\documentclass[pdflatex,sn-mathphys-num]{sn-jnl}% Math and Physical Sciences Numbered Reference Style

\usepackage{lineno}
\usepackage[english]{babel}

\usepackage{paralist}%
\usepackage{appendix}
\usepackage{siunitx}
\let\SInum\num % renamed to SInum, otherwise it will cause "Command '\num' already defined!"
\let\num\relax

\usepackage{pifont}
\usepackage{mathtools}

\usepackage{graphicx}

\usepackage{fancyvrb} % for Verbatim
\usepackage[dvipsnames]{xcolor}

\usepackage[T1]{fontenc}

\usepackage{url}

\usepackage{pgfplots} %  
\pgfplotsset{compat=1.18}
\usepackage[]{subfig}
\usepackage{rotating} % provides sidewaysfigure and sidewaystable environments for full-page rotated figures/tables

\usepackage{geometry}
\usepackage{amsmath,amssymb}
\usepackage{algorithm}
\usepackage{algpseudocode}
\usepackage{etoolbox}

\usepackage{tikz}

\usetikzlibrary{tikzmark, fit, calc}
\makeatletter
\newcommand*{\algrule}[1][\algorithmicindent]{%
  \makebox[#1][l]{%
    \hspace*{.1em}% <------------- This is where the rule starts from
    \vrule height .75\baselineskip depth .25\baselineskip
  }
}
\newcount\ALG@printindent@tempcnta
\def\ALG@printindent{%
    \ifnum \theALG@nested>0% is there anything to print
    \ifx\ALG@text\ALG@x@notext% is this an end group without any text?
    \else
    \unskip
    \ALG@printindent@tempcnta=1
    \loop
    \algrule[\csname ALG@ind@\the\ALG@printindent@tempcnta\endcsname]%
    \advance \ALG@printindent@tempcnta 1
    \ifnum \ALG@printindent@tempcnta<\numexpr\theALG@nested+1\relax
    \repeat
    \fi
    \fi
}

\patchcmd{\ALG@doentity}{\item[]\nointerlineskip}{}{}{} % no spurious vertical space
\makeatother
\algrenewcommand\algorithmicindent{1.1em}

\def\BibTeX{{\rm B\kern-.05em{\sc i\kern-.025em b}\kern-.08em
    T\kern-.1667em\lower.7ex\hbox{E}\kern-.125emX}}
\usepackage{hyperref}
\hypersetup{pdfborder={0 0 0}}
\usepackage{listings}
\usepackage{textcomp} % upquote
\usepackage{lstautogobble}
\usepackage[moderate,tracking=normal]{savetrees}
\usepackage[inline]{enumitem}
\setlist[enumerate]{leftmargin=*}
\usepackage{cleveref}

\def \varg      {\_}

\theoremstyle{thmstyleone}%
\theoremstyle{thmstyletwo}%
\newtheorem{example}{Example}[section]%
\newtheorem{remark}{Remark}%

\theoremstyle{thmstylethree}%
\newtheorem{definition}{Definition}%
\usepackage[color]{zed}
\DeclareMathSymbol{\zseqcat}{\mathbin}{AMSa}{"61}  % \frown
\def \cat {\mathbin{\raise 0.8ex\hbox{$\mathchar\zseqcat$}}}

\newcommand{\setV}{\mathbb{V}}
\newcommand{\real}{\mathbb{R}}
\newcommand{\rational}{\mathbb{Q}}
\newcommand{\ratfunc}{\mathbb{F}}
\newcommand{\set}[1]{\left\{#1\right\}}
\newcommand{\card}{{\#}}
\newcommand{\union}{\cup}
\newcommand{\inter}{\cap}
\usepackage{graphbox} % provide option align=c to \includegraphics
\usepackage{float} % for the H option to place the figures HERE
\hypersetup{
    colorlinks,
    linkcolor={red!50!black},
    citecolor={blue!50!black},
    urlcolor={blue!80!black}
}

\usepackage{multirow} % 
\usepackage{multicol}
\usepackage{booktabs}

\newcommand{\partially}{\rlap{\checkmark}\ \ {\scriptsize{?}}}

\DeclareFontFamily{U}{mathc}{}
\DeclareFontShape{U}{mathc}{m}{it}%
{<->s*[1.03] mathc10}{}

\DeclareMathAlphabet{\mathscr}{U}{mathc}{m}{it}

\definecolor{MetaColor}{rgb}{1,0,0}%
\definecolor{GreyColor}{rgb}{0.5,0.5,0.5}%
\newcommand{\meta}[1]{\mathsf{\color{GreyColor}{#1}}}

\newcounter{tRule}%[section]
\newenvironment{TRule}[2][]{
  \begin{figure*}\centering
    \begin{minipage}{.98\textwidth}
      \noindent\rule{\linewidth}{.1pt}
      \refstepcounter{tRule}
      \textbf{Rule~\thetRule.~#1\ {$#2$}}
      \\
      \noindent\rule{\linewidth}{.1pt}
      \small
      \begin{displaymath}
      }
      {\end{displaymath}\medskip\noindent\rule{\linewidth}{.1pt}	\end{minipage}\end{figure*}}
\newcommand{\lsem}{\lbrack\mkern-3mu\lbrack}
\newcommand{\rsem}{\rbrack\mkern-3mu\rbrack}

\DeclarePairedDelimiter{\sem}{\lsem}{\rsem}
\newcommand{\semL}[2]{\sem{#1}_{\mathscr{#2}}}

\newcolumntype{t}{>{\mathsf\bgroup}l<{\egroup}}
\newcommand{\metablockbegin}{\left(\begin{array}{t}}

\DeclareFontFamily{OMX}{MnSymbolE}{}
\DeclareFontShape{OMX}{MnSymbolE}{m}{n}{
  <-6>  MnSymbolE5
  <6-7>  MnSymbolE6
  <7-8>  MnSymbolE7
  <8-9>  MnSymbolE8
  <9-10> MnSymbolE9
  <10-12> MnSymbolE10
  <12->   MnSymbolE12}{}
\DeclareFontShape{OMX}{MnSymbolE}{b}{n}{
  <-6>  MnSymbolE-Bold5
  <6-7>  MnSymbolE-Bold6
  <7-8>  MnSymbolE-Bold7
  <8-9>  MnSymbolE-Bold8
  <9-10> MnSymbolE-Bold9
  <10-12> MnSymbolE-Bold10
  <12->   MnSymbolE-Bold12}{}

\DeclareSymbolFont{MnSymbol}{OMX}{MnSymbolE}{m}{n}
\SetSymbolFont{MnSymbol}{bold}{OMX}{MnSymbolE}{b}{n}

\DeclareMathDelimiter{\langlebar}{\mathopen}{MnSymbol}{'152}{MnSymbol}{'152}
\DeclareMathDelimiter{\ranglebar}{\mathclose}{MnSymbol}{'157}{MnSymbol}{'157}
\DeclarePairedDelimiter{\blot}{\langlebar}{\ranglebar}

\newcommand{\bmapsto}{\mkern -1mu{\rightsquigarrow}\mkern -1mu} 

\usepackage{lstlinebgrd}
\definecolor{altgreen}{RGB}{220,235,220}
\newcommand{\CodeNote}[1]{%
  \hfill\hspace{-0.5em}{\scriptsize\normalfont\itshape #1}%
}
\usepackage[colorinlistoftodos]{todonotes}
\usepackage{marginnote}

\usepackage{xparse} % to split C1.10 into C1 and 10
\hypersetup{
  colorlinks,
  linkcolor={blue!50!black},
  citecolor={blue!50!black},
  urlcolor={blue!80!black}
}

\usepackage[skins]{tcolorbox}
\usepackage[subfigure]{tocloft}
\usepackage{fmtcount}
\usepackage{imakeidx}
\makeindex[title=Index of Changes, name=changes]

\ifdefined\CHANGES



\newlistof{changes}{s}{\listchangename}	
\newcommand{\changed}[2][?]{%
	\refstepcounter{changes}%
	\linkdest{\thechanges}%
	{\color{blue}#2}\todo{#1}%
	\addcontentsline{s}{changes}{\protect\numberline{\thechanges}#1}%
}%

\newtcolorbox{echanged}[1]{enhanced,title=#1, sharp corners,colback=blue!20,
  attach boxed title to top right=
  {xshift=0mm,yshift=0mm},
  boxed title style={size=small,colback=blue},
  size=minimal,
  finish={
    \refstepcounter{changes}
    \addcontentsline{s}{changes}
    {\protect\numberline{\thechanges}#1}
  }
}
\fi
\newcommand{\notinpaper}[1]{%
  \index[changes]{Comment \textbf{C\expandafter\sortgref#1\sortgref} not reflected in the text.\removecomma|HIDE}%
}

\newrobustcmd{\removecomma}[1]{}
\newcommand{\HIDE}[1]{}

\newcommand{\C}[1]{%
  \index[changes]{Comment \textbf{C\expandafter\sortgref#1\sortgref}|BH{\arabic{changes}}}{C#1}%
}%

\newcommand{\Cn}[2]{%
  \index[changes]{Comment \textbf{RC#1.\expandafter\sortgref#2\sortgref}|BH{\arabic{changes}}}{RC#1.#2}%
}%

\def\sortgref#1\sortgref{%
  \ignoresort{\ifnum#1<10 00\else\ifnum #1<100 0\fi\fi#1}#1%
}
\protected\def\ignoresort#1{}
\ifdefined\CHANGES	

\else
\newcommand{\listchangename}{List of Changes}
\newlistof{changes}{s}{\listchangename}	
\newcommand{\changed}[2][?]{#2}

\newtcolorbox{echanged}[1]{size=minimal}	

\fi
\newcommand{\ad}[1]{\textsf{#1}} % font for elements in activity diagrams
\newcommand{\PRISM}[1]{PRISM} % for the PRISM tool 
\newcommand{\Prism}[1]{Prism} % for the Prism language 

\newcommand{\mykeyword}[1]{\emph{#1}}

\newcommand{\myline}[1]{on line \texttt{\#}\texttt{#1}}
\newcommand{\mylines}[2]{on lines \texttt{\#}\texttt{#1-#2}}
\newcommand{\mylinestwo}[2]{on lines \texttt{\#}\texttt{#1} and \texttt{\#}\texttt{#2}}
\newcommand{\mylinesthree}[3]{on lines \texttt{\#}\texttt{#1}, \texttt{\#}\texttt{#2}, and \texttt{\#}\texttt{#3}}

\newcommand{\method}{QANAD}
\robustify{\method}

\newcommand{\lstinprism}[1]{\lstinline[language=PRISM,columns=fullflexible,breaklines=true,basicstyle=\normalsize\ttfamily,postbreak=]{#1}}
\newcommand{\lstinprop}[1]{\lstinline[language=ADProperty,columns=fullflexible,breaklines=true,basicstyle=\normalsize\ttfamily,postbreak=]{#1}}
\lstdefinelanguage{Epsilon}{
  morekeywords={
      if, then, else, operation, var, and, or, not, new,
      @lazy, rule, transform, to, self, in, for, while, void,
      int, double, removeAt, remove, add, instanceof,
  },
  sensitive=true, % keywords are not case-sensitive
  morecomment=[l]{//}, % l is for line comment
  morecomment=[is]{/*}{*/}, % s is for start and end delimiter
  morestring=[b]" % defines that strings are enclosed in double quotes
} %

\definecolor{eclipseBlue}{RGB}{42,0.0,255}
\definecolor{eclipseGreen}{RGB}{63,127,95}
\definecolor{eclipsePurple}{RGB}{127,0,85}

\definecolor{epsilonred}{rgb}{0.6,0,0} % for strings
\definecolor{epsilongreen}{rgb}{0.25,0.5,0.35} % comments
\definecolor{epsilonpurple}{rgb}{0.5,0,0.35} % keywords
\definecolor{epsilondocblue}{rgb}{0.25,0.35,0.75}

\lstdefinelanguage{PRISM}{
  morekeywords={
      dtmc, mdp, ctmc, const, int, float, bool, global, module, init, endmodule, true, false, double,
      rewards, endrewards, formula, label
  },
  sensitive=true, % keywords are not case-sensitive
  morecomment=[l]{//}, % l is for line comment
  morecomment=[is]{/*}{*/}, % s is for start and end delimiter
  morestring=[b]" % defines that strings are enclosed in double quotes
} %

\definecolor{eclipseBlue}{RGB}{42,0.0,255}
\definecolor{eclipseGreen}{RGB}{63,127,95}
\definecolor{eclipsePurple}{RGB}{127,0,85}
 
\lstdefinelanguage{ADProperty}{
  morekeywords={
      reaches, at, terminated, successfully, failed, terminated, on, fail
  },
  sensitive=true, % keywords are not case-sensitive
  morecomment=[l]{//}, % l is for line comment
  morecomment=[is]{/*}{*/}, % s is for start and end delimiter
  morestring=[b]" % defines that strings are enclosed in double quotes
} %

\begin{document}

\title[Quantitative Analysis from Activity Diagrams]{{QANAD: A Model-Driven Framework for Quantitative Analysis \\and Synthesis of Activity Diagrams via Probabilistic Model Checking}}


\author*[1]{\fnm{Kangfeng} \sur{Ye}}\email{kangfeng.ye@york.ac.uk}

\author[1]{\fnm{Fang} \sur{Yan}}\email{fang.yan@york.ac.uk}

\author[2]{\fnm{Simos} \sur{Gerasimou}}\email{simos.gerasimou@cut.ac.cy}

\affil*[1]{\orgdiv{Department of Computer Science}, \orgname{University of York}, \orgaddress{\street{Deramore Lane}, \city{York}, \postcode{YO10 5GH}, \state{North Yorkshire}, \country{United Kingdom}}}

\affil*[2]{\orgdiv{Department of Electrical Engineering, Computer Science and Engineering}, \orgname{Cyprus University of Technology}, \orgaddress{\street{30 Arch. Kyprianos Str.}, \postcode{3036} \city{Limassol}, \country{Cyprus}}}

\abstract{
      Probabilistic model checking is a widely used formal verification technique to automatically verify qualitative and quantitative Quality-of-Service (QoS) 
      properties for probabilistic models, such as reliability, performance, and cost. 
      % in many application domains. 
      However, devising suitable and representative models for systems of interest in probabilistic model checkers, writing corresponding properties, and verifying them requires domain knowledge in probabilistic models, formal verification, and the target probabilistic model checkers. 
      This unavoidably makes probabilistic model checking not easily accessible for researchers and engineers who may lack the required knowledge. 
      Previous work has extended UML activity diagrams with probability and time, along with quality annotations, developed transformations from activity diagrams to probabilistic models which model checkers support, and implemented accompanying tools for automation. The research, however, is fragmented, informal, or not fully open, making its use, extension, and adaptation non-trivial. 
      To mitigate these challenges, we introduce \method, a comprehensive verification framework for activity diagrams, that includes a new profile for probability, time, parameters, and quality annotations, a semantics interpretation of activity diagrams in Markov models variants (DTMCs, MDPs, and CTMCs) and their parametric versions, and a set of {formal} transformation rules from activity diagrams to the PRISM probabilistic modelling language, supported by the leading probabilistic model checkers PRISM and Storm. 
      The theory underpinning \method\ has been realised into transformation \changed[\Cn{2}{2}]{implementation} using model-driven engineering, enabling fully automated {model transformation with automated verification using (parametric) probabilistic model checking and semi-automated synthesis.}
      We demonstrate \method\ on a case study where multiple robots are used for delivery in a hospital environment and their workflows are modelled as activity diagrams. We further evaluated six other examples from the literature demonstrating the applicability of \method.
      Taken together, this work makes noteworthy contributions to the verification of activity diagrams by improving evaluation, extensibility, adaptability, and accessibility.
}
  
  
\maketitle

\section{Introduction}
\label{sec:intro}

Probabilistic model checking is a powerful formal verification technique for the automatic analysis of probabilistic models given a set of qualitative and quantitative Quality-of-Service (QoS) properties, typically expressed in temporal logic variants~\cite{Baier2008,Katoen2016,Kwiatkowska2018}. 
The technique has been widely used for the performance analysis of randomised distributed algorithms~\cite{Kwiatkowska2001,Kwiatkowska2012},
dependability assessment of communication protocols and networks~\cite{Fruth2011,Petridou2013,Dombrowski2016,Mohammad2017}, 
security~\cite{Basagiannis2009,Elboukhari2010}, safety and reliability evaluation~\cite{Gomes2010,Kikuchi2011,Baouya2019,Gleirscher2020}, planning~\cite{Lahijanian2010} and controller synthesis~\cite{gerasimou2015search,gerasimou2021evolutionary,Feng2015}. 

The underlying probabilistic models are typically Markov model variants, such as discrete-time Markov chains (DTMCs)~\cite{Kemeny1976}, Markov decision processes (MDPs)~\cite{Howard1971,Puterman1994}, and continuous-time Markov chains (CTMCs)~\cite{Anderson1991}.  
These models can be represented graphically as diagrams of labelled state transition systems (LTS), as shown in related research~\cite{Lahijanian2010,paterson2018observation,fang2022presto}, which employed transition diagrams to represent simple Markov models (i.e., comprising only a few states and transitions).
Using diagrams, however, to model larger systems comprising many states and transitions is neither convenient nor sustainable. 
To mitigate this challenge, formal notations such as the PRISM language~\cite{kwiatkowska2011prism}, supported by model checkers such as PRISM and Storm~\cite{Hensel2022}, represent systems as a composition of modules, making modelling more accessible provided that stakeholders have adequate background knowledge of these formalisms. 
For stakeholders unfamiliar with probabilistic modelling, understanding and becoming proficient in using these formal notations is time-consuming and challenging.
Unavoidably, the steep learning curve detracts from their engagement with probabilistic model checking, outweighing its potential benefits. 

{This usability challenge is not unique to probabilistic model checking, but has long been recognised as a fundamental obstacle to the adoption of formal methods in engineering practice.
As a result, a line of research commonly referred to as \emph{Hidden Formal Methods} has emerged, aiming to make formal analysis accessible by embedding it behind familiar modelling abstractions and development workflows~\cite{horvath2020model}.
Representative examples include the notion of \emph{hidden models} in model checking~\cite{visser2012hidden} and the ``Secret Ninja'' view of deploying formal methods in industrial settings~\cite{kiniry2008secret}.
In this paradigm, engineers are not required to directly interact with formal notations or verification tools; instead, formal models and analyses are constructed and executed transparently in the background.
% The approach presented in this paper follows the same philosophy by enabling 
In this work, we embrace this philosophy to enable 
probabilistic model checking and synthesis directly from UML activity diagrams, allowing designers to work at the modelling level while formal verification occurs under the hood.}

Established graphical notations, like UML or SysML activity diagrams (ADs)~\cite{Wolny2020}, have been widely adopted across industry and academia to represent the behaviour of complex systems. 
These notations have been extended with probability and time to model systems that exhibit uncertain and time-critical behaviour, such as in robotics, cyber-physical systems, and business processes~\cite{Morozov2014,Silva2017,paterson2018observation,Calinescu2021,fang2022presto,
Gleirscher2023}, 
{providing a loose connection between ADs and formal models as the probabilistic models are typically manually derived}~\cite{Silva2017,Gleirscher2023}. 
% where ADs are loosely connected to the formal models used for analysis.
% For example, probabilistic models are manually derived from ADs~\cite{Silva2017,Gleirscher2023}. 
Nevertheless, the manual transformation of ADs into probabilistic models is a non-trivial and error-prone procedure that demands expertise in both the semantics of the source (ADs) and target (probabilistic) models.

Notwithstanding research advances in establishing links between ADs and formal models~\cite{Jarraya2007,Gallotti2008,Jarraya2010,Debbabi2010a,Ouchani2011,Calinescu2013,Ouchani2013,Baouya2015,Steurer2020,Lima2020}, they either lack formal semantics or enable the derivation of probabilistic models from ADs at a conceptual level~\cite{Calinescu2013,Silva2017,Gallotti2008}, come without any automation or tool support~\cite{Jarraya2010,Baouya2015} and are limited to only a single type of probabilistic models~\cite{Baouya2021}. 
In addition, the verification properties of interest must be manually transformed in the property language supported by the target model checkers. 
{This is a burdensome task as users must read the generated model and manually extract the mapping between the model states and transitions, and their counterparts in ADs.}

In this paper, we present \method,{\footnote{The \method\ source code is publicly available at \url{https://github.com/RandallYe/QANAD}}} a tool-supported approach that addresses these problems 
% through % delivering a comprehensive verification framework for 
and enables the quantitative analysis of ADs using probabilistic model checking and synthesis. 
\method\ is underpinned by a new UML profile, exposing several stereotypes for annotating QoS information like probability, reliability {(i.e., the probability of successful mission conclusion)}, duration, rate, rewards, and properties for AD components, including activities, nodes and edges. 
By leveraging modelling formalisms that software and system engineers are already familiar with, \method, through its UML profile, provides the mechanisms to easily capture the stochastic behaviour of complex systems. 
These quantitative measures can be initialised based on prior knowledge, for example, using domain expertise and information from related systems, or inferred from past system observations~\cite{paterson2018observation}.

\begin{figure}[ht]
    \centering
    \includegraphics[scale=0.39]{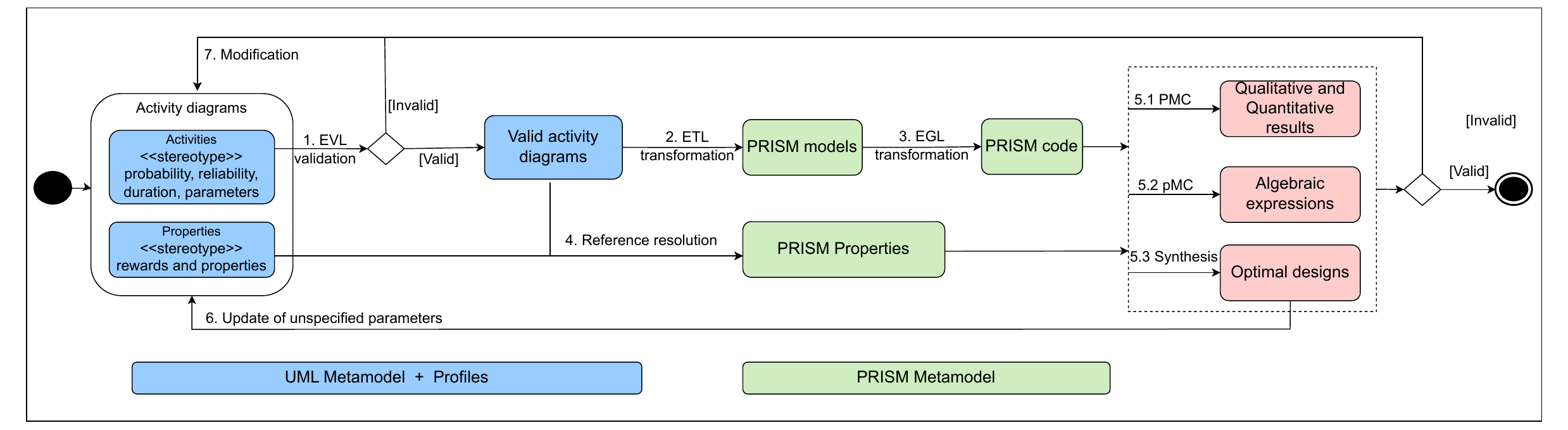}
    \caption{\method\ overview for  the verification of annotated (parametric) activity diagrams for QoS system properties, based on its semantic interpretation in various Markov models (DTMCs, MDPs, and CTMCs) using probabilistic model checking (PMC), parametric model checking (pMC), and synthesis. {Our workflow is automated except for \changed[\Cn{2}{4}]{the} \emph{manual adaptation} of generated PRISM models and properties for synthesis and necessary \emph{modification} of diagrams based on verification results. The details of the figure regarding each step are explained in Section~\ref{subsec:tool}.}}
    \label{fig:our_approach_impl}
\end{figure}

{
\Cref{fig:our_approach_impl} shows an overview of \method, which receives as input an activity diagram whose nodes and edges are annotated with QoS information based on the \method-based UML profile.
\method\ consumes the specified AD and, through model-to-model and model-to-text transformation, produces probabilistic models that conform to the PRISM language~\cite{kwiatkowska2011prism} and which can be loaded and analysed by leading probabilistic model checkers such as PRISM and Storm. 
Accordingly, \method\ supports the specification of ADs that encode diverse information about the target system and which can be transformed into Markovian models that represent discrete-time (DTMC), continuous-time (CTMC) and nondeterministic (MDP) behaviour. If properties are only related to probabilities and rewards (instead of time), DTMCs or MDPs are produced, depending on whether the interleaving behaviour of concurrent flows of an AD or multiple enabled outgoing edges from a decision node is treated as uniform distributions or nondeterminism. 
If time is needed in properties, CTMCs can be used to model continuous-time by exponential distributions.
When verification (PMC) is chosen, \method\ consumes the generated PRISM model and the resolved properties, and uses a probabilistic model checker to derive qualitative (true or false) or quantitative (probabilities or rewards) results. 
In contrast, when  parametric model checking (pMC) is selected, \method\ employs the same pipeline to extract algebraic expressions (such as $p_1 + 3 * p_2$). 
Lastly, when synthesis is chosen, \method\ employs {probabilistic model synthesis engines like EvoChecker~\cite{gerasimou2015search} for PRISM models} to extract optimal parameters (designs) for the QoS system properties.
If the results are not what we expect or are invalid, we can modify the diagram or its annotation and then verify it again until we get the valid results {(that is, all property constraints are satisfied and optimisation objectives are achieved)}. 
{Except \changed[\Cn{2}{3}]{for} the manual adaptation of generated PRISM models and properties for synthesis and manual modification of activity diagrams accordingly based on the results, this workflow is automated using model-based software engineering.}

%The generated PRISM model is then verified against the resolved properties using a probabilistic model checker. 
%The verification results could be qualitative (true or false), quantitative (probabilities or rewards) for probabilistic model checking (PMC), or algebraic expressions (such as $p_1 + 3 * p_2$) for parametric model checking (pMC). 
% When synthesis is invoked, \method\ produces the {PRISM} model to get optimal designs in terms of optimisation objectives (or properties) using tools such as EvoChecker~\cite{gerasimou2015search}, after necessary {manual adaptation} of the generated PRISM model and properties. If the results are not what we expect or are invalid, we can modify the diagram or its annotation and then verify it again until we get the valid results {(that is, all property constraints are satisfied and optimisation objectives are achieved)}. Other work~\cite{Gallotti2008,Baouya2015a} for the verification of activity diagrams that are annotated with probability information, do not support the annotations of rewards, parameters, and properties. They, therefore, are not able to use parametric model checking like what we do to verify parametric models.
}

% an activity diagram annotated with probability, reliability, duration, reward, parameter, properties for verification and a Markov model type for the semantic interpretation of the diagrams based on a profile we build in this work. The profile is presented in \Cref{sec:approach:profile}.

% We illustrate our approach to verify annotated activity diagrams and their context in \Cref{fig:our_approach}. 
% %
% \begin{figure}[t]
%     \centering
%     \includegraphics[scale=0.45]{pics/overall_approach.pdf}
%     \caption{\method\ overview for  the verification of annotated (parametric) activity diagrams for QoS system properties, based on its semantic interpretation in various Markov models (DTMCs, MDPs, and CTMCs) using probabilistic model checking (PMC), parametric model checking (pMC), and synthesis. {Our workflow is automated except \emph{manual adaptation} of generated PRISM models and properties for synthesise and necessary \emph{modification} of diagrams based on verification results.}}
%     \label{fig:our_approach}
% \end{figure}

{The transformation from ADs to PRISM is underpinned by our interpretation of the AD semantics in various Markovian models and is realised using formal rules.}
%Furthermore, unlike related work that generates single module models, \method\ realises a modular approach, producing modular PRISM models that more closely match the semantics of the input ADs. 
{\method\ also supports the annotation of} AD edges with parameters that allow the generation of parametric Markov models and the use of \emph{parametric probabilistic model checking} to \emph{explore} the design space of captured systems and \emph{synthesise} optimal designs. 
This is very useful to support the design process for engineers to start with open ADs (and correspondingly models) with many unspecified parameters and work towards identifying appropriate values for those parameters that satisfy the QoS properties of the system~\cite{gerasimou2015search}. 
Finally, \method\ offers a controlled natural language for specifying basic \emph{atomic propositions (APs)} at the AD-level which are automatically transformed into temporal logic formulae of the underpinning Markov model. 
{
These APs involve additional variables introduced in our transformation to deal with localisation of nodes, termination, and reliability. In our property specification, we introduce constructs to let users specify APs in terms of ADs, and so no knowledge is required from users about these variables.}
Consequently, stakeholders can employ this controlled natural language and focus on system modelling and property specification at the AD-level which is distinct from other work~\cite{Calinescu2013,Silva2017,Gallotti2008} where stakeholders must interpret the generated code to find out the corresponding atomic propositions for the QoS properties of interest.

Our contributions are summarised as follows: 
\begin{enumerate}[label=(\arabic*)]
  \item a \method\ verification framework underpinned by a UML profile {to annotate complex QoS information directly on nodes and edges, and to specify} properties of interest using a controlled natural language {at the AD level};
  {\item the semantics of ADs given in multiple Markov models (DTMCs, CTMCs, MDPs) and realised using formal transformation rules from ADs to the PRISM language; 
    \item the support of parametric ADs, design-space exploration, and synthesis;} 
    \item an open-source \method\ tool, underpinned by model-driven engineering (MDE) principles, that fully automates the workflow {(except minimal manual adaptation for synthesis) and supports single-source multi-formalism}; and
    \item a comprehensive evaluation of our approach using six examples from literature and a realistic case study using multiple robots for material delivery within a hospital facility.
\end{enumerate}
The remainder of this paper is organised as follows. \Cref{sec:background} briefly introduces ADs, Markov models, and the PRISM language. 
In \Cref{sec:motiving_example}, we present a case study from a hospital intralogistics scenario.  %example to use multiple robots for delivery in a logistics scenario within a hospital environment. 
We discuss \method\ and introduce the new UML profile in \Cref{sec:approach}. The semantics interpretation of ADs and transformation rules to PRISM are defined in \Cref{sec:semantics}. 
The open-source \method\ tool is discussed in \Cref{sec:ver_tool}, while the \method\ evaluation is presented in \Cref{sec:cases}. 
We review related work in \Cref{sec:relwork}, and conclude and discuss future work in \Cref{sec:concl}.

\section{Background} 
\label{sec:background}
%Can be removed to save space
%This section briefly introduces UML activity diagrams (\Cref{sec:background:activity}), defines Markov models (\Cref{sec:background:markov}), and describes the PRISM language and probabilistic model checkers (\Cref{sec:background:prism}).

\subsection{UML Activity Diagrams}
\label{sec:background:activity}

\begin{figure*}
    \centering
    \includegraphics[scale=0.45]{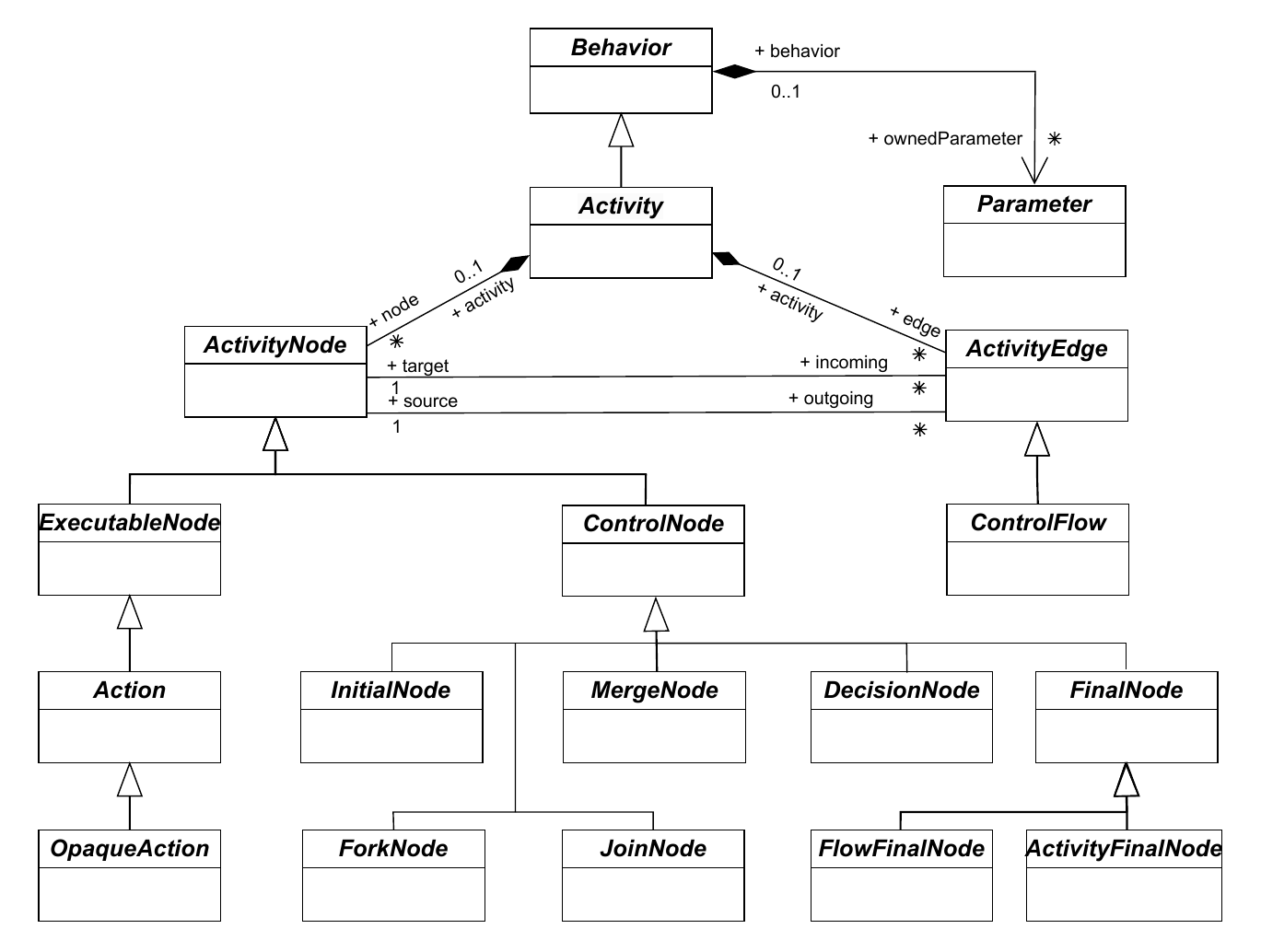}
    \caption{Activity diagram metamodel adapted from~\cite{omguml2017}.}
    \label{fig:activity_metamodel}
\end{figure*}

We consider activity diagrams that are compliant with UML Version 2.5.1.\footnote{\url{https://www.omg.org/spec/UML/2.5.1}} 
\Cref{fig:activity_metamodel} shows the activity diagram metamodel, focusing on the diagram elements considered in this work. %including their structure and relation of these elements. 
Interested readers can find a complete analysis of the UML notation in~\cite{omguml2017}.
The concrete syntax of activity diagram elements conforming to the UML notation and employed in this paper is shown in \Cref{fig:activity_notations}.

An \ad{Activity} is considered to be a \ad{Behavior}, encoded as a graph of \ad{ActivityNode}s interconnected by \ad{ActivityEdges}. 
%Control nodes specify sequencing of executable nodes via control flow edges.
%Object nodes hold data that is input to and output from executable nodes, and moves across object flow edges.
Activities are essentially what are commonly called ``control and data flow'' models. Such models of computation are inherently concurrent, as any sequencing of activity node execution is modelled explicitly by \ad{ActivityEdges}, and no ordering is mandated for any computation not explicitly sequenced~\cite{omguml2017}.
Furthermore, an \ad{Activity} can have \ad{Parameters} that provide the ability to pass values into and out of \ad{Behavior} executions.
When an \ad{Activity} is invoked, \ad{Parameter}s with direction \emph{in} may be provided. 
The \ad{defaultValue} of a \ad{Parameter} can be set to a concrete value of a certain type (e.g., real type) or be left undefined.

% A \ad{Behavior} may have \ad{Parameter}s that provide the ability to pass values into and out of \ad{Behavior} executions.
% Therefore, an \ad{Activity} can have \ad{Parameter}s.
% When an \ad{Activity} is invoked, \ad{Parameter}s with direction ``in'' may be provided. %, as constrained by the multiplicity of those \ad{Parameter}s.
% The \ad{defaultValue} of a \ad{Parameter} can be set to a concrete value of a certain type (e.g., real type) or be left undefined.
%\ad{Parameter}s may also be marked as streaming which allows \ad{Parameter}s' values to be passed into and out of an \ad{Activity} execution any time during its course, rather than just on invocation and completion~\cite{omguml2017}.
%An \ad{Activity} can also have \ad{Variables}.

\begin{figure}
    \centering
    \includegraphics[scale=0.45]{pics/activity_notations_r1.png}
    \caption{Activity diagram notations.}
    \label{fig:activity_notations}
\end{figure}
 
Likewise, \ad{ActivityNodes} can model the individual steps in the behaviour specified by an \ad{Activity}.
An \ad{ActivityEdge} is a directed connection between two \ad{ActivityNode}s along which control or data may flow, between the source and target \ad{ActivityNodes}.
% An \ad{ActivityEdge} can be a \ad{ControlFlow} or an \ad{ObjectFlow}. 
%Tokens are used for describing the execution of an \ad{Activity} but are not explicitly modeled in an \ad{Activity}. 
%An object token is a container for a value that flows over \ad{ObjectFlow} edges. 
%A control token affects the execution of \ad{ActivityNode}s, but does not carry any data, and flows only over ControlFlow edges~\cite{omguml2017}.
%An ActivityEdge may have a guard, which is a ValueSpecification that is evaluated for each token offered to the edge. 
%An ActivityEdge without a guard is equivalent to one with a guard that evaluates to true for every token.
%The notation of a control flow is shown in \Cref{fig:control_flow_notation}.
%The weight property dictates the minimum number of tokens that must traverse the edge at the same time. It is a ValueSpecification that is evaluated every time a new token is offered by the source ActivityNode.
%
% In this work, we implemented the \ad{ControlFlow}. 
A \ad{ControlFlow} is an \ad{ActivityEdge} that only passes control and is used to explicitly encode the sequence of execution for \ad{ActivityNodes}. 
The target \ad{ActivityNode} cannot start execution until the source \ad{ActivityNode} completes its execution.
%A control flow is an arrowed line connecting two actions.%~\cite{omguml2017}. 
%The notation is shown in \Cref{fig:control_flow_notation}.
%
%
An \ad{ActivityNode} can be either an \ad{ExecutableNode} or a \ad{ControlNode}. 
All concrete kinds of \ad{ExecutableNodes} are \ad{Actions} with an \ad{OpaqueAction} specifying the system behaviour and which can be encoded in textual concrete syntax. 
% There are three kinds of \ad{ActivityNode}s including \ad{ExecutableNode}s, \ad{ControlNode}s, and \ad{ObjectNode}s.
% The \ad{ObjectNode} is not yet supported in our approach and will be considered in the future. 
% All concrete kinds of \ad{ExecutableNodes} are \ad{Actions} with an 
% Further, an \ad{OpaqueAction} is an \ad{Action} whose specification may be given in a textual concrete syntax. %~\cite{omguml2017}. 
% We use \ad{OpaqueAction}s to represent the actual behaviour of the system.
%\textcolor{red}{In this work, we explored Action and \ad{ControlNode}.}
%An \ad{Action} can also be a \ad{CallBehaviorAction}, an \ad{ AcceptEventAction}, or a \ad{SendSignalAction}. These are not included in our approach presented in this paper but are considered for future extension.
%In SysML, there are four kinds of actions including \ad{Action}, \ad{CallBehaviorAction},\ad{ AcceptEventAction}, \ad{SendSignalAction}. 
%There notations are shown in \Cref{fig:sysml_action_notation}.

A \ad{ControlNode} manages the flow between other nodes in an Activity and its subclasses include  \ad{InitialNodes}, \ad{FinalNodes}, \ad{ForkNodes}, \ad{JoinNodes}, \ad{MergeNodes}, and \ad{DecisionNodes}.
More specifically, an \ad{InitialNode} has no incoming \ad{ActivityEdges} and acts as a starting point for executing an \ad{Activity}. 
If an \ad{Activity} has multiple \ad{InitialNodes}, then invoking the \ad{Activity} starts multiple concurrent  control flows, one per \ad{InitialNode}.
The outgoing \ad{ActivityEdge}s of an \ad{InitialNode} must all be \ad{ControlFlows}.

% An \ad{InitialNode} is a \ad{ControlNode} that acts as a starting point for executing an \ad{Activity}. An \ad{Activity} may have more than one \ad{InitialNode}. If an \ad{Activity} has more than one \ad{InitialNode}, then invoking the \ad{Activity} starts multiple concurrent  control flows, one for each \ad{InitialNode}.
% An \ad{InitialNode} shall not have any incoming \ad{ActivityEdge}s.
% The outgoing \ad{ActivityEdge}s of an \ad{InitialNode} must all be \ad{ControlFlow}s. 

A \ad{FinalNode} is a \ad{ControlNode} at which the flow of an \ad{Activity} stops; thus, it has no outgoing \ad{ActivityEdge}s.
A \ad{FinalNode} accepts all flows to it on its incoming \ad{ActivityEdges}.
There are two kinds of \ad{FinalNodes}:
(1) a \ad{FlowFinalNode} that terminates a flow but does not affect other flows in the \ad{Activity}; and 
(2) an \ad{ActivityFinalNode} that stops all flows in an \ad{Activity}. 
If an Activity owns several \ad{ActivityFinalNodes}, then the first accepting a flow terminates the execution of the Activity, including the execution of any other \ad{ActivityFinalNodes}.

Another important \ad{ControlNode} is the \ad{ForkNode} which splits a flow into multiple concurrent flows. A \ad{ForkNode} has exactly one incoming \ad{ActivityEdge} and may have multiple outgoing \ad{ActivityEdges}.
%If the incoming edge is a \ad{ControlFlow}, then all outgoing edges shall be \ad{ControlFlow}s and, if the incoming edge is an \ad{ObjectFlow}, then all outgoing edges shall be \ad{ObjectFlow}s.
Similarly, a \ad{JoinNode} is a \ad{ControlNode} that synchronises multiple flows, and has exactly one outgoing \ad{ActivityEdge} but can have multiple incoming \ad{ActivityEdges}. 
%If any of the incoming edges of a \ad{JoinNode} are \ad{ObjectFlow}s, the outgoing edge must be an \ad{ObjectFlow}. Otherwise, the outgoing edge must be a \ad{ControlFlow}.
A \ad{MergeNode}  brings together multiple flows without synchronisation, and has exactly one outgoing \ad{ActivityEdge} but may have multiple incoming \ad{ActivityEdges}.

A \ad{DecisionNode} is a \ad{ControlNode} that chooses between outgoing flows, and has at least one and at most two incoming \ad{ActivityEdges}, and at least one outgoing \ad{ActivityEdge}. 
If two incoming edges exist, then one shall be identified as the \ad{decisionInputFlow}, and the other termed the primary incoming edge. 
If the \ad{DecisionNode} has only one incoming edge, then it is the primary incoming edge. %If the primary incoming edge of a \ad{DecisionNode} is a \ad{ControlFlow}, then all outgoing edges shall be \ad{ControlFlow}s and, if the primary incoming edge is an \ad{ObjectFlow}, then all outgoing edges shall be \ad{ObjectFlow}s.
If any of the outgoing edges of a \ad{DecisionNode} have guards, then these are evaluated for each incoming flow.%token. 

In summary, an \ad{InitialNode} or \ad{Action} with multiple outgoing edges, or multiple \ad{InitialNode}s, or a \ad{ForkNode} will create multiple concurrent flows. A \ad{JoinNode} brings multiple incoming concurrent flows into one flow. A \ad{FlowFinalNode} terminates current concurrent flow, while a \ad{ActivityFinalNode} destroys all concurrent flows of the owning activity. Other control nodes such as \ad{OpaqueAction}, \ad{MergeNode}, and \ad{DecisionNode} do not change their concurrent flows.

\subsection{Markovian Models}
\label{sec:background:markov}
We consider three Markov models for activity diagrams: discrete-time Markov chains (DTMCs)~\cite{Katoen2016}, continuous-time Markov chains (CTMCs)~\cite{Baier1998}, and Markov decision processes (MDPs)~\cite{kwiatkowska_stochastic_2007}. 
{In DTMCs and MDPs, time is discrete and each transition takes one unit of time. Time in CTMCs, however, is continuous, and the probability of an outgoing transition from a state follows an exponential distribution. 
Another difference between these Markov model variants is that DTMCs and CTMCs are deterministic, while MDPs support nondeterminism.}
%{In DTMCs and MDPs, time is discrete and each transition takes one unit of time. From each state, the choice of successor states follows one discrete probability distribution for DTMCs or multiple distributions for CTMCs. Time in CTMCs, however, is continuous. When a transition leaves the source state follows an exponential distribution }%We define the \emph{Markov Chains}, and these Markov models below.

%\begin{definition}[Markov chain and Markov property]
%  Let $S$ be a countable set (named \emph{state space}), and let $X$ be a sequence of random variables taking values from $S$. $X$ is a \emph{Markov chain} if
%  \begin{align*}
%    & P\left(X_{n+1}=s_{n+1} | X_{0}=s_{0}, X_{1}=s_{1}, \cdots, X_{n}=s_{n}\right) 
%    = P\left(X_{n+1}=s_{n+1} | X_{n}=s_{n}\right)
%  \end{align*}
%  for all $n \geq 0$ and $s_{0}, s_{1}, \cdots s_{n+1} \in S$.  
  %\qed
%\end{definition}

%In other words, the probability of the future system states $s_{n+1}$ depends only on its current state $s_n$, and is independent of all its past states. 
%This is called the \emph{Markov property}, signifying that a Markov model is  \emph{memoryless}.

% A DTMC is a sequence of discrete random variables $X_1, X_2, X_3, \ldots$ with the Markov property. 
%The possible values of the $X_i$s form a countable set $S$ called the state space of the chain.
% The formal DTMC definition is given below.
\begin{definition}[Discrete-Time Markov chains]
  \label{def:dtmc}
  A DTMC is a tuple $\left(S, s_{init}, P, L\right)$ where
  \begin{itemize}
  \item $S$ is a non-empty and countable set of states;
  \item $s_{init} \in S$ is an initial state; 
  \item $P: S \times S \fun [0,1]$ is a transition probability matrix such that $\sum\limits_{s' \in S}^{} P(s, s') = 1$ for all $s \in S$; 
  \item $L: S \fun 2 ^ {AP}$ is a labelling function where {$AP$ is the set of atomic propositions that are valid in that state.}
  \end{itemize}   
  %\qed
\end{definition}

% In DTMCs, each transition consumes one discrete time unit. Transitions in CTMCs, however, can occur at any real-valued time instant. The probability of a transition taken at time $t$ is modelled by the exponential distribution $\lambda e^{-\lambda t}$ where $\lambda$ is a rate parameter. The expected value of $t$ is $\frac{1}{\lambda}$. We define CTMCs below.

\begin{definition}[Continuous-Time Markov chains]
  \label{def:ctmc}
  A CTMC is a tuple $\left(S, s_{init}, R, L\right)$ where
  \begin{itemize}
  \item $S$ is a non-empty and countable set of states;
  \item $s_{init} \in S$ is an initial state; 
  \item $R: S \cross S \fun \real_{\geq 0}$ is a transition rate matrix; 
  \item $L: S \fun 2 ^ {AP}$ is a labelling function.% where $AP$ denotes a set of atomic propositions.
  \end{itemize}   
  %\qed
\end{definition}

Each CTMC state $s$ can have multiple outgoing transitions with different rates. The exit rate of $s$, $E(s)$, is equal to $\sum_{s' : S}^{}~R(s, s')$. 
Thus, the probability of leaving $s$ within $[0,t]$ is the cumulative distribution function (CDF), {$1- e^{-E(s) \cdot t}$}, of the exponential distribution with rate $E(s)$, and the expected time (also called the \emph{mean duration}) in $s$ is {$1 / E(s)$}. A race condition~\cite{kwiatkowska_stochastic_2007} arises at $s$ if there is more than one state $s':S$ such that $R(s, s') > 0$.
The choice for the outgoing transition 
% But which transition is taken from $s$? The choice 
is probabilistic and based on the probability, $R(s,s')/E(s)$ of the transition from $s$ to $s'$.

% MDPs are an extension of DTMCs to allow nondeterministic choice. A formal definition of an MDP is given as follows.
\begin{definition}[Markov decision processes]
  \label{def:mdp}
  A MDP is a tuple $\left(S, s_{init}, Act, Steps, L\right)$ where
  \begin{itemize}
  \item $S$ is a non-empty and countable set of states;
  \item $s_{init} \in S$ is an initial state; 
  \item $Act$ is a set of actions;
  \item $Steps: S \times Act \times S \fun [0,1]$ is a transition probability function such that $\sum\limits_{s' \in S}^{} Steps(s, \alpha, s') \in \{0, 1\}$ for all $s \in S$ and $\alpha \in Act$.
  \item $L: S \fun 2 ^ {AP}$ is a labelling function.% where $AP$ denotes a set of atomic propositions.
  \end{itemize}   
  %\qed
\end{definition}
% Unlike DTMCs where $P(s)$ for each state $s$ is a distribution (the probabilities for all its target states sum to 1), $Steps(s)$ in an MDP is a set of distributions which is indexed by an action $\alpha$. 
% We note that every transition in a DTMC or MDP takes one unit of (discrete) time.

{
States and transitions of these Markovian models can be associated with rewards or cost.
\begin{definition}[Cost and rewards]
    For a DTMC or a CTMC, its reward structure is a pair $(R_s, R_t)$ where $R_s: S \fun \real_{\geq 0}$ is a state reward function associating a state with a non-negative real number reward,
    and $R_t: S \cross S \fun \real_{\geq 0}$ is a transition reward function associating a transition with a reward. 

    However, for a MDP, its transition reward function $R_t: S \cross Act \fun \real_{\geq 0}$ associates rewards with actions, instead of individual states in a DTMC or CTMC.
    % $R_t$ is also called a transition reward matrix.
\end{definition}
}
{In this work, costs are treated as a special case of rewards, where a cost is represented as a reward value (e.g., using a negative reward or a dedicated cost reward structure), following the standard interpretation adopted in probabilistic model checking.}

%\simos{Nothing is mentioned up to this point about specifying and verifying properties over the probabilistic models. Since, we use these properties later in the paper, their semantics should be briefly and formally presented here. eg something like the following
%``The properties to verify are expressed in temporal logic over these atomic propositions, e.g., prob- abilistic temporal tree logic (PCTL) [9], [40] for DTMCs
%and continuous stochastic logic (CSL) [3], [5] for CTMCs. State-of-the-art probabilistic model checkers (PRISM [52], Storm [22]) implement efficient PMC algorithms for the veri- fication of complex Markov chain models and properties [51].''}

{
\subsubsection{Parametric Markovian models}
We introduce polynomials, rational functions, and instantiations of rational functions over parameters before defining parametric Markovian models. Similar definitions have already been given in~\cite{Hahn2011,Chen2013,Baier2020,Junges2024}.
\begin{definition}[Polynomials]
    A polynomial $P$ in a finite set of $n$ variables or indeterminates over $\real$, $\setV = \set{x_1, x_2, \ldots, x_n}$, can be expressed as a weighted sum of finite terms or monomials.
\begin{align*}
    & P(x_1,x_2,\ldots,x_n) = \sum_{i_1\in I_1;i_2\in I_2;\ldots;i_n\in I_n}^{} c_{i_1,i_2,\ldots,i_n}\cdot x_1^{i_1} \cdot x_2^{i_2} \cdot \ldots \cdot x_n^{i_n} \mbox{ \quad where } c_{i_1,i_2,\ldots,i_n} \in \rational
\end{align*}
where $I_i$ (for $i \in \set{1..n}$) is a set of rational numbers and of type $\power \nat$, and $c_{i_1,i_2,\ldots,i_n}$ is a constant rational number or coefficient (can be treated as a constant function over indices $\set{i_1,i_2,\ldots,i_n}$) for a monomial whose power part is $x_1^{i_1} \cdot x_2^{i_2} \cdot \ldots \cdot x_n^{i_n}$. A $P$ of $m$ terms is defined as $\card\set{i_1\in I_1;i_2\in I_2;\ldots;i_n\in I_n | c_{i_1,i_2,\ldots,i_n} \neq 0} = m$ (that is, the number of non-zero coefficients is $m$). The \emph{degree} of $P$ is defined as $max\ \set{i_1\in I_1;i_2\in I_2;\ldots;i_n\in I_n | i_1 + i_2 + \ldots + i_n}$ (that is, the maximum number of the sum of the power exponents). For a monomial, if its coefficient is 1, it is called a \emph{primitive monomial}.

We use $\rational[\setV]$ to denote the set containing all polynomials in $\setV$.
\end{definition}

%One polynomial example is given below.
\begin{example}[Polynomials]
    $P_1(x_1,x_2) = 3x_1^2 x_2 + 5 x_1 + 2 x_2$ has three terms and its degree is 3. It can be expressed as 
    \begin{align*}
        & P_1(x_1,x_2) = \sum_{i_1 \in \set{0,1,2}; i_2 \in \set{0,1}}^{} c_{i_1,i_2} \cdot x_1^{i_1} \cdot x_2^{i_2} \mbox{ \quad where } c_{i_1,i_2} =  
        \begin{cases}
            3 & i_1=2 \land i_2=1\\
            5 & i_1=1 \land i_2=0\\
            2 & i_1=0 \land i_2=1\\
            0 & \mbox{otherwise}
        \end{cases}
    \end{align*}
\end{example}

By using polynomials, we can define rational functions.
\begin{definition}[Rational functions]
    A rational function is a fraction of two polynomials in $\setV$, $f(x_1,x_2,\ldots,x_n) = P_1(x_1,x_2,\ldots,x_n) / P_2(x_1,x_2,\ldots,x_n)$ where $P_1(x_1,x_2,\ldots,x_n) \in \rational[\setV]$, $P_2(x_1,x_2,\ldots,x_n) \in \rational[\setV]$, and $P_2(x_1,x_2,\ldots,x_n) \neq 0$. We use $\ratfunc[\setV]$ to represent the set containing all such rational functions in $\setV$.
\end{definition}

Polynomials and relational functions can also be instantiated.
\begin{definition}[Instantiations]
    A partial function $\mu: \setV \pfun \real$ is an instantiation of a finite set $\setV$ of parameters. The set $\setV$ in $\mu$ is \emph{fully} instantiated if $\dom\ \mu = \setV$ (that is, every parameter has an associated value from $\real$) or \emph{partially} instantiated in $\mu$ if $\dom\ \mu \subset \setV$ (that is, at least one parameter is not associated with a value or \emph{open} or \emph{indetermined}).
    For a polynomial $P(x_1,x_2,\ldots,x_n)$ or $P[\setV]$, $P(\mu)$ is an instantiation of $P$ and $P(\mu) = P(x_1,x_2,\ldots,x_n)[\mu'(x_1),\mu'(x_2),\ldots,\mu'(x_n)/x_1,x_2,\ldots,x_n]$ where $\mu'(x_i) = \IF \left(\exists e \bullet (x_i \mapsto e) \in \mu\right) \THEN \mu(x_i) \ELSE x_i$. In other words, an instantiation is a substitution of parameters such as $x_i$ in $P$ by 
    \begin{enumerate*}[label={(\arabic*)}]
        \item its corresponding value $\mu(x_i)$ if $\mu(x_i)$ is defined (that is, there exists a value $e$ in the codomain of $\mu$ such that $x_i$ is mapped to $e$ by $\mu$), or 
        \item $x_i$ itself otherwise (that is, nothing changed for this parameter or not instantiated).
    \end{enumerate*}
    For a rational function, $f(\mu)$ is an instantiation of the fraction of $P_1(\mu) / P_2(\mu)$. 
\end{definition}

We can define a parametric DTMC using polynomials and rational functions.
\begin{definition}[Parametric DTMCs]
  \label{def:pdtmc}
  A parametric DTMC (pDTMC) in $\setV$ is a tuple $\mathcal{M} = \left(\setV, S, s_{init}, \mathbf{P}, L\right)$ where
  \begin{itemize}
  \item $\setV$ is a finite non-empty set of parameters or variables over $\real$;
  \item $S$, $s_{init}$, and $L$ are the same as those in \Cref{def:dtmc}.
  \item $\mathbf{P}: S \times S \fun \left(\ratfunc[\setV] \union [0,1]\right)$ is a transition probability function or matrix where a transition could be associated with a rational function from $\ratfunc[\setV]$ or a real number from 0 to 1 inclusive (or a unit interval). %such that $\sum\limits_{s' \in S}^{} P(s, s') = 1$ for all $s \in S$; 
  \end{itemize}   
  %\qed
\end{definition}

A pDTMC can be instantiated using instantiation.
\begin{definition}[Instantiation of a pDTMC]
  \label{def:pdtmc:inst}
  For a pDTMC $\mathcal{M}$ defined in~\Cref{def:pdtmc} and an instantiation function $\mu:\setV \pfun \real$, we define $\mathcal{M}_{\mu} = \left(\setV \setminus \dom\ \mu, S, s_{init}, \mathbf{P}(\mu), L\right)$. If $\dom\ \mu = \setV$, then $\mathcal{M}$ is fully instantiated and a \emph{valid} DTMC is derived if it is stochastic (see \Cref{def:dtmc}). 
\end{definition}

For a parametric CTMC, $R$ in \Cref{def:ctmc} becomes $\mathbf{R}: S \times S \fun \left(\ratfunc[\setV] \union \real_{\geq 0}\right)$. For a parametric MDP, $Steps$ in \Cref{def:mdp} becomes $Steps: S \times Act \times S \fun \left(\ratfunc[\setV] \union [0,1]\right)$. We omit their definitions and instantiations for simplicity.

The properties of these Markovian models or their parametric counterparts are expressed in variants of temporal logic over atomic propositions, e.g., the probabilistic computation tree logic (PCTL)~\cite{Hansson1994} or PCTL*~\cite{Aziz1995,Bianco1995,Baier1998} (a combination of PCTL and the linear temporal logic (LTL)~\cite{Pnueli1977}) for DTMCs and MDPs, and the continuous stochastic logic (CSL)~\cite{Aziz1996} for CTMCs. State-of-the-art probabilistic model checkers (e.g., PRISM and Storm) implement efficient algorithms to verify properties for these complex (parametric) Markovian models. PRISM can verify unbounded until, steady-state probabilities, reachability reward, and steady-state reward for parametric DTMCs and CTMCs; and unbounded until and reachability rewards for parametric MDPs.\footnote{\url{https://www.prismmodelchecker.org/manual/RunningPRISM/ParametricModelChecking}.} Storm can also compute exact solution functions for parametric Markovian models~\cite{Junges2024}. The model checking results are {mappings from regions or subsets of input parameter valuations to rational functions or algebraic expressions over these input parameters (such as transition probabilities)} that satisfy given properties.
}

\subsection{PRISM Language and Probabilistic Model Checkers}
\label{sec:background:prism}

PRISM~\cite{kwiatkowska2011prism} is a probabilistic model checker for verifying various Markov models, including DTMCs, CTMCs, and MDPs.
The input language of PRISM, called the PRISM language, is based on the formalism of reactive modules~\cite{Alur1999}. The semantics of PRISM is discussed in~\cite{PRISM2008}, including how a system module is constructed from multiple input modules and how the system module is semantically interpreted as a DTMC, MDP, or CTMC model.

\begin{figure}
  \centering
\begin{lstlisting}[language = PRISM,literate={\&}{{\ttfamily\&}}1, linewidth=0.98\linewidth,numbersep=6pt]
dtmc // model type: mdp, ctmc
// constants
const double p1;
const double p2;
const int N;
const int MaxDamages;
// global variable
global b : [0..MaxDamages] init 0;
// a module for the left-right controller
module LeftRight 
  x : [-N..N] init 0; // a local variable
  [] (x=-N)&(b<MaxDamages) -> (x'=x+1)&(b'=b+1);
  [] (x= N)&(b<MaxDamages) -> (x'=x-1)&(b'=b+1);
  [move] (x>-N)&(x<N) -> p1:(x'=x-1)+(1-p1):(x'=x+1);
  [] ((x=-N)|(x=N))&(b=MaxDamages) -> true;
endmodule
// a module for the up-down controller
module UpDown
  y : [-N..N] init 0;
  [] (y=-N)&(b<MaxDamages) -> (y'=y+1)&(b'=b+1);
  [] (y= N)&(b<MaxDamages) -> (y'=y-1)&(b'=b+1);
  [move] (y>-N)&(y<N) -> p2:(y'=y-1)+(1-p2):(y'=y+1);
  [] ((y=-N)|(y=N))&(b=MaxDamages) -> true;
endmodule
// rewards
rewards "battery_consumption"
  [move] true : 1;
endrewards
\end{lstlisting}
\caption{{The PRISM model of a two-dimensional random walk: four {constants}, one global variable, two modules (one local variable and four commands for each module), and one transition reward.}}
  \label{fig:prism_model_trw}
\end{figure}

\Cref{fig:prism_model_trw} illustrates a PRISM example modelling a variation of a two-dimensional random walk within a restricted area. This walk decides its next destination by flipping two different coins simultaneously: one coin for the left-right dimension with probability $p1$ moving towards left and one coin for the up-down dimension with probability $p2$ moving downwards. The restricted area is a $2N \times 2N$ square. The initial position of the walk is at the centre of the area at location (0,0).

The PRISM example in \Cref{fig:prism_model_trw} is a DTMC model, given by the model type \lstinprism{dtmc} \myline{1}. 
%Other supported types include \lstinprism{mdp} and \lstinprism{ctmc}. 
The model declares four {constants} \lstinprism{p1}, \lstinprism{p2}, \lstinprism{N}, and \lstinprism{MaxDamages} \mylines{3}{6}. Among them, \lstinprism{p1} and \lstinprism{p2} are of type \lstinprism{double} for real numbers and they correspond to probabilities $p1$ and $p2$ for the controllers. \lstinprism{N} and \lstinprism{MaxDamages} are of type \lstinprism{int} for integer numbers. \lstinprism{N} denotes the distance of the area border (wall) from its centre in each dimension. We use {\lstinprism{MaxDamages} to limit the maximum damage level to the walk (so it must be repaired manually before it can carry on its duty again) where we assume one unit of damage is caused if the walk reaches the area border, hits the wall, and bounces back.}

A global variable \lstinprism{b} is defined \myline{8} over a range from 0 to \lstinprism{MaxDamages} and initialised to 0. The variable is used to record the number of times that the walk {hits the wall (that is, the units of damage). This is useful to determine when the walk terminates and cannot move further.} %estimate battery consumption because we assume the walk will cost 2 units of battery levels to turn around and move back one cell towards the centre when it reaches the boundary. 

Global variables in PRISM can be read and updated by any module. This example contains two modules: \lstinprism{LeftRight} \mylines{10}{16} and \lstinprism{UpDown} \mylines{18}{24} to model the left-right controller and the up-down controller respectively. Each module has a \emph{local} variable \lstinprism{x} \myline{11} (or \lstinprism{y} \myline{19}) to record the current position of the walk in each dimension. The variables are defined over the range from \lstinprism{-N} to \lstinprism{N}, corresponding to the area in each dimension. Local variables can be read by all modules but written only by the module where they have been defined. So \lstinprism{x} (or \lstinprism{y}) can only be updated in \lstinprism{LeftRight} (or \lstinprism{UpDown}). %Updating global variables in parallel is forbidden to avoid race condition.

The behaviour of each controller is modelled by the \emph{commands} in the corresponding module: four commands in each module. A command in PRISM defines a collection of transitions from its source states to its destination states with augmented probabilities with the form: 
%\\ \lstinprism{[e] g -> p1:u1 + ... + pn:un;}
\begin{lstlisting}[language=prism,frame=,numbers=none,basicstyle=\small\ttfamily]
  [e] g -> p1:u1 + ... + pn:un;
\end{lstlisting}
The optional \emph{action} \lstinprism{e} is used for synchronisation with other modules, %if these modules have commands with the same actions, 
such as commands \myline{14} and \myline{22} with the same action \lstinprism{move}. Therefore, the two commands must be executed at the same time. 
{This does not imply two separate movements occurring independently; instead, it represents a modelling abstraction in which one transition corresponds to a combined update of both dimensions in a single step of the random walk.}
For other commands without an action, or with an action but no commands with the same action in other modules, called an \emph{independent} action, they proceed independently. 

A \emph{state} is a valuation of all variables (global and local) in
the model. The state space $S$ of a model is all valuations of variables and contains all states. The guard condition \lstinprism{g} in a command identifies a subset $S_g$ of $S$ that satisfies \lstinprism{g}. $S_g$ is also the source states of the transitions defined by the command. For example, the condition \lstinprism{(x=-N)&(b<MaxDamages)} in the command \myline{12} characterises a subset of states in which \lstinprism{x} is equal to \lstinprism{-N} and \lstinprism{b} is less than \lstinprism{MaxDamages}. 

A collection of updates is given on the right side of a command, such as $n$ updates from \lstinprism{p1:u1} to \lstinprism{pn:un} where these probabilities must sum to 1. Each update is a pair of a probability \lstinprism{pi} and a set of assignments \lstinprism{ui}, and defines the probability \lstinprism{pi} of the transition going from a source state $s$ in $S_g$ to a destination state $s'$ which is specified by \lstinprism{ui} based on $s$. For example, the two updates (\lstinprism{p1:x'=x-1} and \lstinprism{(1-p1):x'=x+1}) \myline{14} denote that the walk moves to its left (\lstinprism{x'=x-1}) with probability \lstinprism{p1} and to its right (\lstinprism{x'=x+1}) with probability \lstinprism{(1-p1)}.% \emph{when \lstinprism{x} is between \lstinprism{-N} and \lstinprism{N}}.

If the probability of an update is omitted, this implies it is assumed to be 1.0. The updates in the commands \mylinestwo{12}{13} are such examples. The \lstinprism{true} \mylinestwo{15}{23} is a special update denoting a skip assignment (nothing changed) with probability 1. Since no state is changed, if these commands are enabled, they can be always taken, representing self-loop transitions.

Given this PRISM model, we can specify quantitative properties\footnote{\url{https://www.prismmodelchecker.org/manual/PropertySpecification/}} such as \lstinprism{P=? [F<=150 (b=MaxDamages)]}, enquiring about  the probability (\lstinprism{P=?}) of finally (\lstinprism{F}) \lstinprism{b} being equal to \lstinprism{MaxDamages} within 150 units of (discrete) time (\lstinprism{<=150}). 
A \emph{reward} or \emph{cost}, named \lstinprism{battery_consumption}, is defined \mylines{26}{28}, which contains one transition reward \myline{27}, denoting every \lstinprism{move} action (for synchronisation between the commands \mylinestwo{14}{22}) costs one unit (\lstinprism{1}) of battery. This allows us to specify reward-based properties such as \lstinprism{R=? [F (b=MaxDamages)]}, denoting the expected cost of battery (\lstinprism{R=?}) when finally \lstinprism{b} equal to \lstinprism{MaxDamages}. %Because the walk costs 2 units of battery levels every time to move back from the boundary, the total expected cost of battery, in this case, should add 2*\lstinprism{MaxDamages} to the above result.

Using the PRISM or Storm~\cite{Hensel2022} model checkers, we can automatically verify Markov models captured in the PRISM notation. For example, we verified that the results of properties \lstinprism{P=? [F<=150 (b=MaxDamages)]} and \lstinprism{R=? [F (b=MaxDamages)]} are 0.239 and 160.8 when the constants \lstinprism{p1}, \lstinprism{p2}, \lstinprism{N}, and \lstinprism{MaxDamages} are set to \lstinprism{0.5}, \lstinprism{0.6}, \lstinprism{10} and \lstinprism{30} respectively.

\section{Running Example: Hospital Intralogistics Mission}
\label{sec:motiving_example}

\begin{figure}
    \centering
    \begin{minipage}{0.5\textwidth}
        \centering
    \includegraphics[scale=0.15]{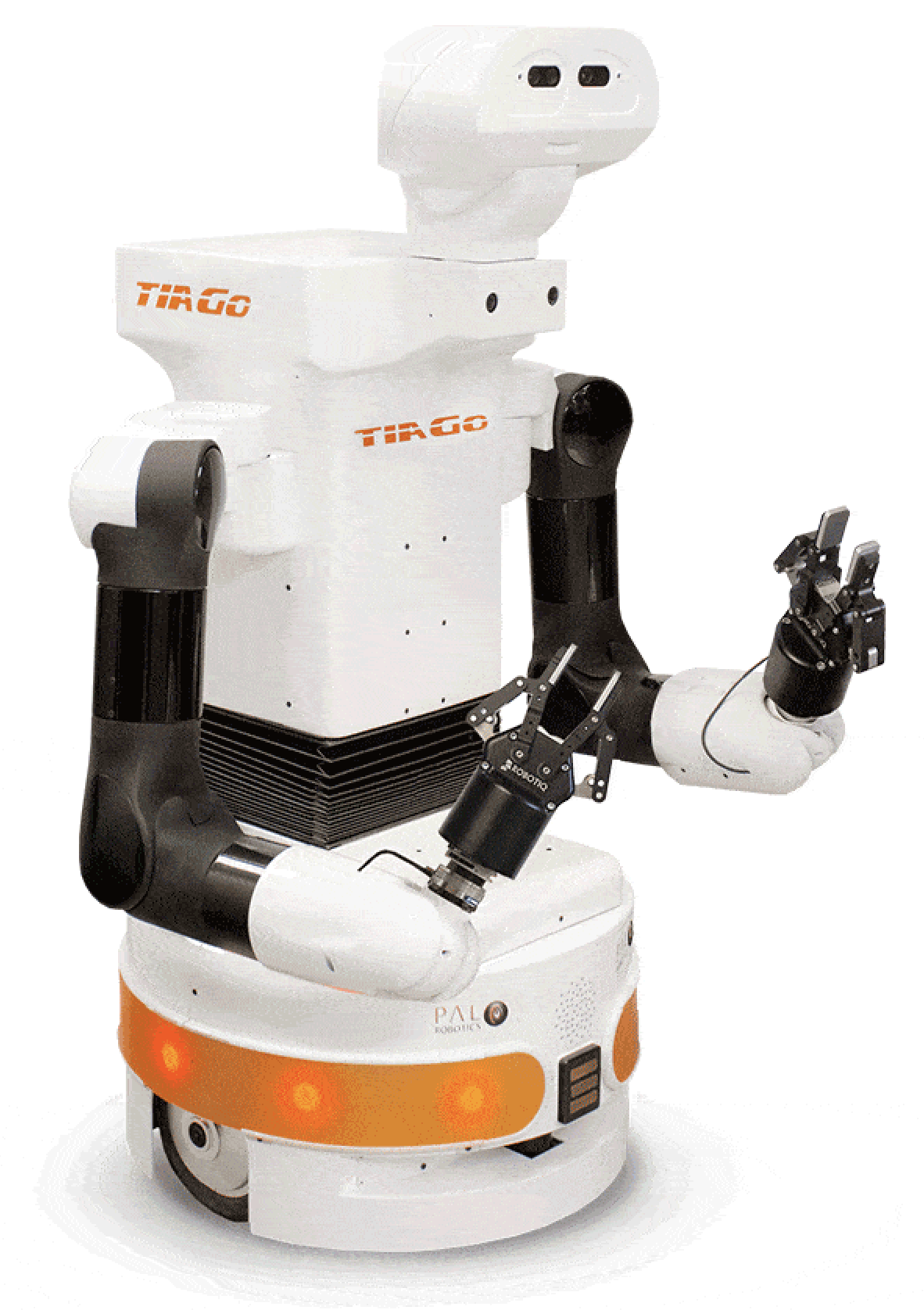}
    \includegraphics[scale=0.15]{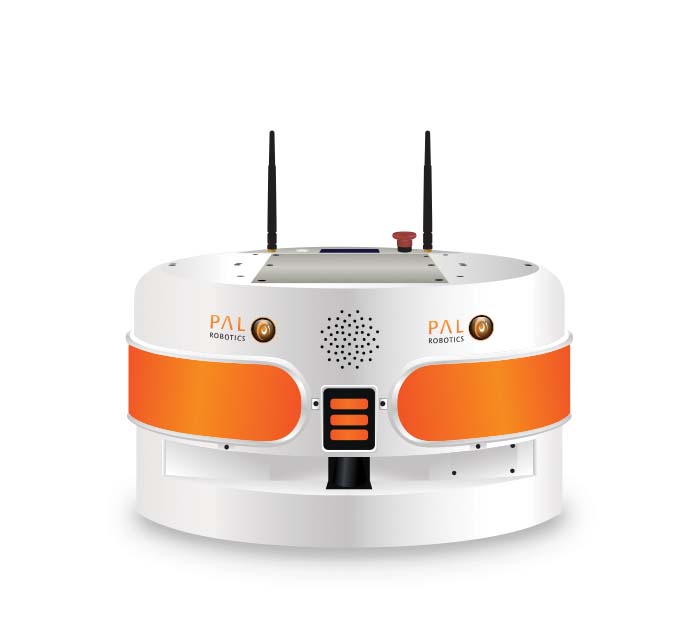}
    \caption[]{Tiago and Tiago base robots.}
    %\footnote{\url{https://pal-robotics.com/robots/tiago/}.}}
    \label{fig:tiago}
    \end{minipage}\hfill
    \begin{minipage}{0.5\textwidth}
        \centering
    \includegraphics[scale=0.40]{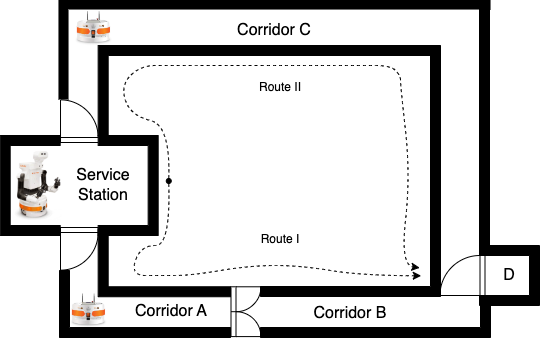}
    \caption[]{Hospital floorplan.}
    \label{fig:floorplan}
    \end{minipage}
\end{figure}

We consider an intralogistics mission within a hospital environment that involves three robots from the PAL Robotics\footnote{\url{https://pal-robotics.com}}: a handling robot equipped with a robotic arm and two mobile robots that can deliver goods, such as medicine and small equipment to their destination (\Cref{fig:tiago}). 
% a Tiago robot (\Cref{fig:tiago} left) and two Tiago base robots (\Cref{fig:tiago} right) from PAL robotics\footnote{\url{https://pal-robotics.com/}} for transporting goods, such as medicine. 
The handling robot is positioned at the service station, it can grasp objects and place them on the mobile robots, which transport the objects from the service station to the requested destinations. 

We consider a scenario where an urgent request has been received, requiring the delivery of  medicine to Room D within 10 minutes. The mobile robots are working in a dynamic environment and are subject to various kinds of uncertainty, including increased traversal time due to obstacles, physical attacks on their sensors, unexpected battery depletion, and trapped behind closed doors. 
Evidently, these situations may result in a failed or delayed delivery. 
\emph{This urgent request demands a delivery with a 99.9\% probability of success.}

As shown in \Cref{fig:floorplan}, there are two routes from the service station to Room D:
\\ $\bullet$ Route I: service station, Corridor A (CA), door connecting corridors A and B (DAB),  Corridor B (CB), Room D 
\\ $\bullet$ Route II: the service station, Corridor C (CC), Room D. 

Route I is clearly shorter than Route II. Route II is usually busy and potentially has more moving elements, like people and trolleys. Usually, it takes less time to reach Room D by following Route I than Route II. 
However, Route I is not as reliable as Route II because there are two doors between corridors A and B. 
Although at least one door is normally open so that the mobile robots can pass through, both doors might be accidentally closed making it impossible for any mobile robot to continue its mission.

Based on domain expertise and statistical analysis of past deliveries, this request for high reliability cannot be achieved using a single mobile robot for delivery by following either route. 
Executing the mission using both mobile robots is therefore more desirable. 
In this plan, each mobile robot transports the same medicine to Room D by following different routes to minimise the probability of a failed or delayed delivery and maximise its reliability (thus, offering enhanced quality of service). 

The stakeholders are interested in analysing the set of QoS requirements shown in Table~\ref{table:reqs}.
{In the next section, we will introduce the profile first and then present the modelling of this example using activity diagrams in \Cref{fig:pal_use_case}.}

\begin{table}[t]
\caption{Quality-of-service requirements for the hospital intralogistics mission} \label{table:reqs}
\begin{tabular}{lll}
\toprule
R1 & What is the probability of successful delivery through either routes within 10 minutes?\\
R2 & What is the probability of successful delivery through both routes within 10 minutes?\\
R3 & What is the probability of a failed delivery (no successful delivery) within 10 minutes?\\ 
R4 & What is the probability of mission termination (either success or failure) within 10 minutes?\\
R5 & The average attempts on passing DAB within 10 minutes\\
R6 & Parametric model checking: the probability of successful delivery through either Route I or Route II. \\ 
\bottomrule
\end{tabular}
\end{table}

\section{\method\ Overview}
\label{sec:approach}
% \method, whose high-level overview is shown in \Cref{fig:our_approach}, receives as input an activity diagram annotated with QoS information (e.g., probability, reliability, duration, cost) 

\subsection{\method\ UML Profile and Stereotypes}
\label{sec:approach:profile}

\begin{figure*}[t]
    \centering
    \includegraphics[scale=0.3]{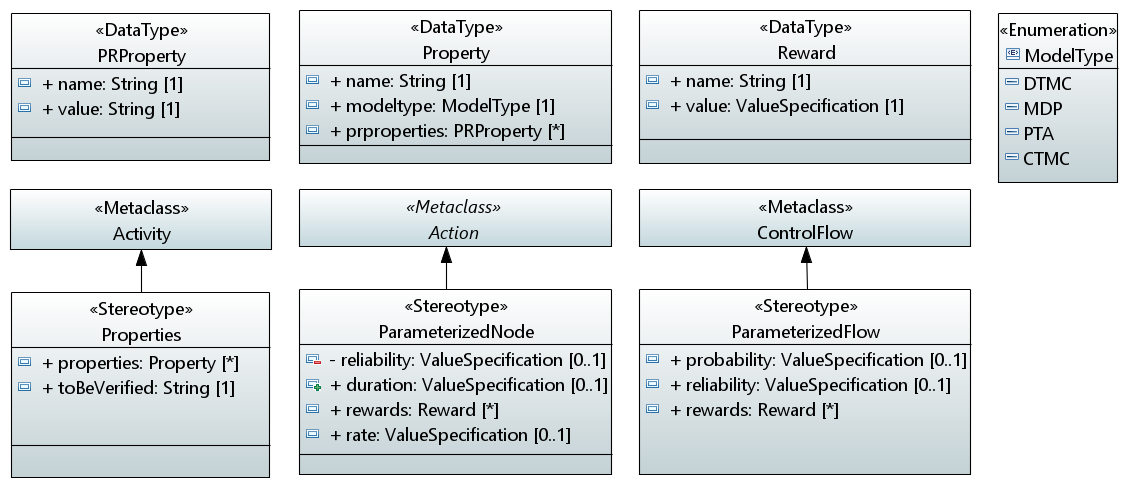}
    \caption{Profile and stereotypes}
    \label{fig:profile}
\end{figure*}

{The MARTE (Modelling and Analysis of Real-Time and Embedded systems) profile\footnote{\url{https://www.omg.org/omgmarte/}} is popularly used for modelling real time and embedded systems. It is also suitable for our work. We, however, create a new profile to make it more lightweight and easy to adapt for our purpose to facilitate the tool implementation.} 
We defined a profile including three stereotypes for \ad{ControlFlow}, \ad{Action}, and \ad{Activity}, respectively, as shown in \Cref{fig:profile}.

%{\mycomment{SysML has the probability and rate feature, but we didn't use it.}}

An \ad{Activity} can be annotated with properties through the stereotype ``Properties'' whose attributes are the properties of ``Property'' type and ``toBeVerified'' of String type.
An activity can have multiple groups of properties. Each group is of ``Property'' type, which has three attributes including ``name'' of the property group, the type of the models ``modeltype'', which can be DTMC, MDP, CTMC, or PTA (not supported now but defined for the future extension), and ``prproperties'', which is a set of PRISM properties of ``PRProperty'' type. Each property of ``PRProperty'' type has a name and a value. The value is a string in restricted natural language to present the content of the property.
%For example, in the PAL use case, when defining the first property listed in \Cref{sec:motiving_example:pal_property}, we annotate the activity diagram with the ``Properties'' stereotype by creating a group of properties.
%The type of the group is ``Property'', the group's name is ``PAL\_ctmc'', and its model type is CTMC. Then, for this group, we can first add a property of ``PRProperty'' type, whose name is ``either success'' and the value is ``P=? [ F<=t  (PAL reaches at PAL::R1DeliveredRoomD) $|$ (PAL reaches at PAL::R2DeliveredRoomD)]''.
%Then, we can continue to add the rest of the properties of  \Cref{sec:motiving_example:pal_property} into the ``PAL\_ctmc'' group.
%Furthermore, we can add a second group of properties, named e.g., ``PAL\_dtmc'', and its model type is DTMC.

We allow an activity diagram to be transformed into PRISM models of DTMC, CTMC, or MDP types.
This is determined by the attribute ``toBeVerified'' of the stereotype ``Properties''.
Before the activity diagram is transformed into PRISM models, we need to set the ``toBeVerified'' as the corresponding group of the properties that we aim to verify first, e.g., ``PAL\_ctmc''. By reading the ``modeltype'' of ``PAL\_ctmc'', which is CTMC, the PRISM models of a CTMC type are generated.

An \ad{Action} can be annotated with reliability, duration, rate, and multiple rewards through the stereotype ``ParameterizedNode''.{\footnote{The font in  \Cref{fig:profile} for \ad{Action} is italicised, while the fonts for other elements are not. We double-checked this font difference and found that it is set by the tool we are using. We did not find a way to make it consistent, but it should be treated as the same.}}
%For example, the \ad{Action} \ad{Prepare} in \Cref{fig:pal_use_case} is annotated with the reliability \textit{r\_prep} and the duration \textit{d\_prep}.
The reliability, duration, and rate are of \ad{ValueSpecification} type.
The reward is of type ``Reward'', which has the attributes of name of String type and value of \ad{ValueSpecification} type.

Similarly, a \ad{ControlFlow} can be annotated with probability, reliability, and multiple rewards through the stereotype ``ParameterizedFlow''.
%For example in \Cref{fig:pal_use_case}, the outgoing edges \ad{E11} and \ad{E12} of \ad{DecisionNode} \ad{D2} are both annotated with the probabilities \textit{p\_d\_12} and \textit{(1-p\_d\_12)}.
%The edges \ad{E17} and \ad{E18} are annotated with the rewards.
All the annotated nodes and edges are marked with ``ParameterizedNode'' and ``ParameterizedFlow''.%, respectively, as shown in \Cref{fig:pal_use_case}.

\subsection{PAL use case}
\label{sec:approach:pal_use_case}

% \begin{figure*}
%     \centering
%     \includegraphics[width=\linewidth]{pics/pal_0129.pdf}
%     \caption{Activity diagram of the hospital intralogistics robotic mission.}
%     \label{fig:pal_use_case}
% \end{figure*}

\begin{sidewaysfigure}
    \centering
    \includegraphics[width=\textheight]{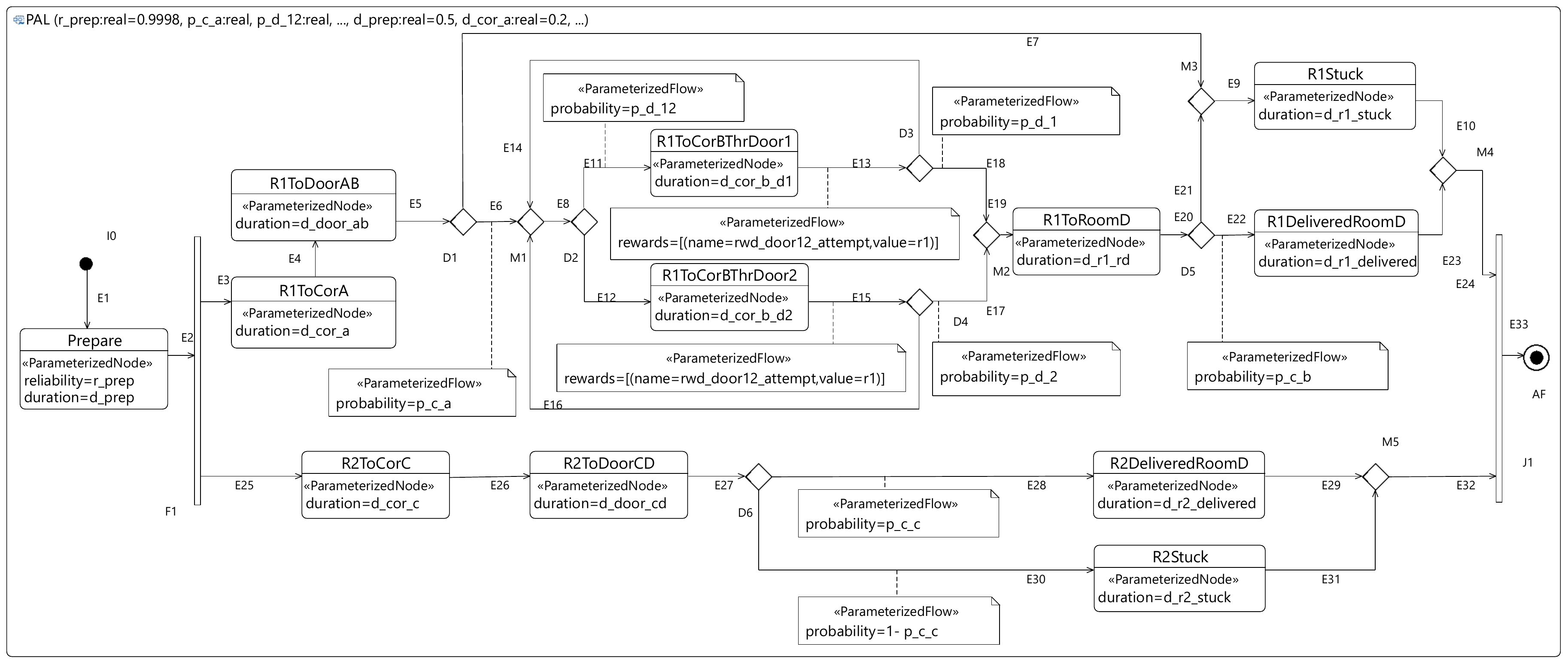}
    \caption{{Activity diagram of the hospital intralogistics robotic mission.}}
    \label{fig:pal_use_case}
\end{sidewaysfigure}
 %\simos{\\ I am thinking of removing the activity diagram from this section to keep it self-contained and include it later in the following section as many concepts have not been introduced yet\\}
 %\toremove{
\Cref{fig:pal_use_case} shows the activity diagram for this hospital intralogistic scenario comprising a sequence of actions (or subtasks) with support for choice, iteration, and concurrency. 

% We model the scenario as an activity diagram in \Cref{fig:pal_use_case}. 
% The diagram contains an activity named \ad{PAL} which specifies the behaviour of this delivery scenario as a sequence of actions (or subtasks) with support for choice, iteration, and concurrency. 
The activity has 20 input parameters (of which five are displayed) which appear inside the parenthesis after the name \ad{PAL}. For example, \ad{r\_prep:real=0.9998} declares a parameter, named \ad{r\_prep}, of type \ad{real} numbers, with a default value \ad{0.9998}. So the parameter, indeed, is \emph{specified} like a constant. Another parameter \ad{p\_c\_a:real}, without a default value, is a loosely specified or \emph{unspecified} parameter 
%whose value is instantiated through the analysis (instead of the design stage like \ad{r\_prep}).
whose value should be fixed during design through analysis.
In this example, six parameters, including \ad{p\_c\_a}, \ad{p\_c\_b}, \ad{p\_c\_c}, \ad{p\_d\_12}, \ad{p\_d\_1}, and \ad{p\_d\_2}, are left unspecified when the activity diagram is initially built and should be fixed later on.
Some other parameters, such as the duration of the preparation \ad{d\_prep}, whose values are affected by the environment, are assumed during design time and will be updated with real-time data during runtime.
{These parameters are QoS annotations for the model, representing reliability and duration of particular actions and probabilities of different decision choices. They could be set to concrete values according to prior knowledge and updated according to new observations from simulation or daily operations.} 

The activity starts its execution from its \ad{InitialNode} on the left of the diagram. Then, its control flows via a \ad{ControlEdge} \ad{E1} to the action \ad{Prepare}, denoting the preparation for the delivery, including picking and placing medicine (by the Tiago) on Base 1 and 2. On average, it takes \ad{d\_prep} minutes for the preparation where \ad{d\_prep} is an input parameter of the activity and is associated with the duration annotation of the action. %We note that the annotation and the parameter are omitted in the diagram for the sake of the presentation. 
The action is also subject to potential rare failures like the medicine shortage or the inability of the Tiago robot to pick and place the object to one of the delivery robots. 
We annotate the action with a very high \ad{reliability} (\ad{r\_prep}=0.9998), in addition to the duration \ad{d\_prep}.  {We set this high reliability because this use case requires the 99.9\% probability of delivery success (see \Cref{sec:motiving_example}). }
These annotations are introduced in our work and are presented in \Cref{sec:approach:profile}.

\begin{remark}[Modelling parameters] \label{rmk:parameter}
    An activity, as a kind of \ad{Behavior} in UML, can have input, output, input-output, and result parameters. Our work presented here only considers input parameters, such as \ad{d\_prep}. %Indeed, the \ad{PAL} activity in \Cref{fig:pal_use_case} contains 20 input parameters. 
    {Input parameters are read-only and will not be updated inside the activity.}
In UML, parameters are handled through an \ad{ActivityParameterNode}, a kind of \ad{ObjectNode}. An \ad{ActivityParameterNode} is associated with one parameter of the activity and appears on the border of the activity in the diagram. The \ad{ActivityParameterNode} then is connected to the nodes that are contained in the activity via \ad{ObjectFlow}s. Such one node is \ad{DecisionNode}. The \ad{ObjectFlow} from the \ad{ActivityParameterNode} is connected to the \ad{decisionInputFlow} of the \ad{DecisionNode}, so the parameter can be referred to, for example, in the guards of the \ad{ActivityEdge}s from the \ad{DecisionNode}. 
This (UML) way to use parameters, however, is quite heavy for modelling and also transformation later because we need to create many \ad{ActivityParameterNode}s and \ad{ObjectFlow}s and connect them together. If a parameter is referred to many times within the activity, it is not easy to draw a well-placed and readable diagram (because of so many flows). 

An alternative to \ad{ObjectFlow}s for passing data in UML is \ad{Variable}s which hold values and can be indirectly (in the sense of no explicit \ad{ObjectFlow}s) used in various \ad{VariableAction}s to read or write values to or from these variables. But similar to \ad{ActivityParameterNode}s, there are more \ad{VariableAction}s and \ad{ObjectFlow}s to be created within the activity. It is also not convenient for modelling. 
 
For this reason, our approach presented here simplifies the modelling of parameters through direct references (using \ad{RefToPara(p)}) to parameters \ad{p} in \ad{ValueSpecification} (for specifying values in UML), specifically in \ad{Expression}s. For example, the reliability annotation of the \ad{Prepare} action is \ad{r\_prep}, which, indeed, is an expression \ad{RefToPara(r\_prep)} (but hidden) in our diagram. We discuss parameters in \Cref{sec:semantics:transformation:parameter} in details.

{In short, parameters are modelled using the owned parameters of $\meta{Activity}$, instead of \ad{ActivityParameterNode}s. Parameters are referred to using \ad{RefToPara} in the \ad{ValueSpecification} part of new stereotypes in our profile, such as probability values in \ad{ParameterizedFlow}. A parameter can be specified (that is, its default value is set) or unspecified. Unspecified parameters can become specified by setting their default values based on parametric model checking and synthesis results (although this step is manual in our approach). 
}
\qed
\end{remark}

After the preparation, we use a \ad{ForkNode} \ad{F1} to create two concurrent flows: one for Base 1 to follow Route I and another for Base 2 to follow Route II.

For Base 1, it moves to CA, denoted as \ad{R1ToCorA} with duration \ad{d\_cor\_a}, passes through CA to DAB, denoted as \ad{R1ToDoorAB} with a success probability \ad{p\_c\_a} (annotated on \ad{E6} from the \ad{DecisionNode} \ad{D1}) to get to DAB and a potential failure of probability (\ad{1-p\_c\_a}, annotated on \ad{E7}) to the \ad{MergeNode} \ad{M3} which is further linked to the action \ad{R1Stuck} for the delivery failure of Base 1. If a move to DAB is successful, denoted as \ad{E6} to the \ad{MergeNode} \ad{M1}, the robot probabilistically (the \ad{DecisionNode} \ad{D2}) chooses DA (\ad{E11}) with probability \ad{p\_d\_12} or DB (\ad{E12}) with probability \ad{1-p\_d\_12}. If the robot chooses DA, then it may pass through DA successfully with probability \ad{p\_d\_1} (annotated on \ad{E18}) or fail with probability \ad{1-p\_d\_1} (annotated on \ad{E14}). This attempt, denoted as \ad{R1ToCorBThrDoor1}, takes on average \ad{d\_cor\_b\_d1} units of time. If the robot chooses DB, it is similar but with success probability \ad{p\_d\_2} (on \ad{E17}) and failure probability \ad{1-p\_d\_2} (on \ad{E16}), and duration \ad{d\_cor\_b\_d2}. In case of a failure in either case, the robot keeps retrying the process through the control flow \ad{E14} or \ad{E16} back to \ad{M1}. In case of a successful pass (\ad{E17} or \ad{E18}), the robot moves to CB and tries to pass through CB, denoted as \ad{R1ToRoomD} with duration \ad{d\_r1\_rd}, with successful probability \ad{p\_c\_b} on \ad{E22} to \ad{R1DeliveredRoomD} or failure probability (\ad{1-p\_c\_b}) on \ad{E21} to \ad{M3} (further to \ad{R1Stuck}). Both a successful delivery (\ad{R1DeliveredRoomD}) and a failed delivery (\ad{R1Stuck}) will be merged by the \ad{MergeNode} \ad{M4} and then the flow comes to the \ad{JoinNode} \ad{J1}. The outgoing edges \ad{E13} and \ad{E15} from \ad{R1ToCorBThrDoor1} and \ad{R1ToCorBThrDoor2} are annotated with rewards respectively to record the attempts of passing \ad{DAB}.

For Base 2, it moves to \ad{CC}, denoted as \ad{R2ToCorC} with duration \ad{d\_cor\_c}, and passes through CC, denoted as \ad{R2ToDoorCD} with duration \ad{d\_door\_cd} and successful probability \ad{p\_c\_c} (on \ad{E28}) to \ad{R2DeliveredRoomD} or failure probability \ad{1-p\_c\_c} (on \ad{E30}) to \ad{R2Stuck}. Similarly, both flows are merged by \ad{M5} and then the new flow reaches J1 for synchronisation with the flow for Base 1. After the synchronisation, the flow reaches the \ad{ActivityFinal} \ad{AF} and so the activity is terminated.
%}
{
\begin{figure}\centering
  \subfloat[Multiple groups of properties and which one to be verified]{\label{fig:pal_use_case:properties:1} 
  \begin{minipage}[b]{.48\linewidth}
    \includegraphics[width=\textwidth]{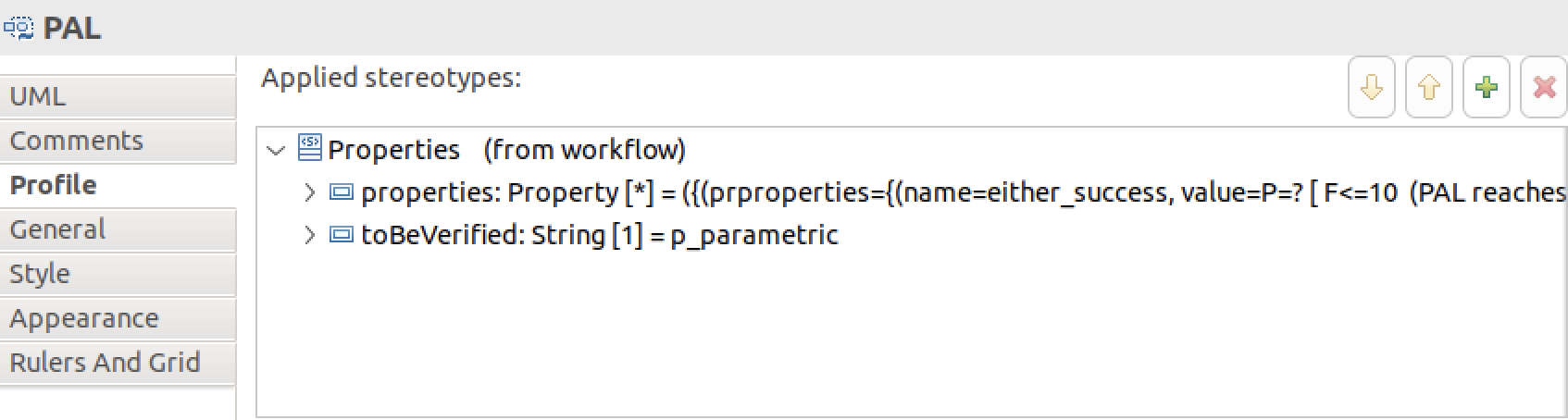}
  \end{minipage}
}
\ 
\subfloat[Four groups for DTMC, MDP, CTMC, and CTMC-parametric]{\label{fig:pal_use_case:properties:2} 
\begin{minipage}[b]{.48\linewidth}
    \includegraphics[width=\textwidth]{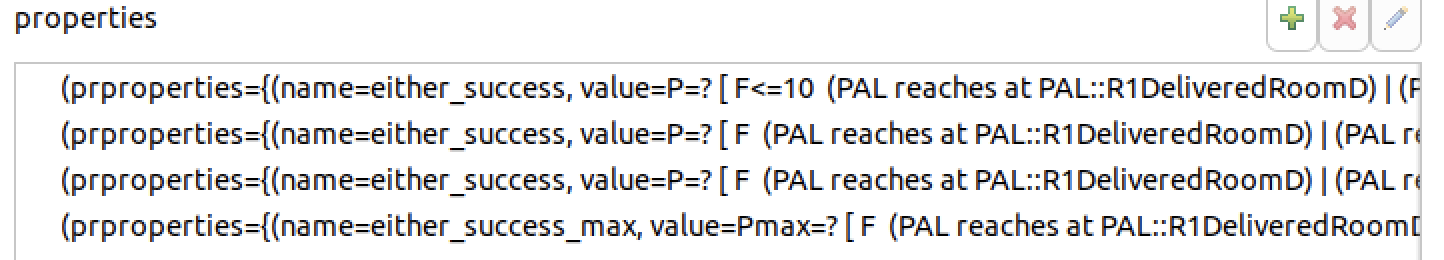}
\end{minipage}
}
\\
  \subfloat[The group for CTMC-parametric: a set of four properties]{\label{fig:pal_use_case:properties:3} 
  \begin{minipage}[b]{.48\linewidth}
    \includegraphics[width=\textwidth]{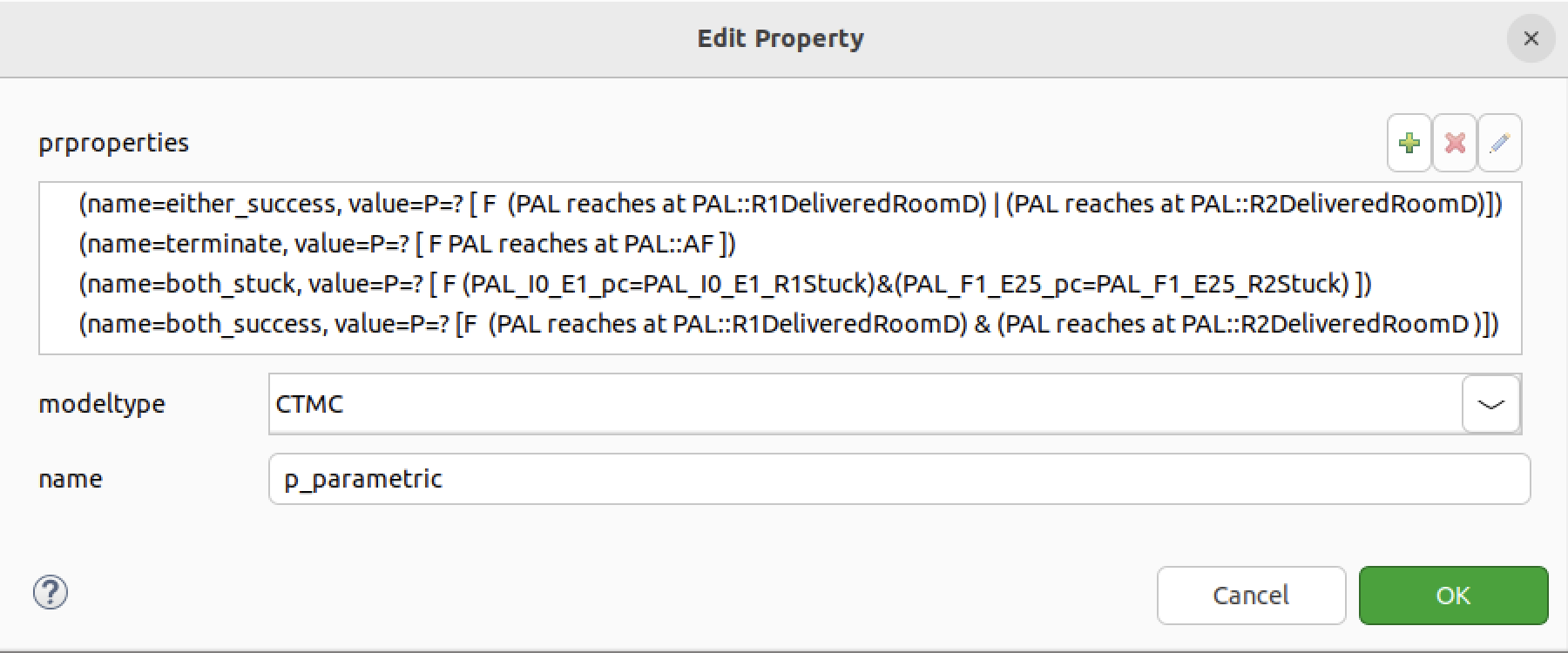}
  \end{minipage}
}
\ 
\subfloat[Each property can be named and edited]{\label{fig:pal_use_case:properties:4}
\begin{minipage}[b]{.48\linewidth}
    \includegraphics[width=\textwidth]{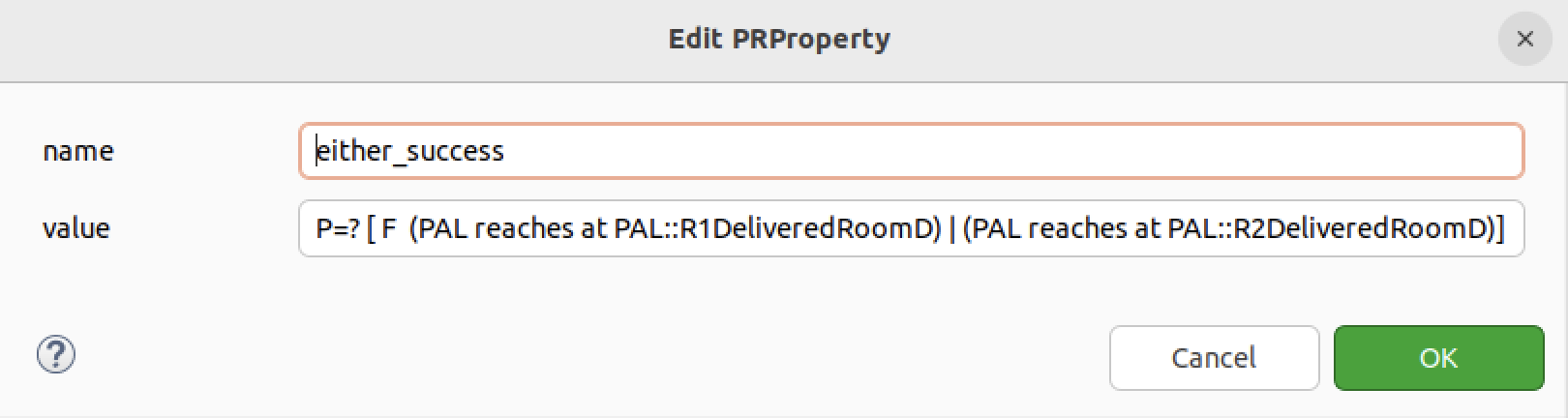}
\end{minipage}
}
\caption{Properties of the hospital intralogistics robotic mission.}\label{fig:pal_use_case:properties}
\end{figure}

The properties in \Cref{table:reqs}, as well as the type of Markov models to be analysed, are also annotated in the activity diagram, illustrated in \Cref{fig:pal_use_case:properties}, so they are part of the model. An activity can have multiple groups of annotated properties (e.g., four named groups shown in \Cref{fig:pal_use_case:properties:2}) for the configurations of different Markovian models and features. For the verification, however, one group must be specified in \ad{toBeVerified} (e.g., \ad{p\_parametric} in \Cref{fig:pal_use_case:properties:1}). A group contains a set of properties to be verified using the same model type (see \Cref{fig:pal_use_case:properties:3} where \ad{p\_parametric} contains four properties to be verified using CTMC), and each property has a name and a value for its specification (see \Cref{fig:pal_use_case:properties:4}). 
%For example, R4 is captured in the diagram as \lstinprop{P=? [F PAL reaches at PAL::AF]}, denoting the probability of the activity \ad{PAL} finally reaching its \ad{ActivityFinal} \ad{AF}. The syntax of properties is discussed in \Cref{sec:semantics:property}.
For example, R1 is captured in the diagram as \lstinprop{P=? [F (PAL reaches at PAL::R1DeliveredRoomD)|(PAL reaches at PAL::R2DeliveredRoomD)]}, denoting the probability of the activity \ad{PAL} finally reaching the \ad{OpaqueAction} \ad{R1DeliveredRoomD} or \ad{R2DeliveredRoomD}. The syntax of properties is discussed in \Cref{sec:semantics:property}.
{We note that our tool will automatically generate the corresponding properties for the PRISM model from these property annotations.}
}

\section{Semantics of activity diagrams}
\label{sec:semantics}

{This section gives the semantics of ADs in PRISM. First, we present the assumptions and well-formedness conditions for ADs in \Cref{sec:semantics:wfc} in order to make our semantics applicable. Then we introduce the general (non-probabilistic and token-based) semantic interpretation of ADs in \Cref{sec:semantics:token}. The probabilistic semantics of extended activity diagrams (with probability, rate or duration annotations) is defined using transformation rules from ADs to PRISM, specified in \Cref{sec:semantics:transformation}. 
{
  Our transformation supports a subset of AD nodes or features, including \ad{Activity}, 
  \ad{InitialNode}, 
  \ad{OpaqueAction}, 
  \ad{ForkNode}, 
  \ad{DecisionNode}, 
  \ad{MergeNode}, 
  \ad{JoinNode}, 
  \ad{FlowFinalNode}, 
  \ad{ActivityFinalNode}, 
  {input parameters}, 
  and {extensions including} reliability and duration in CTMC.
}

Afterwards, \Cref{sec:semantics:property} discusses how to specify properties for verification in our activity diagrams and their interpretation in the property specification language in PRISM, including the transformation of rewards.}

%\subsection{Overview}

\subsection{Semantics of activity diagrams}
\label{sec:semantics:ad}

\subsubsection{Assumptions and well-formedness conditions}
\label{sec:semantics:wfc}
We list the conditions that need to be satisfied by an AD for our approach to be applicable. We have defined below \changed[\Cn{1}{1}]{a complete set} of these conditions \changed[\Cn{1}{1}]{that our transformation rules and implementation rely on}.

%\paragraph{General.}
\begin{enumerate}[label={{wfc-{\arabic*}}}]
    %\item An activity to be analysed must have at least one InitialNode.
    %\item An activity to be analysed must have at least one ActivityFinalNode?
   \item {There is only one activity in an AD and no other embedded activities (therefore no \ad{CallBehaviourAction} used).} \label{wfc:activity:root}
   \item {The activity contains no event-based actions, such as \ad{AcceptEventAction} and \ad{SendSignalAction}.} \label{wfc:activity:no_event}
   \item {The activity contains no \ad{ObjectNode}s and \ad{ObjectFlow}s.} \label{wfc:activity:no_object}
   \item {The activity is initiated only by \ad{InitialNode}s, and not by other actions.} \label{wfc:activity:only_initialnode}
   \item {The activity can have read-only input parameters and no other parameters.} \label{wfc:activity:input_params}
    \item All nodes and edges must have their names, and their names are unique. \label{wfc:node_edges:has_names}
    %\item The name of a node or an edge must be \verb+[a-zA-z]([a-zA-z][0-9]_)*+, and so blank space is not allowed.
    \item {The weight property of any \ad{ActivityEdge} is assumed to be 1, meaning only one token can traverse an edge at any given time.} \label{wfc:one_token}
    \item There is exactly one incoming edge to an action. If there are more, we can use a {\ad{JoinNode}} before the action to merge all incoming edges. \label{wfc:in_edge_action:one}
    \item There is exactly one outgoing edge from an action. If there are more, we can use a \ad{ForkNode} after the action to fork all outgoing edges. \label{wfc:out_edge_action:one}

    \item The incoming alternate flows of a \ad{MergeNode} are mutually exclusive, and so they are not active at the same time. \label{wfc:mergenode:exclusive}
    \item The outgoing edges from a \ad{ForkNode} reach exclusive target nodes. \label{wfc:out_edge_forknode:exclusive_targets}
    \item  \label{wfc:out_edge_forknode:not_loops} {The outgoing edges from a \ad{ForkNode} cannot loop back the \ad{ForkNode} directly or indirectly without its outgoing flows reduced by a \ad{JoinNode}; otherwise concurrent flows are dynamically created with the increasing number of flows.}
    \item The outgoing edges from a \ad{DecisionNode} or a \ad{ForkNode} cannot have associated rewards. \label{wfc:out_edge_decision:no_reward}
        %\begin{itemize}
        %    \item This is because rewards in PRISM are based on the current states or the source states of a transition, not the destination states.
        %\end{itemize}
    %\item The name in a reference expression RefToPara must be an input parameter of the activity, whose type shall be the same as the expression.
    \item The reliability annotation of an \ad{OpaqueAction} must be of type $\mathbb{R}$ with its value between 0 and 1. \label{wfc:reliability:real:0:1}
%\end{enumerate}

%\paragraph{DecisionNode}

%\begin{enumerate}[label={\textbf {WFC-DEC-{\arabic*}}}]
    \item {There is at least one outgoing edge from a \ad{DecisionNode}.} \label{wfc:out_edge_decision:not_zero} 
    \item Each outgoing edge from a \ad{DecisionNode} can be guarded or probabilistic, but not both. \label{wfc:out_edge_decision:guard_or_prob}
    \item %\checkmark 
    The outgoing edges from a \ad{DecisionNode} are either all guarded or all probabilistic, but not mixed. \label{wfc:out_edge_decision:all_guard_or_all_prob}
   \item %\checkmark 
   The probability on any edge from a \ad{DecisionNode} is of type $\mathbb{R}$ with its value between 0 and 1. \label{wfc:out_edge_decision:prob:real:0:1}
    \item %\checkmark 
    The probabilities on the outgoing edges of any probabilistic \ad{DecisionNode} sum to 1. \label{wfc:out_edge_decision:prob:sum:1}
%\end{enumerate}
%
%\paragraph{CTMC}
%If analysed using the CTMC model, 
%\begin{enumerate}[label={\textbf {WFC-CTMC-{\arabic*}}}]
    \item If analysed as a CTMC model, any \ad{OpaqueAction} must be annotated with a positive duration or non-negative rate, but not both. \label{wfc:ctmc:action:duration_or_rate}
    %\item The duration or rate annotation of an action must be a type of $\mathbb{R}$ or $\num$. \label{wfc:ctmc:action:durantion_rate:number}
\end{enumerate}

\subsubsection{General interpretation}
\label{sec:semantics:token}
The behaviour of an \ad{Activity} is virtually and implicitly described by \mykeyword{tokens}. There are two types of tokens: \mykeyword{control} and \mykeyword{object}. A control token does not carry any data and only affects the execution of \ad{ActivityNode}s, while an object token carries data. A control token can only flow over \ad{ControlFlow} edges, pass through \ad{ControlNode}s, and be accepted by \ad{ExecutableNode}s while an object token can only flow over \ad{ObjectFlow} edges and be accepted by \ad{ObjectNode}s. 

\ad{ControlNode}s, with exceptions for \ad{InitialNode}s and \ad{ForkNode}s, do not hold tokens but just pass them through. An \ad{InitialNode} will be placed with a single control token when its parent activity begins to execute. The token will be duplicated and offered to all its outgoing edges. A \ad{ForkNode} has exactly one incoming edge and the token offered to the node will be offered to all its outgoing edges, similar to the \ad{InitialNode}. A \ad{JoinNode} synchronises all its flows by waiting for a token available on each incoming edge. Then a token is offered to its only outgoing edge. A \ad{MergeNode} also brings together multiple flows but without synchronisation. Every token offered to its incoming edge is offered to its outgoing edge. A \ad{FlowFinalNode} simply destroys all the tokens offered in its incoming edge, while an \ad{ActivityFinalNode} stops the execution of the parent activity (by destroying all tokens held in the activity) when a token is offered to its incoming edge. If a token reaches a \ad{DecisionNode}, it is offered to one of the outgoing edges of the node. Which edge would be offered depends on its guard condition being evaluated to true, if the \ad{DecisionNode} is guarded, or its probability (so the choice is probabilistic), if the \ad{DecisionNode} is probabilistic. {If the guard conditions of multiple outgoing edges are evaluated to be true at the same time, UML specifies only one edge will accept the token but does not specify which one. Here, we specify that which edge to be offered is nondeterministically chosen.}

\ad{Action}s are the only kind of \ad{ExecutableNode}. This work considers actions with only one incoming edge and one outgoing edge. An action accepts a token from its incoming edge, holds a token during its execution, and offers a token to its outgoing edge upon the completion of its execution. 

\paragraph{Assumptions} Our work presented in this paper considers the \ad{weight} property of any \ad{ActivityEdge} being 1, so there is only one token traversing the edge at the same time. Our semantics and transformation discussed later are dedicated to \ad{ControlNode}s, \ad{ControlFlow}s, and \ad{ExecutableNode}s. The support of \ad{ObjectNode}s and \ad{ObjectFlow}s is part of our future work, but our approach has taken their support into account. 

\begin{example}[Token-based semantic interpretation of PAL use case]\label{ex:pal_token}
    Consider the activity \ad{PAL} in \Cref{fig:pal_use_case}, its execution starts with placing a single token on the \ad{InitialNode} \ad{I0}. The token is passed through its outgoing flow \ad{E1} to the action \ad{Prepare}. Then \ad{Prepare} starts to execute and offers a token to its outgoing edge \ad{E2} to the \ad{ForkNode} \ad{F1} after it completes the execution. \ad{F1} offers the token to the two outgoing edges \ad{E3} and \ad{E25} at the same time. In other words, two \emph{concurrent flows} are created. The \emph{upper flow} represents the behaviour of Base 1 and the \emph{lower flow} is for that of Base 2.

  The token in \ad{E3} controls the movement of Base 1 through actions \ad{R1ToCorA} and \ad{R1ToDoorAB}, and then to a \ad{DecisionNode} \ad{D1} which is probabilistic. The token from \ad{E5} will have probability \ad{p\_c\_a} (annotated on \ad{E6}) or \ad{1-p\_c\_a} (implicit on \ad{E7}) offered to \ad{E6} or \ad{E7} respectively. The token from \ad{E7} is used to execute \ad{R1Stuck} (representing Base 1 gets stuck) via the \ad{MergeNode} \ad{M3} (through \ad{E9}). The token from \ad{E6} is passed to the \ad{DecisionNode} \ad{D2} via the \ad{MergeNode} \ad{M1} (through \ad{E8}). \ad{D2} allows Base 1 to probabilistically choose \ad{R1ToCorBThrDoor1} (with probability \ad{p\_d\_12}) or \ad{R1ToCorBThrDoor2} (with probability \ad{1-p\_d\_12}). Both the actions could succeed or fail, which is modelled by the \ad{DecisionNode}s \ad{D3} and \ad{D4}. The successful probabilities are \ad{p\_d\_1} (annotated on \ad{E18}) and \ad{p\_d\_2} (annotated on \ad{E17}), and the failure probabilities are their complementary on \ad{E14} and \ad{E16}. Either success takes the control flow to \ad{R1ToRoomD}. Either failure takes the control flow back to \ad{M1}, {which represents a probabilistic retry loop for Base~1, where repeated attempts are possible but eventual resolution may occur through a successful execution.}

  \ad{R1ToRoomD} may succeed with probability \ad{p\_c\_b} (on \ad{E22}) to \ad{R1DeliveredRoomD} or fail with \ad{1-p\_c\_b} (on \ad{E21}) to \ad{R1Stuck} via \ad{M3}. Then the two alternative flows are merged in the \ad{MergeNode} \ad{M4}, which further offers a token to the \ad{JoinNode} \ad{J1}.

  The token from \ad{F1} to \ad{E25} controls the movement of Base 2, and finally the control flow to \ad{J1} through \ad{E32}. We omit the details of this concurrent flow here.

  After both tokens on \ad{E24} and \ad{E32} are available to \ad{J1}, \ad{J1} \emph{synchronises both flows into a control flow} moving to the \ad{ActivityFinalNode} \ad{AF}, and then the activity \ad{PAL} is terminated here. 
\qed
\end{example} 

\subsection{Transformation of Activity to PRISM}
\label{sec:semantics:transformation}

{\subsubsection{Overview}
\label{sec:semantics:transformation:overview}
Our approach aims to verify an activity in activity diagrams. %This activity is called the \emph{root activity} though it may call other activities through \ad{CallBehaviourAction}. 
In general, our semantics interpretation of this activity in PRISM is a mapping from each concurrent flow in the activity to a PRISM module. While concurrent flows in ADs are dynamically created and destroyed, their corresponding PRISM modules are not. Instead, in our approach, the dynamic concurrent flows are implemented in PRISM as \emph{dynamic module state changes} where each module is associated with a special \emph{program counter} variable or \emph{pc} to record its current corresponding flow state (or the location of a token in that flow). All the possible values of \emph{pc} are all the possible locations (or nodes) of the token in that flow. We additionally number these locations based on the order of the token passing through. And these numbers start from 0 for each module. There is an extra value -1 (named INACTIVE) to represent the inactive state of a flow. And INACTIVE is the default value of \emph{pc} in a module that corresponds to an initially non-existing flow. 
{By this way, we simulate the dynamics in concurrent flow creation and destruction through dynamically activating and inactivating PRISM modules using pc variables (though these modules are statically created during transformation). } {Our approach, however, is not able to deal with the dynamics where concurrent flows are continuously created in a loop. For example, the outgoing flows of a \ad{ForkNode} are looped back to the incoming edge of the \ad{ForkNode} through \ad{DecisionNode}s and \ad{MergeNode}s. These situations are ruled out in our current work by~\ref{wfc:out_edge_forknode:not_loops}.}

Concurrency, synchronisation, or alternation between the flows are realised in PRISM through synchronisation between the corresponding modules. To deal with the termination of the activity by an \ad{ActivityFinalNode} from any flow, we use a Boolean \emph{to\_be\_terminated} variable in each module to denote whether its corresponding flow is to be terminated by an \ad{ActivityFinalNode} or not. If this is the case, the variable is set to true and next the termination will propagate to each module through a synchronisation of all modules using a \emph{terminate} action or label. At the same time, an extra Boolean variable \emph{terminated} in the corresponding module to the first concurrent flow will be set to true, denoting the activity is also terminated.

\begin{example}[Overview of the PRISM semantics for the PAL activity]\label{ex:pal_prism_sem}
    For example, the PAL activity in~\Cref{fig:pal_use_case} has two concurrent flows and its semantics in PRISM will have two modules of which each corresponds to a flow and has a local \emph{pc} variable. The lower flow, however, is not initially active and is only activated after \ad{F1}. This inactive state is represented in PRISM as a constant INACTIVE (whose value is -1) for the initial value of \emph{pc} in that corresponding module. The \emph{pc} is only updated with another \emph{pc} for the \emph{upper flow} at the same time through synchronisation when \ad{F1} is active. After that, both modules (corresponding to the two flows) are active until a \ad{JoinNode} \ad{J1} is encountered where a synchronisation happens between the two modules to inactivate one module (by setting its \emph{pc} to INACTIVE) and leave another active. Subsequently, the both (active and inactive) modules are terminated due to the \ad{ActivityFinalNode} \ad{AF}.
\end{example}

% We present the PRISM semantics, interpreted in discrete-time Markov models (DTMCs, MDPs, and CTMCs) for the activity (\Cref{sec:semantics:transformation:activity}) and \ad{ActivityNode}s including \ad{InitialNode} (\Cref{sec:semantics:transformation:initialnode}), \ad{OpaqueAction} (\Cref{sec:semantics:transformation:opaqueaction}), \ad{ForkNode}  (\Cref{sec:semantics:transformation:forknode}), \ad{DecisionNode}  (\Cref{sec:semantics:transformation:decisionnode}), \ad{MergeNode}  (\Cref{sec:semantics:transformation:mergenode}), \ad{JoinNode}  (\Cref{sec:semantics:transformation:joinnode}), \ad{ActivityFinalNode}  (\Cref{sec:semantics:transformation:activityfinal}), input parameters  (\Cref{sec:semantics:transformation:parameter}), and reliability (\Cref{sec:semantics:transformation:reliability}), then its interpretation in CTMCs in \Cref{sec:semantics:transformation:duration:ctmc}. The semantics for some nodes such as \ad{JoinNode}, \ad{FlowFinalNode}, \ad{ActivityFinalNode}, reliability, and CTMC are briefed here and detailed in Appendix~\ref{sec:appendix:semantics}.
We present the PRISM semantics, interpreted in discrete-time Markov models (DTMCs, MDPs, and CTMCs) for the activity and \ad{ActivityNode}s including \ad{InitialNode}, \ad{OpaqueAction}, \ad{ForkNode}, \ad{DecisionNode}, \ad{MergeNode}, \ad{JoinNode}, \ad{ActivityFinalNode}, input parameters, reliability, then its interpretation in CTMCs. {For each node, we discuss its semantics informally with templates first, and then present its semantics formally using transformation rules. The informal or formal semantics for some nodes} %such as \ad{JoinNode}, \ad{FlowFinalNode}, \ad{ActivityFinalNode}, reliability, and CTMC 
are briefed here and detailed in Appendix~\ref{sec:appendix:semantics}. We summarise in~\Cref{tab:prism:semantics:nodes} where the semantics of different nodes or features are given. %and where their semantics are implemented.

\begin{table}[ht]
  \caption{Where are the semantics of different node types informally discussed with templates, and formally presented with rules}
\label{tab:prism:semantics:nodes}\centering

\begin{tabular}{@{}llll@{}}
    \toprule
    Node or Feature & \Cref{sec:semantics:transformation} & Appendix \\%& Implementation \\
    \midrule
    \ad{Activity} & Informal and formal in \Cref{sec:semantics:transformation:activity} & - \\ %& Algorithms \ref{algorithm:pre-process} and \ref{algorithm:Transformation} \\
    \ad{InitialNode} & Informal and formal in \Cref{sec:semantics:transformation:initialnode} & - \\%& Algorithms \ref{algorithm:pre-process} and \ref{algorithm:Transformation} \\
    \ad{OpaqueAction} & Formal in \Cref{sec:semantics:transformation:opaqueaction} & Informal in Appendix~\ref{sec:appendix:semantics:opaqueaction} \\%& Algorithms \ref{algorithm:TraverseNds} and \ref{algorithm:Transformation} \\
    \ad{ForkNode} & Formal in \Cref{sec:semantics:transformation:forknode} & Informal in Appendix~\ref{sec:appendix:semantics:forknode} \\%& Algorithms \ref{algorithm:TraverseNds} and \ref{algorithm:Transformation} \\
    \ad{DecisionNode} & Formal in \Cref{sec:semantics:transformation:decisionnode} & Informal in Appendix~\ref{sec:appendix:semantics:decisionnode} \\%& Algorithms \ref{algorithm:TraverseNds} and \ref{algorithm:Transformation} \\
    \ad{MergeNode} & Formal in \Cref{sec:semantics:transformation:mergenode} & Informal in Appendix~\ref{sec:appendix:semantics:mergenode} \\ %& Algorithms \ref{algorithm:MrgProc} and \ref{algorithm:Transformation} \\
    \ad{JoinNode} & Formal in \Cref{sec:semantics:transformation:joinnode} &  Informal in Appendix \ref{sec:appendix:semantics:joinnode} \\%& Algorithms \ref{algorithm:TraverseNds} and \ref{algorithm:Transformation} \\
    \ad{FlowFinalNode} & - & Informal and formal in Appendix \ref{sec:appendix:semantics:flowfinal} \\%& Algorithms \ref{algorithm:TraverseNds} and \ref{algorithm:Transformation} \\
    \ad{ActivityFinalNode} & Formal in \Cref{sec:semantics:transformation:activityfinal} & Informal in Appendix \ref{sec:appendix:semantics:activityfinal} \\%& Algorithms \ref{algorithm:TraverseNds} and \ref{algorithm:Transformation} \\
    Input parameters & Formal and informal in \Cref{sec:semantics:transformation:parameter} & - \\%& Algorithms \ref{algorithm:TraverseNds} and \ref{algorithm:Transformation} \\
    Reliability & Brief in \Cref{sec:semantics:transformation:reliability} & Formal and informal in Appendix \ref{sec:appendix:semantics:reliability} \\%& Algorithms \ref{algorithm:TraverseNds} and \ref{algorithm:Transformation} \\
    Duration and CTMCs & Brief in \Cref{sec:semantics:transformation:duration:ctmc} & Formal and informal in Appendix \ref{sec:appendix:semantics:duration:ctmc} \\%& Algorithms \ref{algorithm:TraverseNds} and \ref{algorithm:Transformation} \\
    \bottomrule
\end{tabular}
\end{table}
}

\subsubsection{Activity}
\label{sec:semantics:transformation:activity}
\begin{figure*}
    \centering
    \begin{lstlisting}[language=PRISM,linewidth=0.98\linewidth]
dtmc // S1: model type: dtmc, mdp, ctmc
// S2: constants
const int INACTIVE = (-1);
...
// S3: formulas
formula {an}::to_be_terminated = ...;
...
// S4: rewards
...
// S5: The main module for the main concurrent flow
// S5.1: Constants (ActivityNodes along the flow)
const int {mn}::{n1} = 0;
...
// S5.2: module definition
module {an}::{nn}::{o1n}
  {mn}::pc : [-1..{numnodes}] init 0;
  {an}::terminated : bool init false;
  ...
  [{an}::terminate] ({an}::to_be_terminated)& (!({an}::terminated)) -> ({mn}::pc'=INACTIVE)& ({an}::terminated'=true);
  [] ({mn}::pc=INACTIVE)& ({an}::terminated)& (!{an}::to_be_terminated) -> true;
  ...
endmodule
// S6: Other modules for other concurrent flows
// S6.1: Constants 
// S6.2: module definition
module {an}::{nn}::{on}
    ...
endmodule
...
\end{lstlisting}
    \caption{Sketch of the PRISM model for an activity}
    \label{fig:transform_activity}
\end{figure*}
The semantics of an activity is a PRISM model sketched in \Cref{fig:transform_activity}. We note that an element inside braces such as \lstinprism{\{an\}} and \lstinprism{\{mn\}} in the sketch is a placeholder and will be expanded to the real name of that element in an AD during transformation. A placeholder is context-specific and the same placeholder in different contexts such as nodes, concurrent flows, edges etc. could be expanded to different names. Here \lstinprism{an} is short for \lstinprism{activity.name} (that is, the name of the activity in an AD) and \lstinprism{\{an\}} will be expanded to a string ``PAL'' for the AD in~\Cref{fig:pal_use_case}. 
Other short names used in this paper include: 
%\begin{enumerate}[label={(\arabic*)}]
\begin{itemize}
    \item \lstinprism{mn} is short for \lstinprism{modname} (the name of the PRISM module containing the placeholder);
    \item \lstinprism{nn}, \lstinprism{n1}, \lstinprism{n2}, and \lstinprism{ni} are short for \lstinprism{node.name}, \lstinprism{node1.name}, \lstinprism{node2.name}, and \lstinprism{nodei.name}; 
    \item \lstinprism{in}, \lstinprism{i1n}, \lstinprism{i2n}, and \lstinprism{inn} are short for \lstinprism{ie.name}, \lstinprism{ie1.name}, \lstinprism{ie2.name}, and \lstinprism{ien.name}; 
    \item \lstinprism{on}, \lstinprism{o1n}, \lstinprism{o2n}, and \lstinprism{onn} are short for \lstinprism{oe.name}, \lstinprism{oe1.name}, \lstinprism{oe2.name}, and \lstinprism{oen.name}; 
    \item \lstinprism{otn}, \lstinprism{o1tn}, \lstinprism{o2tn}, and \lstinprism{ontn} are short for \lstinprism{oe.target.name}, \lstinprism{oe1.target.name}, \lstinprism{oe2.target.name}, and \lstinprism{oen.target.name}.
\end{itemize}
%\end{enumerate}
In general, a transformed model, as shown in \Cref{fig:transform_activity}, consists of six sections from \textsf{S1} to \textsf{S6}.
{%\raggedright
\begin{enumerate}[label={\textbf {S{\arabic*}}}]
    \item \myline{1} gives the model type: dtmc, mdp, or ctmc. \label{activity:sketch:S1}
    \item \mylines{2}{4} declares model-wide {constants}, such as \lstinprism{INACTIVE} whose value is \lstinprism{-1}. We may have other {constants} defined here such as a default mean duration \lstinprism{mean_duration} for a CTMC model. \label{activity:sketch:S2} 
    \item \mylines{5}{7} declares formulas. We illustrate one formula \lstinprism{\{an\}::to_be_terminated} used to denote whether the activity is to be terminated by an \ad{ActivityFinalNode} or not. Its definition will be shown later when discussing the transformation rule for \ad{ActivityFinalNode}s in \Cref{sec:semantics:transformation:activityfinal}. We note that \lstinprism{\{an\}} is a placeholder and will be expanded to the name of the activity during transformation. For the formula name, we use \lstinprism{::} to separate different segments in a qualified name and \lstinprism{::} will be replaced by \lstinprism{_} in the later stage of our transformation to get a valid name in the PRISM language. \label{activity:sketch:S3} 
    \item \mylines{8}{9} defines rewards. \label{activity:sketch:S4} 
    \item \mylines{10}{22} defines the main module containing \textbf{S5.1}, the definition of {constants} for this module, and \textbf{S5.2}, a module definition. Every activity or corresponding PRISM model has a main module, which corresponds to the first concurrent flow created by the first outgoing edge \lstinprism{oe1} from the first \ad{InitialNode} {node}. The name of the module \myline{15}, \lstinprism{\{an\}::\{nn\}::\{o1n\}}, reflects it. The \lstinprism{\{mn\}::pc} variable for the module is defined \myline{16} where \lstinprism{\{mn\}} denotes the name of the module and \lstinprism{\{numnodes\}} denotes the number of nodes along the flow. The initial value of this variable is 0, instead of \lstinprism{INACTIVE} or -1, which means the main module is active by default. The constants on \textbf{S5.1} correspond to the \ad{ActivityNode}s along the concurrent flow and their values are set to ascending numbers starting from 0 (for the first node). These constants represent all possible values of the \lstinprism{pc} variable, {and the value for a specific node \lstinprism{ni} is specified through the constant \lstinprism{\{mn\}::\{ni\}}, such as the one \myline{12}}. The Boolean \lstinprism{\{an\}::terminated} variable \myline{15} is used to denote whether the activity has been terminated or not ({to enforce that, under the current assumption that an activity instance is executed only once, a terminated activity is not terminated again; supporting multiple invocations via {\ad{CallBehaviourAction}s (\ad{CBA}s)} is left for future work}). The first segment of the variable name is \lstinprism{\{an\}} to indicate this variable will be accessed by other modules, and so other modules can evaluate this variable by the activity name (they all know it) without checking its module name (they do not know it if we use \lstinprism{\{mn\}} here). The command \myline{19} has an action label which is used to synchronise the similar commands in other modules to terminate the activity by terminating all modules for all the concurrent flows of the activity when the activity is going to be terminated (\myline{19}) but not yet terminated (\myline{19}). The termination will make the module inactive (\myline{19}) and mark the activity terminated (\myline{19}). The command \myline{20} corresponds to an absorbing state, when the main module of the activity is inactive (\myline{20}) and the activity has terminated (\myline{20}) and is not going to be terminated (\myline{20}), whose only transition is a self-loop. This is simply to make the PRISM model stochastic, in other words, the transition probabilities for each row sum to 1 in the probability matrix. \label{activity:sketch:S5} 
    \item \mylines{23}{28} defines the modules for the other concurrent flows. \label{activity:sketch:S6} 
\end{enumerate} 
}
{
The semantics of an activity is a PRISM model, defined and generated using a formal transformation rule given in Rule~\ref{ad_prism:activity}. We follow the notations used in~\cite{Ye2022} to define rules and their syntax is summarised in~\cite[Table 3]{Ye2022}. These notations are similar to those in the Z specification language~\cite{Spivey1992,Woodcock1996}.

\begin{TRule}[Activity]{\meta{
  \semL{ act:Activity_{ad} }{A}: Model_{pr} \cross string = }}
  \label{ad_prism:activity}
  %\quad
  % Model_{pr}\left(consts_{rp} \cup const_{ct}, globals_{rp} \cup globals_{ct}, modules_{ct}\right) \\
  \meta{\left(\blot*{
    \begin{array}{t}
      modelType \bmapsto pmt, 
      constants \bmapsto \sigma'.consts, 
      formulas \bmapsto \set{fmterm, fmfail} \union fmrate, \\
      rewards \bmapsto \sigma'.rewards,
      %globals \bmapsto \{\},
      modules \bmapsto mods'
    \end{array}
  }_{Model_{pr}}, props \right)} \\
  \meta{\bf where}\\\quad
  \begin{array}{t}
    \meta{p \defs act.getProperty()}\quad
    \meta{pmt \defs toPrismModelType(p.modeltype)}\quad\\
    \meta{constinac \defs prmConstant(\text{``INACTIVE''}, -1)}\quad
    \meta{constparas \defs \semL{act}{Pr}} \quad \\
    \meta{constdur \defs \IF\ pmt.isCTMC() \THEN\ \set{prmRealConstant(\text{``mean\_duration''}, \varg)} \ELSE\ \emptyset}\\
    \meta{{\sigma_0} \defs \left(\set{constinac} \union constparas \union constdur, \emptyset, \emptyset\right)} \quad \\ % \bot means the current mod is not defined.
    % \meta{initnodes \defs \set{n : act.getAllNodes() | n \in InitialNode}}\\
    %\meta{constmods = \semL{act, \varg}{I}^{\sigma_0} \limg initnodes \rimg }\quad \\
    \meta{\sigma' \defs \semL{act}{IN}^{\sigma_0}}\quad 
    \meta{mods \defs \ran\ ~\left(\sigma'.modmap\right)} \quad 
    \meta{mods' \defs \left(\lambda m @ m.varpcTopUpdatedToLastUsed()\right)\limg mods \rimg} \quad \\
    \meta{exprterm \defs prmOrBoolExpr\left(\set{m: mods' | m.vartbterm() \neq null @ prmVarRefExpr(id(m, \text{``to\_be\_terminated''}))}\right)}\\
    \meta{fmterm \defs prmFormula(id(act, \text{``to\_be\_terminated''}), expterm)}\\
    \meta{exprfail \defs prmOrBoolExpr\left(\set{m: mods' | m.vartbfail() \neq null  @ prmVarRefExpr(id(m, \text{``to\_be\_failed''}))}\right)}\\
    \meta{fmfail \defs prmFormula(id(act, \text{``to\_be\_failed''}), expfail)}\\
    \meta{fmrate \defs \left(
        \begin{array}{t}
          \IF\ pmt.isCTMC() \THEN\ \\\quad 
            \set{prmFormula(\text{``default\_rate''}, prmDivExpr(1.0, prmVarExpr(\text{``mean\_duration''})))} \\
          \ELSE\ \emptyset
        \end{array}
      \right)}\\
    \meta{props \defs refResolve(act, \sigma', p.prproperties)}\\
  \end{array}
\end{TRule}

Generally, a rule characterises a function on elements of the AD metamodel. A rule definition is composed of a declaration, a body, and a $\meta{where}$ clause. The declaration gives the function name, its parameters (names and types), and its return type. In Rule~\ref{ad_prism:activity}, the semantics function for an activity is $\meta{\semL{\varg}{A}}$, which has one parameter $\meta{act}$ of type $\meta{Activity_{ad}}$ (that is, a type class in the AD metamodel) and returns a pair whose first component is a PRISM model of type $\meta{Model_{pr}}$ (that is, a type class in the PRISM metamodel, detailed in~\cite[Section~4.2]{Ye2022}) and whose second component is the PRISM properties of type $\meta{string}$ (a general type, from neither of the the two metamodels). In the rest of the paper, we assume that the parameters of a function have their types in the AD metamodel and its return has the type in the PRISM metamodel. Therefore, we omit subscripts in the subsequent rules for simplicity.

The body of Rule~\ref{ad_prism:activity} is a PRISM model and property pair, in which the model is an object of the type $\meta{Model}$ with its components (including the model type, constants, formulas, rewards, and modules) instantiated to corresponding values as defined in the $\meta{where}$ clause. For example, its component $\meta{modelType}$ is instantiated to $\meta{pmt}$, which is translated from $\meta{p.modeltype}$ by a function $\meta{toPrismModelType}$ where $\meta{p}$ is the annotated property (of type \ad{Property} in our profile~\Cref{fig:profile}) of the activity $\meta{act}$, retrieved through the $\meta{getProperty}$ function. 

For each rule, we introduce a transformation context $\meta{\sigma}$, an object of a schema or record type $\tau$ where $$\tau \defs {\blot*{consts: \power~Constant_{pr}, modmap: \left(ActivityNode \cross ActivityEdge \pfun Module_{pr} \right), rewards: \power~Rewards_{pr}}}$$ 
That is, $\tau$ contains three fields $\meta{consts}$ (a set of PRISM constants) and $\meta{modmap}$ (a partial function mapping an activity node and edge pair to a PRISM module), and $\meta{rewards}$ (a set of rewards). The $\meta{modmap}$ denotes the associated module for a given node and one of its incoming or outgoing edge. We also define $\meta{modmap[m'/m]}$ to be a new map with $\meta{m}$ replaced by a new module $\meta{m'}$. A record can be instantiated (initialised to be empty here) using the following syntax: $\sigma_0 \defs {\blot*{consts \bmapsto \emptyset, modmap \bmapsto \emptyset, rewards \bmapsto \emptyset}_{\tau}}$. We use a simpler syntax $\meta{\left(\emptyset, \emptyset, \emptyset\right)}$ in our rules to denote it. A component in the record can be accessed using its name, such as $\meta{\sigma_0.consts}$.

In Rule~\ref{ad_prism:activity}, $\meta{\sigma_0}$ is the initial context, only containing constants: $\meta{constinac}$, a PRISM constant with its name $\meta{\text{``INACTIVE''}}$ and associated value $\meta{-1}$, constructed by $\meta{prmConstant}$ (that is, a function to construct a PRISM object of type $\meta{Constant}$); $\meta{constparas}$, a set of PRISM constants resulting from the semantics of input parameters of the activity by Rule~\ref{ad_prism:input_paras}; $\meta{constdur}$, a single constant, named $\meta{\text{``mean\_duration''}}$, if the AD is treated as a CTMC model, and empty otherwise.

The function $\meta{\semL{act}{IN}^{\sigma_0}}$ (defined in Rule~\ref{ad_prism:initial_nodes} in~\Cref{sec:semantics:transformation:initialnode}) gives the semantics of \ad{InitialNode}s in $\meta{act}$ based on $\meta{\sigma_0}$. 
{This represents that, under the current semantics where activities are assumed to start from \lstinprism{InitialNode}s, the flows (or the modules) are active by default at its \lstinprism{InitialNode}s.}
It returns an updated context $\meta{\sigma'}$, which contains all the elements to construct the PRISM model. The $\meta{mods}$ is the range $\meta{\ran\ ~}$ of the $\meta{modmap}$ in $\meta{\sigma'}$, which contains all the PRISM modules for $\meta{act}$. The $\meta{mods'}$, defined through relational image $\meta{\varg\limg\varg\rimg}$, contains each module $\meta{m}$ from $\meta{mods}$ but with the top value of the range type for its pc variable updated, which is detailed in \Cref{sec:semantics:transformation:initialnode}. The $\meta{exprterm}$ is a disjunction of a set of boolean expressions, in which each corresponds to a reference to a PRISM variable $\meta{id(m, \text{``to\_be\_terminated''})}$ in a module $\meta{m}$ from $\meta{mods'}$ if such a variable $\meta{m.vartbterm()}$ exists. Here, we use $\meta{id(m, n)}$ to construct a unique string name from a named object or its name $\meta{n}$ in its named container object or the name $\meta{m}$, and the set comprehension format $\meta{\set{v:T_v | P(v) @ e(v)}}$ to characterise such a set. Then $\meta{fmterm}$ is a PRISM formula with name and its value being $\meta{exprterm}$. And $\meta{exprfail}$ and $\meta{fmfail}$ are similar. The $\meta{fmrate}$ is a set, containing only one formula to denote the default rate, which is the result of 1 divided by $\meta{\text{``mean\_duration''}}$, if for a CTMC model, or empty otherwise. The use of these formulas will be discussed in \Cref{sec:semantics:transformation:activityfinal}, \Cref{sec:semantics:transformation:reliability}, and \Cref{sec:semantics:transformation:duration:ctmc}.

The $\meta{props}$ is the PRISM properties generated from a collection $\meta{p.prproperties}$ of to-be-verified properties (of type $\meta{PRProperty}$, as exemplified in~\Cref{fig:pal_use_case:properties:3} for the PAL use case) through resolving references to elements in the activity using $\meta{refResolve}$ (see \Cref{sec:semantics:property} for more details).

%Finally, the body of the function defined in Rule~\ref{ad_prism:activity} has the components of the PRISM model instantiated to these definitions in the clause.
}

\begin{example}[PAL use case - PRISM model]
  We show the corresponding (simplified) PRISM model in {\Cref{fig:transform_activity_pal_use_case}}, according to Rule~\ref{ad_prism:activity}, where we consider the Markov model CTMC. The complete PRISM model can be found online.\footnote{\url{https://github.com/RandallYe/QANAD/blob/master/Examples/PAL_use_case/prism_models/p_ctmc/pal.prism}} We omit a detailed explanation here because the model can be easily mapped to the general sketch in \Cref{fig:transform_activity}.
\qed
\end{example}

\subsubsection{InitialNode}
\label{sec:semantics:transformation:initialnode}
Suppose there are $n$ \ad{InitialNode}s and each node has $oe(i)$ (for $i \in \{1..n\}$) outgoing edges. Then $\sum_{i=1}^{n} oe(i)$ concurrent flows will be created. Hence the same number of new PRISM modules are created. One of them becomes the main module as illustrated in \Cref{fig:transform_activity}, but it does not matter which one is the main module. 
%We show one of the other modules below whose name is identified by the \ad{InitialNode} name \lstinprism{\{nn\}} and one of its outgoing edge name \lstinprism{\{on\}} followed after the activity name.
%\begin{lstlisting}[language=PRISM,linewidth=0.45\textwidth]
%const int {mn}::{nn} = 0;
%module {an}::{nn}::{on}
%  {mn}::pc : [-1..{numnodes}] init 0;
%   ...
%endmodule
%\end{lstlisting}
%We also note that the initial value of the \lstinprism{pc} variable is also 0 which is just the value of the {constant} \lstinprism{\{mn\}::\{nn\}} \myline{1}. 
%This means the concurrent flows from all \ad{InitialNode}s are enabled initially.

\begin{TRule}[InitialNodes]{\meta{\quad~
  \semL{act : Activity}{IN}^{\sigma}: \tau =}}
  \label{ad_prism:initial_nodes}
  \quad \meta{\semL{act, \set{oe_1.target \mapsto m_1} \oplus nmods_o}{Ns}^{\sigma'}} \\
  \meta{\bf where}\\\quad
  \begin{array}{t}
    \meta{initnodes \defs \set{n : act.getAllNodes() | n \in InitialNode}}\\
    \meta{noemap \defs \left(\lambda n @ \left(\lambda oe @ n \mapsto oe \right) \limg act.getAllOEdges(n) \rimg \right)\limg initnodes \rimg }\quad \\
    \meta{noe_1 \in noemap \land \set{noe_1} \inter noe_o = \emptyset \land \set{noe_1} \union noe_o = noemap \land (n_1, oe_1) = noe_1} \quad \\
    \meta{constsmod_1 \defs \semL{act, n_1, oe_1}{IO}} \quad 
    \meta{m_1 \defs (constsmod_1.2).setAsMain()} \quad \\
    \meta{modoe \defs \left(\lambda (n, oe) @ \semL{act, n, oe}{IO}\right)}\quad 
    \meta{constmods_o \defs modoe \limg noe_o \rimg} \quad 
    \meta{ consts_o = \bigcup \{r:constmods_o @ r.1\} }\quad \\
    \meta{modmap_o \defs \left(\lambda (n,oe) @ \set{(n, oe) \mapsto modoe((n, oe)).2, (oe.target, oe) \mapsto modoe((n, oe)).2}\right)\limg noe_o \rimg} \quad \\
    \meta{\sigma' \defs \left(
        \begin{array}{t}
            \sigma.consts \union constsmod_1.1 \union consts_o,\ \sigma.modmap \oplus \set{noe_1 \mapsto m_1, (oe_1.target, oe_1) \mapsto m_1} \oplus \bigcup modmap_o, \\ 
            \sigma.rewards
        \end{array}
        \right)} \quad \\
    \meta{nmods_o \defs \left(\lambda (n,oe) @ oe.target \mapsto modoe(n, oe).2\right) \limg noe_o\rimg}
    %\meta{(mods, \sigma_r) \defs \semL{act, nmods}{Ns}^{\sigma}}
  \end{array}
\end{TRule}

{The semantics of \ad{InitialNode}s are defined in Rule~\ref{ad_prism:initial_nodes} by the function $\meta{\semL{act}{IN}^{\sigma}}$ which has one parameter (the activity $\meta{act}$) and returns a new context of type $\meta{\tau}$. Basically, this rule creates new modules for these nodes, updates the context, and then visits all the reachable nodes from these nodes under the new context.

  The body of the function is the transformation of a collection of target \ad{ActivityNode}s from all \ad{InitialNode}s through their outgoing edges by $\meta{\semL{act, \varg}{Ns}}$ (see Rule~\ref{ad_prism:activitynodes}) under the new context $\meta{\sigma'}$. The context is defined in the $\meta{where}$ clause. The $\meta{initnodes}$ denotes all the nodes of $\meta{act}$ that are $\meta{InitialNode}$. The $\meta{noemap}$ is a set of $\meta{(n, oe)}$ pairs where $\meta{n}$ is a node from $\meta{initnodes}$ and $\meta{oe}$ is one of its outgoing edges (retrieved through $\meta{act.getAllOEdges(n)}$). We use a condition in the third line in the clause to instantiate an arbitrary pair $\meta{(n_1, oe_1)}$, whose corresponding PRISM module (by $\meta{\semL{act, n_1, oe_1}{IO}}$, defined in Rule~\ref{ad_prism:initialnode}) is treated as the main module. That function $\meta{\semL{\varg}{IO}}$ returns a PRISM module and corresponding constants, encapsulated in a pair. The first or second component of the pair can be accessed using $\meta{constsmod_1.1}$ and $\meta{constsmod_1.2}$. Then the resulted module ($\meta{constsmod_1.2}$) is set to be main through the $\meta{setAsMain}$ function and named $\meta{m_1}$.

  The remaining $\meta{(n, oe)}$ pairs are denoted as $\meta{noe_o}$. Similarly, each pair corresponds to a module and a set of constants, resulted from the function $\meta{modoe}$. So, $\meta{constmods_o}$ is a set of such results for $\meta{noe_o}$. The $\meta{consts_o}$ is all constants for them, defined using generalised union $\meta{\bigcup}$. The $\meta{modmap_o}$ is a set of mappings from a $\meta{(n,oe)}$ pair to its corresponding module, and mappings from the target $\meta{oe.target}$ of $\meta{oe}$ and $\meta{oe}$ pair to this module. This denotes the corresponding modules for these \ad{InitialNode}s and the directly reachable nodes from them. Finally, the updated context $\meta{\sigma'}$ contains new constants and module mappings for $\meta{noemap}$ where $\meta{\oplus}$ is a relation override operator.   

  The second argument of $\meta{\semL{act, \varg}{Ns}}$ in the rule body is a set of mappings from a node, to be visited, to its corresponding PRISM module, including the mapping for $\meta{noe_1}$ and those $\meta{nmods_o}$ for $\meta{noe_o}$.

\begin{TRule}[InitialNode and its outgoing edge]{\\\meta{\quad~
  \semL{act : Activity, n: InitialNode, oe : ActivityEdge}{IO}: \power~Constant \cross Module =}}
  \label{ad_prism:initialnode}
  \quad \meta{\left(\set{constinit, constnext}, {mod}\right)}
  \\
  \meta{\bf where}\\\quad
  \begin{array}{t}
      \meta{mn \defs id(act, oe)} \;\;\;\;
      \meta{nn \defs id(mn, n)} \;\;\;\;
      \meta{constinit\defs prmConstant(nn, 0)} \;\;\;\;
      \meta{constnext \defs prmConstant(id(mn, oe.target), 1)} \\
      \meta{mod \defs prmNewModule(act, mn, nn, constinit, constnext, 1)} \\
  \end{array}
\end{TRule}

Rule~\ref{ad_prism:initialnode} defines $\meta{\semL{act, n, oe}{IO}}$, which returns a new module for the $\meta{InitialNode}$ $\meta{n}$ and its outgoing edge $\meta{oe}$. The body of the rule is a pair composed of a set of two constants ($\meta{constinit}$ and $\meta{constnext}$) and a new module $\meta{mod}$ which is defined in the clause where $\meta{mn}$ is the module name, $\meta{nn}$ is the name for $\meta{n}$, $\meta{constinit}$ is a constant (associated with value 0) denoting the initial state of the module at $\meta{n}$, and $\meta{constnext}$ is a constant (associated with value 1) representing the target of $\meta{oe}$. Then $\meta{mod}$ is defined by a function $\meta{prmNewModule}$ (in Rule~\ref{ad_prism:newprismmodule}) with the module name $\meta{mn}$ , the action label $\meta{nn}$ for its first command, the initial state $\meta{constinit}$ for its pc variable, and the target state $\meta{constnext}$, along with the $\meta{lastused}$ number $\meta{1}$ to define constants (so the next available number to be used to define a constant will be $\meta{lastused+1}$). 

\begin{TRule}[New PRISM Module]{\\\meta{\quad~
  prmNewModule\left(
    \begin{array}{t}
      act: Activity, mn: string, actlbl: string, \\
      init: Constant_{pr}, next: Constant_{pr}, lastused: int
    \end{array}
  \right): Module =}}
  \label{ad_prism:newprismmodule}
  %\quad \meta{prmModDef\left(mn, \{varpc, vartbterm, vartbfail\}, \{cmd\}, lastused\right)} \\
  \quad \meta{prmModDef\left(mn, \{varpc\}, \{cmd\}, lastused\right)} \\
  \meta{\bf where}\\\quad
  \begin{array}{t}
      %\meta{varpc \defs prmVarDeclRangeInit(id(mn, \text{``pc''}), -1, numnodes, init)} \\
      % numnodes is set -1 initially and will be updated at the last to the lastused 
      \meta{varpc \defs prmVarDeclRangeInit(id(mn, \text{``pc''}), -1, -1, init)} \\
      %\meta{vartbterm \defs prmVarDeclBool(id(mn, \text{``to\_be\_terminated''}), false)} \\
      %\meta{vartbfail \defs prmVarDeclBool(id(mn, \text{``to\_be\_failed''}), false)} \\
      \meta{grd1 \defs prmEqBoolExpr(varpc, init)} \\
      \meta{grd2 \defs prmNotBoolExpr(prmVarRefExpr(id(act, \text{``to\_be\_terminated''})))}\\
      \meta{grd \defs prmAndBoolExpr(grd1, grd2)} \\
      \meta{upd \defs prmUpdate(\{prmAssign(varpc, constnext)\})} \\
      \meta{cmd \defs prmCommand(prmAction(actlbl), grd, \{upd\})}
  \end{array}
\end{TRule}

The function $\meta{prmNewModule}$ in Rule~\ref{ad_prism:newprismmodule} uses $\meta{prmModDef}$ to construct a PRISM module from its name $\meta{mn}$, a set of local variables (only a $\meta{varpc}$ here), a set of commands (only a $\meta{cmd}$ here), and the $\meta{lastused}$ number. We note that $\meta{lastused}$ is not part of the PRISM metamodel, but introduced in the paper to facilitate the formalisation of rules. In the $\meta{where}$ clause, $\meta{varpc}$ is a pc variable of a range type whose bottom and top are both -1, initialised to the value associated with the constant $\meta{init}$. Here, the top is also set to -1 because initially it is unknown when a module is created and will be updated later by $\meta{varpcTopUpdatedToLastUsed}$ in Rule~\ref{ad_prism:activity} after all nodes are visited. 

The command $\meta{cmd}$ is constructed by $\meta{prmCommand}$ with a PRISM action from the action label, a guard $\meta{grd}$ (a boolean expression), and a set of updates (only $\meta{upd}$ here). The $\meta{grd}$ is a conjunction of $\meta{grd1}$ and $\meta{grd2}$, denoting a state where $\meta{varpc}$ is equal to $\meta{init}$ and no module is to be terminated. In the rest of the paper, we use $\meta{prmBoolExprPCAndNotTBTerm(pc, init)}$ to construct $\meta{grd}$ and omit the definition of $\meta{grd1}$ and $\meta{grd2}$.
}

\begin{example}[PAL use case - InitialNode]
  The main module of the example is \lstinprism{PAL::I0::E1} sketched in \Cref{fig:transform_activity_pal_use_case:initialnode}. 
\begin{figure}
\centering
\begin{lstlisting}[language=PRISM,linewidth=0.98\linewidth]
module PAL::I0::E1
  PAL::I0::E1::pc : [-1..19] init 0;
  PAL::I0::E1::to_be_terminated : bool init false;
  PAL::I0::E1::to_be_failed : bool init false;
  PAL::terminated : bool init false;
  [PAL::I0::E1::I0] ((PAL::I0::E1::pc=PAL::I0::E1::I0))& 
    (!(PAL::to_be_terminated)) -> default_rate: (PAL::I0::E1::pc'=PAL::I0::E1::Prepare);
...
endmodule
\end{lstlisting}
\caption{\ad{InitialNode} part of the PRISM model for PAL use case.}
\label{fig:transform_activity_pal_use_case:initialnode}
\end{figure}
\qed
\end{example}

{
\begin{TRule}[ActivityNodes with their modules]{\\\meta{\quad~
  \semL{act : Activity, nmods: ActivityNode \pfun Module_{pr}}{Ns}^{\sigma}: \tau =}}
  \label{ad_prism:activitynodes}
  \quad
  \meta{
    \IF\ nms_o \neq \emptyset\ \THEN\ \semL{act, nms_o[m_1'/m_1]}{Ns}^{\sigma_1}\ \ELSE\ \sigma_1
  }
  \\
  \meta{\bf where}\\\quad
  \begin{array}{t}
    \meta{nm_1 \in nmods \land \set{nm_1} \inter nms_o = \emptyset \land \set{nm_1} \union nms_o = nmods \land (n_1, m_1) = nm_1} \\
    \meta{\sigma_1 \defs \semL{act, n_1, m_1}{N}^{\sigma}}\quad 
    % get the new module from \sigma_1 for m_1
    \meta{\exists ne @ ne \mapsto m_1 \in \left(\sigma.modmap~\rres~\set{m_1}\right) \land m_1' = \sigma_1.modmap(ne)} \\
    %\meta{\sigma_o \defs \semL{act, nms_o[m_1'/m_1]}{Ns}^{\sigma_1}} % the further map will use the updated m_1' module
  \end{array}
\end{TRule}

Rule~\ref{ad_prism:activitynodes} defines the function $\meta{\semL{act, nmods}{Ns}^{\sigma}}$ to transform a set of \ad{ActivityNode}s along with their corresponding modules, recorded in a partial function $\meta{nmods}$. This rule transforms an arbitrary node $\meta{n_1}$ (with its module $\meta{m_1}$) first by a function $\meta{\semL{act, n_1, m_1}{N}^{\sigma}}$ (defined in Rule~\ref{ad_prism:activitynode}) which returns an updated context $\meta{\sigma_1}$. Then we use an existential quantification to establish that $\meta{m_1'}$ is the updated $\meta{m_1}$ module in $\meta{\sigma_1}$ where $\meta{\rres}$ is a range restriction operator. The remaining nodes are within $\meta{nms_o}$ but with their corresponding modules updated to $\meta{m_1'}$ (if they are $\meta{m_1}$) by a substitution $\meta{nms_o[m_1'/m_1]}$. The body of the rule recursively transforms the remaining nodes under the new context $\meta{\sigma_1}$ until $\meta{nms_o}$ is empty. 

\begin{TRule}[ActivityNode]{\\\meta{\quad~
  \semL{act : Activity, n: ActivityNode, mod: Module_{pr}}{N}^{\sigma}: \tau =}}
  \label{ad_prism:activitynode}
  \quad
  \meta{
    \left(
      \begin{array}{tt}
        \meta{\IF\ n \in OpaqueAction \THEN\ \semL{act, n, mod}{Op}^{\sigma}}\quad 
         & \meta{\ELSE \IF\ n \in ForkNode \THEN\ \semL{act, n, mod}{Fk}^{\sigma} }\\
         \meta{\ELSE \IF\ n \in DecisionNode \THEN\ \semL{act, n, mod}{Ds}^{\sigma} }\quad
         & \meta{\ELSE \IF\ n \in MergeNode \THEN\ \semL{act, n, mod}{Mg}^{\sigma} }\\
         \meta{\ELSE \IF\ n \in JoinNode \THEN\ \semL{act, n, mod}{Jn}^{\sigma} }\quad
         & \meta{\ELSE \IF\ n \in ActivityFinalNode \THEN\ \semL{act, n, mod}{Af}^{\sigma} }\\
         \meta{\ELSE \IF\ n \in FlowFinalNode \THEN\ \semL{act, n, mod}{Ff}^{\sigma} }\quad
         & \meta{\ELSE\ \sigma}
      \end{array} 
    \right)
  }
  % \\
  % \meta{\bf where}\\\quad
  % \begin{array}{t}
  %   \meta{initnodes \defs \set{n : act.getAllNodes() | n \in InitialNode}}\\
  %   \meta{oe \defs n.getOEdge()} \quad
  % \end{array}
\end{TRule}

Rule~\ref{ad_prism:activitynode} defines $\meta{\semL{act , n, mod}{N}^{\sigma}}$ to transform a single node $\meta{n}$ with its corresponding module $\meta{m}$. It calls different functions (defined in the following sections) based on the type of $\meta{n}$.
}

\subsubsection{OpaqueAction}
\label{sec:semantics:transformation:opaqueaction}
We consider the \ad{OpaqueAction} with one incoming edge (\ref{wfc:in_edge_action:one}) and one outgoing edge (\ref{wfc:out_edge_action:one}), such as \ad{Prepare} and \ad{R1ToCorA} in \Cref{fig:pal_use_case}.
{If an \ad{OpaqueAction} has multiple incoming edges, a \ad{JoinNode} can be added before it to synchronise concurrent flows into one. Similarly, if an \ad{OpaqueAction} has multiple outgoing edges, a \ad{ForkNode} can be added after it. }
{This consideration is a simplifying assumption rather than a fundamental constraint, and its relaxation is left to future work.}

{
The semantics of a \ad{OpaqueAction} is informally discussed in Appendix~\ref{sec:appendix:semantics:opaqueaction} and formally generated by the function $\meta{\semL{act, n, mod}{Op}^{\sigma}}$ which is defined in Rule~\ref{ad_prism:opaqueaction}.
\begin{TRule}[OpaqueAction]{\\\meta{\quad~
  \semL{act : Activity, n: OpaqueAction, mod: Module_{pr}}{Op}^{\sigma}: \tau =}}
  \label{ad_prism:opaqueaction}
  \quad
\meta{
  \semL{act, oe.target, mod'}{N}^{\sigma'}
} \\
  \meta{\bf where}\\\quad
  \begin{array}{t}
      \meta{oe \defs act.getOEdge(n)} \quad
      \meta{constnext \defs prmConstant(id(mod, oe.target), mod.lastused+1)}\\
      % \meta{grd_2 \defs prmNotBoolExpr(prmVarRefExpr(id(act, \text{``to\_be\_terminated''})))}\\
      % \meta{grd \defs prmAndBoolExpr(grd_1, grd_2)} \\
      \meta{grd \defs prmBoolExprPCAndNotTBTerm(mod.varpc(), prmVarRefExpr(id(mod, n)))} \\
    \meta{assigns \defs \left(
        \begin{array}{t}
          \IF\ oe.target \in ActivityFinalNode \ \THEN\\
          \quad \set{prmAssign(mod.varpc(), constnext), prmAssign(mod.vartbterm(), true)} \\
          \ELSE \set{prmAssign(mod.varpc(), constnext)}\\
        \end{array}
    \right)} \\
      \meta{upd \defs prmUpdate(assigns)} \\
      \meta{cmd \defs prmCommand(prmAction(id(act, n)), grd, \{upd\})} \\
      \meta{mod' \defs mod.addCmdsAndIncLastUsed(\{cmd\})} \quad \\
      \meta{grd_1 \defs prmEqBoolExpr(mod.varpc(), prmVarRefExpr(id(mod, n))} \\
      \meta{rewards \defs \left(\lambda r @ prmReward(r.name, prmAction(id(act, n)), grd_1, \semL{r.value}{e})\right) \limg oe.rewards\rimg } \\
      \meta{\sigma' \defs \left(\sigma.consts \union \set{constnext}, \sigma.modmap[mod'/mod] \oplus \set{\left(oe.target, oe\right) \mapsto mod'}, \sigma.rewards \union rewards\right)} \\
  \end{array}
\end{TRule}
Basically, this function adds a constant $\meta{constnext}$ (associated with a state denoting the target $\meta{oe.target}$ of the outgoing edge $\meta{oe}$ from $\meta{n}$) and a command $\meta{cmd}$ to the corresponding module $\meta{mod}$ (to $\meta{n}$), and then transforms $\meta{oe.target}$ based on the updated module $\meta{mod'}$ and context $\meta{\sigma'}$ as shown in the body of the rule. Particularly, $\meta{constnext}$ is associated to the next available number $\meta{mod.lastused+1}$ because $\meta{mod.lastused}$ is already used. The guard $\meta{grd}$ of $\meta{cmd}$ specifies the state where the pc variable of $\meta{mod}$ is equal to a reference to a constant whose name is $\meta{id(mod,n)}$, denoting a transition from $\meta{n}$ in the AD. The only update $\meta{upd}$ of $\meta{cmd}$ has an implicit probability 1.0 (so omitted here) and a set $\meta{assigns}$ of assignments which sets the pc variable to $\meta{constnext}$ and additionally sets the module's $\meta{vartbterm}$ (that is, $\meta{\text{``to\_be\_terminated''}}$) to true if the target is a $\meta{ActivityFinalNode}$ (which is further discussed in \Cref{sec:semantics:transformation:activityfinal}). 
In the rest of the paper, we use $\meta{prmAssignsWithAFCheck}$ to construct such a set $\meta{assigns}$ with the check if the target node is a \ad{ActivityFinalNode} or not. 
The new module $\meta{mod'}$, therefore, contains $\meta{cmd}$ along with the increased $\meta{lastused}$, which is updated by $\meta{addCmdsAndIncLastUsed}$ from $\meta{mod}$. 

If $\meta{oe}$ is annotated with rewards, the set $\meta{oe.rewards}$ is not empty. Each reward $\meta{r}$ is transformed to a PRISM reward by $\meta{prmReward}$ with its name, a corresponding action, a guard $\meta{grd_1}$, and its corresponding value $\meta{\semL{r.value}{e}}$ (the transformation of an expression). So, $\meta{rewards}$, defined through relational image, is a set of the PRISM rewards.

The new context $\meta{\sigma'}$, therefore, contains the new $\meta{constnext}$, the updated module $\meta{mod'}$ through substitution, a new module mapping $\meta{(oe.target, oe) \mapsto mod'}$, and the new  $\meta{rewards}$.
}

\begin{example}[PAL use case - OpaqueAction]
  The command that corresponds to the \ad{OpaqueAction} \ad{R1ToCorA} is shown in \Cref{fig:transform_activity_pal_use_case:opaqueaction}.
  %\mylines{43}{44} in \Cref{fig:transform_activity_pal_use_case}.

\begin{figure}
\centering
\begin{lstlisting}[language=PRISM,linewidth=0.98\linewidth]
module PAL::I0::E1
  ...
  [PAL::I0::E1::R1ToCorA] ((PAL::I0::E1::pc=PAL::I0::E1::R1ToCorA))& (!(PAL::to_be_terminated)) -> ((1.0)/d_cor_a): (PAL::I0::E1::pc'=PAL::I0::E1::R1ToDoorAB);
...
endmodule      
\end{lstlisting}
\caption{\ad{OpaqueAction} part of the PRISM model for PAL use case.}
\label{fig:transform_activity_pal_use_case:opaqueaction}
\end{figure}
\qed
\end{example}

\subsubsection{ForkNode}
\label{sec:semantics:transformation:forknode}

{
A \ad{ForkNode} has multiple outgoing edges. For one of the outgoing edges, its flow is regarded as the same flow as the incoming edge, and so the same PRISM module. 

The semantics of a \ad{ForkNode} is informally discussed in Appendix~\ref{sec:appendix:semantics:forknode} and formally generated by the function $\meta{\semL{act, n, mod}{Fk}^{\sigma}}$ which is defined in Rule~\ref{ad_prism:forknode}.
\begin{TRule}[ForkNode]{\\\meta{\quad~
  \semL{act : Activity, n: ForkNode, mod: Module_{pr}}{Fk}^{\sigma}: \tau =}}
  \label{ad_prism:forknode}
\quad \meta{
  \semL{act, nmods}{Ns}^{\sigma'}
  } \\
  \meta{\bf where}\\\quad
  \begin{array}{t}
      \meta{oes \defs act.getAllOEdges(n)} \quad
      \meta{oe_1 \in oes \land \set{oe_1} \inter oe_o = \emptyset \land \set{oe_1} \union oe_o = oes } \quad\\
      \meta{constoe_1\defs prmConstant(id(mod, oe_1.target), mod.lastused + 1)} \quad\\
      %\meta{grd_1 \defs prmEqBoolExpr(mod.varpc(), prmVarRefExpr(id(mod, n))} \\
      %\meta{grd_2 \defs prmNotBoolExpr(prmVarRefExpr(id(act, \text{``to\_be\_terminated''})))}\\
      %\meta{grd \defs prmAndBoolExpr(grd_1, grd_2)} \\
      \meta{grd \defs prmBoolExprPCAndNotTBTerm(mod.varpc(), prmVarRefExpr(id(mod, n)))} \\
      %\meta{upd \defs prmUpdate(\{prmAssign(mod.varpc(), constoe_1)\})} \\
      \meta{upd \defs prmUpdate(prmAssignsWithAFCheck(mod, oe_1, constoe_1))} \\
      \meta{cmd_1 \defs prmCommand(prmAction(id(act, n)), grd, \{upd\})} \quad
      \meta{mod' \defs mod.addCmdsAndIncLastUsed(\{cmd_1\})} \quad \\
      \meta{constoe_o \defs \lambda oe @ prmConstant(id(oe.target), 0))} \quad
      \meta{constsoe_o \defs constoe_o \limg oe_o\rimg} \quad \\
      \meta{modoe_o \defs \left(\lambda oe @ prmNewModule(act, id(act, oe), id(act, n), constinac, constoe_o(oe), 0\right)}\quad 
      \meta{modsoe_o \defs modoe_o \limg oe_o\rimg} \quad \\
      \meta{modmap' \defs \left(\lambda oe @ (oe.target,oe) \mapsto modoe_o(oe)\right)\limg oe_o\rimg} \quad \\
      \meta{\sigma' \defs \left(
              \begin{array}{l}
                  \sigma.consts \union \set{constoe_1} \union constsoe_o, \\
                  \sigma.modmap[mod'/mod] \oplus \set{\left(oe_1.target, oe_1\right) \mapsto mod'} \oplus modmap', \sigma.rewards 
              \end{array}
            \right)} \\
      \meta{nmods \defs \left(\lambda oe @ oe.target \mapsto modoe_o(oe)\right) \limg oe_o\rimg} %\\
%      \meta{(mods, \sigma_r) \defs \semL{act, nmods}{Ns}^{\sigma'}}
  \end{array}
\end{TRule}
In general, the function picks an arbitrary $\meta{oe_1}$ from the outgoing edges $\meta{oes}$ of $\meta{n}$ as the edge which has the same flow as that of the incoming edge. So, the module for $\meta{oe_1.target}$ is also $\meta{mod}$. Similar to Rule~\ref{ad_prism:opaqueaction}, a new constant $\meta{constoe_1}$ is added and a new command $\meta{cmd_1}$ is added to $\meta{mod}$ to get an update module $\meta{mod'}$, because of $\meta{oe_1}$.

The other outgoing edges are recorded in $\meta{oe_o}$. Each $\meta{oe}$ of them will create a new concurrent flow, and so a new PRISM module. The first state for the pc variable in such a new module corresponds to the target of $\meta{oe}$, associated with a constant $\meta{constoe_o(oe)}$ whose value is 0. The $\meta{constsoe_o}$, therefore, denotes a set of such constants for $\meta{oe_o}$. Similarly, $\meta{modsoe_o}$ is a set of such modules, created by $\meta{prmNewModule}$. In particular, the initial state for the pc variable in such a module is $\meta{constinac}$, and so the module is inactive initially.

The new context $\meta{\sigma'}$, therefore, contains these new constants, an update module $\meta{mod'}$, and a set of new mappings, including the one $\meta{{\left(oe_1.target, oe_1\right) \mapsto mod'}}$ for $\meta{oe_1}$ and others $\meta{modmap'}$ for $\meta{oe_o}$. 

The result of $\meta{\semL{act, n, mod}{Fk}^{\sigma}}$ is $\meta{\semL{act, nmods}{Ns}^{\sigma'}}$, recursively visiting all the target nodes from $\meta{n}$ along with their corresponding modules (recorded in $\meta{nmods}$) under the new context. 
}

\begin{example}[PAL use case - ForkNode]
%  We show part of the PRISM model that corresponds to the \ad{ForkNode} \ad{F1} in \Cref{fig:transform_activity_pal_use_case:forknode}. 
  The commands that correspond to the \ad{ForkNode} \ad{F1} are shown in \Cref{fig:transform_activity_pal_use_case:forknode}.
\begin{figure}
\centering
\begin{lstlisting}[language=PRISM,linewidth=0.98\linewidth]
module PAL::I0::E1
  ...
  [PAL::F1] ((PAL::I0::E1::pc=PAL::I0::E1::F1))& (!(PAL::to_be_terminated)) -> default_rate: (PAL::I0::E1::pc'=PAL::I0::E1::R1ToCorA);
...
endmodule

module PAL::F1::E25
  ...
  [PAL::F1] ((PAL::F1::E25::pc=INACTIVE))& (!(PAL::to_be_terminated)) -> default_rate: (PAL::F1::E25::pc'=PAL::F1::E25::R2ToCorC);
  ...
endmodule      
\end{lstlisting}
\caption{\ad{ForkNode} part of the PRISM model for PAL use case.}
\label{fig:transform_activity_pal_use_case:forknode}
\end{figure}    
\qed
\end{example}

\subsubsection{DecisionNode}
\label{sec:semantics:transformation:decisionnode}

{
Different from the \ad{ForkNode} and \ad{JoinNode}, all outgoing edges of a \ad{DecisionNode} are in the same concurrent flow, and so in the same module. 

The semantics of a \ad{DecisionNode} is informally discussed in Appendix~\ref{sec:appendix:semantics:decisionnode} and formally generated by the function $\meta{\semL{act, n, mod}{Ds}^{\sigma}}$ which is defined in Rule~\ref{ad_prism:decisionnode}.
\begin{TRule}[DecisionNode]{\\\meta{\quad~
  \semL{act : Activity, n: DecisionNode, mod: Module_{pr}}{Ds}^{\sigma}: \tau =}}
  \label{ad_prism:decisionnode}
  \quad \meta{\semL{act, nmods}{Ns}^{\sigma'}}\\
  \meta{\bf where}\\\quad
  \begin{array}{t}
      \meta{oes \defs act.getAllOEdges(n)} \quad
      \meta{consts \defs prmConstantSets(mod, oes, mod.lastused)} \quad\\ % a set of constants of which each corresponds to a target node of a oe. 
      % \meta{consts \defs \left(\lambda oe @ prmConstant(id(mod, oe.target), lastused + 1)\right)\limg oes\rimg } \quad\\
      \meta{grd \defs prmBoolExprPCAndNotTBTerm(mod.varpc(), prmVarRefExpr(id(mod, n)))} \\
      %\meta{upds \defs \left(\lambda oe @ prmUpdate(oe.prob, \set{prmAssign(mod.varpc(), prmVarRefExpr(id(mod,oe.target)))})\right)\limg oes\rimg } \\
      \meta{assigns \defs \left(\lambda oe @ prmAssignsWithAFCheck(mod, oe, prmVarRefExpr(id(mod,oe.target)))\right)} \\
      \meta{upds \defs \left(\lambda oe @ prmUpdate(\semL{oe.prob}{e}, assigns(oe))\right)\limg oes\rimg } \\
      \meta{cmd \defs prmCommand(prmAction(id(mod, n)), grd, upds)} \quad \\
      \meta{mod' \defs \left(mod.addCmds(\{cmd\})\right).incLastUsed(\#~oes)} \quad \\
      \meta{modmap' \defs \left(\lambda oe @ (oe.target,oe) \mapsto mod'\right)\limg oes\rimg} \quad \\
      \meta{\sigma' \defs \left(\sigma.consts \union consts, \sigma.modmap[mod'/mod] \oplus modmap', \sigma.rewards\right)} \\
      \meta{nmods \defs \left(\lambda oe @ oe.target \mapsto mod'\right) \limg oes\rimg} %\\
      %\meta{(mods, \sigma_r) \defs \semL{act, nmods}{Ns}^{\sigma'}}
  \end{array}
\end{TRule}
The outgoing edges from $\meta{n}$ are denoted as $\meta{oes}$. Each edge has a corresponding constant for its target node. So, there are $\meta{\#~oes}$ constants, created by $\meta{prmConstantSets}$, starting from $\meta{mod.lastused}$. The new module $\meta{mod'}$, therefore, has its $\meta{lastused}$ increased by $\meta{\#~oes}$, in addition to a command $\meta{cmd}$. This command has $\meta{\#~oes}$ updates ($\meta{upds}$). Each is associated with a probability $\meta{\semL{oe.prob}{e}}$ annotated on the corresponding edge $\meta{oe}$. Apart from these described, the updated context and the (target node, module) mappings are usual.
}

\begin{example}[PAL use case - DecisionNode]
  The command that correspond to the \ad{DecisionNode} \ad{D1} is shown in \Cref{fig:transform_activity_pal_use_case:decisionnode}.
\begin{figure}
\centering
\begin{lstlisting}[language=PRISM,linewidth=0.98\linewidth]
module PAL::I0::E1
  ...
  [PAL::I0::E1::D1] ((PAL::I0::E1::pc=PAL::I0::E1::D1))& (!(PAL::to_be_terminated)) -> (default_rate*p_c_a): (PAL::I0::E1::pc'=PAL::I0::E1::M1) + (default_rate*((1.0)-p_c_a)): (PAL::I0::E1::pc'=PAL::I0::E1::M3);
...
endmodule      
\end{lstlisting}
\caption{\ad{DecisionNode} part of the PRISM model for PAL use case.}
\label{fig:transform_activity_pal_use_case:decisionnode}
\end{figure}    
\qed
\end{example}

\subsubsection{MergeNode}
\label{sec:semantics:transformation:mergenode}
A \ad{MergeNode} combines multiple incoming alternate flows into one outgoing flow, but without synchronisation like the \ad{JoinNode}. We require the incoming alternate flows are mutually exclusive, and so they are not active at the same time, by \ref{wfc:mergenode:exclusive}.

{
The semantics of a \ad{MergeNode} is informally discussed in Appendix~\ref{sec:appendix:semantics:mergenode} and formally generated by the function $\meta{\semL{act, n, mod}{Mg}^{\sigma}}$ which is defined in Rule~\ref{ad_prism:mergenode:1} if all the incoming flows are in the same concurrent flow (then the semantics of the \ad{MergeNode} is just a transition to the target of its outgoing edge) or in Rule~\ref{ad_prism:mergenode:2} otherwise. 

{%In our approach, the treatment of a \ad{MergeNode} is determined during the transformation process rather than being determined solely by the syntactic structure of the \ad{MergeNode}.
%When processing a \ad{MergeNode}, the algorithm checks whether its incoming edges have been mapped to the same PRISM module or to different modules representing concurrent flows.
%This information is readily available from the maintained module mapping $\sigma.modmap$.
  To check if all incoming edges belong to the same module $\meta{mod}$, we specify the number of modules corresponding to all incoming edges of $\meta{n}$ using $\meta{\#~\left(\ran~\left(nes~\dres~\sigma.modmap\right)\right)}$ where the module map $\meta{\sigma.modmap}$ is domain-restricted $\meta{\dres}$ to $\meta{nes}$ (all that pairs from $\meta{n}$ to its incoming edges $\meta{es}$). 
If all incoming edges belong to the same module, the merge can be handled locally within that module. So Rule~\ref{ad_prism:mergenode:1} just adds a constant, a command, and updates the context. 
Otherwise, a new module is introduced for the outgoing flow, and synchronisation is used to transfer control from the active incoming module to this new module, as defined in Rule~\ref{ad_prism:mergenode:2}. 
The requirement that incoming flows are mutually exclusive ensures that no deadlock is introduced by our approach. 
}

\begin{TRule}[MergeNode: all incoming edges from the same concurrent flow]{\\\meta{\quad~
  \semL{act : Activity, n: MergeNode, mod: Module_{pr}}{Mg}^{\sigma}: \tau =}}
  \label{ad_prism:mergenode:1}
  \quad \meta{\semL{act, oe.target, mod'}{N}^{\sigma'}}\\
  \meta{\bf where}\\\quad
  \begin{array}{t}
      \meta{ies \defs act.getAllIEdges(n)} \quad
      \meta{nes \defs \left(\lambda e @ (n, e)\right)\limg ies \rimg} \quad
      \meta{\#~\left(\ran~\left(nes~\dres~\sigma.modmap\right)\right) = 1} \quad\\
      \meta{oe \defs getOEdge(act, n)} \quad
      \meta{constnext \defs prmConstant(id(mod, oe.target), mod.lastused + 1)} \quad\\
      \meta{grd \defs prmBoolExprPCAndNotTBTerm(mod.varpc(), prmVarRefExpr(id(mod, n)))} \\
      %\meta{grd_1 \defs prmEqBoolExpr(mod.varpc(), prmVarRefExpr(id(mod, n))} \\
      %\meta{grd_2 \defs prmNotBoolExpr(prmVarRefExpr(id(act, \text{``to\_be\_terminated''})))}\\
      %\meta{grd \defs prmAndBoolExpr(grd_1, grd_2)} \\
      %\meta{upd \defs prmUpdate(\{prmAssign(mod.varpc(), constnext)\})} \\
      \meta{upd \defs prmUpdate(prmAssignsWithAFCheck(mod, oe, constnext))} \\
      \meta{cmd \defs prmCommand(prmAction(id(act, n)), grd, upd)} \quad \\
      \meta{mod' \defs mod.addCmdsAndIncLastUsed(\{cmd\})} \quad\\
      \meta{\sigma' \defs \left(\sigma.consts \union \set{constnext}, \sigma.modmap[mod'/mod] \oplus \set{(oe.target, oe) \mapsto mod'}, \sigma.rewards\right)} %\\
      %\meta{(nmods, \sigma_r) \defs \semL{act, oe.target, mod', lastused+1}{N}^{\sigma'}}
  \end{array}
\end{TRule}
\begin{TRule}[MergeNode: not all incoming edges from the same concurrent flow]{\\\meta{\quad~
  \semL{act : Activity, n: MergeNode, mod: Module_{pr}}{Mg}^{\sigma}: \power~Constant \cross \tau =}}
  \label{ad_prism:mergenode:2}
  \quad \meta{\semL{act, oe.target, omod'}{N}^{\sigma'}}\\
  \meta{\bf where}\\\quad
  \begin{array}{t}
      \meta{ies \defs act.getAllIEdges(n)} \quad
      \meta{nes \defs \left(\lambda ie @ (n, ie)\right)\limg ies \rimg} \quad
      \meta{\#~\left(\ran~\left(nes~\dres~\sigma.modmap\right)\right) > 1} \quad\\
      \meta{grd_1 \defs \left(\lambda mod @ prmEqBoolExpr(mod.varpc(), prmVarRefExpr(id(mod, n))\right)} \\
      \meta{grd_2 \defs prmNotBoolExpr(prmVarRefExpr(id(act, \text{``to\_be\_terminated''})))}\\
      \meta{grd \defs \left(\lambda m @ prmAndBoolExpr(grd_1(m), grd_2)\right)} \\
      %\meta{grd \defs prmBoolExprPCAndNotTBTerm(mod.varpc(), prmVarRefExpr(id(mod, n)))} \\
      \meta{upd \defs \left(\lambda m @ prmUpdate(\set{prmAssign(m.varpc(), constinac)})\right)} \\
      \meta{an \defs id(act, n))} \quad
      \meta{mod_e \defs \left(\lambda ie @ \sigma.modmap((n,ie))\right)} \quad\\
      \meta{cmd \defs \left(\lambda ie @ prmCommand(prmAction(id(an, ie.source)), grd(mod_e(ie)), \set{upd(mod_e(ie))})\right)} \quad \\
      \meta{imods \defs \left(\lambda ie @ (mod_e(ie), mod_e(ie).addCmds(cmd(ie)))\right) \limg ies \rimg} \\
      % \\
      %\meta{mn \defs id(act, n)} \quad 
      \meta{oe \defs getOEdge(act, n)} \quad
      \meta{constnext \defs prmConstant(id(an, oe.target), 0)} \quad\\
      %\meta{varpc \defs prmVarDeclRangeInit(id(an, \text{``pc''}), -1, numnodes, init)} \\
      %\meta{grd_1 \defs prmEqBoolExpr(varpc, prmVarRefExpr(constinit))} \\
      %\meta{grd_2 \defs prmNotBoolExpr(prmVarRefExpr(id(act, \text{``to\_be\_terminated''})))}\\
      %\meta{grd \defs prmAndBoolExpr(grd_1, grd_2)} \\
      %\meta{omod \defs prmModDef(an, \set{varpc}, cmds)} \quad \\
      \meta{omod \defs \left(prmNewModule(act, id(act,oe), id(act, n), constinac, constnext, 0\right)}\quad \\
      \meta{grd' \defs prmBoolExprPCAndNotTBTerm(omod.varpc(), constinac)} \\
      \meta{assigns \defs prmAssignsWithAFCheck(omod, oe, constnext)}\\
      \meta{cmds \defs \left(\lambda ie @ prmCommand(prmAction(id(an, ie.source)), grd', \set{prmUpdate(assigns)})\right)\limg ies\rimg} \quad \\
      \meta{omod' \defs omod.addCmds(cmds)} \quad \\
      % replaced every ie.mod() by its corresponding new modules.
      \meta{\sigma' \defs \left(\sigma.consts \union \set{constnext}, \sigma.modmap[imods] \oplus \set{(oe.target, oe) \mapsto omod'}, \sigma.rewards\right)} %\\
  \end{array}
\end{TRule}
In Rule~\ref{ad_prism:mergenode:2}, for each incoming edge $\meta{ie}$, a command $\meta{cmd(ie)}$ will be added to its corresponding module $\meta{mod_e(ie)}$ to make the module inactive ($\meta{constinac}$). The $\meta{imods}$ is a set of mappings from the old modules to the updated modules. For the outgoing edge of $\meta{n}$, a new module $\meta{omod}$ is created and is initially inactive. And a set $\meta{cmds}$ of commands is added to $\meta{omod}$ where each command synchronises with the corresponding command $\meta{cmd(ie)}$ in the incoming module $\meta{mod_e(ie)}$ to inactivate the incoming module and activate the outgoing module. We note that $\meta{\sigma.modmap[imods]}$ denotes the substitution of the old modules in $\meta{\sigma}$ for the updated modules as given in $\meta{imods}$.
}

\begin{example}[PAL use case - MergeNode]
  The commands that correspond to the \ad{DecisionNode} \ad{D3} and the \ad{MergeNode} \ad{M2} are shown % {below.} %\mylines{48}{50} and \mylines{51}{52} 
  in \Cref{fig:transform_activity_pal_use_case:mergenode}.
%  We show part of the PRISM model that corresponds to the \ad{MergeNode} \ad{M2} in \Cref{fig:transform_activity_pal_use_case:mergenode}.
%%
\begin{figure}
\centering
\begin{lstlisting}[language=PRISM,linewidth=0.98\linewidth]
module PAL::I0::E1
  ...
  [PAL::I0::E1::D3] ((PAL::I0::E1::pc=PAL::I0::E1::D3))& (!(PAL::to_be_terminated)) -> (default_rate*p_d_1): (PAL::I0::E1::pc'=PAL::I0::E1::M2) + (default_rate*((1.0)-p_d_1)): (PAL::I0::E1::pc'=PAL::I0::E1::M1);
  [PAL::I0::E1::D4] ((PAL::I0::E1::pc=PAL::I0::E1::D4))& (!(PAL::to_be_terminated)) -> (default_rate*p_d_2): (PAL::I0::E1::pc'=PAL::I0::E1::M2) + (default_rate*((1.0)-p_d_2)): (PAL::I0::E1::pc'=PAL::I0::E1::M1);
  [PAL::I0::E1::M2] ((PAL::I0::E1::pc=PAL::I0::E1::M2))& (!(PAL::to_be_terminated)) -> default_rate: (PAL::I0::E1::pc'=PAL::I0::E1::R1ToRoomD);
...
endmodule      
\end{lstlisting}
\caption{\ad{MergeNode} part of the PRISM model for PAL use case.}
\label{fig:transform_activity_pal_use_case:mergenode}
\end{figure}
\qed
\end{example}

\subsubsection{JoinNode}
\label{sec:semantics:transformation:joinnode}
Different from the \ad{ForkNode} which creates new concurrent flows (modules), a \ad{JoinNode} destroys the concurrent flows on its $n$ incoming edges, or makes the corresponding modules inactive. 
{
The semantics of a \ad{JoinNode} is informally discussed in Appendix~\ref{sec:appendix:semantics:joinnode} and formally generated by the function $\meta{\semL{act, n, mod}{Jn}^{\sigma}}$ which is defined in Rule~\ref{ad_prism:joinnode}. 
\begin{TRule}[JoinNode]{\\\meta{\quad~
  \semL{act : Activity, n: JoinNode, mod: Module_{pr}}{Jn}^{\sigma}: \tau =}}
  \label{ad_prism:joinnode}
  %\quad \meta{\left(\set{constnext}, \set{imod_1} \union imods_o \right)} \\
  \quad \meta{\semL{act, oe.target, mod_1'}{N}^{\sigma'}}\\
  \meta{\bf where}\\\quad
  \begin{array}{t}
      \meta{ies \defs act.getAllIEdges(n)} \quad
      \meta{oe \defs act.getOEdge(n)} \quad
      \meta{ie_1 \in ies \land ie_1 \inter ie_o = \emptyset \land ie_1 \union ie_o = ies } \quad\\
      \meta{an \defs id(act, n)} \quad
      \meta{mod_1 \defs mod_e(ie_1)} \quad
      \meta{constnext \defs prmConstant(id(an, oe.target), mod_1.lastused + 1)} \quad\\
      %\meta{grd_1 \defs \left(\lambda mod @ prmEqBoolExpr(mod.varpc(), prmVarRefExpr(id(mod, n))\right)} \\
      %\meta{grd_2 \defs prmNotBoolExpr(prmVarRefExpr(id(act, \text{``to\_be\_terminated''})))}\\
      %\meta{grd \defs \left(\lambda mod @ prmAndBoolExpr(grd_1(mod), grd_2)\right)} \\
      \meta{grd \defs \left(\lambda m @ prmBoolExprPCAndNotTBTerm(m.varpc(), prmVarRefExpr(id(m, n)))\right)} \\
      %\meta{upd_1 \defs \left(prmUpdate(\set{prmAssign(mod_1.varpc(), constnext)})\right)} \\
      \meta{upd_1 \defs prmUpdate(prmAssignsWithAFCheck(mod_1, oe, constnext))} \\
      \meta{cmd_1 \defs prmCommand(prmAction(an), grd(mod_1), \set{upd})} \quad \\
      \meta{mod_1' \defs mod_1.addCmdsAndIncLastUsed(cmd_1)} \\
      % \\
      % For upd_o, it is not necessary to check for AF because their next node is this JoinNode
      \meta{upd_o \defs \left(\lambda m @ prmUpdate(\set{prmAssign(m.varpc(), \sigma.constinac())})\right)} \\
      %\meta{cmd_o \defs \left(\lambda ie @ \left(\LET\ m=\sigma.modmap((n, ie))@ prmCommand(prmAction(an), grd(m), \set{upd_o(m)})\right)\right)} \quad \\
      \meta{cmd \defs \left(\lambda ie @ prmCommand(prmAction(an), grd(mod_e(ie)), \set{upd_o(mod_e(ie))})\right)} \quad \\
      \meta{imods_o \defs \left(\lambda ie @ (mod_e(ie), mod_e(ie).addCmds(\set{cmd(ie)}))\right) \limg ie_o \rimg} \\
      % replaced every ie.mod() by its corresponding new modules.
      \meta{\sigma' \defs \left(\sigma.consts \union \set{constnext}, \left(\sigma.modmap[mod_1'/mod_1]\right)[imods_o] \oplus \set{(oe.target, oe) \mapsto mod_1'}, \sigma.rewards\right)} %\\
  \end{array}
\end{TRule}
The outgoing edge $\meta{oe}$ of $\meta{n}$ will have the same flow as an arbitrary incoming edge $\meta{ie_1}$, and so they have the same module $\meta{mod_1}$. A command $\meta{cmd_1}$ (associated with $\meta{oe}$) is added to it and the updated module is  $\meta{mod_1'}$. 

In addition, for each other incoming edge $\meta{ie}$ from $\meta{ie_o}$, a command $\meta{cmd(ie)}$ is added to its corresponding module $\meta{mod_e(ie)}$ to inactivate the module.
Finally, the context is updated and the transformation of the target node of $\meta{oe}$ under the updated context continues.
}

\begin{example}[PAL use case - JoinNode]
  The commands that corresponds to the \ad{JoinNode} \ad{J1} are shown in \Cref{fig:transform_activity_pal_use_case:joinnode}.
\begin{figure}
\centering
\begin{lstlisting}[language=PRISM,linewidth=0.98\linewidth]
module PAL::I0::E1
  ...
  [PAL::J1] ((PAL::I0::E1::pc=PAL::I0::E1::J1))&(! (PAL::to_be_terminated))(*@\CodeNote{JoinNode (S5.2.8)}@*)
    -> default_rate:(PAL::I0::E1::pc'=INACTIVE);
...
endmodule

module PAL::F1::E25
  ...
  [PAL::J1] ((PAL::F1::E25::pc=PAL::F1::E25::J1))&(! (PAL::to_be_terminated))(*@\CodeNote{JoinNode (S5.2.8)}@*)
    -> default_rate:(PAL::F1::E25::pc'=PAL::F1::E25::AF)&(PAL::F1::E25::to_be_terminated'=true);/*
  ...
endmodule      
\end{lstlisting}
\caption{\ad{JoinNode} part of the PRISM model for PAL use case.}
\label{fig:transform_activity_pal_use_case:joinnode}
\end{figure}

\qed
\end{example}

\subsubsection{ActivityFinalNode}
\label{sec:semantics:transformation:activityfinal}
An activity can have more than one \ad{ActivityFinalNode}. If any of these nodes is reached, then the activity (including all its concurrent flows) will be terminated.

{
  In PRISM, we implement its semantics using three types of variables: a model-wide $\meta{\text{``to\_be\_terminated''}}$ formula, a module-wide local variable $\meta{\text{``to\_be\_terminated''}}$ for each module whose corresponding flow contains at least one $\meta{ActivityFinalNode}$ node,  a local variable $\meta{\text{``terminated''}}$ in the main module. 
The semantics of an \ad{ActivityFinalNode} is informally discussed in Appendix~\ref{sec:appendix:semantics:activityfinal} and formally generated by the function $\meta{\semL{act, n, mod}{Af}^{\sigma}}$ (defined in Rule~\ref{ad_prism:activityfinal:main}) if $\meta{mod}$ is the main module and in Rule~\ref{ad_prism:activityfinal:not_main} otherwise.

\begin{TRule}[ActivityFinalNode - if in the main module]{\\\meta{\quad~
  \semL{act : Activity, n: ActivityFinalNode, mod: Module_{pr}}{Af}^{\sigma}: \tau =}}
  \label{ad_prism:activityfinal:main}
  \quad \meta{\left(\sigma.consts, \sigma.modmap[mod'/mod], \sigma.rewards\right)} \\
  \meta{\bf where}\\\quad
  \begin{array}{t}
    \meta{ mod.isMain() \land mod.varterm() = null \land mod.vartbterm() = null} \quad\\
    % add a termiated variable. 
      \meta{at \defs id(act, \text{``terminated''})} \quad 
      \meta{varat \defs prmBoolVarDecl(at, false)} \quad \\
      \meta{modtbt \defs id(mod, \text{``to\_be\_terminated''})} \quad
      \meta{vartbt \defs prmBoolVarDecl(modtbt, false)} \quad \\
      \meta{atbt \defs id(act, \text{``to\_be\_terminated''})} \quad
      \meta{actlbl \defs prmAction(id(act, \text{``terminate''}))} \quad\\
      \meta{grd_1 \defs prmVarRefExpr(atbt)}\quad 
      \meta{grd_2 \defs prmNotBoolExpr(prmVarRefExpr(at))}\\
      \meta{grd \defs prmAndBoolExpr(grd_1, grd_2)} \\
      \meta{upd \defs prmUpdate(\set{prmAssign(mod.varpc(), constinac), prmAssign(varat, true), prmAssign(vartbt, false)})} \\
      \meta{cmd \defs prmCommand(actlbl, grd, \set{upd})} \quad \\
      \meta{mod' \defs \left(mod().addVars(\set{varat, vartbt})\right).addCmds(\set{cmd})} \\
      %\meta{\sigma' \defs \left(\sigma.consts, \sigma.modmap[mod'/mod]\right)} %\\
      % \\
      % the module contains the node n
      %\meta{mod_2 \defs getMod(\sigma, n, n.getIEdge())} \\
      %\meta{mtbt \defs id(mod_2.name, \text{``to\_be\_terminated''})} \quad
      %\meta{constnext \defs prmConstant(id(mod_2.name, n), lastused + 1)} \quad\\
      %\meta{upd_2 \defs prmUpdate(\set{prmAssign(mod_2.varpc(), constnext), prmAssign(mod_2.varterm(), true)})} \\
      %\meta{grd_2 \defs prmVarRefExpr(atbt)}\\
      %\meta{upd_2' \defs prmUpdate(\set{prmAssign(mod_2.varpc(), constinac), prmAssign(mod_2.varterm(), false)})} \\
      %\meta{cmd_2 \defs prmCommand(actlbl, grd_2, \set{upd_2'})} \quad \\
      %\meta{mod_2' \defs mod_2().addCmds(cmd_2)}
      %% all other modules that do not contain a ActivityFinalNode 
      %\meta{mods_o \defs getOtherMods(\sigma, n, n.getIEdge())} \quad \\
      %\meta{mods_o' \defs \left(\lambda m @ m.addCmds(prmCommand(actlbl, grd_1, prmUpdate(\set{prmAssign(m.varpc(), constinac)})))\right)\limg mods_o\rimg} \quad
  \end{array}
\end{TRule}
If $\meta{mod}$ is the main module (that is, $\meta{mod.isMain()}$), two local variables $\meta{varat}$ and $\meta{vartbt}$  are added if they do not exist in the module (that is, $\meta{mod.varterm() = null \land mod.vartbterm() = null}$). Additionally, one command $\meta{cmd}$ is also added to inactivate this module (by assigning a value to its pc variable $\meta{varpc}$) and mark this module is terminated by setting $\meta{varat}$ to true, and no longer to be terminated by setting $\meta{vartbt}$ to false (see $\meta{upd}$), if this module is to be terminated ($\meta{grd_1}$) but not yet terminated ($\meta{grd_2}$). The body of the rule simply returns the updated module mapping because an $\meta{ActivityFinalNode}$ has no outgoing edge.

\begin{TRule}[ActivityFinalNode - if not in the main module]{\\\meta{\quad~
  \semL{act : Activity, n: ActivityFinalNode, mod: Module_{pr}}{Af}^{\sigma}: \tau =}}
  \label{ad_prism:activityfinal:not_main}
  \quad \meta{\left(\sigma.consts, \sigma.modmap[mod'/mod], \sigma.rewards\right)} \\
  \meta{\bf where}\\\quad
  \begin{array}{t}
    \meta{ \lnot mod.isMain() \land mod.vartbterm() = null} \quad\\
    % add a termiated variable. 
      \meta{modtbt \defs id(mod, \text{``to\_be\_terminated''})} \quad
      \meta{vartbt \defs prmBoolVarDecl(modtbt, false)} \quad \\
      \meta{atbt \defs id(act, \text{``to\_be\_terminated''})} \quad
      \meta{actlbl \defs prmAction(id(act, \text{``terminate''}))} \quad\\
      \meta{grd \defs prmVarRefExpr(atbt)}\quad 
      \meta{upd \defs prmUpdate(\set{prmAssign(mod.varpc(), constinac), prmAssign(vartbt, false)})} \\
      \meta{cmd \defs prmCommand(actlbl, grd, \set{upd})} \quad 
      \meta{mod' \defs \left(mod().addVars(\set{vartbt})\right).addCmds(\set{cmd})} \\
      %\meta{\sigma' \defs \left(\sigma.consts, \sigma.modmap[mod'/mod]\right)} %\\
      % \\
      % the module contains the node n
      %\meta{mod_2 \defs getMod(\sigma, n, n.getIEdge())} \\
      %\meta{mtbt \defs id(mod_2.name, \text{``to\_be\_terminated''})} \quad
      %\meta{constnext \defs prmConstant(id(mod_2.name, n), lastused + 1)} \quad\\
      %\meta{upd_2 \defs prmUpdate(\set{prmAssign(mod_2.varpc(), constnext), prmAssign(mod_2.varterm(), true)})} \\
      %\meta{grd_2 \defs prmVarRefExpr(atbt)}\\
      %\meta{upd_2' \defs prmUpdate(\set{prmAssign(mod_2.varpc(), constinac), prmAssign(mod_2.varterm(), false)})} \\
      %\meta{cmd_2 \defs prmCommand(actlbl, grd_2, \set{upd_2'})} \quad \\
      %\meta{mod_2' \defs mod_2().addCmds(cmd_2)}
      %% all other modules that do not contain a ActivityFinal 
      %\meta{mods_o \defs getOtherMods(\sigma, n, n.getIEdge())} \quad \\
      %\meta{mods_o' \defs \left(\lambda m @ m.addCmds(prmCommand(actlbl, grd_1, prmUpdate(\set{prmAssign(m.varpc(), constinac)})))\right)\limg mods_o\rimg} \quad
  \end{array}
\end{TRule}

If $\meta{mod}$ is not the main module (see Rule~\ref{ad_prism:activityfinal:not_main}), one local variables $\meta{vartbt}$ is added if it does not exist in the module (that is, $\meta{mod.vartbterm() = null}$). Similarly, a constant and a command are added. 

We note that there are similar commands in other modules to terminate them if their corresponding flows do not have an $\meta{ActivityFinalNode}$ (because such a node in other flows can also terminate the whole activity) but these modules have no local $\meta{vartbt}$ variable (so the command will not update it to true too). We omit the definition of such rules for simplicity. The new commands from these rules have the same action label $\meta{actlbl}$ and their guards do not specify a particular pc state. So, these commands in different modules will synchronise to terminate all modules simultaneously after any one of $\meta{ActivityFinalNode}$ nodes is reached.

We describe the semantics below using the PAL use case.
}

\begin{example}[PAL use case - ActivityFinalNode]
The termination part of the use case, shown in \Cref{fig:transform_activity_pal_use_case:activityfinal}
 includes a formula \lstinprism{PAL::to_be_terminated} which is a disjunction of two Boolean \lstinprism{to_be_terminated} variables from both modules, a Boolean \lstinprism{PAL::terminated} variable, and two synchronisation commands to terminate both modules (that is, it makes both modules inactive) at the same time. 
  
\begin{figure}
\centering
\begin{lstlisting}[language=PRISM,linewidth=0.98\linewidth]
formula PAL_to_be_terminated = (PAL_I0_E1_to_be_terminated) | (PAL_F1_E25_to_be_terminated);

module PAL::I0::E1
  PAL::I0::E1::pc : [-1..19] init 0;
  PAL::I0::E1::to_be_terminated : bool init false;
  PAL::terminated : bool init false;
  ...
  [PAL::terminate] (PAL::to_be_terminated)& (!(PAL::terminated)) -> default_rate: (PAL::I0::E1::pc'=INACTIVE)& (PAL::terminated'=true)&(PAL::I0::E1::to_be_terminated'=false);
...
endmodule

module PAL::F1::E25
  PAL::F1::E25::pc : [-1..7] init INACTIVE;
  PAL::F1::E25::to_be_terminated : bool init false;
  ...

  [PAL::terminate] PAL::to_be_terminated -> default_rate: (PAL::F1::E25::pc'=INACTIVE)& (PAL::F1::E25::to_be_terminated'=false);
endmodule      
\end{lstlisting}
\caption{\ad{ActivityFinalNode} part of the PRISM model for PAL use case.}
\label{fig:transform_activity_pal_use_case:activityfinal}
\end{figure}

  The \lstinprism{to_be_terminated} variable denotes whether its owning module is to be terminated or not. It is required for a module whose corresponding concurrent flow contains one or more \ad{ActivityFinalNode}s, such as \lstinprism{PAL::F1::E25::to_be_terminated} due to the \ad{AF} node in \Cref{fig:pal_use_case}. We note that there is another \lstinprism{to_be_terminated} in the first module. This is not because of an \ad{ActivityFinalNode}, but due to the reliability annotation of \ad{Prepare}, which will be discussed in \Cref{sec:semantics:transformation:reliability}. The formula means whether there is any module to be terminated. The commands for edges that are not related to termination need to test this formula in their guard conditions to ensure the encoded transitions can only be taken if no module is to be terminated (that is, \lstinprism{!PAL::to_be_terminated}). If the \ad{AF} node is reached following the edge \ad{E33} from the node \ad{J1}, then the activity is to be terminated (that is, \lstinprism{PAL::F1::E25::to_be_terminated'=true}). In this case, the formula is true. The only commands that can be executed are two synchronising on the event \lstinprism{PAL::terminate}, updating the \lstinprism{pc} variables of both modules to \lstinprism{INACTIVE}, and setting \lstinprism{PAL::terminated} to true which denotes the activity has been terminated.
%  The termination part of the PRISM model (corresponding to the \ad{ActivityFinal} \ad{AF}) is shown in \Cref{fig:transform_activity_pal_use_case:termination}. We note that the PAL use case only has one \ad{ActivityFinal}, but there are two \lstinprism{to_be_terminated} variables in both modules. The variable \myline{3} is actually due to the reliability annotation of \ad{Prepare}, which will be discussed in \Cref{sec:semantics:transformation:reliability}. The variable \myline{14}, indeed, is due to the termination by \ad{AF}.
%
% \begin{figure}
% \centering

% \caption{Termination part of the PRISM model for PAL use case.}
% \label{fig:transform_activity_pal_use_case:termination}
% \end{figure}    
\qed
\end{example}

% \subsubsection{FlowFinalNode}
% \label{sec:semantics:transformation:flowfinal}
% The semantics of a \ad{FlowFinalNode}, shown in \Cref{fig:trans:flowfinal}, is to make the module (or the flow) inactive by setting \ad{pc} to \lstinprism{INACTIVE} \myline{6}. 

%\subsubsection{\textbf{CallBehaviourAction}}

%\subsubsection{\textbf{ObjectNode and ObjectFlow}}

%\subsubsection{\textbf{AcceptEventAction} and \textbf{SendSignalAction}}

\subsubsection{Input Parameters}
\label{sec:semantics:transformation:parameter}
An \ad{Activity} can have parameters in UML, as illustrated in the diagram for the PAL use case in \Cref{fig:pal_use_case}. We consider input parameters. They can be used in any expression in the activity, such as probability, duration, and rate annotations. There are 20 parameters in the PAL use case of which 7 are used as transition probabilities, 1 is for rewards, and 12 are as durations of actions. The expressions that use these parameters are polynomials: either directly reference to the parameters (primitive monomials with degree 1) or expressions like $1-p\_c\_b$ (a polynomial with two terms). Particularly, 6 parameters (as probability) are not instantiated in this diagram. 

In UML, \ad{ActivityParameterNode}s are required to connect the parameters of an activity to other \ad{ActivityNode}s through \ad{ObjectFlow}s. For example, an input parameter can be specified in an \ad{ActivityParameterNode}, then a \ad{ObjectFlow} is connected from the \ad{ActivityParameterNode} to the \ad{decisionInputFlow} of a \ad{DecisionNode}. By this way, the value of the parameter is made available to the guards or the probabilities of the \ad{DecisionNode}. If we need to use the parameter in several places, then several \ad{ObjectFlow}s are required by UML. This is not convenient for modelling practice. For this reason, we have simplified the reference to a parameter \ad{p} by introducing a special construct \ad{RefToPara(p)} in the expression syntax. Therefore, \ad{ActivityParameterNode}s and \ad{ObjectFlow}s are not required. 

{
  Input parameters of an activity are transformed using the function $\meta{\semL{act}{Pr}}$ in Rule~\ref{ad_prism:input_paras}. Each parameter $\meta{p}$ corresponds to one constant in PRISM.
\begin{TRule}[Input Parameters]{\\\meta{\quad~
  \semL{act : Activity}{Pr}: \power~Constant =}}
  \label{ad_prism:input_paras}
  \quad 
  \meta{\left(\lambda p @ prmConstantWithType(p.name, p.type, p.defaultValue)\right)\limg act.getAllInputParameters()\rimg} 
  
  %\meta{consts} \\
  %\meta{\bf where}\\\quad
  %\begin{array}{t}
  %  \meta{params \defs act.getAllInputParameters()} \quad
  %  \meta{consts \defs \left(\lambda p @ prmConstantWithType(p.name, p.type, p.defaultValue)\right)\limg params \rimg} \quad
  %\end{array}
\end{TRule}
}

% For each parameter, a {constant} is declared in PRISM, as show below.
% \begin{lstlisting}[language=PRISM,linewidth=0.98\linewidth]
%   const <parameter.type> {parameter.name};
%   \end{lstlisting}
% The constant is of type \lstinprism{<parameter.type>}, the corresponding PRISM type to the type \lstinprism{parameter.type} of the parameter, has the same name as the parameter name \lstinprism{\{parameter.name\}}. 
% 
% If the parameter has a default value, the constant has also a default value \lstinprism{<parameter.defaultValue>}.
% \begin{lstlisting}[language=PRISM,]
% const <{parameter.type}> {parameter.name} = <parameter.defaultValue>;
% \end{lstlisting}
% 
% Then in the expressions where the parameter is referred to by \ad{RefToPara(parameter.name)}, its corresponding expression in PRISM is just the parameter name: \lstinprism{\{parameter.name\}}.

\begin{example}[PAL use case - input parameters]
  We have shown the part of the PRISM model that corresponds to the input parameters in \Cref{fig:transform_activity_pal_use_case:input_para}.
\begin{figure}
\centering
\begin{lstlisting}[language=PRISM,linewidth=0.98\linewidth]
const double p_c_a; (*@\CodeNote{Input Parameters (S5.2.10)}@*)
const double p_d_12;
.../*
const double p_d_1;
const double p_d_2;
const double p_c_c;
const double p_c_b;*/
const double r_prep = (0.9998);
const double r1 = (1.0);
const double d_prep = (0.5);
const double d_cor_a = (0.2);
.../*
\end{lstlisting}
\caption{The input parameter part of the PRISM model for PAL use case.}
\label{fig:transform_activity_pal_use_case:input_para}
\end{figure}
\qed
\end{example}

\subsubsection{Reliability}
\label{sec:semantics:transformation:reliability}
An action can be annotated with reliability $p$, a real value between 0 and 1, to denote the probability $p$ of a successful execution of this service (represented by the action). So the failure probability of the service is $1-p$. There is an implicit \ad{DecisionNode} {with two outgoing edges: one with probability $p$ to the destination of the outgoing edge of the annotated action and one with probability $1-p$ to an \ad{ActivityFinalNode}} here. In activity diagrams, we also do not explicitly show an \ad{ActivityNode} for the failure to facilitate the modelling. \emph{Semantically, if an action fails, the activity owning this action fails.} This is similar to the \ad{ActivityFinalNode}. We refer to Appendix \ref{sec:appendix:semantics:reliability} for more details of its semantics.

\begin{example}[PAL use case - reliability]
  \label{ex:pal:reliability}
  The reliability annotation of \ad{Prepare} in the use case, shown in \Cref{fig:transform_activity_pal_use_case:reliability} 
includes a formula \lstinprism{PAL::I0::E1::to\_be\_failed}, a Boolean \lstinprism{to_be_failed} variable, and one command. The command has two updates: one with probability \lstinprism{1-r\_prep} for the failure and one with probability \lstinprism{r\_prep} for the success. We note that \lstinprism{(1/d_prep)} in the command corresponds to the rate of this node and will be discussed in \Cref{sec:semantics:transformation:duration:ctmc}. On failure, the module becomes inactive, and both \lstinprism{to\_be\_failed} and \lstinprism{to_be_terminated} are set to true. Afterwards, the activity is terminated (that is, all modules are terminated).

\begin{figure}
\centering
\begin{lstlisting}[language=PRISM,linewidth=0.98\linewidth]
formula PAL_to_be_failed = PAL_I0_E1_to_be_failed;

module PAL::I0::E1
  PAL::I0::E1::to_be_terminated : bool init false;
  PAL::I0::E1::to_be_failed : bool init false;
  ...
  [PAL::I0::E1::Prepare] ((PAL::I0::E1::pc=PAL::I0::E1::Prepare))& 
    (! (PAL::to_be_terminated)) -> (((1.0)/d_prep)*((1.0)-r_prep)):(PAL::I0::E1::pc'=INACTIVE)& 
    (PAL::I0::E1::to_be_failed'=true)& (PAL::I0::E1::to_be_terminated'=true) + 
    (((1.0)/d_prep)*r_prep):(PAL::I0::E1::pc'=PAL::I0::E1::F1);
...
endmodule
\end{lstlisting}
\caption{The reliability part of the PRISM model for PAL use case.}
\label{fig:transform_activity_pal_use_case:reliability}
\end{figure}
\qed
    %We show part of the PRISM model that corresponds to the reliability annotation \ad{r\_prep} on \ad{Prepare} \mylines{12}{19} in \Cref{fig:transform_activity_pal_use_case:reliability} where the first update corresponds to failure and the second update corresponds to success.

%\begin{figure}
%\centering
%\begin{lstlisting}[language=PRISM,linewidth=0.98\linewidth]
%module PAL::I0::E1
%  ...
%  [PAL::I0::E1::Prepare] ((PAL::I0::E1::pc=PAL::I0::E1::Prepare))& (!(PAL::to_be_terminated)) -> (((1.0)/d_prep)*((1.0)-r_prep)): (PAL::I0::E1::pc'=INACTIVE)& (PAL::I0::E1::to_be_failed'=true)& (PAL::I0::E1::to_be_terminated'=true) + (((1.0)/d_prep)*r_prep): (PAL::I0::E1::pc'=PAL::I0::E1::F1);
%...
%endmodule      
%\end{lstlisting}
%\caption{Reliability part of the PRISM model for PAL use case.}
%\label{fig:transform_activity_pal_use_case:reliability}
%\end{figure}
%
\end{example}

\subsubsection{Duration and CTMCs}
\label{sec:semantics:transformation:duration:ctmc}
An action can be annotated with a duration $d$, or a rate $r$. Both duration and rate are real numbers. We require all actions must be annotated with either a duration or a rate (see \ref{wfc:ctmc:action:duration_or_rate}) if the activity diagram is to be analysed as a CTMC model. For a DTMC or MDP model, the duration and rate annotations are simply ignored and the corresponding well-formedness condition is not enforced. For other \ad{ControlNode}s, we assume they are instantaneous. In other words, they will not take time. This corresponds to a rate of $\infty$. However, it is not possible to have a rate of $\infty$ in PRISM or other model checkers because they require bounded integer numbers. 
For a CTMC model in PRISM, every update of a command has an associated rate. If a rate is omitted in a PRISM command, then a rate $1.0$ is assumed. 
{As a result, we need to choose a default rate or duration for all these other nodes. Ideally, this rate should be much larger than the biggest rate of all actions to minimise the impact on time from these nodes. However, this will make a PRISM model more stiff (that is, a model with widely different time scales), and so the model checking time could be dramatically increased. 
  So a trade-off between the result accuracy and performance needs to be considered. {In the PAL example, the biggest rate is $1/d\_cor\_a=5$ and our default rate is $1/mean\_duration=1000$. So the difference in the rate is 200 times. We suggest a minimum difference of 100 times and then doing experiments on different default rates using both the PRISM and Storm model checkers to choose the best default rate (to balance between model checking time and accuracy).}

We introduce a parameter \lstinprism{mean\_duration} in our transformation (so we can choose different values during the transformation) as the default duration for the other nodes. It is a constant in the resultant PRISM model (0.001 \myline{3} in \Cref{fig:transform_activity_pal_use_case} is an example). We define a formula \lstinprism{default_rate} \myline{23}. This is the default rate we define for these nodes. However, one node may involve a synchronisation between multiple commands. For example, the \ad{F1} in \Cref{fig:pal_use_case} will result in a synchronisation between the two commands \mylines{41}{42} and \mylines{63}{64} in \Cref{fig:transform_activity_pal_use_case}. If both commands have a \lstinprism{default_rate}, the final rate will be their product: \lstinprism{default_rate*default_rate} according to the PRISM semantics. In the PRISM manual,\footnote{\url{https://www.prismmodelchecker.org/manual/ThePRISMLanguage/Synchronisation}} it recommends to make one command \emph{active} with the desirable rate and other commands \emph{passive} with rate 1. That is why we also define a \lstinprism{default_rate} \myline{23}. In our semantics interpretation of a \ad{ForkNode}, \ad{JoinNode}, \ad{MergeNode}, or \ad{ActivityFinalNode}, only one command in a synchronisation will have a \lstinprism{default_rate} and other commands have a \lstinprism{default_rate}. Therefore, their final rates are all the same: just \lstinprism{default_rate}.
}

We refer to \Cref{sec:appendix:semantics:duration:ctmc} for more details.

\begin{example}[PAL use case - duration and CTMC]
  Most commands have a default rate (or a passive rate) while the command for \ad{Prepare} {(see \Cref{ex:pal:reliability})} has a rate \lstinprism{1/d_prep} where \lstinprism{d_prep} is the duration annotation of the action \ad{Prepare}.
\qed
\end{example}

\subsection{Properties}
\label{sec:semantics:property}
As shown in \Cref{fig:profile}, activities can be annotated with groups of properties. Each group specifies a Markov model type (DTMC, CTMC, or MDP) and a set of properties to be verified based on the model type. While the verification for the CTMC model takes all annotations into account, that for the DTMC and MDP, however, ignores the duration or rate annotations. 

The specified properties are given in PCTL*~\cite{Aziz1995,Bianco1995,Baier1998}, a combination of the probabilistic computation tree logic (PCTL)~\cite{Hansson1994} and the linear temporal logic (LTL)~\cite{Pnueli1977}, for DTMCs and MDPs, and in CSL~\cite{Aziz1996,Baier1999}, an extension of PCTL for CTMCs. We use the PRISM Property Specification (PRPS)\footnote{\url{https://www.prismmodelchecker.org/manual/PropertySpecification}} to specify PCTL* and CSL properties because we use the PRISM and Storm model checkers for verification and both use the property specification.

%The basic elements in both PCTL* and CSL are atomic propositions (AP). AP in the Markov models for activity diagrams is specified in Eq.~(\ref{eqn:APS}). Then in \Cref{sec:semantics:transformation:activityfinal} and~\ref{sec:semantics:transformation:reliability}, we introduce a \lstinprism{terminated} variable, a \lstinprism{to_be_failed} variable for each concurrent flow having actions annotated with reliability, and a \lstinprism{to_be_terminated} variable for each concurrent flow passing through \ad{ActivityFinalNode}s or the corresponding module having the \lstinprism{to_be_failed} variable. We assume there are $r$ modules having 
%\lstinprism{to_be_failed} and $q$ modules having \lstinprism{to_be_terminated}. Then we define AP for activity diagrams in our PRISM semantics as follows, where $B=\{true, false\}$.
%
%\begin{align}
%  APS &\defs \label{eqn:aps}\\
%  & APS_a \union
%   \left\{b:B \bullet \text{\lstinprism{terminated}}=b  \right\} \union \label{eqn:aps:terminated}\\
%      & \bigcup\left\{i: 1..q \bullet \left\{b:B \bullet \text{\lstinprism{to_be_terminated}}_i=b \right\}\right\} \union \label{eqn:aps:to_be_terminated}\\
%      & \bigcup\left\{i: 1..r \bullet \left\{b:B \bullet \text{\lstinprism{to_be_failed}}_i=b \right\}\right\} \label{eqn:aps:to_be_failed}
%\end{align}

We note that some variables (including \lstinprism{pc}, \lstinprism{terminated}, \lstinprism{to_be_terminated}, and \lstinprism{to_be_failed}) in PRISM models are introduced in our semantic and not in the original activity diagrams. It is inconvenient to specify properties in terms of these variables. For this reason, we introduce several ways to specify AP of properties in activity diagrams directly, as shown in \Cref{fig:semantics:properties:ap}.

\begin{figure}
\centering
\begin{lstlisting}[language=ADProperty,linewidth=0.98\linewidth]
{an} reaches at {an}::{nn}
{an} terminated successfully
{an} terminated on fail
{an} failed
\end{lstlisting}
\caption{The ways to specify APs of properties.}
\label{fig:semantics:properties:ap}
\end{figure}

{
\begin{TRule}[Property]{\meta{
  refResolve (act:Activity, \sigma: \tau, props: \power~PRProperty) : \power~string = }}
  \label{ad_prism:property}
  \quad
  {\text{Recursively call $\meta{refAPResolve}$ to resolve references in the four AP syntaxes, other parts remain unchanged}}\\
  \meta{\bf where}\\\quad
  \begin{array}{t}
    \meta{refAPResolve(\text{an reaches at an::nn}) \defs \left(id(an, pc) = id(an, nn) \right)}\\
    \meta{refAPResolve(\text{an terminated successfully}) \defs \left(id(an, \text{``terminated''}) \& !id(an, \text{``to\_be\_failed''}) \right)} \\
    \meta{refAPResolve(\text{an terminated on fail}) \defs \left(id(an, \text{``terminated''}) \& id(an, \text{``to\_be\_failed''} \right)}\\
    \meta{refAPResolve(\text{an failed}) \defs id(an, \text{``to\_be\_failed''}) }\\
  \end{array}
\end{TRule}
Rule~\ref{ad_prism:property} defines reference resolution for the above defined four syntaxes.
}
The first syntax specifies an activity \lstinprop{reaches at} one of its nodes, and it will be transformed to \lstinprism{\{mn\}::pc = \{mn\}::\{nn\}} where \lstinprism{\{mn\}} is the corresponding module name for the concurrent flow involving the node. 
An activity \lstinprop{terminated successfully} \myline{2} specifies the activity has terminated without a failure. This is transformed to the conjunction of \lstinprism{\{an\}::terminated} and \lstinprism{!\{an\}::to_be_failed} where we use the formula to denote no failure occurs in any module. 
An activity \lstinprop{terminated on fail} \myline{3} specifies the activity has terminated with a failure. This corresponds to the conjunction of \lstinprism{\{an\}::terminated} and \lstinprism{\{an\}::to_be_failed}.
An activity \lstinprop{failed} \myline{4} specifies the activity has failed, and is transformed to \lstinprism{\{an\}::to_be_failed}.
{This is different from \lstinprop{terminated on fail} because an activity could be failed, but not yet terminated.}

{In short, we use the four syntaxes above to specify atomic propositions and use the PRPS to syntactically specify other parts of properties. The four syntaxes are automatically generated into APs in the PRPS and other parts are already in the PRPS. Therefore, a complete PRPS file will be generated.}

\begin{example}[PAL use case - properties]
    We show the corresponding PRISM property for the termination property R4 discussed in \Cref{table:reqs} as \lstinprop{P=? [F PAL_F1_E25_pc = PAL_F1_E25_AF]}.
%\begin{lstlisting}[language=ADProperty,frame=none,numbers=none]
%  P=? [F PAL_F1_E25_pc = PAL_F1_E25_AF]
%\end{lstlisting}
The complete PRISM properties can be found online.\footnote{\url{https://github.com/RandallYe/QANAD/blob/master/Examples/PAL_use_case/prism_models/p_ctmc/pal.props}}
\qed
\end{example}

\subsubsection{Rewards}
\ad{ActivityEdge}s can be annotated with rewards except edges from a \ad{DecisionNode} or a \ad{ForkNode}. Each reward has a name and a real number value. It corresponds to a transition reward in PRISM.

\begin{lstlisting}[language=PRISM,,linewidth=0.98\linewidth]
rewards 
  {reward.name} [{mn}::{edge.source.name}] ({mn}::pc={mn}::{edge.source.name}) : <reward.value>;
endrewards
\end{lstlisting}
The action label and the guard \myline{2} of the reward correspond to the source node of the edge, \lstinprism{\{edge.source.name\}}.

\begin{example}[PAL use case - rewards]
  The PRISM rewards {are shown in \Cref{fig:transform_activity_pal_use_case:rewards}}, which correspond to reward annotations on the edges \ad{E13} and \ad{E15} in \Cref{fig:pal_use_case}.
  \begin{figure}
      \centering
\begin{lstlisting}[language=PRISM,linewidth=0.98\linewidth]
rewards "reward_door12_attempt"
  [PAL_I0_E1_R1ToCorBThrDoor1] ((PAL_I0_E1_pc=PAL_I0_E1_R1ToCorBThrDoor1)) : r1;	
	[PAL_I0_E1_R1ToCorBThrDoor2] ((PAL_I0_E1_pc=PAL_I0_E1_R1ToCorBThrDoor2)) : r1;	
endrewards
\end{lstlisting}
\caption{The rewards part of the PRISM model for PAL use case.}
\label{fig:transform_activity_pal_use_case:rewards}
\end{figure}
\qed
\end{example}

{

\subsection{Soundness of the model transformation}
\label{sec:soundness}

The soundness of the model transformation is established by demonstrating that the generated PRISM model faithfully preserves the \textbf{token-based operational semantics} of the source AD as explained in~\Cref{sec:semantics:token}. This argument rests on three structural pillars that map the dynamic behaviour of tokens to the reactive modules of the PRISM language.

\begin{enumerate}
    \item \textbf{Mapping concurrent flows to PRISM modules.} In our transformation, each concurrent flow within an activity is mapped to a unique PRISM module. This structural mapping ensures that the independent execution and parallel nature of activities are maintained. While PRISM modules are statically allocated during the transformation, their dynamic activation simulates the creation and destruction of flows in the original diagram.

    \item \textbf{Token representation via program counters.} In our semantics, the presence and location of a control token within a flow are represented by a local program counter (\texttt{pc}) variable in each module.
    \begin{itemize}
        \item \textit{Token location mapping:} all the potential token locations (at any \ad{ActivityNode}) in a flow are mapped to distinguished non-negative numbers in each PRISM module, representing different nodes along this flow.
        \item \textit{Token existence:} a flow possessing a token is considered \textbf{active}, with its \texttt{pc} set to such a non-negative number.
        \item \textit{Token absence:} a flow without a token is represented by the {\texttt{INACTIVE}} state. 
        \item \textit{Token passage:} Rules~\ref{ad_prism:activity} to~\ref{nfm_prism:opaqueaction:reliability:ctmc} ensure that transitions between \texttt{pc} values (and between \texttt{INACTIVE} and active states) correspond exactly to the movement of control tokens across edges and nodes.
    \end{itemize}

    \item \textbf{Coordination and synchronisation.} The logic governing \texttt{ControlNodes} is preserved through PRISM`s synchronisation labels, which enforce collective state updates across modules.
    \begin{itemize}
      \item \textit{\ad{InitialNode}:} Rule~\ref{ad_prism:initial_nodes} creates a PRISM module (using Rule~\ref{ad_prism:newprismmodule}) for each outgoing edge from \ad{InitialNode}s and sets the initial value of the $\meta{varpc}$ to 0 (Rule~\ref{ad_prism:initialnode}). So this module is initially active. This corresponds to one token placed on each outgoing edge in the AD. Therefore, this rule preserves the \ad{InitialNode} semantics. 
      \item \textit{\ad{ForkNode}:} Rule~\ref{ad_prism:forknode} ensures that $n-1$ new PRISM modules are created and initially inactive if a \texttt{ForkNode} has $n$ outgoing edges and the flow control is simultaneously passed (from its incoming module) to $n$ modules by synchronising the incoming module with multiple outgoing modules to update their $\meta{varpc}$ to their corresponding active states. This represents $n$ tokens are duplicated and placed on the outgoing edges, which is the semantics of the \ad{ForkNode}.
      \item \textit{\ad{JoinNode}:} Rule~\ref{ad_prism:joinnode} synchronises $n$ incoming modules on the node and simultaneously inactivates $n-1$ modules by setting their $\meta{varpc}$ to \lstinprism{INACTIVE} and passes the control to the destination of its outgoing edge by updating another module's $\meta{varpc}$. This requires all participating modules to reach the \texttt{JoinNode} state before the outgoing flow can trigger. Upon synchronisation, all but one module are reset to \texttt{INACTIVE}, effectively ``merging'' the tokens. The semantics of \ad{JoinNode}, nevertheless, are preserved.
        \item \textit{Global Termination:} The global effect of an \texttt{ActivityFinalNode} is soundly maintained via Rules~\ref{ad_prism:activityfinal:main} and \ref{ad_prism:activityfinal:not_main}. By using a model-wide formula ($\meta{to\_be\_terminated}$) and the $\meta{terminate}$ action label, the transformation ensures that reaching an \texttt{ActivityFinalNode} destroys all tokens globally by resetting every module's \texttt{pc} to \texttt{INACTIVE} simultaneously.
        \item \textit{\ad{DecisionNode}:} Rule~\ref{ad_prism:decisionnode} ensures that the node corresponds to a transition in the owning module to probabilistically reach their targets with the same probabilities as those annotated in the outgoing edges from the node. And the token is passed to one of its targets, which preserves the \changed[\Cn{2}{7}]{token-based} semantics.
        \item \textit{\ad{MergeNode}:} If all incoming edges to a \ad{MergeNode} are in the same flow, Rule~\ref{ad_prism:mergenode:1} does not create a new module and just creates a new transition for each incoming edge in the owning module to pass the control to the target of the outgoing edge. Only one of such transitions can be enabled at the same time because they share the same $\meta{varpc}$ (provided they are in the same flow) but have the different source states (corresponding to the different source nodes to this \ad{MergeNode} in the AD). Otherwise, Rule~\ref{ad_prism:mergenode:2} creates a new (outgoing) module which is initially inactive. Upon the synchronisation of one of the incoming modules and the outgoing module, the incoming module becomes inactive and the outgoing module is simultaneously set to an active state. For both cases, no token is destroyed and no new token is created. So the number of tokens is maintained. This is semantically consistent with the token interpretation of the AD. 
        \item Other \ad{ActivityNode}s: Rule~\ref{ad_prism:opaqueaction} for \textit{\ad{OpaqueAction}} passes the control and Rule~\ref{nfm_prism:flowfinalnode} for \textit{\ad{FlowFinalNode}} inactivates the flow. They are simple and preserve their original \changed[\Cn{2}{7}]{token-based} semantics.
    \end{itemize}
\end{enumerate}

Through these mappings, the state transitions and probabilistic behaviors in the generated PRISM code are shown to be a reflection of the \changed[\Cn{2}{7}]{original token-based semantics} of the source activity diagrams \changed[\Cn{2}{8}]{informally and intuitively.} 

}

Next, we present the implementation of a tool to automate the transformation and facilitate verification using PRISM. 

\section{Verification and tool support}
\label{sec:ver_tool}

In this section, we first describe, in \Cref{subsec:alg}, \changed[\Cn{2}{2}]{the implementation of} the semantics transformation discussed in \Cref{sec:semantics:transformation}, then the tool support for our approach in \Cref{subsec:tool}.

\subsection{Implementation of the transformation}\label{subsec:alg}

In this section, we \changed[\Cn{2}{2}]{describe the implementation of the} transformation from an activity diagram to a PRISM model.
The transformation consists of two subsequent phases: the pre-processing of an activity diagram and the transformation from the pre-processed activity diagram to the PRISM model.

{
\begin{remark}[Differences in implementation from transformation rules] \label{rmk:rules_algorithm}
    We note that our implementation of the transformation from ADs to PRISM here is different from our presentation in the rules in \Cref{sec:semantics:transformation}. For a node in the rule, we either use its corresponding module from the parameter $\meta{mod}$ (such as Rule~\ref{ad_prism:opaqueaction}), or create new modules (such as Rule~\ref{ad_prism:forknode}), for the subsequent nodes. At the same time, this node is transformed to PRISM under its input context. However, in the \changed[\Cn{2}{2}]{implementation}, we split the transformation into two phases: the pre-processing phase to decide the node and module mappings, and the transformation phase to transform nodes, for the sake of the performance. There are mainly due to two considerations. Firstly, a rule defines a pure function (without side effects) where we use substitutions to update modules and the transformation context, and so there are many substitutions of whole modules and mappings in these rules, which is not efficient for implementation because we can use pointers or references to objects in imperative programming languages for an in-place update of the objects. Secondly, the transformation of a node is completely independent from that of another node as long as their node and module mappings are known. Therefore, in the two-phase approach, the transformation of different nodes can be conducted in parallel, which can largely improve the efficiency of the \changed[\Cn{2}{2}]{implementation}.
  For these reasons, our implementation is not one-to-one mappings from transformation rules. Instead, we use the rules as a reference to develop our \changed[\Cn{2}{2}]{implementation} algorithms.
\qed
\end{remark}
}

\subsubsection{Pre-processing}\label{subsubsec:alg-preproess}

\changed[\Cn{2}{2}]{
The pre-processing phase scans each node in an activity to determine which PRISM module (corresponding to which concurrent flow) the node belongs to. Starting from the \ad{InitialNode}s of the \ad{Activity}, a breadth-first traversal visits the rest of the nodes and builds a set of mappings \textit{modMaps} from (node, edge) pairs to PRISM modules. The traversal handles each node type as follows:

\begin{itemize}
    \item For each \ad{InitialNode}, a new PRISM module is created for each of its outgoing edges, establishing the initial module allocation (see \Cref{sec:semantics:transformation:initialnode}).

    \item An \ad{OpaqueAction} inherits the module of its predecessor, since it has exactly one incoming and one outgoing edge by well-formedness conditions~\ref{wfc:in_edge_action:one} and~\ref{wfc:out_edge_action:one} (see \Cref{sec:semantics:transformation:opaqueaction}).

    \item A \ad{ForkNode} retains the current module for the first outgoing branch and creates a new module for each subsequent branch, reflecting the creation of concurrent flows (see \Cref{sec:semantics:transformation:forknode}).

    \item A \ad{JoinNode} is processed only after the module of at least one incoming branch has been determined. Since a \ad{JoinNode} has multiple incoming edges potentially associated with different modules, the algorithm randomly selects one incoming edge and retrieves the module of its source node. If that module exists, the \ad{JoinNode} is processed; otherwise, it is deferred by re-enqueueing (see \Cref{sec:semantics:transformation:joinnode}).

    \item A \ad{DecisionNode} propagates its module to all outgoing branches (see \Cref{sec:semantics:transformation:decisionnode}).

    \item A \ad{MergeNode} requires checking whether its incoming edges originate from multiple modules and whether all source nodes of its incoming edges have been visited. These conditions determine whether a new PRISM module is needed. Temporary dummy modules are introduced when not all predecessors have been visited, and are resolved once the traversal completes (see \Cref{sec:semantics:transformation:mergenode}).

    \item \ad{FlowFinalNode} and \ad{ActivityFinalNode} are simply marked as visited.
\end{itemize}
}

We illustrate the pre-processing phase with the PAL use case below.

\begin{example}[Pre-processing of PAL use case]\label{ex:pal_preprocessing}
The application of the pre-processing to the PAL use case in \Cref{fig:pal_use_case} results in two PRISM modules. According to \textit{modMaps} shown in \Cref{fig:pal_alg1_result},
\begin{figure}
    \centering
    \begin{minipage}{0.46\textwidth}
        \centering
\begin{lstlisting}[basicstyle=\footnotesize\ttfamily,
  frame=trbl, % draw a frame at the top, right, left and bottom of the listing
  frameround=tttt, % make the frame round at all four corners
  framesep=4pt, % quarter circle size of the round corners
  numbers=left, % show line numbers at the left
%  numberstyle=\tiny\ttfamily, % style of the line numbers
  numberstyle=\scriptsize\ttfamily, % style of the line numbers
  numbersep=5pt,
  linewidth=\linewidth,
  xleftmargin=1.5em,
  breaklines=true,
  ]
{((D1, E5), PAL::I0::E1),
((D2, E8), PAL::I0::E1),
((D3, E13), PAL::I0::E1),
((D4, E15), PAL::I0::E1),
((D5, E20), PAL::I0::E1),
((F1, E2), PAL::I0::E1),
((I0, E1), PAL::I0::E1),
((J1, E24), PAL::I0::E1),
((M1, E14), PAL::I0::E1),
((M1, E16), PAL::I0::E1),
((M1, E6), PAL::I0::E1),
((M2, E17), PAL::I0::E1),
((M2, E18), PAL::I0::E1),
((M3, E21), PAL::I0::E1),
((M3, E7), PAL::I0::E1),
((M4, E10), PAL::I0::E1),
((M4, E23), PAL::I0::E1),
\end{lstlisting}
\end{minipage}\hspace{2em}%\hfill
    \begin{minipage}{0.46\textwidth}
        \centering
\begin{lstlisting}[basicstyle=\footnotesize\ttfamily,
  frame=trbl, % draw a frame at the top, right, left and bottom of the listing
  frameround=tttt, % make the frame round at all four corners
  framesep=4pt, % quarter circle size of the round corners
  numbers=left, % show line numbers at the left
%  numberstyle=\tiny\ttfamily, % style of the line numbers
  numberstyle=\scriptsize\ttfamily, % style of the line numbers
  numbersep=5pt,
  linewidth=\linewidth,
  xleftmargin=1.5em,
  breaklines=true,
  firstnumber=18]
((Prepare, E1), PAL::I0::E1),
((R1DeliveredRoomD, E22), PAL::I0::E1),
((R1Stuck, E9), PAL::I0::E1),
((R1ToCorA, E3), PAL::I0::E1),
((R1ToCorBThrDoor1, E11), PAL::I0::E1),
((R1ToCorBThrDoor2, E12), PAL::I0::E1),
((R1ToDoorAB, E4), PAL::I0::E1),
((R1ToRoomD, E19), PAL::I0::E1),
((AF, E33), PAL::F1::E25),
((D6, E27), PAL::F1::E25),
((J1, E32), PAL::F1::E25),
((M5, E29), PAL::F1::E25),
((M5, E31), PAL::F1::E25),
((R2DeliveredRoomD, E28), PAL::F1::E25),
((R2Stuck, E30), PAL::F1::E25),
((R2ToCorC, E25), PAL::F1::E25),
((R2ToDoorCD, E26), PAL::F1::E25)}
\end{lstlisting}
    \end{minipage}
    \caption{The resulting mappings for the application of pre-processing to PAL use case.}
    \label{fig:pal_alg1_result}
\end{figure}
the nodes \ad{R2ToCorC}, \ad{R2ToDoorCD}, \ad{D6}, \ad{R2DeliveredRoomD},
\ad{R2Stuck},
\ad{M5},
\ad{J1},
and \ad{AF} (\mylines{26}{34}) will be allocated to one module (\lstinprism{PAL::F1::E25}) and other nodes to another module (\lstinprism{PAL::I0::E1}). We also note that \ad{J1} is also associated with the module \lstinprism{PAL::I0::E1}.

The order of visited nodes, recorded in \textit{visited}, is
$\langle$\ad{I0},
\ad{Prepare},
\ad{F1},
\ad{R1ToCorA},
\ad{R2ToCorC},
\ad{R1ToDoorAB},
\ad{R2ToDoorCD},
\ad{D1},
\ad{D6},
\ad{R2DeliveredRoomD},
\ad{R2Stuck},
\ad{M5},
\ad{J1},
\ad{AF},
\ad{M3},
\ad{R1Stuck},
\ad{M4},
\ad{M1},
\ad{D2},
\ad{R1ToCorBThrDoor1},
\ad{R1ToCorBThrDoor2},
\ad{D3},
\ad{D4},
\ad{M2},
\ad{R1ToRoomD},
\ad{D5}, and
\ad{R1DeliveredRoomD}$\rangle$.
We note that after \ad{D1} on the first concurrent flow (FCF) is visited, the algorithm visits all nodes (\ad{D6},
\ad{R2DeliveredRoomD},
\ad{R2Stuck},
\ad{M5},
\ad{J1},
and \ad{AF}) on the second concurrent flow (SCF) first before it revisits the nodes (starting from \ad{M3}) on FCF. This is because the nodes from \ad{D1} on FCF (shown in \Cref{fig:pal_use_case}) are \ad{M1} and \ad{M3} of which both are \ad{MergeNode}. {Because only one of the incoming edges for both \ad{MergeNode}s (that is, \ad{E7} for \ad{M3} and \ad{E6} for \ad{M1}) has been visited, we cannot decide if a new module is required for either. In this case, visiting both \ad{MergeNode}s is delayed until the other flow (SCF) is visited.} After that, the set \textit{visited} contains the two \ad{MergeNode}s only. However, visiting both the nodes cannot make any progression because \ad{M3} has another incoming edge \ad{E21} to be visited but \ad{E21} is dependent on the flow from \ad{M1}, and \ad{M1} has other two incoming edges \ad{E14} and \ad{E16} to be visited but they are also dependent on the flow from \ad{M1} (due to the two loops between \ad{M1} and \ad{D3} (and \ad{D4}). In our algorithm, we will create a dummy module (DMod3) for the outgoing edge of \ad{M3} to let the process continue. So the next visited nodes after \ad{M3} are \ad{R1Stuck} and \ad{M4} (whose situation is the same as \ad{M3} and a dummy module DMod4 is created). Because \ad{J1} has been visited, the process completes the visit of this branch and continues to visit \ad{M1}.
Now the only node to be visited in \textit{tbvisited} is \ad{M1}.

A new dummy module DMod1 is created for the outgoing edge \ad{E8} of \ad{M1} to let the process continue. After both the edges \ad{E14} and \ad{E16} are visited where their source nodes \ad{D3} and \ad{D4} are associated with the same dummy module DMod1 because the nodes \ad{D2} and two \ad{OpaqueAction}s between \ad{M1} and \ad{D3} (or \ad{D4}) will not create a new module. With the help of dummy modules, the algorithm visits all the nodes in the diagram but leaves three dummy modules (DMod3, DMod4, and DMod1, created when visiting \ad{M3}, \ad{M4}, and \ad{M1}). These dummy modules are then resolved: the basic idea of the resolution is to check how many different modules are from all the incoming edges of a \ad{MergeNode}. For \ad{M1}, there are two: one (Mod1) corresponding to \ad{D1} and one dummy module (DMod1) corresponding to \ad{D3} and \ad{D4}. In this case, we simply replace the dummy module (DMod1) with the module (Mod1) for \ad{D1} in \textit{modMaps} because this module (Mod1) fulfils all pre-processing rules for these nodes.

For \ad{M3}, the module for the source node \ad{D1} of its incoming edge \ad{E7} is Mod1 and that for the source node \ad{D5} of its incoming edge \ad{E21} is also Mod1 (was DMod1 but replaced with Mod1 when resolving \ad{M1}, discussed previously). In this case, its outgoing dummy module (DMod3) is also replaced with Mod1 because there is only one module for all its incoming edges and its outgoing edge should use the same module. For \ad{M4}, it is a similar case. Finally, all three dummy modules are replaced with Mod1.
\qed
\end{example}

\subsubsection{Transformation}\label{subsubsec:transformation}

\changed[\Cn{2}{2}]{
Given the node-to-module mappings (\textit{modMaps}) from the pre-processing, the transformation phase generates the PRISM commands, synchronisation labels, rewards, and properties for each node according to its type. The mappings are first grouped by modules so that all nodes sharing a module are processed together. For each node, a PRISM constant is created to represent the node's location within its module, and then PRISM commands are generated according to the node type:

\begin{itemize}
    \item \textit{InitialNode}: a regular command and possible edge rewards (see \Cref{sec:semantics:transformation:initialnode}).

    \item \textit{OpaqueAction}: a regular command, or a reliability-related command if annotated with a reliability value; additionally, edge and node rewards are generated if present (see Sections~\ref{sec:semantics:transformation:opaqueaction} and~\ref{sec:semantics:transformation:reliability}).

    \item \textit{ForkNode}: a synchronisation command for the target node of the first outgoing edge, which shares the same module as the fork node (see \Cref{sec:semantics:transformation:forknode}).

    \item \textit{JoinNode}: a synchronisation command if the target node of its outgoing edge is in the same module; otherwise, an inactivation synchronisation command to set the module state to ``INACTIVE'' (see \Cref{sec:semantics:transformation:joinnode}).

    \item \textit{DecisionNode}: a regular command for each outgoing edge, or a probabilistic command if the branches are annotated with probabilities (see \Cref{sec:semantics:transformation:decisionnode}).

    \item \textit{MergeNode}: similar to \textit{JoinNode} but with a regular command instead of a synchronisation command, plus possible edge rewards (see \Cref{sec:semantics:transformation:mergenode}).

    \item \textit{FlowFinalNode}: an inactivation command to terminate the flow (see \Cref{sec:appendix:semantics:flowfinal}).

    \item \textit{ActivityFinalNode}: commands to terminate the whole behaviour of the activity diagram (see \Cref{sec:semantics:transformation:activityfinal}).
\end{itemize}

\noindent After all nodes are processed, parameters and property specifications are generated to complete the PRISM model.
An excerpt of the PRISM code generated from the transformed PRISM models is available in ~\Cref{fig:transform_activity_pal_use_case}.
}

\subsection{Implementation workflow and tool support}\label{subsec:tool}
%{\mycomment Describe the general workflow of our transformation implementation on Eclipse with Papyrus, Epsilon, Ant etc. with a diagram. }
The detailed workflow to implement our approach in an Eclipse-based tool, QANAD,\footnote{Available at \url{https://github.com/RandallYe/QANAD/}} for automation is shown in \Cref{fig:our_approach_impl}. The tool uses the Eclipse Modeling Framework (EMF).\footnote{\url{www.eclipse.org/modeling/emf/}}
The implementation exploits the development environments of Eclipse Papyrus\footnote{\url{https://eclipse.dev/papyrus/}}~\cite{lanusse2009papyrus} for system modelling, the Eclipse Epsilon\footnote{\url{https://eclipse.dev/epsilon/}}~\cite{kolovos2008epsilon} for PRISM code generation through model-to-model and model-to-text transformation, and the probabilistic model checkers (e.g., PRISM, Storm) for verifying the system properties. 
The open-source
implementation of our approach, full experimental results summarised in the next section, the case studies
used for its evaluation, and the tool setup instructions and requirements are available on our project webpage.\footnote{\url{https://github.com/RandallYe/QANAD}}

% {Figure~\ref{fig:our_approach_impl} presents a \emph{tool-level implementation workflow} in QANAD, including the concrete Eclipse components and the artefacts exchanged between them.
% In contrast, Figure~6 provides a \emph{conceptual, method-level view} of our overall approach, abstracting away from tooling details.}

{As shown in \Cref{fig:our_approach_impl}, the verification workflow comprises a set of task-oriented steps, from modelling and validation to transformation, property generation, and model checking.
To make the boundary of automation explicit, we summarise for each step whether it is automated by QANAD or requires human intervention in~\Cref{tab:automation}.}
\begin{table}[t]
\centering
\small
{\begin{tabular}{p{0.05\linewidth} l p{0.25\linewidth}}
\toprule
Step & Task (inputs $\rightarrow$ outputs) & Automation \\
\midrule
0 & Model AD and annotate it in (Papyrus) $\rightarrow$ UML/EMF model & Manual \\
1 & Validate well-formedness using EVL $\rightarrow$ validation report & Automated \\
2 & Valid AD $\rightarrow$ PRISM EMF model & Automated \\
3 & PRISM EMF model $\rightarrow$ \texttt{.prism}; property model $\rightarrow$ \texttt{.props}/\texttt{.pctl} & Automated \\
4 & Resolve property stereotypes and references $\rightarrow$ property model & Automated \\
5 & Model checking / parametric checking / synthesis $\rightarrow$ results & Automated + human decision \\
6 & Update selected parameters back to the AD model $\rightarrow$ the updated model & Manual (currently) \\
7 & Revise model upon failures $\rightarrow$ the updated model; and re-run & Manual + re-run automated\\
\bottomrule
\end{tabular}}
\caption{Workflow tasks and automation in QANAD.}
\label{tab:automation}
\end{table}

% These activities include \textit{Step 0} - modelling of the activity diagrams with some parameters being kept unspecified; \textit{Step 1} - validation of the activity diagrams; \textit{Step 2} - the transformation of activity diagrams into PRISM models and generation of formal properties;  \textit{Step 3} - generation of PRISM code from PRISM models;  \textit{Step 4.1} -  verification of the PRISM code in probabilistic model checker, e.g., PRISM, Storm; \textit{Step 4.2} -  parametric model checking of the PRISM code when we need to acquire an algebiac expression; \textit{Step 4.3} - synthesis of models to obtain the optimal properties and the corresponding parameters which are not specified at \textit{Step 0};  \textit{Step 5} - update of the unspecified parameters in the activity diagram; \textit{Step 6} - model modification, if EVL validation or PRISM verification fails, we need to modify the UML activity diagrams and rerun the process.
% We used Epsilon model management languages to implement the activities from \textit{Step 1} to \textit{Step 3}.

\paragraph{Step 0 - Modelling of the activity diagrams.}
%{\mycomment We also need to briefly mention why we chose Papyrus and emphasize our approach can be easily adapted to other UML modelling tools.}
%The system behaviour is first modelled using activity diagrams. % based on the UML metamodel and the pre-defined profile.
In this work, we use Papyrus as the modelling tool for activity diagrams. Papyrus is a leading open-source UML modelling tool developed under the Eclipse Foundation and driven by the PolarSys Initiative and the Papyrus Industry Consortium.
The reasons to use Papyrus are that the tool provides an integrated, user-consumable environment for supporting UML and related modelling languages such as SysML; it supports UML modelling as EMF models, and also offers advanced support for UML profiles. 
Other UML modelling tools that support EMF models and profiling can also be used in our workflow.
During the modelling, we create activity diagrams to model systems and also properties the systems shall satisfy through property stereotypes in the profile.

\paragraph{Step 1 - EVL Validation and modification of the activity diagrams} 
We use Epsilon Validation Language (EVL)\footnote{\url{www.eclipse.org/epsilon/doc/evl/}} to validate the activity diagrams against a set of predefined well-formedness conditions. %These models comply with the UML 2.5 metamodel. 
A subset of these well-formedness conditions is listed in \Cref{sec:semantics:wfc}.
There are three sources of the conditions: (1) the constraints for activity diagrams modelling, e.g.,~\ref{wfc:node_edges:has_names} ``each edge shall have a name''; \textcolor{black}{(2) the conditions imposed from Markov models, e.g.,~\ref{wfc:out_edge_decision:prob:sum:1} ``the probabilities of all the outgoing edges from a node shall sum to one''; and (3) the constraints on the content of the property, e.g., the name of the activity diagram used in the property should be correct. 
}
If the validation of an activity diagram reports a failure, we will need to modify the diagram accordingly based on the error messages provided.

\paragraph{Step 2 - ETL Transformation of activity diagrams into PRISM models} 
After activity diagrams are validated, %there are two steps of transformation. Firstly, 
we use Epsilon Transformation Language (ETL)\footnote{\url{www.eclipse.org/epsilon/doc/etl/}} to implement the model-to-model transformation from the activity diagram models to PRISM models according to the transformation rules defined in \Cref{sec:semantics:transformation} and the corresponding algorithms in \Cref{subsec:alg}. Both the source and target models of the transformation are in the form of EMF models and are supported by their metamodels respectively. The UML 2.5 metamodel is provided by an Eclipse project UML2\footnote{\url{https://projects.eclipse.org/projects/modeling.mdt.uml2}} and the PRISM metamodel\footnote{Available at \url{https://github.com/RandallYe/QANAD/blob/master/eclipse_workspace/AD2PRISM_Transformation_workspace/SysML_ActivityDiagram2PRISM/metamodels/sesame_prism.ecore}} is an extension of the metamodel presented in~\cite{Ye2022} to deal with CTMCs. 
%The algorithms of the ETL transformation for \textit{Step 2.1} are given in \Cref{subsec:alg}.
% As the activity diagram is stereotyped with system properties to be verified, within the ETL transformation, these properties written in a controlled natural language are also resolved into formal assertions that can be parsed in the probabilistic model checkers.

\paragraph{Step 3 - EGL Transformation from PRISM models to PRISM code } 
We use Epsilon Generation Language (EGL)\footnote{\url{www.eclipse.org/epsilon/doc/egl/}} to implement the model-to-text transformation from the PRISM models to PRISM code. 
Here, the PRISM model denotes the EMF model and the PRISM code is the textual representation of the EMF model and accepted by the PRISM or Storm model checker. 
%As the generation of PRISM code from the PRISM models is straightforward, no corresponding algorithms are provided in this paper.
Although we use Epsilon for transformation, other transformation languages like ATL~\cite{Jouault2008} and QVT~\cite{ObjectManagementGroupOMG2016} can also be used as needed.

\paragraph{Step 4 - Property generation through reference resolution} 
{The activity diagram is stereotyped with system properties to be verified, which are written in a controlled natural language.
Based on these properties and the validated activity diagrams, QANAD automatically derives formal property specifications via reference resolution.
These specifications are consumable by probabilistic model checkers and can be passed directly to Step~5 for verification.}

\paragraph{Step 5 - Model checking} 
\paragraph{Step 5.1 - Probabilistic model checking} 
{
Once the PRISM code and the corresponding property specifications are generated, the tool invokes probabilistic model checkers (e.g., PRISM or Storm) programmatically as part of the end-to-end workflow. In this step, probabilistic model checking is fully automated within QANAD.
Concretely, we use an Ant script to automate model validation, transformation, and property generation to automatically get a model and a property file, which are passed to the configured model checker. 
The verification results may be Boolean (true or false) or numerical values, such as probabilities or rewards. They are automatically stored along with the generated model and property file.
As a result, designers are not required to directly interact with the model checkers or their formal notations.

% {The verification results (e.g., Boolean satisfaction, probabilities, or reward values) are collected automatically and reported back to the Eclipse environment through generated reports and log files, enabling designers to assess verification outcomes without inspecting raw model-checker outputs.}

Overall, QANAD realises a \emph{hidden formal methods} workflow: designers specify systems and properties at the UML level, while all formal verification steps are performed automatically in the background, thereby lowering the barrier to adoption of probabilistic model checking.
Steps~5.2 and~5.3 build on these automatically obtained verification results, where further analysis and synthesis may require design-time decisions.
}

\paragraph{Step 5.2 - Parametric model checking}
If parametric model checking is applied to the generated PRISM code, the verification results are algebraic expressions.
{This step is also executed automatically as part of the QANAD workflow.} 

\paragraph{Step 5.3 - Synthesis of models} 
%From \textit{Step 2}, the properties are resolved as formal assertions from the activity diagram. 
%Some of the properties may conflict in nature, and may need adjustment to reach the optimal definition of the properties.
%On the other hand, the 
The PRISM code from \textit{Step 3} may contain parameters whose values are not specified. 
These parameters shall be fixed based on the design analysis and can be evolved during runtime.
To satisfy the multiple properties, the designers need to verify probabilistic models for numerous instantiations of these parameters to identify a proper set of parameters.
This is a tedious manual process and the obtained models are often suboptimal in terms of their trade-offs between properties~\cite{gerasimou2015search}.
In our approach, we use the synthesis tool EvoChecker~\cite{gerasimou2015search} to automatically optimise the properties and identify the proper set of parameters.
EvoChecker is a search-based synthesis approach that % employs multi-objective optimisation genetic algorithms to 
automates the multi-objective optimisation process. % and improves the  its outcome.
Through EvoChecker, we can obtain several sets of optimised properties as design targets and the corresponding values for the evolvable parameters. 
Then, the engineers shall select the proper set of properties and the constant values according to the specific design decisions. This will be further illustrated in \Cref{sec:cases:results}.

EvoChecker is employed as an external synthesis tool to explore the design space and compute candidate parameter configurations that optimise the specified properties.
In QANAD, this synthesis step is \emph{not fully automated. {The adaptation process currently requires manual effort — identifying evolvable parameters, inserting them into the generated PRISM model, and selecting properties as EvoChecker objectives — all of which represent strictly structural and syntactic modifications that do not alter the underlying behavioural semantics of the models. These manual tasks are entirely amenable to automation in future work, for instance, via a dedicated intermediate EvoChecker adapter.}
Selecting a preferred solution and adapting the activity diagram accordingly requires manual intervention and design-time decisions by engineers.}

\paragraph{Step 6 - Update of the unspecified parameters} 
After the synthesis, we use the selected set of parameters to update the activity models. 
This is a manual process at the moment.
Then, we re-run \textit{Step 1} to \textit{Step 3} to obtain the complete PRISM code, which is ready for verification.

\paragraph{Step 7 - Model Modification} 
If the validation in \textit{Step 1} or the verification in \textit{Step 5.1} or \textit{Step 5.2} fails, we will need to modify the activity diagram and restart the process from \textit{Step 1}. 

%{\mycomment Emphasise our validation, transformation and generation approach starts from the exported UML model (UML 2.5), and so it works for any tools that can export UML-compliant models.}

\section{Evaluation}
\label{sec:cases}

\subsection{Research Questions}
We evaluated the effectiveness and validity of our approach through case studies to answer the following three research questions.

\textbf{RQ1 (Correctness): %How does our approach perform compare to existing solutions?
%Does our approach generate valid PRISM models from activity diagrams?
  {Do the transformation rules and implementation of our approach generate correct PRISM models from activity diagrams {for selected case studies?}}
}
We gave the semantics of activity diagrams in the PRISM language, as presented in \Cref{sec:semantics:transformation}, and implemented it as discussed in \Cref{subsec:alg}.
Therefore, we evaluated {several examples to provide some evidence for the correctness of the transformation and its implementation}.

%\textbf{RQ2 (Scalability): Can our approach be applied to large-scale activity diagrams?}
%{\mycomment We didn't evaluate large-scale activity diagrams now and it is not possible to evaluate scalability now. It might be better to change RQ2 to (Comparison) with other approaches (especially the 2015 approach) in terms of state space, the complexity implication (easy or difficult) for the implementation of tools, what features they can support but we cannot, and what features we can support but they cannot.
%}

\textbf{RQ2 (Efficiency):
Can activity diagrams be verified in the probabilistic model checkers with acceptable overheads in terms of computational resources and time?}
We checked the time efficiency of the PRISM code generation from activity diagrams. 
Also, we analysed the state space size of the generated PRISM code and its impact on the model checking efficiency.

%{
%\mycomment we are not able to evaluate efficiency at this stage as well. Maybe we can change to (extensibility) or (insights) to give insights how our approach can be extended to support other features: CallBehaviourAction, ObjectFlow and ObjectNode, Event and Signals etc.
%}

\textcolor{black}{\textbf{RQ3 (Decision Support): How does our approach support the design process and facilitate design decisions for users?}
This research question aims to explore the practical implications and benefits of our approach in supporting users to make design decisions based on verification results of the generated PRISM code. }

\subsection{Evaluated Examples}

In this work, we used seven case studies (as shown in \Cref{tab:use_case_table}) from different application domains to evaluate the research questions.
Each case study is used to evaluate different features of activity diagrams, but there is also overlap in terms of features among these case studies.
For each case study, we show its name, the configuration of features or annotations in the AD, chosen Markovian models for analysis, the number of nodes in the AD, the number of parameters defined in the activity diagram (and the number of open parameters among them), and the number of concurrent flows.

\begin{table}[ht]
    \caption{{The use cases used for evaluation}}
\label{tab:use_case_table}\centering

\begin{tabular}{@{}llllllll@{}}
    \toprule
    %& \multicolumn{3}{c|}{TTT} & \multicolumn{3}{c|}{TTT}\\
    Example           & Configuration & ModelType  & \#Nodes & \#Parameters & \#Flows %& Note
\\
    \midrule
    \multirow{2}{*}{SD} & Fair coin & DTMC & \multirow{2}{*}{24} & \multirow{2}{*}{1 (1 open parameter)} & \multirow{2}{*}{1} \\
                                   & Parametrised & DTMC &&&\\\midrule
    \multirow{3}{*}{DC} & Parametrised & DTMC & \multirow{3}{*}{21} & \multirow{3}{*}{3 (0 open parameter)} & \multirow{3}{*}{4} %& \multirow{3}{*}{Original PTA}
    \\
    &Parametrised&MDP&&&&\\
    &Duration&CTMC&&&&\\\midrule
    FP & Parametrised-Reward & DTMC & 14 & 11 (11 open parameters) & 1 %& Algebraic expression
    \\\midrule
    TW & Parametrised & CTMC & 13 &8 (0 open parameters) & 1 %& Default duration
    \\\midrule
    IS & Parametrised & CTMC & 16 & 14 (0 open parameters) & 1 %&Default duration 
    \\\midrule
    \multirow{3}{*}{TM} & Reliability & DTMC & \multirow{3}{*}{18} & \multirow{3}{*}{2 (0 open parameters)} & \multirow{3}{*}{1}\\
                                        & Reliability & MDP&&& \\
                                        & Reliability-Duration & CTMC &&&%& Default duration 
                                        \\\midrule
    \multirow{4}{*}{PAL} & Reward  & DTMC & \multirow{4}{*}{27} & \multirow{4}{*}{20 (6 open parameters)} & \multirow{4}{*}{2} \\
                         & Reward  & MDP &&&  \\
                         & Reward  & CTMC  &&&%&   Default duration      
                         \\
                         & Parametrised-Reward  & CTMC  &&&
                         \\
    \bottomrule
\end{tabular}
\end{table}

The six-sided dice (SD)~\cite{knuth1976complexity} models the Knuth's algorithm of simulating a die using a fair coin in an activity diagram. Imagine standing at the root vertex (state 0) of a horizontal tree-like structure whose six leaves are the values of the six sides of a dice. 
At this starting point, you initiate the process by tossing a fair coin. If the result is heads, you move along the upper branch; if tails, you take the lower branch. This process is iteratively repeated based on the outcome of each coin toss. The traversal continues until the final value of the die is determined, i.e., a leaf is reached.
If the coin used is a fair coin, then the probabilities of getting heads and tails are the same, otherwise, the probability for each toss is parametrised. We experimented with both of these two configurations.
\textcolor{black}{The SD activity diagram has the features of reward, parametrised probability, and property annotations, and the decision-merge loop structure. The SD activity has one parameter which is open or indetermined or uninstantiated.}

The digital camera (DC) case study~\cite{Baouya2015a} models the process of using a digital photo-camera device to take a picture in an activity diagram.
%The activity starts by turning on the camera. Then, there are three concurrent flows.
%The first flow is to use the autofocus function of the camera,  followed by a decision checking the status of the availbility of the memory. If the memory is full, the camera is turned off. 
%The second parallel flow is to detect the ambient lighting conditions to further determine the necessity of the flash.
%The third flow allows charging the flash (ChargeFlash) if it is not already charged. The action of picture-taking executes if it is sunny and the memory is not full or the flash (Flash) is needed because of the lack of luminosity. 
\textcolor{black}{The DC activity diagram has the features of duration, parametrised probability and property annotations, and the concurrent flow structure.}

The fruit-picking (FP) case study~\cite{fang2022presto} models the functionality of a fruit-picking robot from the autonomous farming domain.
The robot performs three operations autonomously including:
(1) to position itself in the right location for the fruit picking; (2) to use its arm to pick up the fruit; and (3) when operation 2 is unsuccessful, to decide whether to retry the fruit picking from operation 1 or to terminate the operation.
\textcolor{black}{The FP activity diagram has the features of parametrised probability, reward, and property annotations%, and the parametrised algebraic expression
.}

The travel web (TW) application case study~\cite{paterson2018observation} models a service-based system that is implemented using
six real-world web services: two commercial web
services provided by Thales Group, three free Bing
web services provided by Microsoft, and a free {web service provided by WebServiceX.Net}.
The TW application handles two types of user requests: 
(1) a plan to meet and entertain a visitor arriving by train; and
(2) looking for a possible destination for a day trip by train for the user.
\textcolor{black}{The TW activity diagram has the features of parametrised duration and property annotations, and the decision-merge loop structure.}

The IT support system (IS) case study~\cite{paterson2018observation} models a real-world IT support system deployed at the Federal Institute of Education, Science and Technology of Rio Grande de Norte (IFRN), Brazil.
The system enables the IFRN IT support team to handle
user tickets reporting problems with the institute’s computing systems.
\textcolor{black}{The IS activity diagram has the features of parametrised rate, parametrised probability, and property annotations, and the multiple decision-merge loop structure.}

The travel management (TM) case study~\cite{Gallotti2008}
models a travel management functionality that starts from a travel location, and offers booking services and notifications.
\textcolor{black}{The TM activity diagram has the features of parametrised duration, parametrised reliability, and property annotations, the concurrent flow structure and the decision-merge loop structure.}

The PAL example presented in \Cref{sec:motiving_example} is also used for evaluation in this section.

Among the seven case studies, SD and FP are verified as DTMC models in the experiment; TW and IT are verified as CTMC models; DC, TM and PAL are all verified as three types of models, including DTMC, MDP, and CTMC models.

\subsection{Evaluation methodology and results}
\label{sec:cases:results}

%{\mycomment evaulation by using six use cases, having consistent result; by comparing the manual model and auto models. need to analyse  the difference in the numbers of the states and transitions between manual and auto models.}
%{\mycomment Compare our verification results with those in the literature or from other source. If the results are different, we explore other approaches (manual implementation in PRISM using a verification approach from literature, manually draw a LTS and verify the results using mathematics directly without PRISM), or contact the authors to clarify their results.}

We used the seven case studies listed in \Cref{tab:use_case_table} to evaluate our approach. The activity diagram, the generated PRISM code and properties, and verification results for each case study can be found online.\footnote{\url{https://github.com/RandallYe/QANAD/tree/master/Examples}}

\textbf{RQ1 (Correctness)}. 
To address the first question regarding {correctness of the transformation implementation}, we compare the verification results of our automatically generated PRISM models with those reported in the literature across five case studies, excluding the DC and PAL examples. The verified properties and the corresponding results, obtained using the PRISM model checker, are presented in \Cref{tab:use_case_result_table}. We note that {the concrete values of the QoS annotations in these models are from corresponding literature or references} and all these examples are verified to be deadlock-free.

The verification results for DC and PAL are not used for the comparison, as we lack available referenced results for these cases. The literature source~\cite{Baouya2015a} for the DC example focuses on PTA models, while we generated DTMC, MDP, and CTMC models from the same DC example. Consequently, the verification results are not directly comparable for the DC case.
Also, the PAL use case is not sourced from published literature so it has no available referenced results.

\begin{table*}[ht]
\caption{The comparison of our verification (actual) results with the (expected) results from literature for evaluated use cases.}
\label{tab:use_case_result_table}\centering
\renewcommand{\arraystretch}{1.2} % Adjust the value as needed
\setlength\tabcolsep{.9mm}
\resizebox{\textwidth}{!}{\begin{tabular}{@{}lllm{7.5cm}m{2.0cm}lll@{}}
    \toprule
    %& \multicolumn{3}{c|}{TTT} & \multicolumn{3}{c|}{TTT}\\
    Use case       & \#St & \#Tr    & Property & CTime(s)
& Actual result  & Exp Result & Delta   \\
  \hline
 \multirow{3}{*}{SD-DTMC}&\multirow{3}{*}{25} &\multirow{3}{*}{32} &\lstinprop{P=?  [F Six\_dice reaches at Six\_dice::x] for x in \{O1,O2,O3,O4,O5,O6\}}  &%\lstinprop{0.025, 0.027, 0.015, 0.015, 0.017, 0.013} &0.167 &	0.167 &0\\
 \lstinprop{0.013 - 0.027} &0.167 &	0.167 &0\\

  &&&\lstinprop{R\{"reward\_flip"\}=?[F Six\_dice reaches at Six\_dice::F0]} &0.016 &3.67	 &3.67& 0\\
 \hline

 \multirow{7}{*}{SD-para} &\multirow{3}{*}{25} &\multirow{3}{*}{32}&\lstinprop{P=? [F Six\_dice reaches at Six\_dice::O1]} 	 &0.033 & $p^2/(p + 1)$ &	the same &0\\
   && &\lstinprop{P=? [F Six\_dice reaches at Six\_dice::O2]}	 &0.012&$p^2/(p + 1)$ &	the same& 0\\
    &&&\lstinprop{P=? [F Six\_dice reaches at Six\_dice::O3]} &0.008&	$(p - p^2)/(p + 1)$ &	the same&  0\\
   &&&\lstinprop{P=? [F Six\_dice reaches at Six\_dice::O4]} &0.021&$(p^2-p^3)/(p^2-p+1)$ &	the same& 0\\
    &&&\lstinprop{P=? [F Six\_dice reaches at Six\_dice::O5]}	 &0.018&$(p^3-2p^2+p)/(p^2-p+1)$	 &the same& 0\\
    &&&\lstinprop{P=? [F Six\_dice reaches at Six\_dice::O6]}	 &0.016&$(3p^2-p^3-3p+1)/(p^2-p+1)$ &	the same& 0\\
    &&&\lstinprop{R\{"reward\_flip"\}=? [Six\_dice reaches at Six\_dice::F0]} &0.057&$(p^4-5p^3+4p^2+p-3)/(p^4 - p^3 + p-1)$ &the same& 0
 \\
 %&~~~~~? [Six\_dice reaches at Six\_dice::F0]	 & && \\
 \hline
 
 \multirow{3}{*}{FP} & \multirow{3}{*}{15}& \multirow{3}{*}{18}&\lstinprop{P=? [F "picking success"]} & 0.049&
 $(\alpha p_1 \beta p_2 -\alpha p_1)/(\alpha p_1 \beta p_2 p_3 -1)$
 %($\alpha$\textasteriskcentered$p1$\textasteriskcentered$\beta$\textasteriskcentered$p2$ + (-1)\textasteriskcentered$\alpha$\textasteriskcentered$p1$)	
 &the same& 0\\
%     & &/($\alpha$\textasteriskcentered$p1$\textasteriskcentered$\beta$\textasteriskcentered$p2$\textasteriskcentered$p3$ + (-1))            & &\\
&& &\lstinprop{R\{"time"\} =? [F "done"]}&0.063&
 $-(\alpha p_1 \beta p_2 t_3 + t_1 + \alpha p_1 t_2) / (\alpha p_1 \beta p_2 p_3 -1)$
 &the same& 0\\
 &&&\lstinprop{R\{"energy"\} =? [F "done"]}	&0.021&
 $-(\alpha p_1 \beta p_2 e_3 + e_1 + \alpha p_1 e_2) / (\alpha p_1 \beta p_2 p_3 -1)$
 &the same& 0\\
%                    &&/($\alpha$\textasteriskcentered $p1$\textasteriskcentered $\beta$\textasteriskcentered $p2$\textasteriskcentered $p3$ + (-1))& & 
\hline
\multirow{2}{*}{IS}  & \multirow{2}{*}{17}  & \multirow{2}{*}{22}  & \lstinprop{P=? [F<=100 "complete"]}	& 0.007 &0.4394%0.439417949

 &	0.4395 %0.439465776
&1e-4  %4.8×1e-5
 \\
 &&&\lstinprop{P=? [(!"reopen" \& !"addInfo") U<=100 "complete"]}	&0.005&0.4296 (duration=0.01)  %0.429642643

 &	0.4297  %0.429680054
&  1e-4 %3.7×1e-5
\\
%                    &   ~~~~~~~~~~U<=100 ``complete"]&(duration = 0.01)&&\\
\hline

\multirow{4}{*}{TW}  & \multirow{4}{*}{14}  &\multirow{4}{*}{16}  &\lstinprop{P=? [F<=1 "complete"]}	&0.056 &	0.4147  %0.414695030
% 0.4146950095517946
& 0.4147  %0.414695033
& 0  %3e-9、
\\
&&&\lstinprop{P=? [!"arrival" U<=1 "complete"]/(1-p1)}	&0.038&0.3625 
 %0.362525671 
 & 0.3625  %0.362525673	
 & 0  %2e-9
 \\
 &&&\lstinprop{P=? [F<=1 "complete"] - 2*(1-P=? [F=3 "complete"])}	&0.192&0.2664 (duration=1e-5)
 %0.266407563 
 & 0.2664   %0.266407592	
 &0  % 29e-9
 \\
 %                   &~~~~~~~~~~-2\textasteriskcentered(1-P=? [ F=3 "complete" ])&(duration = 1e-8)&&\\
\hline

\multirow{7}{*}{TM-DTMC} &\multirow{7}{*}{39} &\multirow{7}{*}{73} & \lstinprop{P=? [F (TM terminated successfully)]} &0.062& 0.775 & 	0.775  &  0\\
&&&\lstinprop{filter(max, P=? [F (TM terminated successfully)], (TM reaches at TM::checkSchedule))}&0.063&0.824	 &0.824 & 0\\
                        %Result: 0.8242755073293442
%                        &\qquad(TM reaches at TM::checkSchedule))	 & &&\\
&&&\lstinprop{filter(max, P=? [F (TM terminated successfully)], (TM reaches at TM::requestMeeting))} &0.038&0.950 &	0.950 & 0\\
                        %Result: 0.9507920738075207
%                        & \qquad(TM reaches at TM::requestMeeting))&&&\\
\hline
\multirow{4}{*}{TM-MDP} & \multirow{4}{*}{39} &\multirow{4}{*}{73} &\lstinprop{filter(max, Pmax=? [F (TM terminated successfully)], (TM reaches at TM::makeCall))}&0.027 &	0.932 &	0.932 & 0\\
%                         & \qquad(TM reaches at TM::makeCall))}&&&\\
&&&\lstinprop{filter(min, Pmin=? [F (TM terminated successfully)], (TM reaches at TM::makeCall))} &0.086&0.841	 &0.841 & 0\\
%                        & \qquad(TM reaches at TM::makeCall)) &&&\\
\hline

\multirow{2}{*}{TM-CTMC} &\multirow{2}{*}{39} &\multirow{2}{*}{73} &\lstinprop{P=? [F=t (TM terminated successfully)]} for (t =  100)&0.049& 0.775 & N/A & N/A \\
&&&\lstinprop{P=? [F=t (TM terminated successfully)]} for (t = 40 )&0.018&0.688 (duration=0.001)	&N/A &N/A
\\

\hline
\multirow{3}{*}{DC-DTMC} %!E [F "deadlock"]&	true	&true&N/A\\
&\multirow{3}{*}{226}&\multirow{3}{*}{514}& \lstinprop{A [F dc reaches at dc::AF]}&0.013 &	true	&true&N/A\\
&&& \lstinprop{P=?[F dc reaches at dc::AF]}&0.01&	1.0	&N/A&N/A\\ 
&&& \lstinprop{P=? [F (dc terminated successfully)]}&0.003	&1.0	&N/A&N/A\\
\hline

\multirow{3}{*}{DC-MDP} &\multirow{3}{*}{226} &\multirow{3}{*}{514} %& !E [F "deadlock"]&	true	&true&N/A\\
& \lstinprop{A [F dc reaches at dc::AF]}&0.013 &true	&true&N/A\\
%& Pmin=?[F dc reaches at dc::AF]&	1.0	&N/A&N/A\\ 
%& Pmax=?[F dc reaches at dc::AF]&	1.0	&N/A&N/A\\ 
&&& \lstinprop{Pmin=? [F (dc terminated successfully)]}&0.009	&1.0	&N/A&N/A\\
&&& \lstinprop{Pmax=? [F (dc terminated successfully)]}&0.01&	1.0	&N/A&N/A\\\hline

\multirow{3}{*}{DC-CTMC} & \multirow{3}{*}{226} & \multirow{3}{*}{514} & \lstinprop{P=? [F<=10 dc reaches at dc::AF]}&0.001 &0.822	&N/A &N/A\\   
&&&\lstinprop{P=? [F<=10 ((dc reaches at dc::M1)|(dc reaches at dc::AutoFocus))\&(dc reaches at dc::TakePicture)]}  &0.919& 0.195 (duration=0.1) &N/A &N/A\\ 
%Result: 0.23199566646020925
%&~~~~~dc::AutoFocus)) $\And$ (dc reaches at dc::TakePicture)]&(duration=0.1)&&\\
\hline
\multirow{5}{*}{PAL-CTMC} & \multirow{5}{*}{126} &\multirow{5}{*}{275} &\lstinprop{P=? [F<=10 "Either delivered"]}&0.017&	0.999	&N/A&N/A\\
%0.9991309971762218  duration = 0.001 using predefined parameters
%Result: 0.9992147988542862 using evochecker result with p_c_a changed to 0.989
%Evochecker config:
%const double p_c_a = (0.918);
%const double p_d_1 = (0.947);
%const double p_d_2 = (0.833);
%const double p_c_c = (0.987);
%const double p_c_b = (0.996);

%0.9991315149570098 when duration = 0.0001, t= 810.971 seconds.
%if set d to 0.00001 or less, overflow

&&& \lstinprop{P=? [F<=10 "Both delivered"]}&0.031&	0.178	&N/A &N/A\\
%Result: 0.17570663358120214 using evochecker result with p_c_a changed to 0.989

%0.1777062172823482 d=0.001 using predefined parameters
%& P=? [F<=10 "Robot1 delivered"]& 0.977	&N/A &N/A\\
%0.9774598638673966 d=0.001
%& P=? [F<=10 "Robot2 delivered"]&	0.970	&N/A &N/A\\
%0.9700251744530175 d=0.001
&&& \lstinprop{P=? [F<=10 "Both stuck"]}&0.017&	3.61×1e-5	&N/A &N/A\\
%Result: 6.50949961614106E-5 using evochecker result with p_c_a changed to 0.989
%3.61555652409591E-5  d=0.001 using predefined parameters
&&& \lstinprop{P=? [F<=10 "terminate"]}&0.022&	0.968	&N/A &N/A\\
%Result: 0.9675359377118761 using evochecker result with p_c_a changed to 0.989
%0.9676623136663314  using predefined parameters
%0.9992286926587634 (t=20)

&&& \lstinprop{R\{"rwd\_door12\_attempt"\}=? [C<=10]}&0.022&	1.03 (duration=0.001)	&N/A &N/A\\
%Result: 1.1110119434781025 using evochecker result with p_c_a changed to 0.989
%Result: 1.0300013393333454  using predefined parameters
%&&(duration = 0.001)&&\\

\hline
\multirow{1}{*}{PAL-para} &\multirow{1}{*}{126} &\multirow{1}{*}{275} &\multirow{1}{*}{\lstinprop{P=? [F "Either delivered"]}} &0.291&	$4.999*\left(
\begin{array}{l}
    p\_c\_a*p\_c\_b +p\_c\_c - \\
    p\_c\_a*p\_c\_b*p\_c\_c
\end{array}
\right)/5$ &the same &\multirow{1}{*}{0}\\
%\multirow{2}{*}{PAL-para} & \multirow{2}{*}{\lstinprop{P=? [F "Either delivered"]}} &	4.999\textasteriskcentered ( $p\_c\_a$ \textasteriskcentered $p\_c\_b$ +$p\_c\_c$&the same &\multirow{2}{*}{0}\\
%&&	 - $p\_c\_a$ \textasteriskcentered $p\_c\_b$ \textasteriskcentered $p\_c\_c$ )/ 5	& &\\\hline
\hline

PAL-DTMC &126&275& \lstinprop{P=? [F "Either delivered"]}&0.054&	0.9996	&N/A &N/A\\   
%Result: 0.9996054046923334 using evochecker result with p_c_a changed to 0.989
%0.9997 using predefined parameters
\hline
\multirow{2}{*}{PAL-MDP} & \multirow{2}{*}{126} &\multirow{2}{*}{275} &\lstinprop{Pmax=? [F "Either delivered"]}&0.073&	0.9997	&N/A &N/A\\
%max = Result: 0.9996906263783023 (+/- 4.806698372661344E-9 estimated; rel err 4.808185898546574E-9)
&&& \lstinprop{Pmin=? [F "Either delivered"]}&0.056&0.9997	&N/A &N/A\\
%min: Result: 0.9996906266744479 (+/- 1.9636994037508517E-9 estimated; rel err 1.964307107973251E-9)
\bottomrule
\end{tabular}}
\end{table*}

For SD, we compared the results of our approach with the results obtained from the PRISM code online.\footnote{\url{https://www.prismmodelchecker.org/casestudies/dice.php}} Seven properties are verified, and we obtain the same results from our PRISM code as the ones online. 
For the parametrised variant of SD, we use a biased coin instead of a fair coin and changed the probability of getting a head from $0.5$ to a parameter \ad{p} of the activity. We also adapted the PRISM code from online manually to make it a parametric model and used PRISM for the parametric model checking. The results obtained from our tool are the same as those from the manual adaptation. 

For FP, we compared our results with the ones from the literature. Three properties are verified, and the results are in the form of an algebraic expression. 
We obtain the same results as the ones in the literature.

For IS and TW, we obtained the PRISM code from the authors and compared our results with the ones of the authors' PRISM code. The results for both IS and TW are consistent. However, for the IS use case, both our results and the ones from the PRISM code provided by the authors are different from the results listed in the literature.
The reason is that a different set of probability values was actually used to generate the figures of verification results in the literature. 
Both PRISM code of IS and TW are CTMC models, therefore in our models the control nodes (e.g., \ad{DecisionNode}, \ad{MergeNode}, etc.) are assigned a default duration of a small value. The control node duration for IS is \textit{0.01}, and \textit{1e-5} 
for TW. \textcolor{black}{The values are selected to closely match the results in the literature.}
%when the duration is lower than these value, the PRISM model checking overflows.
As the PRISM code we compared with is manually created according to the transition system, there is a small discrepancy in the verification result of the time-based properties due to the duration of the control nodes. 
Since we set the default duration of the \ad{ControlNode}s as small as possible, {the observed discrepancy is negligible for the considered use cases, as shown in the ``Delta'' column of \Cref{tab:use_case_result_table}.}
When the numerical values are extended to four decimal places, the magnitude of the discrepancy is on the order of \textit{1e-4} for the IS use case, and \textit{0} for TW use case.
When the numerical values are extended to four decimal places, the magnitude of the discrepancy {for the considered settings} is on the order of \textit{1e-4} for the IS use case, and \textit{0} for the TW use case.

%we can conclude that our approach's results are consistent with the literature.

For TM, we verified the property for DTMC, MDP, and CTMC. The property is to evaluate the probability that TM terminates successfully.
We first compared our results with those reported in the literature. 
Due to the absence of explicit probability that should be assigned to the outgoing edges of decision nodes in the activity diagram from the literature, we assumed that the decision nodes have the same probability for each outgoing edge.
With this assumption, we obtained the same results as the ones in the literature for the properties of DTMC and MDP.

Regarding TM's CTMC models, since probabilities evolve over time and the literature does not provide a concrete CTMC model, we were unable to fully reproduce the results. Moreover, no exact probability values are provided at specific time points—only the probability-time curve is presented in the literature. Therefore, we marked ``Not Applicable'' (N/A) for both "Exp Result" and ``Delta'' because no precise values are available from the literature. However, we can estimate the values by referencing the probability-time curve to make a rough comparison.
We checked our CTMC model against the termination property from t = 10 to t = 200 at 10-second intervals. Starting from t = 100, the probability values are consistent with those reported in the literature.
When t $\leq$ 100, our result is slightly different from the values in the probability-time curve presented in the literature. For example, at time t=40, we observed a rough difference of 0.034 between our result and the estimated value from the curve.
%One exception in TM is the verification result of the properties for the CTMC models.
%the probability value for the ``long run'' in the literature is 0.775, which is the same as our result. %when \textit{t $>$ 100}. %, with precision up to three decimal places. 
%So we claim that we obtained the same result for this property when \textit{t} is large enough to have a steady probability.
%As the literature provides the graph of results for \textit{0} \textit{$\leq$}  \textit{t} \textit{$\leq$} \textit{70}, \Cref{fig:Comparison_of_TM-CTMC_fig5_VS_auto} compares our results with the literature's for this time range and reveals a notable discrepancy. The maximum difference is \textit{0.195} when time is \textit{15}.
The reason for the discrepancy may lie in the default duration assigned to the \ad{ControlNode}s, e.g., \ad{MergeNode}s, \ad{DecisionNode}s.
Since the lack of the PRISM code and the details in the literature, we could not know the default duration used to obtain the verification results in the literature.
% \begin{comment}
% \begin{figure}[ht]
%     \centering
%     \includegraphics[scale=0.45]{pics/Comparison_of_TM-CTMC_fig5_VS_auto.png}
%     \caption{Comparison of TM-CTMC verification results between literature and our approach.}
%     \label{fig:Comparison_of_TM-CTMC_fig5_VS_auto}
% \end{figure}
%     
% \end{comment}

\begin{figure}
    \centering
    \includegraphics[scale=0.35]{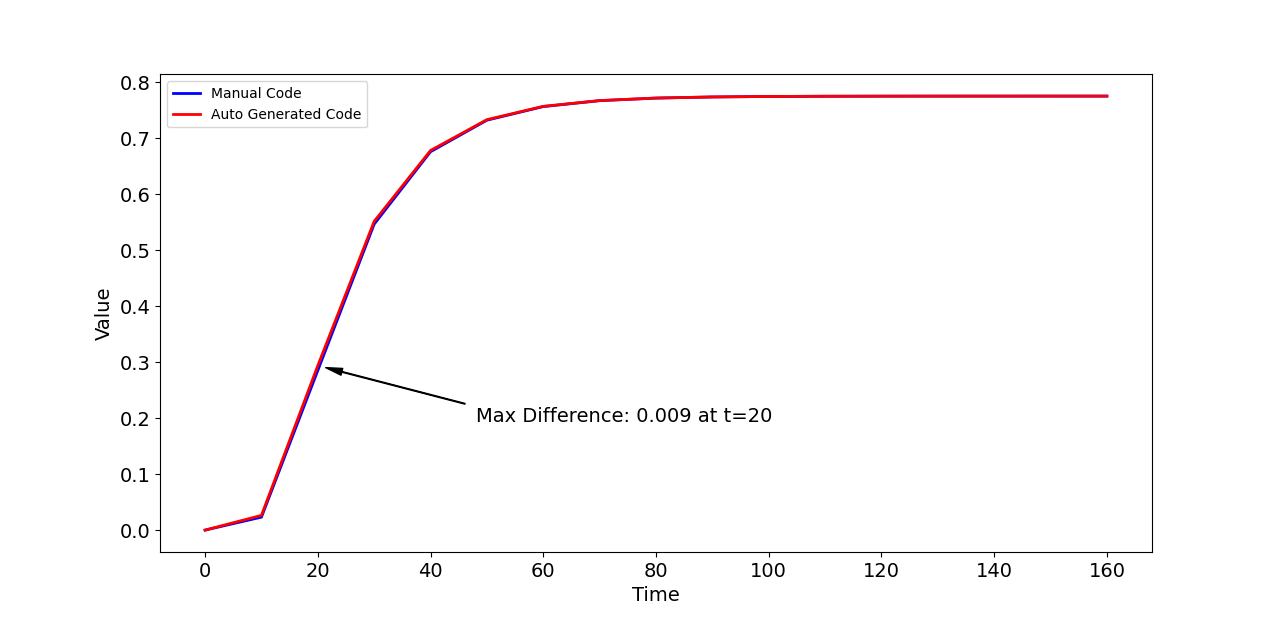}
    \caption{Comparison of TM-CTMC verification results of approach in~\cite{Baouya2015a} and our approach.}
    \label{fig:Comparison_of_TM-CTMC_autoVSmanual}
\end{figure}

%\begin{comment}
%\begin{figure}
%\centering
%\begin{subfigure}{0.2\textwidth}
%  \centering
% \frame{ \includegraphics[width=1.0\linewidth]{pics/Comparison_of_TM-CTMC_verification_results.png}}
%  \caption{Comparison between literature and our approach.}
%  \label{fig:sub1}
%\end{subfigure}
%
%\begin{subfigure}{0.2\textwidth}
%  \centering
%  \frame{ \includegraphics[width=1.0\linewidth]{pics/Comparison_of_TM-CTMC_autoVSmanual.png}}
%  \caption{Comparison between manual approach and our approach.}
%  \label{fig:sub2}
%\end{subfigure}
%
% \caption{Comparison of TM-CTMC verification results}
%    \label{fig:Comparison_of_TM-CTMC_verification_results}
%\end{figure}
%    
%\end{comment}

To further evaluate the correctness of our transformation for TM (CTMC), we conducted three manual activities to compare with the results from the automatically generated code using our tool.

The first is to manually implement the transformation from the TM activity diagram to the PRISM CTMC code according to the semantics in \Cref{sec:semantics}. 
This is to ensure our implementation of the transformation is correct.

Secondly,  we manually constructed the TM PRISM code following the approach in~\cite{Baouya2015a} which declares a set of Boolean variables (one variable for each node in an AD) and uses MSM as the input language for PRISM so the transformation results in one module in PRISM.
If a collection of nodes in the AD is active, then the corresponding variables are set to \textit{true}, and others are set to \textit{false}.
In contrast, our automated transformation uses the module approach to generate two modules for TM, and to declare one \lstinprop{pc} variable for each module.
As our automatic code includes multiple modules, additional states and transitions are introduced compared to the manual code, serving the purpose of module synchronisation.
The comparison of our approach (using automatically generated code) and the manual implementation by the approach~\cite{Baouya2015a} is shown in \Cref{fig:Comparison_of_TM-CTMC_autoVSmanual}.
There is a slight discrepancy due to the extra states and transitions which is within an acceptable range.
When {$t>120$}, the discrepancy is less than {1e-6}.
When {$t \leq 120$}, the maximum value of the discrepancy is {0.009}. 
Therefore, we claim that our transformation implementation is correct.

{Thirdly, we manually drew an LTS, computed the transition rate matrix, and calculated the bounded termination probability using MATLAB and a Python program (written by us and without using the model checking tools) based on the matrix exponential.}
The calculated results are consistent with the results from our \method\ too. 
{Therefore, we believe our approach and results for TM-CTMC are correct.}

For DC, we annotated actions in the AD with mean duration instead of minimum and maximum execution time (interpreted in PTA) in the original literature~\cite{Baouya2015a} and verified three types of properties for DTMC, MDP, and CTMC. The default duration is set to \textit{0.1} for the CTMC model. From the verification results for DTMC and MDP, we conclude that the activity diagram is always terminated (because the result is true), and {it almost surely terminates independently from the choices of the scheduler/policy,} {meaning that the minimum probability of termination over all schedulers is equal to~1, as indicated by the outcome of the \lstinprop{Pmin=?} property.}

For PAL, three types of properties are verified including
properties for DTMC, MDP, and CTMC. We also conducted the parametric model checking for one property. Five properties are verified for the CTMC model. The default duration is set to {0.001}. 
The property ``the probability of successful delivery through either
Route I or Route II'' is checked also for both DTMC and MDP models. 
This property should be presented as \lstinprop{P=? [F (PAL reaches at PAL::R1DeliveredRoomD)|(PAL reaches at PAL::R2DeliveredRoomD)]}.
To save space in the table, we use the label ``Either delivered'' to replace the state property \lstinprop{(PAL reaches at PAL::R1DeliveredRoomD)|(PAL reaches at PAL::R2DeliveredRoomD)}.
The same is applied to the labels in other three properties for CTMC models.
The PAL use case and the DC use case are used to demonstrate that the transformation can go through.  
We conducted a manual review of the PRISM code generated for these two use cases to assess the correctness of the transformation.

Because each case study carries different features in activity diagrams, we can conclude that our approach is capable of generating the correct PRISM code for activity diagrams through the evaluation of these case studies.

%\textcolor{red}{Table~\ref{tab:use_case_result_table} also presents the complexity of the PRISM models, quantified by the number of states and transitions, as well as the model checking time for each property across different models. The checking of models tends to be  complete within approximately 10 milliseconds.
%However, we do not observe a direct correlation between the scale of the PRISM models and the model checking time. For example, 
% in DTMC model checking: the checking time for DC's eventual termination is 0.003 seconds, whereas for the TM example, it is 0.062 seconds, despite TM having only 39 states and 73 transitions.}
{Table~\ref{tab:use_case_result_table} also presents the size of the generated PRISM models in terms of the number of states (\#St) and transitions (\#Tr), as well as the model checking time (CTime) for each property across different models. The DC example is the largest and the PAL is the second largest. While the verification of most properties is finished within 0.1s, there are three exceptions. The verification of the third property of TW takes 0.192s because we use a small default duration (1e-5) to get the exact same result (0.2664) as that from the literature. The second property of DC-CTMC was verified over 0.9s and the parametric model checking of PAL-para takes 0.291s.
}

\begin{table*}[ht]
\caption{The CPU time of PRISM code generation for use cases}
\label{tab:cpu_time}\centering
\renewcommand{\arraystretch}{1.2} % Adjust the value as needed
\setlength\tabcolsep{.9mm}
\resizebox{\textwidth}{!}{\begin{tabular}{c*{10}{c}c}
\toprule
\multirow{2}{*}{Activity} & \multicolumn{10}{c}{Use Case CPU time (ms)} & \multirow{2}{*}{Mean} \\
\cmidrule{2-11}
& SD & Para-SD & DC & FP & TW & IS & TM-DTMC & TM-MDP& TM-CTMC & PAL-CTMC&   (ms) \\
\hline

AD validation& 81.2 & 69.4 & 88.7 & 73.7 & 77.8 & 72.1 & 105.5 & 305.6 & 300.5 & 312.1 & 148.7 \\ \hline

ETL transformation & 214.1 & 166.7 & 375.9 & 242.6 & 364.6 & 303.4 & 275.5 & 216.7 & 191.5 & 240.9 & 259.2 \\ \hline

EGL transformation& 62.0 & 67.4 & 75.5 & 59.2 & 72.8 & 78.2 & 67.5 & 62.3 & 61.5 & 61.6 & 66.8 \\ \hline
Total& 357.2 & 303.4 & 540.0 & 375.4 & 515.2 & 453.7 & 448.4 & 584.5 & 553.5 & \textbf{614.5} & \textbf{474.6} \\ 
\bottomrule 
\end{tabular}
}
\end{table*}
\textbf{RQ2 (Efficiency)}. To address the second research question, concerning the efficiency of activity diagram verification, we assess the time required for model transformation from activity diagrams to PRISM code. The focus here is on the efficiency of the transformation process, rather than the subsequent verification in probabilistic model checkers, as the latter reflects the performance of the model checking tools.

The case studies were conducted on a computer with a \textcolor{black}{1.6GHz Core i5 processor and 8GB of RAM}. The times presented in \Cref{tab:cpu_time} are averages derived from 10 individual runs. The table provides a breakdown of the duration for each use case and corresponding activities. 
Overall, the average CPU time needed to complete the entire process, excluding the verification by model checkers, is 474.6 ms across all case studies. The PAL use case exhibits the lengthiest execution time at 614.5 ms.
%This comprehensive analysis provides insights into the computational efficiency of the activity diagram to PRISM code transformation process, offering a nuanced perspective on the performance characteristics of the proposed approach.

%\begin{comment}
%\begin{table}[ht]
%\caption{The CPU time of PRISM code generation for use cases}
%\label{tab:cpu_time}\centering
%
%\begin{tabular}{@{}|c|c|c|c|c|c|c|c|c|c|c|c|@{}}
%\hline
%\multirow{2}{*}{Activity} & \multicolumn{10}{c|}{Use Case CPU time (ms)} & \multirow{2}{*}{Average} \\
%\cline{2-11}
%& SD & Para-SD & DC & FP & TW & IS & TM-DTMC &TM-MDP & TM-CTMC & PAL&   (ms) \\
%
%\hline
%
%EVL& 81.2 & 69.4 & 88.7 & 73.7 & 77.8 & 72.1 & 105.5 & 305.6 & 300.5 & 312.1 & 148.7 \\ \hline
%
%ETL & 214.1 & 166.7 & 375.9 & 242.6 & 364.6 & 303.4 & 275.5 & 216.7 & 191.5 & 240.9 & 259.2 \\ \hline
%
%EGL & 62.0 & 67.4 & 75.5 & 59.2 & 72.8 & 78.2 & 67.5 & 62.3 & 61.5 & 61.6 & 66.8 \\ \hline
%Total& 357.2 & 303.4 & 540.0 & 375.4 & 515.2 & 453.7 & 448.4 & 584.5 & 553.5 & \textbf{614.5} & \textbf{474.6} \\ 
%
%\hline
%\end{tabular}%
%
%\end{table}
%    
%\end{comment}

To evaluate the efficiency of our approach, we also took into account the size of the generated PRISM code and the corresponding verification time. We compared the code size for the case studies TW, IS, and TM with our results.
In the case studies TW and IS, the PRISM code that we obtained from the authors is manually generated according to the transition systems and therefore does not include any control nodes. 
In our experiments with TW and IS, we first converted these transition systems into activity diagrams, and then applied our approach to obtain the PRISM code. The activity diagrams have more nodes than the states in the original transition systems because we need extra \ad{DecisionNode}s and \ad{MergeNode}s (in addition to actions) in the diagrams to deal with probabilistic choice and multiple transitions into one state.

% \begin{table}[ht]
% \caption{The comparison of sizes of manual PRISM code and automatically generated code}
% \label{tab:sizd_of_model}\centering
% \renewcommand{\arraystretch}{1.3} % Adjust the value as needed
% \begin{tabular}{ccccc}
% \toprule
% \multirow{2}{*}{Use case} & \multicolumn{2}{c}{\#States} & \multicolumn{2}{c}{\#Transitions} \\
% \cmidrule{2-5}
% & Manual & Auto & Manual & Auto  \\
% \hline

% TW & 7 & 14   & 9 & 16 \\
% \hline
% IS & 8 & 16  & 13 & 21 \\
% \hline
% TM & 31 & 39  & 61 & 73 \\
% \bottomrule
% \end{tabular}%

% \end{table}

\begin{table}[t]
\caption{{The comparison of sizes and verification time of manual PRISM code and automatically generated code. 
The verification time corresponds to the total time required to check all properties reported in Table~\ref{tab:use_case_result_table}.
For TM, the total verification time is reported separately for DTMC, MDP and CTMC models.}}
\label{tab:sizd_of_model}\centering
\renewcommand{\arraystretch}{1.3} % Adjust the value as needed
\begin{tabular}{ccccccc}
\toprule
\multirow{2}{*}{Use case} & \multicolumn{2}{c}{\#States} & \multicolumn{2}{c}{\#Transitions}  & \multicolumn{2}{c}{Verification time (s)} \\
\cmidrule{2-7}
& Manual & Auto & Manual & Auto & Manual & Auto  \\
\hline

\multirow{1}{*}{TW} & \multirow{1}{*}{7} & \multirow{1}{*}{14}   & \multirow{1}{*}{9} & \multirow{1}{*}{16} & 0.020 & 0.286 \\
\hline
\multirow{1}{*}{IS} & \multirow{1}{*}{8} & \multirow{1}{*}{17}  & \multirow{1}{*}{13} & \multirow{1}{*}{22}& 0.012 & 0.012\\
\hline
\multirow{3}{*}{TM} & \multirow{3}{*}{31} & \multirow{3}{*}{39}  & \multirow{3}{*}{61} & \multirow{3}{*}{73} & 0.017~(DTMC) & 0.163~(DTMC)\\
&&&&&0.016~(MDP) & 0.113~(MDP)\\
&&&&&0.017~(CTMC) & 0.067~(CTMC)\\
\bottomrule
\end{tabular}%
\end{table}
We use the number of states and transitions to measure the size difference as shown in \Cref{tab:sizd_of_model}. 
As both cases are on a small scale but involve several probabilistic transitions, the states of our PRISM code are twice the size of the manual ones, and the transitions are about 1.7 times larger than the manual size %approximately 70\% more, 
due to the introduction of extra control nodes.
But, their scale is within the same order of magnitude.
%\textcolor{purple}{Furthermore, their CPU time for model checking in PRISM is similar. We analyzed both the time for model construction and for property verification. For both case studies, the CPU time for construction and verification is approximately in the order of milliseconds.} {\mycomment (shall be removed?) not valid for TW because we used d= 1e-8 to have a small delta. the smaller the duration, the longer the execution time.}

%In our semantics, all the control nodes are transformed into states of the PRISM models. Therefore, the size of our PRISM code is larger than the ones manually generated without considering the control nodes.

Despite the increase in state space, the verification time does not increase disproportionately.
As reported in \Cref{tab:sizd_of_model}, the verification time for TW and IS remains low for both manual and automatically generated models.
This suggests that, for these case studies, the increase in state space introduced by our approach does not lead to a noticeable degradation in verification performance.

%\begin{comment}
%\begin{table}[ht]
%\caption{The comparison of sizes of manual PRISM models and automatically generated models}
%\label{tab:sizd_of_model}\centering
%\resizebox{0.48\textwidth}{!}{%
%\begin{tabular}{|c|c|c|c|c|c|c|}
%\hline
%\multirow{2}{*}{Use case} & \multicolumn{3}{c|}{\#State} & \multicolumn{3}{c|}{\#Transition} \\
%\cline{2-7}
%& Manual & Auto & Delta & Manual & Auto & Delta \\
%\hline
%
%TW & 7 & 14 & 100\%  & 9 & 16 & 78\%\\
%\hline
%IS & 8 & 16 & 100\%  & 13 & 21 & 62\%\\
%\hline
%TM & 29 & 39 & 34\%  & 57 & 73 & 28\%\\
%\hline
%\end{tabular}%
%}
%\end{table}
%\end{comment}
For the case study TM, we compared the PRISM code that we manually created following the approach in~\cite{Baouya2015a} with our code.
Because the approach in~\cite{Baouya2015a} also introduces the control node into the PRISM code, the numbers of states and transitions of the two versions, as shown in \Cref{tab:sizd_of_model}, are similar.
The size of our code is slightly bigger because we use the module approach (which involves multiple modules) and so additional variables and transitions are required to deal with termination and failure.
Although the automatically generated models incur higher checking times than the manually constructed ones, the overall verification effort remains within a practical and manageable range. This indicates that the additional structural complexity introduced by our approach does not lead to prohibitive verification costs.

{
The evaluation results in Table~\ref{tab:use_case_result_table} and Table~\ref{tab:sizd_of_model} also provide insights into the scalability characteristics of the generated PRISM models.
The size of the generated models depends not only on the number of elements in an AD, but also on the degree of concurrency. In particular, additional states and transitions are introduced to coordinate concurrent execution flows, termination conditions, and failure propagation. %which may substantially increase the state space for highly concurrent models.
If an activity has $n$ concurrent flows, its $i$th flow (where $i$ is from $1$ to $n$) contains $m_i$ nodes. Then the generated PRISM model has up to $\left(\prod_{i=1}^{n} \left((m_i + 1) * f_i * r_i\right) \right) * f_m$ states where $f_i$ is equal to 2 (corresponding to the boolean variable $\meta{to\_be\_terminated}$) if the flow $i$ contains at least one \ad{ActivityFinal} and equal to 1 otherwise; $r_i$ is equal to 2 (corresponding to the boolean variable $\meta{to\_be\_failed}$) if the flow $i$ contains at least one reliability annotation and equal to 1 otherwise; $f_m$ is 2, corresponding to the boolean $\meta{terminated}$ variable in the main module. This is, however, a pessimistic estimation. For the PAL use case, it has two flows and the generated PRISM model has up to $(21 * 2 * 2) * (9 * 2 * 1) * 2 = 3024$ states. However, it only has 126 reachable states, as shown in \Cref{tab:use_case_result_table}. For the DC example, it has four flows and the generated model has six modules (additional two modules corresponding to \ad{MergeNode}) can have up to $(8*2*1)*(3*1*1)*(7*1*1)*(6*1*1)*(4*1*1)*(3*2*1) *2 = 96768$ states. However, it only has 226 reachable states. Two main reasons for the smaller reachable states are the synchronisation between modules and the place where the synchronisation occurs. It is hard to give a precise bound on the state space. 

%Therefore, it is difficult to derive a general mathematical formula that precisely characterises the number of generated states and transitions. Nevertheless, the results in Table~\ref{tab:use_case_result_table} and Table~\ref{tab:sizd_of_model} indicate that the generated overhead remains manageable for the evaluated case studies. Automatically generated models may contain additional coordination structures compared with manually optimised models, which explains part of the observed verification overhead.

% Compared with traditional encodings that may require up to $2^n$ combinations for $n$ activity nodes, our approach represents execution progress using program-counter-style variables, which helps mitigate excessive state-space explosion in practice.
}

{It is well-known, however, that model checking is subject to the state explosion problem. There are time and space restrictions in how large an activity diagram can be verified using our approach. The seven cases that we evaluate are comparatively small. It is part of our future work to evaluate our approach using larger activity diagrams.}
%\textcolor{orange}{The run time of the PRISM model checking for manual code and automatic code is close.}{\mycomment this is valid for TM, but necessary?}

%The time complexity of our approach is discussed in \Cref{subsec:alg}. 
%We mainly focus on the transformation by ETL from an activity diagram to a PRISM model.
%The validation by EVL and the generation by EGL are generally straightforward; therefore, we do not consider the complexity of their implementation.
%The maximum complexity of the ETL transformation is \(O(n^{3})\) or \(O(n^{2})\) if no guarded edges from \ad{DecisionNode}s are in activity diagrams. 
%The complexity of Algorithm~\ref{algorithm:pre-process} is \(O(n^{2})\).
%The complexity of Algorithm~\ref{algorithm:TraverseNds}, who is invoked by Algorithm~\ref{algorithm:pre-process}, is also \(O(n^{2})\). 
%The complexity of Algorithm~\ref{algorithm:MrgProc} is \(O(1)\).
%The complexity of Algorithm~\ref{algorithm:Transformation} is \(O(n^{3})\).

\textbf{RQ3 (Decision Support)}. To address the third question of design decision support, we look at the benefit of our approach through the whole process in \Cref{fig:our_approach_impl}.

In the phase of activity diagram validation, the automated error-checking of activity diagrams can prevent errors from being passed to the next design phase, which may cause additional difficulties in error identification.
The modelling errors to be reported include the ones related to the syntax of the activity diagram, e.g., the name of an \ad{ActivityEdge} is not given, the errors related to the well-formedness of Markov models, e.g., all probabilistic edges from a \ad{DecisionNode} shall have their value within [0,1], and the errors in the properties, e.g., the name of the activity to be verified is not correct.

To facilitate probabilistic model checking, our approach 
{allows APs in properties to be specified at the activity-diagram level using a designer-oriented property notation, which is automatically translated into the property specification languages required} by the underlying model checking tools (e.g., the PRISM Property Specification\footnote{\url{https://www.prismmodelchecker.org/manual/PropertySpecification}}).
%{As a result, end-users are not required to write properties directly in formal languages such as PRISM, PCTL*, or CSL.}
To specify atomic propositions involving, e.g., \lstinprism{pc} variables, termination, and failure (introduced in our semantics for activity diagrams), we introduce several controlled notations in the properties as shown in \Cref{sec:semantics:property}, e.g., \lstinprop{PAL reaches at PAL::R2Stuck}.
Our approach translates them into formal assertions in PCTL* and CSL.

Our approach supports the analysis of an AD of different model types (e.g., DTMC, MDP, and CTMC) without changing the activity diagrams. 
In our approach, a valid activity diagram can be transformed into three types of Markov models, i.e., DTMC, MDP, and CTMC.
By setting the attribute ``toBeVerified'' of the stereotype ``Properties'' as the name of the property group to be verified, the type of Markov models to be generated is selected accordingly, and the corresponding set of properties is parsed.

Another benefit of our approach to the design process is to get the result of system properties as algebraic expressions to derive the proper values of the system parameters at design time and to evaluate the satisfaction of the system properties at runtime.
Our approach supports generating verification results as algebraic expressions through parametric model checking.
We consider the parameters as two types: \emph{controllable parameters}, whose values must be fixed during the design time to meet the system properties (e.g., the probability \textit{p\_c\_a} in the PAL example), and the \emph{observable parameters}, whose values are assumed during design time and are updated with real-time data collections during runtime (e.g., the duration \ad{d\_prep} of the preparation before the robot sets off).

\begin{figure}
    \centering
    \includegraphics[scale=0.5]{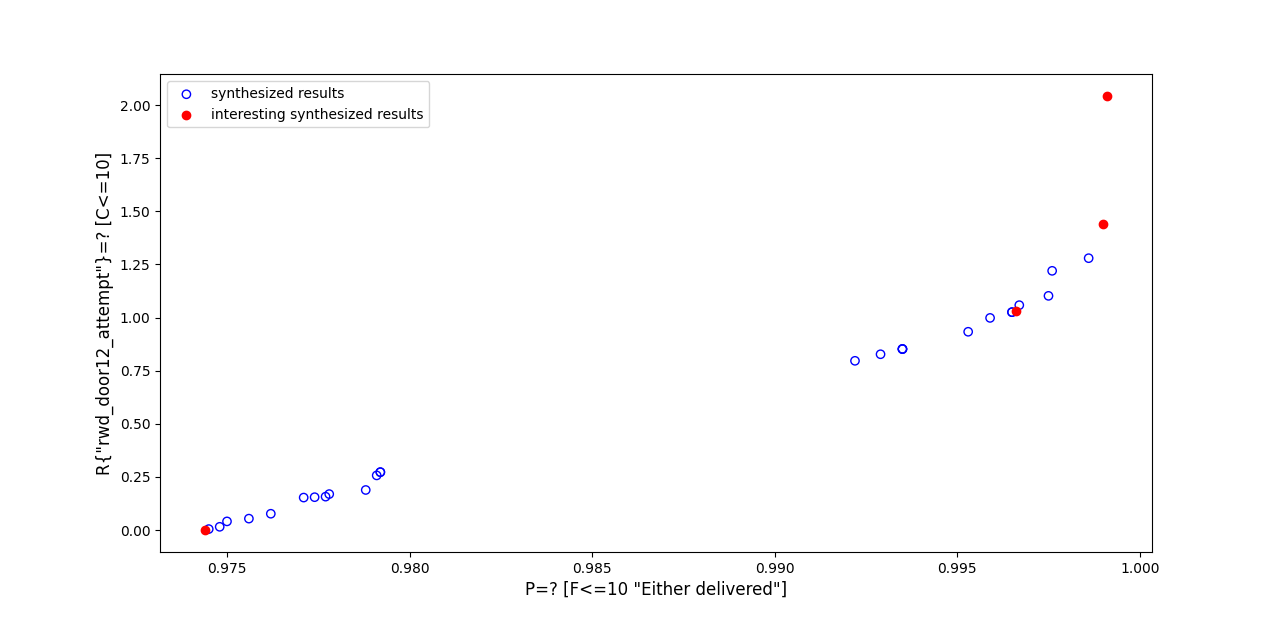}
    \caption{The example of optimal design options for PAL derived by EvoChecker.}
    \label{fig:pal_evocheckre_result}
\end{figure}

For the controllable parameters, after assuming the observable parameters, engineers can obtain the optimised values of the controllable parameters for specified property constraints and optimisation objectives through synthesis.
In our approach, we use the EvoChecker tool to derive the proper values of the controllable parameters and obtain optimised designs. {Firstly, we need to decide QoS constraints and optimisation objectives, which are properties in CSL and are generated by our tool. These properties need to be manually annotated with corresponding marks such as constraints or objectives (either maximising or minimising). Secondly, the generated PRISM model needs to be manually updated to a format that EvoChecker accepts, including prefixing open parameters with \emph{evolve} and specifying their value ranges. Thirdly, EvoChecker is configured to run on this model with the constraints and objectives. Finally, optimisation results are obtained. We note that the manual adaptation of properties and models for EvoChecker can be integrated in our tool to make synthesis fully automated (though we have not integrated it yet).
} 
Taking PAL as an example, if we desire the maximum possibility of either robot's successful delivery within 10 minutes and the minimum cost of ``rwd\_door12\_attempt'', we can obtain a set of design candidates of the probability and reward pairs as shown in \Cref{fig:pal_evocheckre_result}, along with the corresponding set of the values of the controllable parameters, such as \textit{p\_c\_a} and \textit{p\_c\_b}. Then, engineers can choose the designs that fit their requirements better, e.g., the ``interesting synthesised results'' marked as red dots in \Cref{fig:pal_evocheckre_result}.
By then, the algebraic expressions for the properties are fully instantiated.

Furthermore, during runtime, the observable parameters are updated through statistical analysis of the data collected in real-time.
Subsequently, algebraic expressions, serving as key elements during runtime operation, are continually re-evaluated to determine if system properties, especially the safety properties, are satisfied.
In case of a violation, the values of controllable parameters need to be updated to meet the requirements of the properties.

%The capability to support parametric model checking in our approach enabled this 

%\paragraph{Integration with EvoChecker}
%\textcolor{red}{In terms of identifying the system properties, we can make use of EvoChecker tool ~\cite{gerasimou2015search} which is a search-based software engineering approach employing multi-objective optimisation genetic algorithms to automate multi-objective optimisation process and considerably improve its outcome. Through EvoChecker, we can obtain the optimised requirements as design targets. Taking PAL as an example, if we need to have the maximum possibility of either robot's successful delivery and the minimum cost of "rwd\_door12\_attempt", we can obtain a set of requirement candidates of the probability and reward pairs. Then, the engineers can choose from the set the pair that fits the project which is further used to calculate the parameters.}

%, demonstrating the efficiency of the proposed method. Any updates made to the hazard analysis table or the RoboChart models can be reflected in the AC module of the LRE by simply rerunning the generation process.

\subsection{ Threats to Validity}

\textbf{Construct validity} threats may be due to the assumptions made during the modelling of activity diagrams and the definition of probabilistic properties to be verified.
To mitigate this threat, most of the activity diagrams and the corresponding properties used for evaluation (except for PAL) are taken from valid case studies in the literature.

\textbf{Internal validity} threats can arise from the definition of the transformation from activity diagrams to PRISM models and the implementation of the transformation. To mitigate these threats, we conducted a thorough evaluation using the case studies in \Cref{tab:use_case_table} by comparing the verification results of our automatically generated PRISM code with the results in the literature and with the manual calculation results. 

\textbf{External Validity} threats may arise due to the application of our approach to a limited set of systems or scenarios, limiting the generalisability of our approach to systems with significantly different characteristics or domains.
Given that the primary goal of this work is to verify activity diagrams modelled for various systems, it is crucial to ensure coverage of the elements of activity diagrams and consider features from diverse systems and domains.
{To address these concerns, we selected six use cases from the literature based on explicit coverage criteria rather than domain representativeness alone. Specifically, we aim to cover (i) diverse activity diagram structures, including sequential, branching, and concurrent flows; (ii) different types of control nodes and control flows; and (iii) the use of quantitative annotations relevant to probabilistic analysis.}
Each use case's activity diagram incorporates different features with some overlaps among them, collectively covering all the elements and features of the activity diagram semantics discussed in \Cref{sec:semantics}.
In this study, we designed the semantics and transformation processes specifically for \ad{ControlNode}s, \ad{ControlFlow}s, and \ad{ExecutableNode}s.
It is noteworthy that the inclusion of \ad{ObjectNode}s and \ad{ObjectFlow}s is part of our future work.

%\textcolor{red}{Another threat to external validity may arise from the scalability of the approach as the systems in the use cases are of small scale.}

\section{Related work}
\label{sec:relwork}
The topic of using UML diagrams
{ or other graphical notations such as {Business Process Model and Notation (BPMN)}} 
to model probabilistic systems and then applying formal verification techniques to evaluate the qualitative and quantitative properties of these systems is not new. This section discusses previous studies on modelling, verification, and tool automation.
%To the best of our knowledge, our work is the first comprehensive approach for modelling, verification, and automation with the support of various annotations and properties.

UML state machine diagrams, sequence diagrams, and activity diagrams are commonly used to specify systems behaviour. State machines are event-triggered and model reactive systems. Activity diagrams capture the workflow of activities, actions or tasks. The flow is controlled upon the completion of actions, instead of events in state machines. Sequence diagrams capture the interaction of processes in a manner of message sequences. Comparatively, state machines are more complex, and sequence and activity diagrams are less complex and can be easily understood and mastered by more stakeholders in addition to engineers. 

{
    The BPMN is a de-facto standard (ISO/IEC 19510:2013) in business process modelling. Similar to ADs, BPMN also models workflows but specialises for business processes from an end-to-end view, involving stakeholders from different stages (such as business users, analysts, developers etc.), and interactions and (information and data) flows between them. ADs focus more on the modelling of internal behavioural activities within systems or processes.
}

These diagrams and notations are all extended to capture probabilistic behaviour and corresponding verification approaches are developed by several authors. This section discusses the relevant extensions of state machine diagrams and sequence diagrams briefly first, and then focuses on the comparison of the extensions of AD 
{ and BPMN} with our work.

\subsection{State machine diagrams} 
{Gnesi et al.~\cite{Gnesi2000} proposed a stochastic time extension of UML statechart diagrams (SUMLSD) in which a set of clocks is introduced to be associated with both states and transitions. The clocks that are associated with a state will be set to particular values by their individual probability distributions (where clocks are regarded as random variables), as long as the state is entered. The distributions for clocks are general and not restricted to exponential distributions in CTMCs%(because the exponential distribution is the only continuous distribution which is memoryless)
. These clocks will be decreased at the same speed according to an assumed global time.  When a clock reaches 0, it expires. 
A transition can be labelled with a trigger event, an action, and an optional set of clocks. A transition from the current state is enabled only if the event is offered by  the environment and the specified set of clocks expires. SUMLSD also supports composite states, AND or concurrent states, and inter-level transitions.
SUMLSD has an operational semantics given in Stochastic Automata (SA)~\cite{Argenio1999}. SA, in general, cannot be expressed in PTAs or CTMCs (because SA can have multiple clocks with individually more general continuous distributions) and cannot be model-checked by the tools such as PRISM and Storm. Instead, Gnesi et al. used a simulation tool, that supports SPADES~\cite{Argenio1999}, a Stochastic Process Algebra for Discrete-Event Simulation, to analyse SUMLSD diagrams where the diagrams were manually translated to SPADES. SUMLSD has richer semantics than the semantics given to ADs in our approach, but it lacks the support of model-checking as for our approach.
}

Jansen et al.'s P-statecharts~\cite{Jansen2002} is a conservative (so the semantics of existing elements are not changed) probabilistic extension of UML statecharts to allow nondeterministic choice, probabilistic choice, and a priority scheme to resolve nondeterministic choice between transitions in different hierarchies (composite states). The semantics of P-statecharts are given in MDPs, and the P-statecharts models can be verified using the PRISM model checker. {P-statecharts support both AND-nodes and OR-nodes, and can model complex system behaviours. To model systems using P-statecharts properly, users need to understand them very well and to be supported by a good tool, which makes P-statecharts less accessible to engineers, particularly considering that there is no such tool support. %Comparatively, our approach is based on ADs and is easier to model systems with simple workflows (though it is still hard or sometimes not possible to complex systems) and verify them using the integrated tool. 
} 

DAMRTS (Dependability Analysis Models for Real-Time Systems) \cite{Addouche2004,Addouche2005} is a UML profile that extends UML state machines with time and probability. Its semantics is given in probabilistic timed automata (PTA) in PRISM, which allows state machine models to be analysed using PRISM. The authors in~\cite{Baouya2015} use a UML2 profile MARTE (Modelling and Analysis of Real-Time and Embedded systems)\footnote{\url{https://www.omg.org/spec/MARTE/}} to annotate time and probability requirements in UML state machine diagrams. Similarly, its semantics is given in PTA and the annotated models can be analysed using PRISM.

DA-Charts (Dependability Assessment Charts)~\cite{Mustafiz2008}, are also a probabilistic extension of the UML statecharts for capturing dependability requirements such as reliability and safety. The authors implemented a tool in AToM\textsuperscript{3}~\cite{Lara2002} to automatically construct Markov chains and then developed an algorithm for analysis. 

RoboChart~\cite{Miyazawa2019,Ye2022} is a timed and probabilistic domain-specific language for robotics. It is based on a subset of UML state machines but also has a component model with notions of platform, controller, and module to foster reuse for robotic applications. Its probabilistic semantics~\cite{Ye2022} is given in DTMCs and MDPs in PRISM and defined as transformation rules between RoboChart and PRISM. A property language is also designed for RoboChart to specify quantitative properties. An accompanying tool, RoboTool\footnote{\url{https://robostar.cs.york.ac.uk/robotool/}}, is developed to automatically generate semantics for RoboChart models using model-based techniques, and automatically verify the models using PRISM. This work inspired the work presented in this paper. We also use transformation rules to PRISM to define semantics for ADs and implement our tool based on the same Epsilon framework as RoboTool. The PRISM metamodel in our tool, indeed, reuses that in RoboTool with just a small modification for CTMCs. 

{All these discussed statechart-based approaches can model complex system behaviours. %In general, state machines are more expressive than ADs but also more difficult to model systems correctly. 
ADs are simpler to model system behaviours that can be expressed as workflows like a sequence of actions. For such systems, the AD-based approaches like ours are more accessible. For instance, to model the PAL use case in \Cref{fig:pal_use_case}, in RoboChart we need at least two state machines (one for each concurrent flow) and introduce additional events for synchronisation between two state machines with extra concepts like controllers and platforms in its component model. The whole model is not obviously flow-driven so it is not easily linked to the sequence of actions in this use case. Using SUMLSD and P-statecharts could improve the presentation, thanks to their support of composite states and AND-nodes. %However, the main problem with them is the lack of a tool support.
The semantics of these approaches except for SUMLSD are DTMCs, MDPs, or PTAs, which are based on discrete time. Instead, the semantics interpretation of ADs in our approach includes CTMCs, so we can model continuous time. In terms of automation, for some approaches such as SUMLSD, P-statecharts, DAMRTS, and DA-Charts, no tool is available or a tool was developed but not maintained and available now. RoboChart is supported with RoboTool, but it cannot model CTMCs and annotate reliability and rewards directly in its state machines.
} 

\subsection{Sequence diagrams}
pSTAIRS~\cite{Refsdal2005,Refsdal2015} extends STAIRS~\cite{Haugen2003,Haugen2005}, an approach for the use of UML sequence diagrams in step-wise and incremental system development, to allow specification of soft real-time constraints and probability. The formal semantics of pSTAIRS is given as a set of probability obligations and a refinement relation is defined for sequence diagrams for incremental development. Subjective sequence diagrams~\cite{Refsdal2008} extend UML sequence diagrams with probabilistic choice \emph{palt} and further supports a notion of \emph{trust}, a belief (subjective probability) of the trustor on the behaviour of the trustee, to model trust-dependent systems. The analysis includes subjective probability estimates and decisions based on such estimates.

The work in \cite{Tribastone2008} extends UML sequence diagrams to capture timing (the delay) using a rate as a parameter for an exponential distribution. From the diagrams, PEPA~\cite{Hillston1996} (short for Performance Evaluation Process Algebra, a stochastic timed process algebra) descriptions are automatically extracted by an implemented algorithm. PEPA descriptions can be interpreted in the Markovian semantics (CTMCs) or the continuous-state semantics (a system of ODEs).

\subsection{Activity diagrams}
We summarise related work for modelling and verification of ADs in \Cref{tab:related_work}.  

\begin{table*}[ht]
    \caption{Modelling and verification approaches for activity diagrams}
    \label{tab:related_work}\centering
\bgroup
\def\arraystretch{1.0}
\setlength\tabcolsep{.5mm}
    \begin{tabular}{@{}l c ccccc c ccccc c cccc c c c @{}}
        \toprule
        \multirow{2}{*}{Approach} & \phantom{a} & \multicolumn{5}{c}{Activity features} & \phantom{a} & \multicolumn{5}{c}{Modelling annotations} & \phantom{a} & \multicolumn{4}{c}{Verification} & \phantom{a} & \multirow{2}{*}{Auto} & \phantom{a} \\
        \cmidrule{3-7} \cmidrule{9-13} \cmidrule{15-18}
        %& & Probability & Reliability & Rate & Parameters & Rewards & Properties & &
        &
        & Cmn & Param 
        & CBA % CallBehaviorAction
        & EVT & Obj & 
        & Prob & Relb & Rate & Rew & Prop & 
        %Verification
        & Markov & pMC %parametric model checking
        & Syn & NoC &
        & \\
        \midrule
        %Synchronous Timed AD
        STAD~\cite{Jarraya2007} & & \checkmark%\ding{51}
        &  &  &  &  & & \checkmark &  &  &  &  & & DTMC &  &  & 1 & &  & \\
        SEC~\cite{Ouchani2011} & & \checkmark & & \partially & \checkmark & \checkmark & & \checkmark & & & & & & MDP & & & 1 & & & \\
        NuAC~\cite{Ouchani2013} & & \checkmark & & \checkmark & \checkmark & \checkmark & & \checkmark & & & & & & MDP & & & 2 & & \checkmark & \\
        AC~\cite{Jarraya2010} & & \checkmark & & \partially & \partially & & & \checkmark & & & & & & MDP & & & 1 & & & \\ 
        TAC~\cite{Baouya2015a} & & \checkmark & & \partially & \partially & & & \checkmark & & & & & & PTA & & & 1 & & \checkmark & \\
        DAC~\cite{Baouya2021} & & \checkmark & & \partially & \partially & & & \checkmark & & & & & & DTMC & & & 1 & & \checkmark & \\
        MIE~\cite{Debbabi2010a} & & \checkmark & & & & & & \checkmark & & & & & & MDP & \checkmark & & 1 & & & \\
        ATOP~\cite{Gallotti2008} & & \checkmark & & & & & & \checkmark & \checkmark & \checkmark & & & & DMC & & & 1 & & \checkmark & \\
        AD-CSP\cite{Lima2020} & & \checkmark & \checkmark & \checkmark & \checkmark & \checkmark & & & & & & & & CSP & & & 1 & & \checkmark & \\
        Our & & \checkmark & \checkmark &  &  &  & & \checkmark & \checkmark & \checkmark & \checkmark & \checkmark & & DMC & \checkmark & \checkmark & 7 & & \checkmark & \\ 
        \midrule
        \multicolumn{21}{p{.95\linewidth}}{\textbf{Acronym}: 
        Auto: Automation;
        CBA: CallBehaviorAction; 
        Cmn: Common; 
        DMC: DTMC, MDP, and CTMC;
        EVT: Event-based actions, such as \ad{AcceptEventAction} and \ad{SendSignalAction};
        %MDE: Model-driven engineering;
        %NA: Not applicable;
        %NC: Not clear;
        %N-MDE: Not model-driven engineering;
        NoC: Number of case studies;
        Obj: \ad{ObjectNode} and \ad{ObjectFlow};
        Param: Parameter;
        pMC: Parametric model checking;
        Prob: Probability;
        Prop: Property;
        Relb: Reliability;
        Rate: Duration or rate;
        Rew: Reward;
        Syn: Synthesis;
        %TD: To be done (or future work);
        %Y: Yes;
        %YN: Yes and No;
        \partially: partially (supported in some stages but not in all stages), or claimed to support but no further details disclosed.
        } \\
        \bottomrule
\end{tabular}
\egroup
\end{table*}

\cite{Jarraya2007} extends {SysML activity diagrams} (SAD) with time constraints to capture both probabilistic and timed requirements in activities. The semantics of an activity diagram is a DTMC where each module, corresponding to a concurrent flow, has a clock variable of type integer numbers to record the (discrete) time steps passed in actions on this flow. Particularly, the progression of all clock variables is synchronous through a \emph{step} event. The approach then uses PRISM to verify the manually generated DTMC models by an algorithm. Our approach presented here supports the DTMC, MDP, and CTMC semantics where time requirements can be captured in the duration or rate annotations and analysed using CTMCs. Importantly, our approach is fully automatic {in terms of model transformation, PMC, and pMC.}

In \cite{Jarraya2010}, the author designed a formal logic, Activity Calculus (AC), to capture the semantics of SAD. AC has operational semantics which is based on probabilistic transition systems (PTS). From a PTS, an MDP can be derived. The author also gives {an operational semantics to the PRISM language.} So the soundness of a translation from an AD to PRISM can be established by their (AC and PRISM) operation semantics. SAD will be translated to PRISM models for analysis using PRISM. 
Then \cite{Baouya2015a} extends this approach with time, and correspondingly defines Timing Activity Calculus (TAC), giving operational semantics to TAC and PRISM based on Probabilistic Timed Automata (PTA). A Java-based tool was implemented to automate the translation process from AD to PRISM. 
Another work~\cite{Baouya2021} considers the reliability-driven software deployment using parametric probabilistic model checking for SAD that captures software behaviour, where one kind of such parameter is software component failure rate. However how these parameters are captured in SAD is not clear, while our approach uses input parameters of AD to model parameters. The work gives SAD semantics in a logic: Deployment Activity Calculus (DAC), which is mapped to DTMC models in PRISM. An Eclipse plugin is developed to automate the transformation and verification process. 
These studies support a large set of activity features (such as \ad{CallBehaviorAction}, \ad{Events}, \ad{ObjectNodes}) in their AC and TAC, but how these features are mapped to PRISM is not clear (without examples illustrated). When we are investigating our approach, we indeed take into account the support of these features (though they are not implemented). The time in~\cite{Baouya2015a} is real-time clocks and interpreted in PTAs, while the time in our approach is duration or rate and interpreted in CTMCs. They cannot support parameters, the annotations of reliability, reward, and property, and parametric verification and synthesis as we do.

The work in \cite{Debbabi2010a} gives MDP semantics to SAD and can capture multiple instances of executions (so each \ad{ActivityNode} can hold more than one token at the same time) in the semantics. Algorithms have been designed for the translation of SAD to PRISM models, but the process is not automated. Comparatively, our approach tackles one instance of executions and allows only one token at most in each \ad{ActivityNode} at the same time, and our approach is fully automatic {in terms of model transformation, PMC, and pMC}. %This work does not support time, parameters, the annotations of reliability, reward, and property as we do.

AD is also used to capture security requirements for quantitative analysis using probabilistic model checking. \cite{Ouchani2011} presents security attack patterns and scenarios in SAD. They are mapped to MDP models in PRISM. The security requirements are finally verified using PRISM. Interestingly, object nodes and flows are mapped to multiple modules in PRISM where one module simulates a one-place buffer for each \ad{ObjectNode} (to support asynchronous communication). We aim to support object nodes and flows in the future. Our approach is similar to this work to have one PRISM module to simulate a buffer for each \ad{ObjectNode}, but our approach requires two commands for sending data to the buffer and one command to read data from the buffer, instead of two commands for both reading and writing in~\cite{Ouchani2011}. Further, \cite{Ouchani2013} designed a New Activity Calculus (NuAC) to capture the semantics of SAD. The operational semantics for both NuAC and PRISM MDP models are presented to establish the soundness of the translation from SAD to PRISM models. A tool is developed to automate the transformation and verification.

These studies~\cite{Jarraya2007,Jarraya2010,Debbabi2010a,Ouchani2011,Ouchani2013,Baouya2015a,Baouya2021} all use a similar approach to represent \ad{ActivityNode}s in PRISM models. A node \ad{nd}, for example, corresponds to one Boolean variable in PRISM whose definition is (\lstinprism{nd: bool init false;}). That the variable is evaluated to be true (or false) denotes that the corresponding node is holding (or not holding) a token and executing (or not executing). %In these studies except \cite{Ouchani2011}, the corresponding PRISM model is composed of a single module. They support \ad{CallBehaviorAction}, events, and object nodes in their semantics and transformation rules. But there are not enough details and no evaluation of the support. 
If an AD has $n$ nodes, $n$ variables are declared in PRISM. So the state space is $2^n$. For example, the PAL use case in \Cref{fig:pal_use_case} has 27 nodes so the corresponding state space would be $2^{27}=\SInum{134217728}$. Our approach uses a \lstinprism{pc} variable for each concurrent flow to record its state. The PAL use case has two concurrent flows: 20 nodes in one flow and 8 nodes in another flow. We note that in our approach the \ad{JoinNode} \ad{J1} will appear in both flows so the total number of nodes in both flows is 28 instead of 27. An additional \lstinprism{INACTIVE}, whose value is -1, is required. Finally, the two \lstinprism{pc} variables for both concurrent flows could take 21 ($=20+1$) and 9 ($=8+1$) values respectively. We also need four Boolean variables \mylines{32}{34} and \myline{62} in \Cref{fig:transform_activity_pal_use_case} to deal with termination and failure. So the state space size in our approach is $21*9*2^4=\SInum{3024}$, much less than that in their studies. We, however, need to emphasise that the number of reachable states by our approach, which is the number reported in verification results by PRISM, is larger than theirs because of the additional Boolean variables for termination and failures, as illustrated in \Cref{tab:sizd_of_model}.  
Their semantics support DTMCs, MDPs, or PTAs, but not CTMCs as we do. They have not supported the modelling and analysis of reliability, rewards, and parameters too.

ATOP~\cite{Gallotti2008} uses a subset of the MARTE profile to provide quality annotations to ADs for service compositions. The annotations include service reliability, execution time, probability, and invocation attempts. An annotated diagram then is automatically translated to a PRISM model according to the specified Markov model: a DTMC, MDP, or CTMC. Finally, the PRISM model is verified using PRISM. Our work is inspired by ATOP. We have similar annotations and support the same three Markov models as ATOP. But ATOP supports only one \ad{InitialNode} and one \ad{ActivityFinal}, while our approach can handle multiple \ad{InitialNode}s and \ad{ActivityFinal}s. ATOP focuses on the engineering side (workflow, implementation, and evaluation) of the approach, but lacks the most important part: the semantics interpretation. Even no PRISM model is illustrated. The ATOP tool is also not available. Without these, we are not able to have an in-depth evaluation of the approach. For example, because of the lack of these details, we are not able to further analyse the deviation of TM-CTMC between our approach and ATOP. Comparatively, our work provides a detailed semantics interpretation of ADs and transformation rules are defined in \Cref{sec:semantics}.
Our tool is open-source and available online. Our work additionally supports parameters and parametric model checking using PRISM, and the reward annotation to verify reward-related properties. These are not provided by ATOP.

ADs may have unexpected behaviours such as deadlock, for example, caused by all guards on the outgoing transitions of a \ad{DecisionNode} evaluated to be false, or nondeterminism due to overlapped guards on these transitions. Lima et al.~\cite{Lima2020} present a compositional CSP~\cite{Hoare1985,Roscoe2011,Cavalcanti2006} (Communicating Sequential Processes, a process algebra for modelling concurrent systems) semantics for AD. A plugin is implemented in the Astah modelling environment to automate the process from semantics generation to verification by FDR~\cite{GibsonRobinson2014}, a refinement model checker for CSP. The tool also supports traceability between AD and generated CSP specifications. This is used to highlight the problematic flows in the diagrams that cause deadlock or nondeterminism by mapping the counterexamples found by FDR back to the diagrams. This work supports a wide variety of \ad{ActivityNode}s such as input and output parameters, \ad{CallBehaviorAction}, \ad{SendSignalAction}, \ad{AcceptEventAction}, and \ad{ObjectNode} and \ad{ObjectFlow}, but not probability and rate extensions. It can only analyse qualitative properties, while our approach can analyse both qualitative and quantitative properties. For the detection of deadlock, we can implement it by checking the reachability of the final states or measuring the termination probability. %For the detection of nondeterminism, we could check both the maximum and minimum probabilities of interesting properties. If they are not equal, this indicates nondeterminism in the diagrams.  

{
\subsection{BPMN} 
The work in~\cite{Prandi2008} gives the BPMN a semantics in the stochastic extension of COWS~\cite{Prandi2007}. The COWS semantics can be interpreted in CTMCs, enabling the quantitative analysis of BPMN models using PRISM via COWS. The translations from BPMN to COWS and from COWS to PRISM, however, are not formally discussed and automated. 
 
Braghetto et al.~\cite{Braghetto2011} interpreted a BPMN diagram as a compositional Stochastic Automata Network (SAN) model~\cite{Plateau1985}, and a system composed of $n$ BPMN process instances is just $n$ parallel instances of SAN models. The conversion from a BPMN diagram to SAN is formally defined and implemented in a tool for automation. The generated SAN models are analysed using the PEPS tool~\cite{Brenner2007a}. Compared to PRISM whose semantics is not compositional, SAN has an advantage of analysing systems or processes with multiple parallel instances (thanks to its compositionality), as in BPMN models. The main issue of this approach is the lack of the tool support because PEPS is discontinued.

Herbert et al.~\cite{Herbert2012,Herbert2014a} extended the BPMN with nondeterministic probabilistic branching and reward annotations. In their extension, each flow $f_i$ from an exclusive gateway is annotated with a label and probability pair $(l_i,p_i)$. Multiple flows from this gateway could share the same label but their probabilities must sum to 1, denoting the choice between these flows is probabilistic. Then the choice between different labels for such flows from the gateway is nondeterministic. On the contrary, our work here targets ADs with no explicit modelling of nondeterminism. The task nodes in their extension can be annotated with rewards, for example, to capture the time spent to perform a task, or failure rates (interpreted as fail-stop or fail-continue behaviours)~\cite{Herbert2013a,Herbert2014a} used to model unreliability in business processes and subsequently generate fault trees. Our work is also capable of modelling reliability but we only consider the fail-stop behaviour now. Herbert et al. also utilised the BPMN extension for the synthesis of optimal schedules and the optimisation of business processes~\cite{Herbert2015,Herbert2016}. 
%Herbert et al.~\cite{Herbert2013a,Herbert2014a} also considered the generation of fault trees from their BPMN extensions through PRISM. A BPMN task node can be annotated with one of two failure rates interpreted as fail-stop or fail-continue behaviours.
The semantics of extended BPMN models is captured in MDPs to be analysed using PRISM. Time in ADs is \emph{implicit} if interpreted as DTMCs or MDPs, and \emph{explicit} (in annotated rates for exponential distributions) if interpreted as CTMCs. Time in their BPMN extension is captured in rewards, so the properties related to time is the reward properties in PRISM (that is, the expected time). Their extension and semantics cannot model and analyse quality of service in terms of time like we do in TM-CTMC and PAL-CTMC in \Cref{tab:use_case_result_table}.
A comprehensive tool, called SBOAT~\cite{Herbert2014a}, was developed to fully automate their approach for different applications. These provide us insights to use our approach for the similar applications in the future. SBOAT is implemented in .NET and built on top of PRISM. It, however, is not accessible and not open-sourced due to the confidentiality requirement from their collaboration with industry partners. On the contrary, our \method\ is open-sourced and accessible.

Kikuchi~\cite{Kikuchi2014} developed an approach to evaluate the availability and cost of business workflows (described in BPMN) using public SaaS offerings such as e-mail and spreadsheet services. Only a limited subset of nodes, including start, finish, task, and decision nodes, is supported. Other nodes related to merge and parallel are not supported. The semantics of such workflows is given in MDPs and then PRISM is used to analyse these quantitative measurement. No tool, however, is available to automate the semantics generation from the workflows.

Ceballos et al.~\cite{Ceballos2015a,Ceballos2018} considered a subset of BPMN constructs as a normal form to model human activities, proposed an automatic transformation procedure to translate BPMN diagrams to Causal Bayesian Networks to support human decision making. Their approach does not support BPMN diagrams with loops or cycles and no tool is available to automate the transformation, which limits its application. 

Duran et al.~\cite{Duran2018a} extended BPMN diagrams with annotations of time duration for tasks and flows, and probabilities for split (exclusive and inclusive) gateways. Time can be constant or sampled from various probability distributions such as normal and uniform distributions. % to represent the duration of a task or a flow taken.
Then the subset of BPMN is mapped to an executable specification in rewriting logic: PMaude~\cite{Agha2006} to be evaluated using a discrete-event simulation and statistical model checking. A plugin extending VBPMN~\cite{Krishna2017} automates the transformation of a BPMN process to PMaude. The work was further extended to support the annotation and analysis of resource allocation~\cite{Duran2019} for multiple instances. Our approach introduces annotations for time (rate or duration) and probabilities but not the resource allocation  (though it can be modelled as rewards), and automates the transformation. The rate or duration, however, in our work is interpreted in exponential distributions in CTMCs. No other distributions can be supported in CTMCs because the exponential distribution is the only continuous distribution which is memoryless. Different from the approximated analysis in their approach, we use PRISM for precise (parametric) probabilistic model checking.

The approach by Dechsupa et al.~\cite{Dechsupa2022} proposed the annotation of tasks in BPMN with time constraints such as delay, schedule, or stochastic (described as exponential probability distributions), the definition and automation of the transformation rules from BPMN to the coloured generalised stochastic Petri net (CGSPN)~\cite{Gharbi2009}, and the stepwise verification to verify separate components first, then manually abstract these components' behaviours to form their composition, and finally simulate the composition in Snoopy, a tool for the analysis of CGSPN. The analysis process is not fully automated and the verification is based on simulation. Instead, our approach is fully automated (except for synthesis) and is supported by probabilistic model checking (not simulation). 

Falcone et al.~\cite{Falcone2022a} considered runtime verification of BPMN processes. Different from other related work to extend the BPMN notation, the approach works on a subset of executable BPMN processes without the annotations of time and probability. A BPMN model is transformed into a labelled transition system (LTS) using VBPMN. The model is also executed in the Activiti framework\footnote{\url{www.activiti.org}} for multiple times. Runtime information is collected and stored in a database, from which execution traces are extracted. Then the probability for each transition in LTS can be statistically calculated from the traces. With the calculated probability information, the LTS is transformed into a probabilistic transition system (PTS) to be model-checked in CADP~\cite{Garavel2012}, along with properties specified in the original BPMN model. In~\cite{Falcone2024}, the authors further considered the enforcement (via reordering or buffering) to output a list of executable tasks that will not violate specified properties. This procedure is fully automated and can be iterated periodically to support verification at runtime. Comparatively, our current approach focuses on verification at design-time. The support of runtime verification and monitor is part of our future work. Their method can deal with probability but not include other quantitative measurements like failure, time duration, and rewards as in our approach. 
}
{
\subsection{Summary} 
The \method\ framework distinguishes itself from these work by offering a high-level analysis workflow rooted in the ``Hidden Formal Methods'' paradigm. Unlike many existing approaches that require deep expertise in formal notations, \method\ enables engineers to perform complex quantitative analysis while remaining entirely at the modelling level. Our workflow is characterised by several key aspects. 

{
  \textbf{Advanced QoS Annotations.}
The framework introduces a dedicated UML profile that exposes specific stereotypes for annotating complex QoS information directly on nodes and edges. This includes reliability (probability of mission success), duration, rates, and rewards. This level of integrated annotation is often fragmented or missing in existing literature.
}

{
  \textbf{End-to-End Automated Model-Driven Engineering (MDE)}. While previous research often involved fragmented or manual transformations, \method\ provides a fully automated {(in terms of model transformation, PMC, and pMC)} workflow based on MDE principles.
} 

{
\textbf{High-level Integrated Property Specification in Modelling.}
A significant differentiator is how properties are specified. In many existing works, stakeholders must manually transform properties into temporal logic or interpret generated code to map formal states back to the original AD components. 
\method\ offers a CNL for specifying atomic propositions directly at the AD level.
These high-level properties (e.g., ``activity reaches a specific node'') are automatically transformed into formal specifications (PCTL* or CSL) for the model checkers. This allows users to focus on system requirements rather than the underlying state-space variables like program counters (pc).
}

{
\textbf{Single-Source Multi-Formalism Support.}
\method\ allows a single activity diagram to be analysed as different types of stochastic models (DTMC, MDP, or CTMC) depending on the property being verified.
Most previous works are limited to a single type of probabilistic model or do not provide a rigorous, automated mapping for multiple variants.
}

{
  \textbf{Support for Design-Space Exploration and Synthesis.}
 Beyond simple verification, the \method\ workflow integrates parametric model checking (pMC) and synthesis. 
 Users can start with ``open'' activity diagrams where certain QoS values are unknown. The workflow can utilise engines like EvoChecker to extract optimal parameter values that satisfy specific QoS trade-offs. These optimised values can then be manually updated back into the original AD model, supporting an iterative design process.
}

}

\section{Conclusion and future work}
\label{sec:concl}
This work proposes a comprehensive verification framework for systems whose complex behaviours (including probability, reliability, and time) are captured in ADs to address problems related to evaluation, extensibility, adaptability, and accessibility, in current practice and research. 

In our framework, we developed a UML profile to annotate ADs with these advanced features along with cost or rewards associated with activity edges and properties to be verified. The transformation rules from ADs to PRISM, and the corresponding \changed[\Cn{2}{2}]{transformation implementation} play a key role in our approach for automation. Underneath the transformation is our semantic interpretation of ADs as three Markov models. 

To make our work amenable to evaluation, we present our framework in details in this paper, evaluated seven case studies with different features using different Markov models, and make our tool open-source. To make it \emph{extensible}, we use a modular approach to map each concurrent flow in ADs into a module in PRISM, which entitles us to further add more features such as \ad{CallBehaviorAction}, event-based actions, object nodes and flows into our framework using a conservative way (without a change of behaviour for currently supported features). Our research is open and our tool is open-source, which could make our work \emph{adaptable} for other researchers or engineers to apply in other domains or implement in other tools (instead of Eclipse and Epsilon), or use it as a front end to capture their labelled transition systems. To make it \emph{accessible} by engineers, we automated the whole workflow and so users can focus on the modelling and verification parts instead of intermediate steps. 

\subsection{Discussions, limitations, and future work}
\textbf{AD features.} Our approach presented in this paper cannot deal with all features in ADs, and our immediate future work is to extend it in order to overcome current limitations. The work discussed in this paper deals with only one activity, which limits its application to ADs with a hierarchical structure, for example, by using \ad{CallBehaviorAction}. The first extension to our framework is to support \ad{CallBehaviorAction} because this action is very useful in practice to model complex systems. %Thanks to our modular approach for PRISM models, the called activity of an \ad{CallBehaviorAction} is nothing different from the work discussed here for one activity except the pass of control from its caller when called and to the caller when terminated. 
Because a \ad{CallBehaviorAction} action could also call other actions, this would be nested calls. We will use a staged start and termination for each call to allow appropriate interpretation of semantics in PRISM. \ad{CallBehaviorAction} also allows its defined behaviour to be reused for multiple instances. Because PRISM does not support such ``functional-level'' (above module-level in PRISM) abstraction, if an activity is called in several instances, then each instance would be treated differently and must be mapped to PRISM individually.

To introduce and support object nodes and flows in our framework, we need synchronous communication in PRISM. Communication requires a way to exchange data on particular channels. PRISM uses action labels for synchronisation between modules to identify channels and variables for data exchange. %An example is shown in \Cref{fig:prism_comm} to simulate communication $c!1$ and $c?x$ between two modules. 
In addition to a proper synchronous communication, asynchronous communication is also required to implement buffers to support object nodes such as \ad{CentralBufferNode}s. We will use an extra module in PRISM to simulate a buffer for each \ad{ObjectNode}.  %We illustrate such one example in \Cref{fig:prism_comm_buffer}. 
%
% \begin{figure}[ht]
%   \centering
% \begin{lstlisting}[language=PRISM,]
% module M1 // simulate c!1
%   c_data : [0..10] init 0;
%   [inbuf] _ -> (c_data'=1)&_;
%   ... 
% endmodule
% 
% module buffer // simulate one-place buffer
%   idle : bool init false;
%   empty : bool init true;
%   data : [0..10] init 0;
%   
%   [inbuf] (idle=true) -> (empty'=true)& (idle'=false);
%   [] (idle=false)-> (data'=c_data)&(idle'=true)&(empty'=false);
%   [outbuf] (empty=false)&(idle=true) -> (empty'=true);
% endmodule
% 
% module M2 // simulate c?x
%   x : [0..10] init 0;
%   [outbuf] -> (x'=data)&_;
%   ... 
% endmodule
% \end{lstlisting}
%   \caption{An example of an implementation of asynchronous communication in PRISM through a buffer module.}
%   \label{fig:prism_comm_buffer}
% \end{figure}
%
% In this example, \lstinprism{M1} and \lstinprism{M2} exchange message (\lstinprism{1}) through a buffer module \mylines{7}{15}. This buffer module has one Boolean variable \lstinprism{idle} to indicate whether it is busy with accepting input or not, and another Boolean variable \lstinprism{empty} to indicate whether the buffer is empty or not, in addition to a variable \lstinprism{data} to hold the data for exchange in the buffer. The command \myline{12} prepares \lstinprism{buffer} for accepting input and the command \myline{13} actually moves the data for exchange into the buffer. The command \myline{14} removes the data from the buffer and at the same time \lstinprism{M2} stores it in \lstinprism{x}. 
With synchronous and asynchronous communication, we can extend our framework to support object nodes and flows. 
The support of event-based actions, specifically \ad{SendSignalAction} and \ad{AcceptEventAction}, also relies on synchronous communication. %illustrated in \Cref{fig:prism_comm}.

{Our approach assumes maximum one token in each node and cannot support the semantics of ADs with multiple tokens. We will investigate this in line with multiple tokens in Petri nets in the future.}

{Our approach imposes a condition~\ref{wfc:out_edge_forknode:not_loops} to avoid dynamically created concurrent flows. This is one of the limitations of our approach. }
{Generalising the semantics to action nodes with multiple incoming and outgoing edges -- removing \ref{wfc:in_edge_action:one} and \ref{wfc:out_edge_action:one} -- remains an immediate direction for future work.}

{
\textbf{Consistency between the transformation rules and the implementation.} As discussed in~\Cref{rmk:rules_algorithm}, our implementation in algorithms is not directly mapped to the corresponding transformation rules, which results in a gap to establish the soundness of the implementation using the rules. Our immediate future work is to implement a pure functional algorithm (instead of imperative here) to map the rules directly to functions in the  implementation, for example, using the same data types, parameters, bodies, and returns. This entitles us to evaluate the performance differences between the imperative and declarative implementations.
}

\textbf{Other Markovian models.} %\cmt{Discuss modelling limitations introduced by CTMCs where time is the mean duration (using an exponential distribution), and not realtime. Consider other expressive models, such as PTA and Markov automata in Storm.}
In DTMCs and MDPs, time is discrete and encoded into transitions. In CTMCs, time is modelled by exponential distributions (parametrised with rates) for states. {However, we could not capture real-time behaviours (e.g., a system will stay in a state for at least 1 second and up to 3 seconds, or exactly two seconds) using these models.} Probabilistic timed automata (PTAs)~\cite{Norman2012}, another Markov model, extend MDPs with clocks to model real-time behaviour. PTAs support discrete probabilities, continuous time, and nondeterminism. The choice between multiple transitions and the elapse of time are nondeterministic, and the choice between target states, after a transition is chosen, is probabilistic. We will extend our framework to support an annotation (similar to \cite{Baouya2015}) of real-time for \ad{Action}s, interpret the semantics of ADs in PTAs, and define transformation rules from ADs to PTA models in PRISM. 

{Another Markovian model of interest is Markov automata (MAs)~\cite{Eisentraut2010} which combines CTMCs with MDPs. So an MA supports both continuous-time by exponential distributions and nondeterminism. Storm~\cite{Christian2017} is able to verify MA models. We can give ADs the semantics in this more expressive model in the future, and so we can model nondeterministic behaviour in ADs explicitly.}

{For CTMCs, our approach introduces a small mean duration or a passive rate to simulate immediate transitions for control nodes, which results in the common stiffness problem in analysis. In general, there are two ways to address this issue. One is to construct CTMC models without the need of intermediate transitions and another is to use non-Markovian models that support other continuous probability distributions (instead of exponential distributions in CTMCs), such as GSPNs. This is part of our future work to investigate further in this aspect.} 

\textbf{Annotations}. We are also interested in other QoS annotations such as Service Invocations Attempts and Service Degradation Functions as discussed in~\cite{Gallotti2008}. They could be very useful, for example, in our PAL use case if we need to model the maximum number of attempts Base 1 can make to pass through two doors. In our current model shown in \Cref{fig:pal_use_case}, we assume an infinite number of attempts. This is not ideal in practice. With these annotations, we can improve the modelling. We will give a semantics interpretation of these annotations, and define transformation rules.

{\textbf{Property specifications}. Our tool supports basic atomic propositions (APs) in the property specification for the three Markovian models using PCTL* or CSL, as explained in \Cref{sec:semantics:property}, including the APs involving \lstinprism{pc}, \lstinprism{terminated}, \lstinprism{to_be_terminated}, and \lstinprism{to_be_failed}. Our tool assumes no knowledge from users about these variables introduced in the transformation. Instead, we provide a syntax to let them specify APs in terms of ADs, as explained in \Cref{sec:semantics:property} and exemplified in \Cref{fig:pal_use_case:properties}. However, currently we still require users' knowledge in PCTL* and CSL to write property specification other than these APs. This is part of our future work to design a domain-specific property language and develop an accompanying tool.}

%{\textbf{Well-formedness conditions}. \Cref{sec:semantics:wfc} lists all necessary well-formedness conditions to make our transformation valid. These conditions are rather complete. In the future, we aim to add more conditions when we are extending the current work.} 

{\textbf{Case studies}. We have studied seven examples in fifteen different configurations using three different Markovian models. However, these examples contain a limited number of nodes, up to 27 in the PAL example. We will investigate larger activity diagrams from different sources in the future.}

{\textbf{Synthesis integration}. The adaptation of models and properties for synthesis will be integrated into our workflow to make it fully automated in the future.}

For all these extensions, we will implement them in our tool, evaluate case studies from literature, and continue to publish and open our research, the tool, and evaluation results for communities.

\section*{Acknowledgements}
This work has been partially supported by the European Union's Horizon 2020 project: Secure and Safe Multi-Robot Systems (SESAME -- \url{www.sesame-project.org})
% \footnote{%
% SESAME Project: \url{www.sesame-project.org/}%}, 
under grant agreement no. 101017258.

\begin{appendices}
{
  \section{The corresponding PRISM model to the PAL use case}
The generated PRISM model for the PAL use case is sketched in \Cref{fig:transform_activity_pal_use_case}. The complete PRISM model can be found online.\footnote{\url{https://github.com/RandallYe/QANAD/blob/master/Examples/PAL_use_case/prism_models/p_ctmc/pal.prism}} 
}
\begin{figure}

\begin{lstlisting}[
  language=PRISM,
  basicstyle=\ttfamily\scriptsize,
  escapeinside={(*@}{@*)},
  linewidth=\textwidth,
  lineskip=1.3pt,
  breaklines=true,
  breakatwhitespace=false,
  columns=fullflexible,
  keepspaces=true,
  showstringspaces=false,
  postbreak=\mbox{\textcolor{gray}{$\hookrightarrow$}\space},
  linebackgroundcolor={
    \ifnum\value{lstnumber}=3\color{altgreen}\fi
    \ifnum\value{lstnumber}>7 \ifnum\value{lstnumber}<12\color{altgreen}\fi\fi % 8--11
    \ifnum\value{lstnumber}>19 \ifnum\value{lstnumber}<22\color{altgreen}\fi\fi % 20--21
    \ifnum\value{lstnumber}=23\color{altgreen}\fi % 23
    \ifnum\value{lstnumber}>24 \ifnum\value{lstnumber}<29\color{altgreen}\fi\fi % 25--28
    \ifnum\value{lstnumber}>29 \ifnum\value{lstnumber}<32\color{altgreen}\fi\fi % 30--31
    \ifnum\value{lstnumber}=33\color{altgreen}\fi % 33
    \ifnum\value{lstnumber}>34 \ifnum\value{lstnumber}<37\color{altgreen}\fi\fi % 35--36
    \ifnum\value{lstnumber}>40 \ifnum\value{lstnumber}<43\color{altgreen}\fi\fi % 41--42
    \ifnum\value{lstnumber}>44 \ifnum\value{lstnumber}<51\color{altgreen}\fi\fi % 45--50
    \ifnum\value{lstnumber}>52 \ifnum\value{lstnumber}<55\color{altgreen}\fi\fi % 53--54
    \ifnum\value{lstnumber}>59 \ifnum\value{lstnumber}<62\color{altgreen}\fi\fi % 60--61
    \ifnum\value{lstnumber}>62 \ifnum\value{lstnumber}<65\color{altgreen}\fi\fi % 63--64
    \ifnum\value{lstnumber}>66 \ifnum\value{lstnumber}<69\color{altgreen}\fi\fi % 67--68
  }
]
ctmc
const int INACTIVE = (-1);
const double mean_duration = (0.001);(*@\CodeNote{Duration and CTMCs (S5.2.12)}@*)
const int PAL_I0_E1_I0 = (0);(*@\CodeNote{Corresponding nodes in the 1st concurrent flow}@*)
const int PAL_I0_E1_Prepare = (1);/*
const int PAL_I0_E1_F1 = (2);
const int PAL_I0_E1_R1ToCorA = (3);
const int PAL_I0_E1_R1ToDoorAB = (4);
const int PAL_I0_E1_D1 = (5);
const int PAL_I0_E1_M1 = (6);
const int PAL_I0_E1_M3 = (7);
const int PAL_I0_E1_D2 = (8);
const int PAL_I0_E1_R1ToCorBThrDoor1 = (9);
const int PAL_I0_E1_R1ToCorBThrDoor2 = (10);
const int PAL_I0_E1_D3 = (11);
const int PAL_I0_E1_D4 = (12);
const int PAL_I0_E1_M2 = (13);
const int PAL_I0_E1_R1ToRoomD = (14);
const int PAL_I0_E1_D5 = (15);
const int PAL_I0_E1_R1DeliveredRoomD = (16);
const int PAL_I0_E1_M4 = (17);
const int PAL_I0_E1_R1Stuck = (18);*/
...
const int PAL_I0_E1_J1 = (19);
const int PAL_F1_E25_R2ToCorC = (0);(*@\CodeNote{Corresponding nodes in the 2nd concurrent flow (F1)}@*)
const int PAL_F1_E25_R2ToDoorCD = (1);/*
const int PAL_F1_E25_D6 = (2);
const int PAL_F1_E25_R2DeliveredRoomD = (3);
const int PAL_F1_E25_R2Stuck = (4);
const int PAL_F1_E25_M5 = (5);
const int PAL_F1_E25_J1 = (6);*/
...
const int PAL_F1_E25_AF = (7);
const double p_c_a; (*@\CodeNote{Input Parameters (S5.2.10)}@*)
const double p_d_12;
.../*
const double p_d_1;
const double p_d_2;
const double p_c_c;
const double p_c_b;*/
const double r_prep = (0.9998);
const double r1 = (1.0);
const double d_prep = (0.5);
const double d_cor_a = (0.2);
.../*
const double d_cor_c = (0.2);
const double d_door_ab = (0.5);
const double d_door_cd = (2.5);
const double d_cor_b_d1 = (0.5);
const double d_cor_b_d2 = (0.5);
const double d_r1_rd = (1.0);
const double d_r1_stuck = (1.0);
const double d_r2_stuck = (1.0);
const double d_r1_delivered = (0.5);
const double d_r2_delivered = (0.5);
const double t = (10.0);
*/
formula PAL_to_be_terminated = (PAL_I0_E1_to_be_terminated) |(*@\CodeNote{ActivityFinal (S5.2.9)}@*)
  (PAL_F1_E25_to_be_terminated);
formula PAL_to_be_failed = PAL_I0_E1_to_be_failed;(*@\CodeNote{Reliability (S5.2.11)}@*)
formula default_rate = ((1.0)/mean_duration);(*@\CodeNote{Duration and CTMCs (S5.2.12)}@*)

rewards "reward_door12_attempt"(*@\CodeNote{Rewards (S5.3.1)}@*)
  [PAL_I0_E1_R1ToCorBThrDoor1] ((PAL_I0_E1_pc=PAL_I0_E1_R1ToCorBThrDoor1)) : r1;	
	[PAL_I0_E1_R1ToCorBThrDoor2] ((PAL_I0_E1_pc=PAL_I0_E1_R1ToCorBThrDoor2)) : r1;	
endrewards

module PAL::I0::E1(*@\CodeNote{InitialNode (S5.2.3), main module}@*)
  PAL::I0::E1::pc : [-1..19] init 0;
  PAL::I0::E1::to_be_terminated : bool init false;(*@\CodeNote{ActivityFinal (S5.2.9)}@*)
  PAL::I0::E1::to_be_failed : bool init false;(*@\CodeNote{Reliability (S5.2.11)}@*)
  PAL::terminated : bool init false;(*@\CodeNote{ActivityFinal (S5.2.9)}@*)
  [PAL::I0::E1::I0] ((PAL::I0::E1::pc=PAL::I0::E1::I0))& (*@\CodeNote{InitialNode (S5.2.3)}@*)
    (!(PAL::to_be_terminated)) -> default_rate: (PAL::I0::E1::pc'=PAL::I0::E1::Prepare);/*
*/
  [PAL::I0::E1::Prepare] ((PAL::I0::E1::pc=PAL::I0::E1::Prepare))& (*@\CodeNote{Reliability (S5.2.11)}@*)
    (! (PAL::to_be_terminated)) -> (((1.0)/d_prep)*((1.0)-r_prep)):(PAL::I0::E1::pc'=INACTIVE)& 
    (PAL::I0::E1::to_be_failed'=true)& (PAL::I0::E1::to_be_terminated'=true) + 
    (((1.0)/d_prep)*r_prep):(PAL::I0::E1::pc'=PAL::I0::E1::F1);
  [PAL::F1] ((PAL::I0::E1::pc=PAL::I0::E1::F1))&(! (PAL::to_be_terminated)) (*@\CodeNote{ForkNode (S5.2.5)}@*)
    -> default_rate: PAL::I0::E1::pc'=PAL::I0::E1::R1ToCorA);
  [PAL::I0::E1::R1ToCorA] ((PAL::I0::E1::pc=PAL::I0::E1::R1ToCorA))&(*@\CodeNote{OpaqueAction (S5.2.4)}@*)
    (! (PAL::to_be_terminated)) -> ((1.0)/d_cor_a):(PAL::I0::E1::pc'=PAL::I0::E1::R1ToDoorAB);/*

  [PAL::I0::E1::R1ToDoorAB] ((PAL::I0::E1::pc=PAL::I0::E1::R1ToDoorAB))&(! (PAL::to_be_terminated)) -> ((1.0)/d_door_ab):(PAL::I0::E1::pc'=PAL::I0::E1::D1);
*/
  [PAL::I0::E1::D1] ((PAL::I0::E1::pc=PAL::I0::E1::D1))&(*@\CodeNote{DecisionNode (S5.2.6)}@*)
    (! (PAL::to_be_terminated)) -> (default_rate*p_c_a):(PAL::I0::E1::pc'=PAL::I0::E1::M1) 
    + (default_rate*((1.0)-p_c_a)):(PAL::I0::E1::pc'=PAL::I0::E1::M3);/*

  [PAL::I0::E1::M1] ((PAL::I0::E1::pc=PAL::I0::E1::M1))&(! (PAL::to_be_terminated)) -> default_rate:(PAL::I0::E1::pc'=PAL::I0::E1::D2);

  [PAL::I0::E1::M3] ((PAL::I0::E1::pc=PAL::I0::E1::M3))&(! (PAL::to_be_terminated)) -> default_rate:(PAL::I0::E1::pc'=PAL::I0::E1::R1Stuck);

  [PAL::I0::E1::D2] ((PAL::I0::E1::pc=PAL::I0::E1::D2))&(! (PAL::to_be_terminated)) -> (default_rate*p_d_12):(PAL::I0::E1::pc'=PAL::I0::E1::R1ToCorBThrDoor1) + (default_rate*((1.0)-p_d_12)):(PAL::I0::E1::pc'=PAL::I0::E1::R1ToCorBThrDoor2);

  [PAL::I0::E1::R1ToCorBThrDoor1] ((PAL::I0::E1::pc=PAL::I0::E1::R1ToCorBThrDoor1))&(! (PAL::to_be_terminated)) -> ((1.0)/d_cor_b_d1):(PAL::I0::E1::pc'=PAL::I0::E1::D3);

  [PAL::I0::E1::R1ToCorBThrDoor2] ((PAL::I0::E1::pc=PAL::I0::E1::R1ToCorBThrDoor2))&(! (PAL::to_be_terminated)) -> ((1.0)/d_cor_b_d2):(PAL::I0::E1::pc'=PAL::I0::E1::D4);
*/
  [PAL::I0::E1::D3] ((PAL::I0::E1::pc=PAL::I0::E1::D3))&
    (! (PAL::to_be_terminated)) -> (default_rate*p_d_1): PAL::I0::E1::pc'=PAL::I0::E1::M2)
    + (default_rate*((1.0)-p_d_1)):(PAL::I0::E1::pc'=PAL::I0::E1::M1);/*

  [PAL::I0::E1::D4] ((PAL::I0::E1::pc=PAL::I0::E1::D4))&(! (PAL::to_be_terminated)) -> (default_rate*p_d_2):(PAL::I0::E1::pc'=PAL::I0::E1::M2) + (default_rate*((1.0)-p_d_2)):(PAL::I0::E1::pc'=PAL::I0::E1::M1);
*/
  [PAL::I0::E1::M2] ((PAL::I0::E1::pc=PAL::I0::E1::M2))&(*@\CodeNote{MergeNode (S5.2.7)}@*)
    (!(PAL::to_be_terminated)) -> default_rate:(PAL::I0::E1::pc'=PAL::I0::E1::R1ToRoomD);/*

  [PAL::I0::E1::R1ToRoomD] ((PAL::I0::E1::pc=PAL::I0::E1::R1ToRoomD))&(! (PAL::to_be_terminated)) -> ((1.0)/d_r1_rd):(PAL::I0::E1::pc'=PAL::I0::E1::D5);

  [PAL::I0::E1::D5] ((PAL::I0::E1::pc=PAL::I0::E1::D5))&(! (PAL::to_be_terminated)) -> (default_rate*p_c_b):(PAL::I0::E1::pc'=PAL::I0::E1::R1DeliveredRoomD) + (default_rate*((1.0)-p_c_b)):(PAL::I0::E1::pc'=PAL::I0::E1::M3);

  [PAL::I0::E1::R1DeliveredRoomD] ((PAL::I0::E1::pc=PAL::I0::E1::R1DeliveredRoomD))&(! (PAL::to_be_terminated)) -> ((1.0)/d_r1_delivered):(PAL::I0::E1::pc'=PAL::I0::E1::M4);

  [PAL::I0::E1::M4] ((PAL::I0::E1::pc=PAL::I0::E1::M4))&(! (PAL::to_be_terminated)) -> default_rate:(PAL::I0::E1::pc'=PAL::I0::E1::J1);

  [PAL::I0::E1::R1Stuck] ((PAL::I0::E1::pc=PAL::I0::E1::R1Stuck))&(! (PAL::to_be_terminated)) -> ((1.0)/d_r1_stuck):(PAL::I0::E1::pc'=PAL::I0::E1::M4);
*/
  [PAL::J1] ((PAL::I0::E1::pc=PAL::I0::E1::J1))&(! (PAL::to_be_terminated))(*@\CodeNote{JoinNode (S5.2.8)}@*)
    -> default_rate:(PAL::I0::E1::pc'=INACTIVE);/*
*/
  [PAL::terminate] (PAL::to_be_terminated)&(! (PAL::terminated))(*@\CodeNote{ActivityFinal (S5.2.9)}@*)
    -> default_rate:(PAL::I0::E1::pc'=INACTIVE)&(PAL::terminated'=true);/*

  [] ((PAL::I0::E1::pc=INACTIVE))&((PAL::terminated)&(! (PAL::to_be_terminated))) -> default_rate:true;
*/
...
endmodule

module PAL::F1::E25(*@\CodeNote{ForkNode (S5.2.5)}@*)
  PAL::F1::E25::pc : [-1..7] init INACTIVE;
  PAL::F1::E25::to_be_terminated : bool init false;(*@\CodeNote{ActivityFinal (S5.2.9)}@*)
  [PAL::F1] ((PAL::F1::E25::pc=INACTIVE))& (!(PAL::to_be_terminated))(*@\CodeNote{ForkNode (S5.2.5)}@*)
    -> default_rate:(PAL::F1::E25::pc'=PAL::F1::E25::R2ToCorC);/*

  [PAL::F1::E25::R2ToCorC] ((PAL::F1::E25::pc=PAL::F1::E25::R2ToCorC))&(! (PAL::to_be_terminated)) -> ((1.0)/d_cor_c):(PAL::F1::E25::pc'=PAL::F1::E25::R2ToDoorCD);

  [PAL::F1::E25::R2ToDoorCD] ((PAL::F1::E25::pc=PAL::F1::E25::R2ToDoorCD))&(! (PAL::to_be_terminated)) -> ((1.0)/d_door_cd):(PAL::F1::E25::pc'=PAL::F1::E25::D6);

  [PAL::F1::E25::D6] ((PAL::F1::E25::pc=PAL::F1::E25::D6))&(! (PAL::to_be_terminated)) -> (default_rate*p_c_c):(PAL::F1::E25::pc'=PAL::F1::E25::R2DeliveredRoomD) + (default_rate*((1.0)-p_c_c)):(PAL::F1::E25::pc'=PAL::F1::E25::R2Stuck);

  [PAL::F1::E25::R2DeliveredRoomD] ((PAL::F1::E25::pc=PAL::F1::E25::R2DeliveredRoomD))&(! (PAL::to_be_terminated)) -> ((1.0)/d_r2_delivered):(PAL::F1::E25::pc'=PAL::F1::E25::M5);

  [PAL::F1::E25::R2Stuck] ((PAL::F1::E25::pc=PAL::F1::E25::R2Stuck))&(! (PAL::to_be_terminated)) -> ((1.0)/d_r2_stuck):(PAL::F1::E25::pc'=PAL::F1::E25::M5);

  [PAL::F1::E25::M5] ((PAL::F1::E25::pc=PAL::F1::E25::M5))&(! (PAL::to_be_terminated))
    -> default_rate:(PAL::F1::E25::pc'=PAL::F1::E25::J1);
*/
  [PAL::J1] ((PAL::F1::E25::pc=PAL::F1::E25::J1))&(! (PAL::to_be_terminated))(*@\CodeNote{JoinNode (S5.2.8)}@*)
    -> default_rate:(PAL::F1::E25::pc'=PAL::F1::E25::AF)&(PAL::F1::E25::to_be_terminated'=true);/*
*/
  [PAL::terminate] PAL::to_be_terminated -> default_rate:(*@\CodeNote{ActivityFinal (S5.2.9)}@*)
    (PAL::F1::E25::pc'=INACTIVE)&(PAL::F1::E25::to_be_terminated'=false);
...
endmodule      
\end{lstlisting}
\caption{The corresponding PRISM model to PAL use case.{}}
    \label{fig:transform_activity_pal_use_case}
  \end{figure}

\section{Semantics of other nodes}
\label{sec:appendix:semantics}

\subsection{\ad{OpaqueAction}}
\label{sec:appendix:semantics:opaqueaction}
We consider the \ad{OpaqueAction} with one incoming edge and one outgoing edge (such as \ad{Prepare} and \ad{R1ToCorA} in \Cref{fig:pal_use_case}) and its semantics is shown in \Cref{fig:trans:opaqueaction}. 
\begin{figure}\centering
%\lstinputlisting[language = PRISM,literate={\&}{{\ttfamily\&}}1, linewidth=\linewidth,numbersep=6pt]{cases/activityfinal.prism}
\begin{lstlisting}[language=PRISM,literate={\&}{{\ttfamily\&}}1, linewidth=0.98\linewidth,numbersep=6pt]
const int {mn}::{otn} = ?n+1;
module inmod // the module for the incoming edge and not newly created 
  ...
  [{an}::{nn}] ({mn}::pc={mn}::{nn})& (!{an}::to_be_terminated) -> ({mn}::pc'={mn}::{otn});
  ...
endmodule
\end{lstlisting}
\caption{PRISM semantic for \ad{OpaqueAction} with one incoming and one outgoing edge.}\label{fig:trans:opaqueaction}
\end{figure}
For its outgoing edge \lstinprism{oe}, its flow is regarded as the same flow as its incoming edge, and so {the corresponding PRISM command \mylines{4}{5} (to the outgoing edge) is in the same PRISM module (\lstinprism{inmod} \myline{2}, a PRISM model for the incoming edge) as that for its incoming edge}. 

A new {constant} \lstinprism{\{mn\}::\{oe.target.name\}} is defined \myline{1}, corresponding to the target node of the outgoing edge \lstinprism{oe}. The value of the constant is \lstinprism{?n+1}, denoting the currently allocated largest number \lstinprism{?n} (for constants in this module) plus 1, which is the next available number. %The \lstinprism{\{mn\}} here denotes the name of the corresponding module, \lstinprism{curmod} in this case.
The semantics for the action is a command \myline{4}. Its action label is the quantified name to the action. Its guard is a conjunction of two conditions of which the first one denotes the source state (this \ad{OpaqueAction}) and the second one is for the termination of its parent activity and will be explained in details in \Cref{sec:semantics:transformation:activityfinal}. Its only update \myline{4} takes the flow (or the module) to its target node.

\subsection{ForkNode}
\label{sec:appendix:semantics:forknode}

\begin{figure}\centering
\begin{lstlisting}[language = PRISM,literate={\&}{{\ttfamily\&}}1, linewidth=0.98\linewidth,numbersep=6pt]
// The first outgoing edge: reuse the current module
const int {mn}::{o1tn} = ?n+1;
module inmod // the module for the incoming edge and not newly created
  ...
  [{an}::{nn}] ({mn}::pc={mn}::{nn})& (!{an}::to_be_terminated) -> 
    ({mn}::pc'={mn}::{o1tn});
  ...
endmodule
// For the 2nd to nth outgoing edges, (n-1) new modules
const int {mn}::{o2tn} = 0;
module {an}::{nn}::{o2n} //new
  {mn}::pc : [-1..{numnodes}] init INACTIVE;
  [{an}::{nn}] ({mn}::pc=INACTIVE) & (!{an}::to_be_terminated) -> 
    ({mn}::pc'={mn}::{o2tn});
  ...
endmodule
...
// The nth outgoing edge
const int {mn}::{ontn} = 0;
module {an}::{nn}::{onn} //new
  {mn}::pc : [-1..{numnodes}] init INACTIVE;
  [{an}::{nn}] ({mn}::pc=INACTIVE) & (!{an}::to_be_terminated) -> 
    ({mn}::pc'={mn}::{ontn});
  ...
endmodule
\end{lstlisting}
\caption{\ad{ForkNode} with one incoming edge and $n$ outgoing edges.}\label{fig:trans_forknode}
\end{figure}

We consider a \ad{ForkNode} with $n$ outgoing edges and show its PRISM semantics in \Cref{fig:trans_forknode}. For its first outgoing edge \lstinprism{oe1}, its flow is regarded as the same flow as its incoming edge, and so the same PRISM module (\lstinprism{inmod} \myline{3}). A new {constant} \lstinprism{\{mn\}::\{oe1.target.name\}} is defined \myline{2}, corresponding to the target node of the outgoing edge \lstinprism{oe1}. The value of the constant is \lstinprism{?n+1}, denoting the current used largest number \lstinprism{?n} plus 1, which is the next available number. The \lstinprism{\{mn\}} here denotes the name of the corresponding module, \lstinprism{inmod} in this case.  

For the other $n-1$ outgoing edges from \lstinprism{oe2} to \lstinprism{oen}, we create $n-1$ new PRISM modules (\mylines{11}{25}). The \lstinprism{pc} variables declared in these module \mylinestwo{12}{21}, have their initial values be \lstinprism{INACTIVE}, and so the new modules are initially inactive. These modules become active by the similar commands \mylines{13}{14} and \mylines{22}{23}. These commands synchronise with the command \mylines{5}{6} (when the activity is not going to be terminated \mylinesthree{5}{13}{22}) on the action label \lstinprism{\{an\}::\{nn\}} which represents the \ad{ForkNode}. On synchronisation, these commands representing all the outgoing edges of the node are taken simultaneously. These edges take the concurrent flows to their target nodes (the constants \mylinesthree{2}{10}{19}) by the updates \mylinesthree{6}{14}{23}. Eventually, the $n$ flows are active at the same time and among them $n-1$ are new concurrent flows.

\subsection{DecisionNode}
\label{sec:appendix:semantics:decisionnode}

\begin{figure*}\centering
\subfloat[\ad{DecisionNode} with $n$ guarded outgoing edges (including one else)]{\label{fig:trans_decision:guard} 
\begin{minipage}[b]{\linewidth}%[\baselineskip]{6cm}
\begin{lstlisting}[language = PRISM,literate={\&}{{\ttfamily\&}}1, linewidth=0.45\linewidth,firstline=1,lastline=22,numbersep=6pt]
const int {mn}::{o1tn} = ?n+1;
const int {mn}::{o2tn} = ?n+2;
...
const int {mn}::{ontn} = ?n+n;
module inmod
  ...
  [{mn}::{nn}] 
    ({mn}::pc={mn}::{nn})&<g1>& 
    (!{an}::to_be_terminated) -> 
    ({mn}::pc'={mn}::{o1tn});
  [{mn}::{nn}] 
    ({mn}::pc={mn}::{nn})&<g2>&
    (!{an}::to_be_terminated) -> 
    ({mn}::pc'={mn}::{o2tn});
  ...
  [{mn}::{nn}] 
    ({mn}::pc={mn}::{nn})&
    !(<g1>|<g2>|...)& 
    (!{an}::to_be_terminated) -> 
    ({mn}::pc'={mn}::{ontn});
  ...
endmodule
\end{lstlisting}
\end{minipage}
}
\ \ \  
\subfloat[\ad{DecisionNode} with $n$ probabilistic outgoing edges]{\label{fig:trans_decision:prob} 
\begin{minipage}[b]{\linewidth}%[\baselineskip]{6cm}
\begin{lstlisting}[language = PRISM,literate={\&}{{\ttfamily\&}}1, linewidth=0.45\linewidth,firstline=24,lastline=38,numbersep=6pt]
const int {mn}::{o1tn} = ?n+1;
const int {mn}::{o2tn} = ?n+2;
...
const int {mn}::{ontn} = ?n+n;
module inmod
  ...
 [{mn}::{nn}] 
   ({mn}::pc={mn}::{nn})&
    (!{an}::to_be_terminated) -> 
    <oe1.prob>:({mn}::pc' = {mn}::{o1tn})+
    <oe2.prob>:({mn}::pc' = {mn}::{o2tn})+
    ... + 
    <oen.prob>:({mn}::pc' = {mn}::{ontn});
  ...
endmodule
\end{lstlisting}
\end{minipage}
}
\caption{PRISM semantic for \ad{DecisionNode}.}\label{fig:prism_sem_decisionnode}
\end{figure*}

Different from the \ad{ForkNode} and \ad{JoinNode}, all outgoing edges of a \ad{DecisionNode} are in a same concurrent flow, and so in a same module as illustrated in \Cref{fig:trans_decision:guard} for the \emph{guarded} \ad{DecisionNode} where all its outgoing edges have guards, and \Cref{fig:trans_decision:prob} for the \emph{probabilistic} \ad{DecisionNode} where all its outgoing edges have annotated probability values. In both cases, $n$ new {constants} \mylines{1}{4} in both figures are defined to represent the $n$ target nodes of the $n$ outgoing edges from the \ad{DecisionNode}. 

If the node is guarded, then $n$ commands \mylines{7}{16} in \Cref{fig:trans_decision:guard} are defined and each command corresponds to one outgoing edge. In particular, the guard of the command is a conjunction and one of the conjuncts, \lstinprism{<g1>}, corresponds to the guard of the edge \lstinprism{g1}, but it is the translation of \ad{g1} in PRISM. If the guard of one outgoing edge is \lstinprism{else}, then the corresponding conjunct is \lstinprism{!(<g1>|<g2>|...)} \myline{18}, which is the negation of the disjunction of all other edge guards, and so no other guards are true.

If the node is probabilistic, its semantics is encoded in one command \mylines{7}{13} in \Cref{fig:trans_decision:prob}. In particular, there are $n$ updates corresponding to the $n$ outgoing edges. Each update has the probability \lstinprism{<oei.prob>} of the edge \ad{oei} and its assignment updates the \lstinprism{pc} variable to the target of \ad{oei}.

\subsection{MergeNode}
\label{sec:appendix:semantics:mergenode}
A \ad{MergeNode} combines multiple incoming alternate flows into one outgoing flow, but without synchronisation like the \ad{JoinNode}. We require the incoming alternate flows are mutually exclusive, and so they are not active at the same time, by \ref{wfc:mergenode:exclusive}.

If all the incoming flows are in the same concurrent flow, then the semantics of the \ad{MergeNode} is just a transition to the target of its outgoing edge, as shown \myline{4} in \Cref{fig:trans:mergenode:onemodule}.

\begin{figure}\centering
  \subfloat[\ad{MergeNode} with all its incoming edges in a same flow]{\label{fig:trans:mergenode:onemodule} 
  \begin{minipage}[b]{\linewidth}
\begin{lstlisting}[language = PRISM,literate={\&}{{\ttfamily\&}}1, linewidth=0.98\linewidth, firstline=1,lastline=6, numbersep=6pt]
const int {mn}::{otn} = ?n+1;
module inmod
  ...
  [{an}::{nn}] ({mn}::pc={mn}::{nn})& (!{an}::to_be_terminated) -> ({mn}::pc'={mn}::{otn});
  ...
endmodule
\end{lstlisting}
  \end{minipage}
}
\ \ 
\subfloat[\ad{MergeNode} with all its incoming edges not in a same flow]{\label{fig:trans:mergenode:newmodule} 
\begin{minipage}[b]{\linewidth}
\begin{lstlisting}[language = PRISM,literate={\&}{{\ttfamily\&}}1, linewidth=0.98\linewidth, firstline=8,lastline=27, numbersep=6pt]
module inmod_1
  ...
  [{an}::{nn}::{i1n}] ({mn}::pc={mn}::{nn})& (!{an}::to_be_terminated) -> ({mn}::pc'=INACTIVE);
  [{an}::{nn}::{i2n}] ({mn}::pc={mn}::{nn})& (!{an}::to_be_terminated) -> ({mn}::pc'=INACTIVE);
  ...
endmodule
...
module inmod_m
  ...
  [{an}::{nn}::{inn}] ({mn}::pc={mn}::{nn})& (!{an}::to_be_terminated) -> ({mn}::pc'=INACTIVE);
  ...
endmodule
// A new module for its outgoing flow
module {an}::{nn}
  {mn}::pc : [-1..{numnodes}] init INACTIVE;
  [{an}::{nn}::{i1n}] ({mn}::pc=INACTIVE) & (!{an}::to_be_terminated) -> ({mn}::pc'={mn}::{otn});
  ...
  [{an}::{nn}::{inn}] ({mn}::pc=INACTIVE) & (!{an}::to_be_terminated) -> ({mn}::pc'={mn}::{otn});
  ...
endmodule
\end{lstlisting}
\end{minipage}
}
\caption{PRISM semantic for \ad{MergeNode}.}\label{fig:prism_sem_mergenode}
\end{figure}

If not all the incoming flows are from the same concurrent flow, we assume $n$ incoming flows are in $m$ concurrent flows where $n\geq m$. The $m$ modules from \lstinprism{inmod_1} \mylines{1}{6} to \lstinprism{inmod_m} \mylines{8}{12}, as shown in \Cref{fig:trans:mergenode:newmodule}, correspond to $m$ concurrent flows. %We also assume there are two incoming flows (on the incoming edges \ad{ie1} and \ad{ie2}) in the same concurrent flow (or the module \lstinprism{inmod_1}). 
Each incoming edge corresponds to one command in one of the $m$ modules, such as the command \myline{3} for \ad{ie1}, the command \myline{4} for \ad{ie2}, and the command \myline{10} for \ad{ien}. 

We need a new module for the outgoing flow of the \ad{MergeNode}, as shown \mylines{14}{20} in \Cref{fig:trans:mergenode:newmodule}. The new module is initially inactive (\myline{15}). It becomes active only after one of its $n$ commands \mylines{16}{18} synchronises with one of the $n$ commands \mylines{3}{10} for each incoming edge on an action label which is based on the incoming edge name such as \lstinprism{\{an\}::\{nn\}::\{ien.name\}} \mylinestwo{10}{18}. On synchronisation, the corresponding incoming concurrent module becomes inactive (such as \myline{10} in \lstinprism{inmod_m}) and the new module becomes active (\myline{18}). 

We note that the commands in the new module are only enabled when the module is inactive (\lstinprism{\{mn\}::pc =INACTIVE} such as \myline{16}). This means the module would not be able to accept another synchronisation on these action labels if the module is active after one synchronisation. Except the \lstinprism{inmod} that has been inactive after the synchronisation, other \lstinprism{inmod} modules can continue to execute until they reach the \ad{MergeNode}. At that point, they stop at these incoming flows because it is not possible to have the new module ready for another synchronisation. Our assumption of mutually exclusive incoming flows can avoid this issue because there is only one flow active at the same time.  

\subsection{\ad{JoinNode}}
\label{sec:appendix:semantics:joinnode}

\begin{figure}\centering
\begin{lstlisting}[language = PRISM,literate={\&}{{\ttfamily\&}}1, linewidth=0.98\linewidth,numbersep=6pt]
// The module for the 1st incoming edge
const int {mn}::{otn} = ?n+1;
module inmod_ie1
  ...
  [{an}::{nn}] ({mn}::pc={mn}::{nn})& (!{an}::to_be_terminated) -> 
    ({mn}::pc'={mn}::{otn});
  ...
endmodule
// The module for the 2nd incoming edge
module inmod_ie2 
  ...
  [{an}::{nn}] ({mn}::pc={mn}::{nn})& (!{an}::to_be_terminated) -> 
    ({mn}::pc'=INACTIVE);
  ...
endmodule
...
// The module for the nth incoming edge
module inmod_ien 
  ...
  [{an}::{nn}] ({mn}::pc={mn}::{nn})& (!{an}::to_be_terminated) -> 
    ({mn}::pc'=INACTIVE);
  ....
endmodule
\end{lstlisting}
\caption{\ad{JoinNode} with $n$ incoming edges and one outgoing edge}\label{fig:trans_joinnode}
\end{figure}

We consider a \ad{JoinNode} with $n$ incoming edges and show its semantics in \Cref{fig:trans_joinnode}. Different from the \ad{ForkNode} which creates new concurrent flows (modules), a \ad{JoinNode} destroys the concurrent flows on its $n$ incoming edges, and makes the corresponding modules inactive. Similar to the semantics of the \ad{ForkNode}, we would not terminate the module \lstinprism{inmod_ie1} (\mylines{3}{8}) on its first incoming edge and allow the outgoing edge of the \ad{JoinNode} to reuse this module. All other modules (from \lstinprism{inmod_ie2} to \lstinprism{inmod_ien} \mylines{10}{23}) become inactive simultaneously through the synchronisation of the commands \mylinesthree{5}{12}{20} on the action label \lstinprism{\{an\}::\{nn\}}. At the same time, the outgoing edge of the \ad{JoinNode} reaches its target node \lstinprism{\{mn\}::\{otn\}} \myline{6}.

\subsection{\ad{FlowFinalNode}}
\label{sec:appendix:semantics:flowfinal}

\begin{figure}\centering
\begin{lstlisting}[language = PRISM,literate={\&}{{\ttfamily\&}}1, linewidth=0.98\linewidth,numbersep=6pt]
module inmod
  ...
  [{an}::{nn}] ({mn}::pc={mn}::{nn})& (!{an}::to_be_terminated) -> ({mn}::pc'=INACTIVE);
  ...
endmodule
\end{lstlisting}
\caption{\ad{FlowFinalNode}}\label{fig:trans:flowfinal}
\end{figure}

The semantics of a \ad{FlowFinalNode}, shown in \Cref{fig:trans:flowfinal}, is to make the module (or the flow) inactive by setting \ad{pc} to \lstinprism{INACTIVE} \myline{3}. Here, the condition \lstinprism{(!\{an\}::to_be_terminated)} ensures this only happens if no any \ad{ActivityFinalNode} has been reached. If it is, \lstinprism{\{an\}::to_be_terminated} will be set to true, and so next the activity will be terminated (that is, all modules will be inactive). Therefore, it is not necessary to terminate this flow using \ad{FlowFinalNode}.

{
  Rule~\ref{nfm_prism:flowfinalnode} defines the function $\meta{\semL{act, n, mod}{Ff}^{\sigma}}$ for a \ad{FlowFinalNode}. We omit the detailed explanation of this rule.

\begin{TRule}[FlowFinalNode]{\\\meta{\quad~
  \semL{act : Activity, n: FlowFinalNode, mod: Module_{pr}}{Ff}^{\sigma}: \tau =}}
  \label{nfm_prism:flowfinalnode}
  \quad \meta{\left(\sigma.consts, \sigma.modmap[mod'/mod], \sigma.rewards\right)} \\
  \meta{\bf where}\\\quad
  \begin{array}{t}
    % add a termiated variable. 
      \meta{grd \defs prmBoolExprPCAndNotTBTerm(mod.varpc(), prmVarRefExpr(id(mod, n)))}\quad \\
      \meta{upd \defs prmUpdate(\set{prmAssign(mod.varpc(), constinac)})} \\
      \meta{cmd \defs prmCommand(prmAction(id(mod,n)), grd, \set{upd})} \quad \\
      \meta{mod' \defs mod().addCmds(cmd)} 
  \end{array}
\end{TRule}
}

\subsection{\ad{ActivityFinalNode}}
\label{sec:appendix:semantics:activityfinal}
An activity can have more than one \ad{ActivityFinalNode}. We show our PRISM semantics for the termination of the activity through any \ad{ActivityFinalNode} in \Cref{fig:prism_sem:activityfinal} where we assume $m$ concurrent flows contain \ad{ActivityFinalNode}s and other $n$ concurrent flows do not. 

\begin{figure}\centering
\begin{lstlisting}[language = PRISM,literate={\&}{{\ttfamily\&}}1, linewidth=0.98\linewidth,numbersep=6pt]
// Disjunctions of all local to_be_terminated variables
formula {an}::to_be_terminated = inmod_1::to_be_terminated | inmod_2::to_be_terminated | ... | 
  inmod_m::to_be_terminated;
// The main module (assume no ActivityFinal in the flow)
module mainmod
  {mn}::terminated : bool init false;
  ...
  [{an}::terminate] ({an}::to_be_terminated) & (!({an}::terminated)) -> 
    ({mn}::pc'=INACTIVE) & ({an}::terminated'=true);
  ...
endmodule
module inmod_1
  {mn}::to_be_terminated : bool init false;
  ...
  [_] _ -> _: ({mn}::pc'={mn}::{nn}) & ({mn}::to_be_terminated'=true);
  [{an}::terminate] ({an}::to_be_terminated) -> 
    ({mn}::pc'=INACTIVE) & ({mn}::to_be_terminated'=false);
endmodule
module inmod_2
  {mn}::to_be_terminated : bool init false;
  ...
endmodule
...
module inmod_m 
  {mn}::to_be_terminated : bool init false;
  ...
endmodule
// For all other modules without ActivityFinals in their flows
module inmod_other_1
  ...
   [{an}::terminate] {an}::to_be_terminated) -> (mn::pc'=INACTIVE);
  ...
endmodule
...
module inmod_other_n
  ...
endmodule
\end{lstlisting}
\caption{PRISM semantic for \ad{ActivityFinalNode}.}\label{fig:prism_sem:activityfinal}
\end{figure}

For the $m$ flows, their corresponding modules \mylinesthree{12}{19}{24} all have one local variable \lstinprism{\{mn\}:: to_be_terminated} \mylinesthree{13}{20}{25}. This variable denotes that the corresponding flow to the owning module has arrived at an \ad{ActivityFinalNode} (\myline{15}) and so it is going to be terminated (\myline{15}). Because the source node of the edge to this \ad{ActivityFinalNode} is not important (for the illustration here), we omit the action label, the guard, and the probability \myline{15} (denoted as \lstinprism{_}). The formula \mylines{2}{3} is a disjunction of all these variables. In other words, if any one of the variables is true (i.e. any concurrent flow arrives at its \ad{ActivityFinalNode}), then the formula is true. 

Each of the $m$ modules \mylines{12}{27} contains a command like that \mylines{16}{17}. Each of the other $n$ modules \mylines{29}{37} (corresponding to the other $n$ concurrent flows) contains a command like that \myline{31}. The main module \mylines{5}{11} also has a command \mylines{8}{9} for termination. All these commands have the same action label \lstinprism{\{an\}::terminate}, and so all modules need to synchronise on them to terminate at the same time. The termination is only allowed if the activity is not terminated yet (\myline{8}). On termination, all modules become inactive (\mylinesthree{9}{17}{31}), the activity is terminated (\myline{9}), and all local \lstinprism{to_be_terminated} variables are set to \lstinprism{false} (\myline{17}).

In \Cref{fig:prism_sem:activityfinal}, we assume the main module \mylines{5}{11} does not have any \ad{ActivityFinalNode} in its corresponding concurrent flow. If this is not the case, then the module will contain a local \lstinprism{to_be_terminated} variable like the one \myline{13}. Additionally, the command \mylines{8}{9} will contain another assignment like the one \myline{17} to set the local \lstinprism{to_be_terminated} to \lstinprism{false}. 

\subsection{Reliability}
\label{sec:appendix:semantics:reliability}
%An action can be annotated with reliability $p$, a real value between 0 and 1, to denote the probability $p$ of a successful execution of this service (represented by the action). So the failure probability of the service is $1-p$. There is an implicit \ad{DecisionNode} here. In activity diagrams, we also do not explicitly show an \ad{ActivityNode} for the failure to facilitate the modelling. \emph{Semantically, if an action fails, the activity owning this action fails.} This is similar to the \ad{ActivityFinal}. 
\begin{figure}
\begin{lstlisting}[language=PRISM,linewidth=0.98\linewidth]
// Disjunctions of all local to_be_failed variables
formula {an}::to_be_terminated = inmod_1::to_be_terminated | inmod_2::to_be_terminated | ... | 
  inmod_m::to_be_terminated | 
  ... ;
formula {an}::to_be_failed = inmod_1::to_be_failed | inmod_2::to_be_failed | ... | 
  inmod_m::to_be_failed;

const int {mn}::{otn} = ?n+1;
module inmod_1
  {mn}::to_be_terminated : bool init false;
  {mn}::to_be_failed : bool init false;

  [{an}::{nn}] ({mn}::pc={mn}::{nn})& (!{an}::to_be_terminated) -> 
    (<node.reliability>): ({mn}::pc'={mn}::{otn})+ 
    (1 - <node.reliability>): ({mn}::pc'=INACTIVE) & ({mn}::to_be_failed'=true) & ({mn}::to_be_terminated'=true);
  ...
endmodule
...
module inmod_m 
  {mn}::to_be_terminated : bool init false;
  {mn}::to_be_failed : bool init false;
  ...
endmodule
\end{lstlisting}
\caption{PRISM semantic for \ad{OpaqueAction} annotated with reliability.}\label{fig:prism_sem:reliability}
\end{figure}

We informally illustrate the semantics for actions annotated with reliability in \Cref{fig:prism_sem:reliability} and present its formal semantics afterwards. 
In PRISM, we make the failure node explicit by introducing a local variable \lstinprism{\{mn\}:: to_be_failed} (such as \mylinestwo{11}{21}) to each module (like \lstinprism{inmod_1} \mylines{9}{17} and \lstinprism{inmod_m} \mylines{19}{23}) that corresponds to a concurrent flow containing the actions annotated with reliability. We define a formula \lstinprism{\{an\}::to_be_failed} \mylines{5}{6} which is a disjunction of all these local \lstinprism{to_be_failed} variables. Each module also contains a \lstinprism{to_be_terminated} variable \mylinestwo{10}{20} used to indicate a termination. A failure is regarded as a termination of the activity with \lstinprism{\{an\}::to_be_failed} being true.

For an action with the annotation of reliability, its semantics are captured in the command \mylines{13}{15} where two updates denote a probabilistic choice between a successful execution of the service (\lstinprism{pc} updated to the target node of the outgoing edge of the action) with probability equal to the reliability \lstinprism{<node.reliability>} and a failure with probability (\lstinprism{1-<node.reliability>}). The failure is encoded in three assignments \myline{15}. The first assignment makes the module inactive. The second one sets the local \lstinprism{to_be_failed} to true. The third one sets  \lstinprism{to_be_terminated} to true. As discussed in \Cref{sec:semantics:transformation:activityfinal}, all modules will be terminated after that. We note the local \lstinprism{to_be_failed} variable will not be set to false after the termination, and so the formula \lstinprism{\{an\}::to_be_failed} is still true. This formula is not used in the PRISM model, but can be quantified in properties to indicate the occurrence of a failure.

{
  Rule~\ref{nfm_prism:opaqueaction:reliability} defines the function $\meta{\semL{act, n, mod}{Op}^{\sigma}}$ for an \ad{OpaqueAction} that is annotated with reliability (that is, $\meta{n.reliability}$ is not $\meta{null}$). We also assume that $\meta{mod}$ does not contain the two required local variables yet, and so they are added to the updated module $\meta{mod'}$. We omit the detailed explanation of this rule.

\begin{TRule}[OpaqueAction with reliability annotation]{\\\meta{\quad~
  \semL{act : Activity, n: OpaqueAction, mod: Module_{pr}}{Op}^{\sigma}: \tau =}}
  \label{nfm_prism:opaqueaction:reliability}
  \quad \meta{
    \semL{act, oe.target, mod'}{N}^{\sigma'}
  } \\
  \meta{\bf where}\\\quad
  \begin{array}{t}
    \meta{n.reliability \neq null \land mod.vartbfail() = null \land mod.vartbterm() = null} \quad \\
      \meta{modtbf \defs id(mod, \text{``to\_be\_failed''})} \quad 
      \meta{vartbf \defs prmBoolVarDecl(modtbf, false)} \quad \\
      \meta{modtbt \defs id(mod, \text{``to\_be\_terminated''})} \quad
      \meta{vartbt \defs prmBoolVarDecl(modtbt, false)} \quad \\
      \meta{atbt \defs id(act, \text{``to\_be\_terminated''})} \quad
      \meta{oe \defs act.getOEdge(n)} \quad \\
      \meta{constnext \defs prmConstant(id(mod, oe.target), mod.lastused+1)}\\
      \meta{grd \defs prmBoolExprPCAndNotTBTerm(mod.varpc(), prmVarRefExpr(id(mod, n)))} \\

    \meta{assigns_s \defs \left(
        \begin{array}{t}
          \IF\ oe.target \in ActivityFinalNode \ \THEN\\
          \quad \set{prmAssign(mod.varpc(), constnext), prmAssign(mod.vartbterm(), true)} \\
          \ELSE \set{prmAssign(mod.varpc(), constnext)}\\
        \end{array}
    \right)} \\
      \meta{upd_s \defs prmUpdate(\semL{n.reliability}{e}, assigns_s)} \\

    \meta{assigns_f \defs \set{
          \begin{array}{t}
            prmAssign(mod.varpc(), constinac), 
            prmAssign(vartbf, true),
            prmAssign(vartbt, true)
          \end{array}
        }} \\
      \meta{upd_f \defs prmUpdate\left(1 - \semL{n.reliability}{e}, assigns_f\right)} \\
      \meta{cmd \defs prmCommand(prmAction(id(act, n)), grd, \set{upd_r, upd_o})} \\
      \meta{mod' \defs \left(mod().addVars(\set{vartbf, vartbt})\right).addCmdsAndIncLastUsed(\set{cmd})} \\
      \meta{grd_1 \defs prmEqBoolExpr(mod.varpc(), prmVarRefExpr(id(mod, n))} \\
      \meta{rewards \defs \left(\lambda r @ prmReward(r.name, prmAction(id(act, n)), grd_1, \semL{r.value}{e})\right) \limg oe.rewards\rimg } \\
      \meta{\sigma' \defs \left(\sigma.consts \union \set{constnext}, \sigma.modmap[mod'/mod] \oplus \set{\left(oe.target, oe\right) \mapsto mod'}, \sigma.rewards \union rewards\right)} \\
  \end{array}
\end{TRule}
}

\subsection{Duration and CTMCs}
\label{sec:appendix:semantics:duration:ctmc}
We introduce a {constant} \lstinprism{mean_duration} (whose value can be specified during model checking and here 0.0001 \myline{1} is an example) and define a formula \lstinprism{default_rate} for a CTMC model, as illustrated \mylines{1}{2} in \Cref{fig:prism_sem:ctmc} where the semantics for actions with or without reliability annotations, and for other \ad{ControlNode}s are illustrated.

\begin{figure}
  \begin{lstlisting}[language=PRISM,linewidth=0.98\linewidth]
const double mean_duration = 0.0001;
formula default_rate = 1/mean_duration;

module inmod
  ...
  // if an action is not annotated with reliability
  [{an}::{nn}] 
    ({mn}::pc={mn}::{nn})&
    (!{an}::to_be_terminated) ->
    (1/<node.duration>): ({mn}::pc'={mn}::{otn});
  // if an action is annotated with reliability
  [{an}::{nn}] 
    ({mn}::pc={mn}::{nn})&
    (!{an}::to_be_terminated) -> 
    (1/<node.duration>)*(<node.reliability>): ({mn}::pc'={mn}::{otn})+ 
    (1/<node.duration>)*(1-<node.reliability>): ({mn}::pc'=INACTIVE) & ({mn}::to_be_failed'=true) & ({mn}::to_be_terminated'=true);
  // for ControlNodes
  [{an}::{nn}] 
    ({mn}::pc={mn}::{nn})& 
    (!{an}::to_be_terminated) ->
    default_rate: ({mn}::pc'={mn}::{otn});
  // for probabilistic DecisionNode
  [{mn}::{nn}] 
  ({mn}::pc={mn}::{nn})&
   (!{an}::to_be_terminated) -> 
   default_rate*<oe1.prob>: ({mn}::pc' = {mn}::{o1tn})+
   default_rate*<oe2.prob>: ({mn}::pc' = {mn}::{o2tn})+
   ... + 
   default_rate*<oen.prob>: ({mn}::pc' = {mn}::{ontn});
  // for guarded DecisionNode
  [{mn}::{nn}] 
  ({mn}::pc={mn}::{nn})&<g1>& 
  (!{an}::to_be_terminated) -> 
  default_rate: {mn}::pc'={mn}::{o1tn});
  ...
endmodule
  \end{lstlisting}
\caption{PRISM semantic for \ad{Action} annotated with duration and \ad{ActivityNode}s analysed in CTMCs.}\label{fig:prism_sem:ctmc}
\end{figure}

We show the semantics of an action without and with a reliability annotation \mylines{7}{10} and \mylines{12}{16} where each update has a rate \lstinprism{1/<node.duration>} \myline{10} or rates proportional to its reliability \mylinestwo{15}{16}. For any \ad{ControlNode}, it has a \lstinprism{default_rate}, shown \myline{21}. Similarly, the rates (\mylinesthree{26}{27}{29}) for the transitions from a probabilistic \ad{DecisionNode} are proportional to \lstinprism{default_rate}.

{
  Rule~\ref{nfm_prism:opaqueaction:reliability:ctmc} defines the function $\meta{\semL{act, n, mod}{Op}^{\sigma}}$ for an \ad{OpaqueAction} that is annotated with both reliability and duration. We omit the detailed explanation of this rule.

\begin{TRule}[OpaqueAction with reliability and duration annotations when modelled as a CTMC]{\\\meta{\quad~
  \semL{act : Activity, n: OpaqueAction, mod: Module_{pr}}{Op}^{\sigma}: \tau =}}
  \label{nfm_prism:opaqueaction:reliability:ctmc}
  \quad \meta{
    \semL{act, oe.target, mod'}{N}^{\sigma'}
  } \\
  \meta{\bf where}\\\quad
  \begin{array}{t}
    \meta{n.reliability \neq null \land n.duration \neq null \land mod.vartbfail() = null \land mod.vartbterm() = null} \quad \\
      \meta{modtbf \defs id(mod, \text{``to\_be\_failed''})} \quad 
      \meta{vartbf \defs prmBoolVarDecl(modtbf, false)} \quad \\
      \meta{modtbt \defs id(mod, \text{``to\_be\_terminated''})} \quad
      \meta{vartbt \defs prmBoolVarDecl(modtbt, false)} \quad \\
      \meta{atbt \defs id(act, \text{``to\_be\_terminated''})} \quad
      \meta{oe \defs act.getOEdge(n)} \quad \\
      \meta{constnext \defs prmConstant(id(mod, oe.target), mod.lastused+1)}\\
      \meta{grd \defs prmBoolExprPCAndNotTBTerm(mod.varpc(), prmVarRefExpr(id(mod, n)))} \\

    \meta{assigns_s \defs \left(
        \begin{array}{t}
          \IF\ oe.target \in ActivityFinalNode \ \THEN\\
          \quad \set{prmAssign(mod.varpc(), constnext), prmAssign(mod.vartbterm(), true)} \\
          \ELSE \set{prmAssign(mod.varpc(), constnext)}\\
        \end{array}
    \right)} \\
      \meta{upd_s \defs prmUpdate(\semL{\left(1/n.duration\right) * n.reliability}{e}, assigns_s)} \\

    \meta{assigns_f \defs \set{
          \begin{array}{t}
            prmAssign(mod.varpc(), constinac), 
            prmAssign(vartbf, true),
            prmAssign(vartbt, true)
          \end{array}
        }} \\
      \meta{upd_f \defs prmUpdate\left(\semL{\left(1/n.duration\right) * \left(1 - {n.reliability}\right)}{e}, assigns_f\right)} \\
      \meta{cmd \defs prmCommand(prmAction(id(act, n)), grd, \set{upd_r, upd_o})} \\
      \meta{mod' \defs \left(mod().addVars(\set{vartbf, vartbt})\right).addCmdsAndIncLastUsed(\set{cmd})} \\
      %\meta{\sigma' \defs \left(\sigma.consts \union \set{constnext}, \sigma.modmap[mod'/mod] \oplus \set{\left(oe.target, oe\right) \mapsto mod'}\right)} %\\
      \meta{grd_1 \defs prmEqBoolExpr(mod.varpc(), prmVarRefExpr(id(mod, n))} \\
      \meta{rewards \defs \left(\lambda r @ prmReward(r.name, prmAction(id(act, n)), grd_1, \semL{r.value}{e})\right) \limg oe.rewards\rimg } \\
      \meta{\sigma' \defs \left(\sigma.consts \union \set{constnext}, \sigma.modmap[mod'/mod] \oplus \set{\left(oe.target, oe\right) \mapsto mod'}, \sigma.rewards \union rewards\right)} \\
  \end{array}
\end{TRule}
}

\end{appendices}
%\vspace{-2ex}

\newpage
\bibliography{main,BPMN}

@incollection{Braghetto2011,
  title = {Performance {{Evaluation}} of {{Business Processes}} through a {{Formal Transformation}} to {{SAN}}},
  booktitle = {Computer {{Performance Engineering}}},
  author = {Braghetto, Kelly Rosa and Ferreira, Jo{\~a}o Eduardo and Vincent, Jean-Marc},
  editor = {Thomas, Nigel},
  year = {2011},
  volume = {6977},
  pages = {42--56},
  publisher = {Springer Berlin Heidelberg},
  address = {Berlin, Heidelberg},
  doi = {10.1007/978-3-642-24749-1\_5},
  copyright = {http://www.springer.com/tdm},
  isbn = {978-3-642-24748-4 978-3-642-24749-1},
  langid = {english}
}

@inproceedings{Brenner2007a,
  title = {{{PEPS2007}} - {{Stochastic Automata Networks Software Tool}}},
  booktitle = {Fourth {{International Conference}} on the {{Quantitative Evaluation}} of {{Systems}} ({{QEST}} 2007)},
  author = {Brenner, Leonardo and Fernandes, Paulo and Plateau, Brigitte and Sbeity, Ihab},
  year = {2007},
  month = sep,
  pages = {163--164},
  doi = {10.1109/QEST.2007.33},
  url = {https://ieeexplore.ieee.org/document/4338252},
  urldate = {2024-05-27}
}

@inproceedings{Ceballos2015a,
  title = {A {{Probabilistic BPMN Normal Form}} to {{Model}} and {{Advise Human Activities}}},
  booktitle = {Engineering {{Multi-Agent Systems}}},
  author = {Ceballos, Hector G. and {Flores-Solorio}, Victor and Garcia, Juan Pablo},
  editor = {Baldoni, Matteo and Baresi, Luciano and Dastani, Mehdi},
  year = {2015},
  pages = {51--69},
  publisher = {Springer International Publishing},
  address = {Cham},
  doi = {10.1007/978-3-319-26184-3\_4},
  isbn = {978-3-319-26184-3},
  langid = {english}
}

@inproceedings{Ceballos2018,
  author       = {Hector G. Ceballos and Francisco J. Cantu},
  editor       = {Daniel D. Lee and Alexander Steen and Toby Walsh},
  title        = {Discovering Causal Relations in Semantically-Annotated Probabilistic Business Process Diagrams},
  booktitle    = {GCAI-2018, 4th Global Conference on Artificial Intelligence, Luxembourg, September 18-21, 2018},
  series       = {EPiC Series in Computing},
  volume       = {55},
  pages        = {29--40},
  publisher    = {EasyChair},
  year         = {2018},
  doi          = {10.29007/ND7R},
  address      = {Stockport},
}

@article{Dechsupa2022,
  title = {Stepwise {{Verification}} for the {{BPMN With Timed}} and {{Stochastic Process Using}} a {{Colored Generalized Stochastic Petri Net}}},
  author = {Dechsupa, C. and Vatanawood, W. and Thongtak, A.},
  year = {2022},
  journal = {IEEE Access},
  volume = {10},
  pages = {42983--43002},
  issn = {2169-3536},
  doi = {10.1109/ACCESS.2022.3168000},
  url = {https://ieeexplore.ieee.org/document/9758738/},
  urldate = {2024-05-26},
  copyright = {https://creativecommons.org/licenses/by/4.0/legalcode}
}

@article{Duran2018a,
  title = {Stochastic Analysis of {{BPMN}} with Time in Rewriting Logic},
  author = {Dur{\'a}n, Francisco and Rocha, Camilo and Sala{\"u}n, Gwen},
  year = {2018},
  month = dec,
  journal = {Science of Computer Programming},
  volume = {168},
  pages = {1--17},
  issn = {0167-6423},
  doi = {10.1016/j.scico.2018.08.007},
  url = {https://www.sciencedirect.com/science/article/pii/S0167642318303307},
  urldate = {2024-05-26}
}

@inproceedings{Duran2019,
  title = {Analysis of {{Resource Allocation}} of {{BPMN Processes}}},
  booktitle = {Service-{{Oriented Computing}}},
  author = {Dur{\'a}n, Francisco and Rocha, Camilo and Sala{\"u}n, Gwen},
  editor = {Yangui, Sami and Bouassida Rodriguez, Ismael and Drira, Khalil and Tari, Zahir},
  year = {2019},
  pages = {452--457},
  publisher = {Springer International Publishing},
  address = {Cham},
  doi = {10.1007/978-3-030-33702-5\_35},
  isbn = {978-3-030-33702-5},
  langid = {english}
}

@inproceedings{Falcone2022a,
  title = {Probabilistic {{Model Checking}} of {{BPMN Processes}} at {{Runtime}}},
  booktitle = {Integrated {{Formal Methods}}},
  author = {Falcone, Yli{\`e}s and Sala{\"u}n, Gwen and Zuo, Ahang},
  editor = {Ter Beek, Maurice H. and Monahan, Rosemary},
  year = {2022},
  volume = {13274},
  pages = {191--208},
  publisher = {Springer International Publishing},
  address = {Cham},
  doi = {10.1007/978-3-031-07727-2\_11},
  isbn = {978-3-031-07726-5 978-3-031-07727-2},
  langid = {english}
}

@inproceedings{Falcone2024,
  title = {Probabilistic {{Runtime Enforcement}} of {{Executable BPMN Processes}}},
  booktitle = {Fundamental {{Approaches}} to {{Software Engineering}}},
  author = {Falcone, Yli{\`e}s and Sala{\"u}n, Gwen and Zuo, Ahang},
  editor = {Beyer, Dirk and Cavalcanti, Ana},
  year = {2024},
  pages = {56--76},
  publisher = {Springer Nature Switzerland},
  address = {Cham},
  doi = {10.1007/978-3-031-57259-3\_3},
  isbn = {978-3-031-57259-3},
  langid = {english}
}

@article{Gharbi2009,
  title = {Colored Stochastic {{Petri}} Nets for Modelling and Analysis of Multiclass Retrial Systems},
  author = {Gharbi, Nawel and Dutheillet, Claude and Ioualalen, Malika},
  year = {2009},
  month = apr,
  journal = {Mathematical and Computer Modelling},
  volume = {49},
  number = {7},
  pages = {1436--1448},
  issn = {0895-7177},
  doi = {10.1016/j.mcm.2008.11.006},
  url = {https://www.sciencedirect.com/science/article/pii/S0895717709000132},
  urldate = {2024-06-10}
}

@proceedings{Herbert2012,
    author = {Herbert, Luke T. and Sharp, Robin},
    title = {Quantitative Analysis of Probabilistic BPMN Workflows},
    volume = {Volume 2: 32nd Computers and Information in Engineering Conference, Parts A and B},
    series = {International Design Engineering Technical Conferences and Computers and Information in Engineering Conference},
    pages = {509-518},
    year = {2012},
    month = {08},
    doi = {10.1115/DETC2012-70653},
}

@inbook{Herbert2013a,
  title = {Workflow Fault Tree Generation through Model Checking},
  booktitle = {Safety, {{Reliability}} and {{Risk Analysis}}},
  author = {Herbert, Luke and Sharp, Robin},
  year = {2013},
  month = sep,
  edition = {1},
  pages = {2229--2236},
  publisher = {CRC Press},
  address = {London},
  doi = {10.1201/b15938-333},
  collaborator = {Steenbergen, R.D.J.M. and Van Gelder, P.H.A.J.M. and Miraglia, S. and Vrouwenvelder, A.C.W.M.},
  isbn = {978-0-429-22771-4},
  langid = {english}
}

@book{Herbert2014a,
  title = {Specification, {{Verification}} and {{Optimisation}} of {{Business Processes}}: {{A Unified Framework}}},
  shorttitle = {Specification, {{Verification}} and {{Optimisation}} of {{Business Processes}}},
  author = {Herbert, Luke Thomas},
  year = {2014},
  series = {{{DTU Compute PHD-2014}}},
  publisher = {Technical University of Denmark},
  address = {Kgs. Lyngby}
}

@inproceedings{Herbert2015,
  title = {Optimal {{Scheduling}} of {{Stochastic Production Processes Through Model Checking}}},
  booktitle = {Modelling and {{Management}} of {{Engineering Processes}}},
  author = {Herbert, L. and Hansen, Z. N. L. and Sharp, R. and Jacobsen, P.},
  editor = {Schabacker, Michael and Gericke, Kilian and Sz{\'e}lig, Nikoletta and Vajna, S{\'a}ndor},
  year = {2015},
  pages = {187--202},
  publisher = {Springer},
  address = {Berlin, Heidelberg},
  doi = {10.1007/978-3-662-44009-4\_16},
  isbn = {978-3-662-44009-4},
  langid = {english}
}

@article{Herbert2016,
  title = {Restructuring of Workflows to Minimise Errors via Stochastic Model Checking: {{An}} Automated Evolutionary Approach},
  shorttitle = {Restructuring of Workflows to Minimise Errors via Stochastic Model Checking},
  author = {Herbert, L. T. and Hansen, Z. N. L.},
  year = {2016},
  month = jan,
  journal = {Reliability Engineering \& System Safety},
  volume = {145},
  pages = {351--365},
  issn = {0951-8320},
  doi = {10.1016/j.ress.2015.07.002},
  url = {https://www.sciencedirect.com/science/article/pii/S0951832015001921},
  urldate = {2024-05-26}
}

@inproceedings{Kikuchi2014,
  title = {Performance {{Estimation}} for {{Business Workflows}} on {{Public Cloud Offerings Using Probabilistic Model Checker}}},
  booktitle = {2014 {{IEEE}}/{{ACM}} 7th {{International Conference}} on {{Utility}} and {{Cloud Computing}}},
  author = {Kikuchi, Shinji},
  year = {2014},
  month = dec,
  pages = {317--326},
  doi = {10.1109/UCC.2014.41},
  urldate = {2024-05-26},
}

@inproceedings{Krishna2017,
  title = {{{VBPMN}}: {{Automated Verification}} of {{BPMN Processes}} ({{Tool Paper}})},
  shorttitle = {{{VBPMN}}},
  booktitle = {Integrated {{Formal Methods}}},
  author = {Krishna, Ajay and Poizat, Pascal and Sala{\"u}n, Gwen},
  editor = {Polikarpova, Nadia and Schneider, Steve},
  year = {2017},
  pages = {323--331},
  publisher = {Springer International Publishing},
  address = {Cham},
  doi = {10.1007/978-3-319-66845-1\_21},
  isbn = {978-3-319-66845-1},
  langid = {english}
}

@article{Plateau1985,
  title = {On the Stochastic Structure of Parallelism and Synchronization Models for Distributed Algorithms},
  author = {Plateau, Brigitte},
  year = {1985},
  month = aug,
  journal = {ACM SIGMETRICS Performance Evaluation Review},
  volume = {13},
  number = {2},
  pages = {147--154},
  issn = {0163-5999},
  doi = {10.1145/317786.317819}
}

@inproceedings{Prandi2007,
  title = {Stochastic {{COWS}}},
  booktitle = {Service-{{Oriented Computing}} -- {{ICSOC}} 2007},
  author = {Prandi, Davide and Quaglia, Paola},
  editor = {Kr{\"a}mer, Bernd J. and Lin, Kwei-Jay and Narasimhan, Priya},
  year = {2007},
  pages = {245--256},
  publisher = {Springer},
  address = {Berlin, Heidelberg},
  doi = {10.1007/978-3-540-74974-5\_20},
  isbn = {978-3-540-74974-5},
  langid = {english}
}

@inproceedings{Prandi2008,
  title = {Formal {{Analysis}} of {{BPMN Via}} a {{Translation}} into {{COWS}}},
  booktitle = {Coordination {{Models}} and {{Languages}}},
  author = {Prandi, Davide and Quaglia, Paola and Zannone, Nicola},
  editor = {Lea, Doug and Zavattaro, Gianluigi},
  year = {2008},
  pages = {249--263},
  publisher = {Springer},
  address = {Berlin, Heidelberg},
  doi = {10.1007/978-3-540-68265-3\_16},
  isbn = {978-3-540-68265-3},
  langid = {english}
}

@Article{Agha2006,
  author     = {Agha, Gul and Meseguer, Jos{\'e} and Sen, Koushik},
  journal    = {Electronic Notes in Theoretical Computer Science},
  title      = {{{PMaude}}: {{Rewrite-based Specification Language}} for {{Probabilistic Object Systems}}},
  year       = {2006},
  issn       = {1571-0661},
  month      = may,
  number     = {2},
  pages      = {213--239},
  volume     = {153},
  doi        = {10.1016/j.entcs.2005.10.040},
  series     = {Proceedings of the {{Third Workshop}} on {{Quantitative Aspects}} of {{Programming Languages}} ({{QAPL}} 2005)},
  shorttitle = {{{PMaude}}},
  url        = {https://www.sciencedirect.com/science/article/pii/S1571066106002672},
  urldate    = {2024-06-09},
}

@Article{Alur1999,
  author    = {Alur, Rajeev and Henzinger, Thomas A},
  journal   = {{Formal Methods in System Design}},
  title     = {Reactive modules},
  year      = {1999},
  number    = {1},
  pages     = {7--48},
  volume    = {15},
  publisher = {Springer},
}

@article{visser2012hidden,
  title={The hidden models of model checking},
  author={Visser, Willem and Dwyer, Matthew B and Whalen, Michael},
  journal={Software \& Systems Modeling},
  volume={11},
  number={4},
  pages={541--555},
  year={2012},
  publisher={Springer}
}

@inproceedings{kiniry2008secret,
  title={Secret ninja formal methods},
  author={Kiniry, Joseph R and Zimmerman, Daniel M},
  booktitle={International Symposium on Formal Methods},
  pages={214--228},
  year={2008},
  organization={Springer}
}

@inproceedings{lanusse2009papyrus,
  title={{Papyrus UML: an open source toolset for MDA}},
  author={Lanusse, Agnes and Tanguy, Yann and Espinoza, Huascar and Mraidha, Chokri and Gerard, Sebastien and Tessier, Patrick and Schnekenburger, Remi and Dubois, Hubert and Terrier, Fran{\c{c}}ois},
  booktitle={Proc. of the Fifth European Conference on Model-Driven Architecture Foundations and Applications (ECMDA-FA 2009)},
  pages={1--4},
  year={2009},
  organization={Citeseer}
}

@InProceedings{Gallotti2008,
  author       = {Gallotti, Stefano and Ghezzi, Carlo and Mirandola, Raffaela and Tamburrelli, Giordano},
  booktitle    = {Quality of Software Architectures. Models and Architectures: 4th International Conference on the Quality of Software-Architectures, QoSA 2008, Karlsruhe, Germany, October 14-17, 2008. Proceedings 4},
  title        = {{Quality prediction of service compositions through probabilistic model checking}},
  year         = {2008},
  organization = {Springer},
  pages        = {119--134},
}

@article{paterson2018observation,
  title={{Observation-enhanced QoS analysis of component-based systems}},
  author={Paterson, Colin and Calinescu, Radu},
  journal={IEEE Transactions on Software Engineering},
  volume={46},
  number={5},
  pages={526--548},
  year={2018},
  publisher={IEEE}
}

@inproceedings{fang2022presto,
  title={{PRESTO: predicting system-level disruptions through parametric model checking}},
  author={Fang, Xinwei and Calinescu, Radu and Paterson, Colin and Wilson, Julie},
  booktitle={Proceedings of the 17th Symposium on Software Engineering for Adaptive and Self-Managing Systems},
  pages={91--97},
  year={2022}
}

@inproceedings{kolovos2008epsilon,
  title={{The Epsilon transformation language}},
  author={Kolovos, Dimitrios S and Paige, Richard F and Polack, Fiona AC},
  booktitle={{International Conference on Theory and Practice of Model Transformations}},
  pages={46--60},
  year={2008},
  organization={Springer}
}

@Article{Baouya2021,
  author   = {Baouya, Abdelhakim and Mohamed, Otmane Ait and Ouchani, Samir and Bennouar, Djamal},
  journal  = {Expert Systems with Applications: An International Journal},
  title    = {Reliability-driven {Automotive} {Software} {Deployment} based on a {Parametrizable} {Probabilistic} {Model} {Checking}},
  year     = {2021},
  issn     = {0957-4174},
  month    = jul,
  number   = {C},
  volume   = {174},
  doi      = {10.1016/j.eswa.2021.114572},

}

@article{knuth1976complexity,
  title={{The complexity of nonuniform random number generation}},
  author={Knuth, Donald},
  journal={Algorithm and Complexity, New Directions and Results},
  year={1976},
  publisher={Academic Press}
}

@InCollection{kwiatkowska_stochastic_2007,
  author    = {Kwiatkowska, Marta and Norman, Gethin and Parker, David},
  booktitle = {Formal {Methods} for {Performance} {Evaluation}: 7th {International} {School} on {Formal} {Methods} for the {Design} of {Computer}, {Communication}, and {Software} {Systems}, {SFM} 2007, {Bertinoro}, {Italy}, {May} 28-{June} 2, 2007, {Advanced} {Lectures}},
  publisher = {Springer Berlin Heidelberg},
  title     = {Stochastic {Model} {Checking}},
  year      = {2007},
  address   = {Berlin, Heidelberg},
  editor    = {Bernardo, Marco and Hillston, Jane},
  isbn      = {978-3-540-72522-0},
  pages     = {220--270},
  doi       = {10.1007/978-3-540-72522-0_6},

}

@InProceedings{Aziz1995,
  author    = {Aziz, Adnan and Singhal, Vigyan and Balarin, Felice and Brayton, Robert K. and Sangiovanni-Vincentelli, Alberto L.},
  booktitle = {Computer Aided Verification},
  title     = {{It usually works: The temporal logic of stochastic systems}},
  year      = {1995},
  address   = {Berlin, Heidelberg},
  editor    = {Wolper, Pierre},
  pages     = {155--165},
  publisher = {Springer Berlin Heidelberg},
  isbn      = {978-3-540-49413-3},
}

@InProceedings{Bianco1995,
  author    = {Bianco, Andrea and de Alfaro, Luca},
  booktitle = {Foundations of Software Technology and Theoretical Computer Science},
  title     = {Model checking of probabilistic and nondeterministic systems},
  year      = {1995},
  address   = {Berlin, Heidelberg},
  editor    = {Thiagarajan, P. S.},
  pages     = {499--513},
  publisher = {Springer Berlin Heidelberg},
  isbn      = {978-3-540-49263-4},
}

@Article{Baier1998,
  author = {Baier, Christel},
  school = {Habilitation thesis, Fakult{\"a}t f{\"u}r Mathematik \& Informatik, Universit{\"a}t Mannheim},
  title  = {On algorithmic verification methods for probabilistic systems},
  year   = {1998},
}

@techreport{omguml2017,
  title = {{OMG Unified Modeling Language Version 2.5.1}},
  institution = {{OMG}},
  year = {2017}
}

@InProceedings{gerasimou2015search,
  author    = {Gerasimou, Simos and Tamburrelli, Giordano and Calinescu, Radu},
  booktitle = {2015 30th IEEE/ACM International Conference on Automated Software Engineering (ASE)},
  title     = {{Search-Based Synthesis of Probabilistic Models for Quality-of-Service Software Engineering}},
  year      = {2015},
  pages     = {319-330},
  doi       = {10.1109/ASE.2015.22},
}

@InProceedings{Aziz1996,
  author    = {Aziz, Adnan and Sanwal, Kumud and Singhal, Vigyan and Brayton, Robert},
  booktitle = {Computer Aided Verification},
  title     = {Verifying continuous time Markov chains},
  year      = {1996},
  address   = {Berlin, Heidelberg},
  editor    = {Alur, Rajeev and Henzinger, Thomas A.},
  pages     = {269--276},
  publisher = {Springer Berlin Heidelberg},
  isbn      = {978-3-540-68599-9},
}

@InProceedings{Baier1999,
  author    = {Baier, Christel and Katoen, Joost-Pieter and Hermanns, Holger},
  booktitle = {CONCUR'99 Concurrency Theory},
  title     = {Approximative Symbolic Model Checking of Continuous-Time Markov Chains},
  year      = {1999},
  address   = {Berlin, Heidelberg},
  editor    = {Baeten, Jos C. M. and Mauw, Sjouke},
  pages     = {146--161},
  publisher = {Springer Berlin Heidelberg},
  isbn      = {978-3-540-48320-5},
}

@Article{Hansson1994,
  author    = {Hans Hansson and Bengt Jonsson},
  journal   = {Formal Asp. Comput.},
  title     = {{A Logic for Reasoning about Time and Reliability}},
  year      = {1994},
  number    = {5},
  pages     = {512--535},
  volume    = {6},
  bibsource = {dblp computer science bibliography, https://dblp.org},
  doi       = {10.1007/BF01211866},
}

@InProceedings{Pnueli1977,
  author    = {Pnueli, Amir},
  booktitle = {18th Annual Symposium on Foundations of Computer Science (sfcs 1977)},
  title     = {The temporal logic of programs},
  year      = {1977},
  pages     = {46-57},
  doi       = {10.1109/SFCS.1977.32},
}

@inproceedings{kwiatkowska2011prism,
  title={{PRISM 4.0: Verification of probabilistic real-time systems}},
  author={Kwiatkowska, Marta and Norman, Gethin and Parker, David},
  booktitle={Computer Aided Verification: 23rd International Conference, CAV 2011, Snowbird, UT, USA, July 14-20, 2011. Proceedings},
  pages={585--591},
  year={2011},
  organization={Springer}
}

@InProceedings{Addouche2004,
  author       = {Addouche, Nawal and Antoine, Christian and Montmain, Jacky},
  title        = {UML models for dependability analysis of real-time systems},
  booktitle    = {2004 IEEE International Conference on Systems, Man and Cybernetics (IEEE Cat. No. 04CH37583)},
  year         = {2004},
  volume       = {6},
  organization = {IEEE},
  pages        = {5209--5214},
}

@InProceedings{Jansen2002,
  author="Jansen, David N.
  and Hermanns, Holger
  and Katoen, Joost-Pieter",
  editor="Damm, Werner
  and Olderog, Ernst -R{\"u}diger",
  title="A Probabilistic Extension of UML Statecharts",
  booktitle="Formal Techniques in Real-Time and Fault-Tolerant Systems",
  year="2002",
  publisher="Springer Berlin Heidelberg",
  address="Berlin, Heidelberg",
  pages="355--374",
  isbn="978-3-540-45739-8"
}

@InProceedings{Addouche2005,
  author    = {Addouche, Nawal and Antoine, Christian and Montmain, Jacky},
  booktitle = {Computer Safety, Reliability, and Security},
  title     = {Combining Extended UML Models and Formal Methods to Analyze Real-Time Systems},
  year      = {2005},
  address   = {Berlin, Heidelberg},
  editor    = {Winther, Rune and Gran, Bj{\o}rn Axel and Dahll, Gustav},
  pages     = {24--36},
  publisher = {Springer Berlin Heidelberg},
  isbn      = {978-3-540-32000-5},
}

@Article{Ye2022,
  author  = {Ye, Kangfeng and Cavalcanti, Ana and Foster, Simon and Miyazawa, Alvaro and Woodcock, Jim},
  journal = {Softw. Syst. Model.},
  title   = {{Probabilistic modelling and verification using RoboChart and PRISM}},
  year    = {2022},
  number  = {2},
  pages   = {667--716},
  volume  = {21},
  doi     = {10.1007/s10270-021-00916-8},

}

@Article{Mustafiz2008,
  author   = {Mustafiz, Sadaf and Sun, Ximeng and Kienzle, Jörg and Vangheluwe, Hans},
  journal  = {Software \& Systems Modeling},
  title    = {Model-driven assessment of system dependability},
  year     = {2008},
  issn     = {1619-1374},
  month    = oct,
  number   = {4},
  pages    = {487--502},
  volume   = {7},
  doi      = {10.1007/s10270-008-0084-1},
  language = {en},

}

@InProceedings{Lara2002,
  author    = {Lara, Juan de and Vangheluwe, Hans},
  booktitle = {Fundamental Approaches to Software Engineering},
  title     = {AToM3: A Tool for Multi-formalism and Meta-modelling},
  year      = {2002},
  address   = {Berlin, Heidelberg},
  editor    = {Kutsche, Ralf-Detlef and Weber, Herbert},
  pages     = {174--188},
  publisher = {Springer Berlin Heidelberg},
  isbn      = {978-3-540-45923-1},
}

@InProceedings{Baouya2015,
  author    = {Baouya, Abdelhakim and Bennouar, Djamal and Mohamed, Otmane Ait and Ouchani, Samir},
  booktitle = {2015 12th International Symposium on Programming and Systems (ISPS)},
  title     = {A probabilistic and timed verification approach of SysML state machine diagram},
  year      = {2015},
  pages     = {1-9},
  doi       = {10.1109/ISPS.2015.7245001},
}

@Article{Miyazawa2019,
  author   = {Miyazawa, Alvaro and Ribeiro, Pedro and Li, Wei and Cavalcanti, Ana and Timmis, Jon and Woodcock, Jim},
  title    = {{RoboChart}: {Modelling} and verification of the functional behaviour of robotic applications},
  journal  = {Software {\&} Systems Modeling},
  year     = {2019},
  month    = {Jan},
  issn     = {1619-1374},
  day      = {23},
  doi      = {10.1007/s10270-018-00710-z},
}

@InProceedings{Refsdal2005,
  author    = {Refsdal, Atle and Husa, Knut Eilif and Stølen, Ketil},
  booktitle = {Formal {Modeling} and {Analysis} of {Timed} {Systems}},
  title     = {Specification and {Refinement} of {Soft} {Real}-{Time} {Requirements} {Using} {Sequence} {Diagrams}},
  year      = {2005},
  address   = {Berlin, Heidelberg},
  editor    = {Pettersson, Paul and Yi, Wang},
  pages     = {32--48},
  publisher = {Springer},
  series    = {Lecture {Notes} in {Computer} {Science}},
  doi       = {10.1007/11603009_4},
  isbn      = {978-3-540-31616-9},
  language  = {en},
}

@InProceedings{Tribastone2008,
  author    = {Tribastone, Mirco and Gilmore, Stephen},
  booktitle = {2008 {Fifth} {International} {Conference} on {Quantitative} {Evaluation} of {Systems}},
  title     = {Automatic {Translation} of {UML} {Sequence} {Diagrams} into {PEPA} {Models}},
  year      = {2008},
  month     = sep,
  pages     = {205--214},
  doi       = {10.1109/QEST.2008.18},
  url       = {https://ieeexplore.ieee.org/abstract/document/4634973},
  urldate   = {2024-01-05},
}

@Article{Refsdal2015,
  author   = {Refsdal, Atle and Runde, Ragnhild Kobro and Stølen, Ketil},
  journal  = {Journal of Computer and System Sciences},
  title    = {Stepwise refinement of sequence diagrams with soft real-time constraints},
  year     = {2015},
  issn     = {0022-0000},
  month    = nov,
  number   = {7},
  pages    = {1221--1251},
  volume   = {81},
  doi      = {10.1016/j.jcss.2015.03.003},
  url      = {https://www.sciencedirect.com/science/article/pii/S0022000015000252},
  urldate  = {2024-01-05},
}

@Article{Refsdal2008,
  author   = {Refsdal, Atle and Stølen, Ketil},
  journal  = {Science of Computer Programming},
  title    = {Extending {UML} sequence diagrams to model trust-dependent behavior with the aim to support risk analysis},
  year     = {2008},
  issn     = {0167-6423},
  month    = dec,
  number   = {1},
  pages    = {34--42},
  volume   = {74},
  doi      = {10.1016/j.scico.2008.09.003},
  series   = {Special {Issue} on {Security} and {Trust}},
  url      = {https://www.sciencedirect.com/science/article/pii/S0167642308001044},
  urldate  = {2024-01-05},
}

@InProceedings{Haugen2003,
  author    = {Haugen, $\O$ystein and St$\o$len, Ketil},
  booktitle = {«{UML}» 2003 - {The} {Unified} {Modeling} {Language}. {Modeling} {Languages} and {Applications}},
  title     = {{STAIRS} – {Steps} {To} {Analyze} {Interactions} with {Refinement} {Semantics}},
  year      = {2003},
  address   = {Berlin, Heidelberg},
  editor    = {Stevens, Perdita and Whittle, Jon and Booch, Grady},
  pages     = {388--402},
  publisher = {Springer},
  series    = {Lecture {Notes} in {Computer} {Science}},
  doi       = {10.1007/978-3-540-45221-8_33},
  isbn      = {978-3-540-45221-8},
  language  = {en},
}

@Article{Haugen2005,
  author   = {Haugen, $\O$ystein and Husa, Knut Eilif and Runde, Ragnhild Kobro and St$\o$len, Ketil},
  journal  = {Software \& Systems Modeling},
  title    = {{STAIRS} towards formal design with sequence diagrams},
  year     = {2005},
  issn     = {1619-1374},
  month    = nov,
  number   = {4},
  pages    = {355--357},
  volume   = {4},
  doi      = {10.1007/s10270-005-0087-0},
  language = {en},

}

@book{Hillston1996,
  author = {Hillston, Jane},
  title = {A compositional approach to performance modelling},
  year = {1996},
  isbn = {0521571898},
  publisher = {Cambridge University Press},
  address = {USA}
}

@InProceedings{Jarraya2007,
  author    = {Jarraya, Yosr and Soeanu, Andrei and Debbabi, Mourad and Hassaine, Fawzi},
  booktitle = {14th Annual IEEE International Conference and Workshops on the Engineering of Computer-Based Systems (ECBS'07)},
  title     = {{Automatic Verification and Performance Analysis of Time-Constrained SysML Activity Diagrams}},
  year      = {2007},
  pages     = {515-522},
  doi       = {10.1109/ECBS.2007.22},
}

@PhdThesis{Jarraya2010,
  author   = {Jarraya, Yosr},
  school   = {Concordia University},
  title    = {{Verification and validation of UML and SysML based systems engineering design models}},
  year     = {2010},
  note     = {Unpublished},
  url      = {https://spectrum.library.concordia.ca/id/eprint/979278/},
}

@article{Jouault2008,
  title={{ATL: A model transformation tool}},
  author = {Jouault, Fr{\'{e}}d{\'{e}}ric and Allilaire, Freddy and B{\'{e}}zivin, Jean and Kurtev, Ivan},
  doi = {10.1016/j.scico.2007.08.002},
  journal={Science of computer programming},
  volume={72},
  number={1-2},
  pages={31--39},
  year={2008},
  publisher={Elsevier}

}

@misc{ObjectManagementGroupOMG2016,
  author = {{OMG}},
  title = {{MOF Query/View/Transformation Version:1.3}},
  url = {https://www.omg.org/spec/QVT/About-QVT/},
  year = {2016}
}

@Article{Baouya2015a,
  author   = {Abdelhakim Baouya and Djamal Bennouar and Otmane Ait Mohamed and Samir Ouchani},
  journal  = {Expert Systems with Applications},
  title    = {{A quantitative verification framework of SysML activity diagrams under time constraints}},
  year     = {2015},
  issn     = {0957-4174},
  number   = {21},
  pages    = {7493-7510},
  volume   = {42},
  doi      = {https://doi.org/10.1016/j.eswa.2015.05.049},
  url      = {https://www.sciencedirect.com/science/article/pii/S0957417415003851},
}

@Inbook{Debbabi2010a,
  author="Debbabi, Mourad
  and Hassa{\"i}ne, Fawzi
  and Jarraya, Yosr
  and Soeanu, Andrei
  and Alawneh, Luay",
  title="Probabilistic Model Checking of SysML Activity Diagrams",
  bookTitle="Verification and Validation in Systems Engineering: Assessing UML/SysML Design Models",
  year="2010",
  publisher="Springer Berlin Heidelberg",
  address="Berlin, Heidelberg",
  pages="153--166",
  isbn="978-3-642-15228-3",
  doi="10.1007/978-3-642-15228-3\_9",
  url="https://doi.org/10.1007/978-3-642-15228-3\_9"
}

@Article{Lima2020,
  author   = {Lucas Lima and Amaury Tavares and Sidney C. Nogueira},
  journal  = {Science of Computer Programming},
  title    = {{A framework for verifying deadlock and nondeterminism in UML activity diagrams based on CSP}},
  year     = {2020},
  issn     = {0167-6423},
  pages    = {102497},
  volume   = {197},
  doi      = {https://doi.org/10.1016/j.scico.2020.102497},
  url      = {https://www.sciencedirect.com/science/article/pii/S0167642320301064},
}

@Book{Hoare1985,
  author    = {Hoare},
  publisher = {Prentice-Hall Int.},
  title     = {{Communicating Sequential Processes}},
  year      = {1985},
  address   = {Englewood Cliffs, NJ},
}

@BOOK{Roscoe2011,
  title     = {{Understanding Concurrent Systems}},
  author    = {Roscoe, A W},
  publisher = {Springer},
  series    = {Texts in computer science},
  edition   =  2010,
  month     = May,
  year      =  2010,
  address   = {London, England},
  language  = {en}
}

@Inbook{Cavalcanti2006,
  author="Cavalcanti, Ana
  and Woodcock, Jim",
  editor="Cavalcanti, Ana
  and Sampaio, Augusto
  and Woodcock, Jim",
  title="A Tutorial Introduction to CSP in Unifying Theories of Programming",
  bookTitle="Refinement Techniques in Software Engineering: First Pernambuco Summer School on Software Engineering, PSSE 2004, Recife, Brazil, November 23-December 5, 2004 Revised Lectures",
  year="2006",
  publisher="Springer Berlin Heidelberg",
  address="Berlin, Heidelberg",
  pages="220--268",
  isbn="978-3-540-46254-5",
  doi="10.1007/11889229\_6",
}

@InProceedings{GibsonRobinson2014,
  author    = {T.~Gibson-Robinson and P.~Armstrong and A.~Boulgakov and A.~W.~Roscoe},
  booktitle = {Tools and Algorithms for the Construction and Analysis of Systems},
  title     = {{FDR3: A modern refinement checker for CSP}},
  year      = {2014},
  pages     = {187-201},
}

@InProceedings{Ouchani2011,
  author    = {Ouchani, Samir and Jarraya, Yosr and Ait Mohamed, Otmane},
  booktitle = {2011 Ninth Annual International Conference on Privacy, Security and Trust},
  title     = {Model-based systems security quantification},
  year      = {2011},
  pages     = {142-149},
  doi       = {10.1109/PST.2011.5971976},
}

@PhdThesis{Ouchani2013,
  author   = {Ouchani, Samir},
  school   = {Concordia University},
  title    = {{A Security Verification Framework for SysML Activity Diagrams}},
  year     = {2013},
  month    = {September},
  note     = {Unpublished},
  url      = {https://spectrum.library.concordia.ca/id/eprint/977746/},
}

@Article{Hensel2022,
  author   = {Hensel, Christian and Junges, Sebastian and Katoen, Joost-Pieter and Quatmann, Tim and Volk, Matthias},
  journal  = {International Journal on Software Tools for Technology Transfer},
  title    = {{The probabilistic model checker Storm}},
  year     = {2022},
  issn     = {1433-2787},
  month    = {Aug},
  number   = {4},
  pages    = {589-610},
  volume   = {24},
  day      = {01},
  doi      = {10.1007/s10009-021-00633-z},

}

@InProceedings{Silva2017,
  author    = {da Silva, Carlos Eduardo and Saraiva da Silva, José Diego and Paterson, Colin and Calinescu, Radu},
  booktitle = {2017 IEEE/ACM 12th International Symposium on Software Engineering for Adaptive and Self-Managing Systems (SEAMS)},
  title     = {Self-Adaptive Role-Based Access Control for Business Processes},
  year      = {2017},
  pages     = {193-203},
  doi       = {10.1109/SEAMS.2017.13},
}

@Article{Wolny2020,
  author     = {Wolny, Sabine and Mazak, Alexandra and Carpella, Christine and Geist, Verena and Wimmer, Manuel},
  journal    = {Softw. Syst. Model.},
  title      = {{Thirteen years of SysML: a systematic mapping study}},
  year       = {2020},
  issn       = {1619-1366},
  month      = {jan},
  number     = {1},
  pages      = {111–169},
  volume     = {19},
  address    = {Berlin, Heidelberg},
  doi        = {10.1007/s10270-019-00735-y},
  issue_date = {Jan 2020},
  numpages   = {59},
  publisher  = {Springer-Verlag},

}

@Article{Calinescu2021,
  author  = {Calinescu, Radu and Paterson, Colin and Johnson, Kenneth},
  journal = {IEEE Transactions on Software Engineering},
  title   = {{Efficient Parametric Model Checking Using Domain Knowledge}},
  year    = {2021},
  number  = {6},
  pages   = {1114-1133},
  volume  = {47},
  doi     = {10.1109/TSE.2019.2912958},
}

@Article{Morozov2014,
  author   = {Andrey Morozov and Klaus Janschek},
  journal  = {Mechatronics},
  title    = {{Probabilistic error propagation model for mechatronic systems}},
  year     = {2014},
  issn     = {0957-4158},
  number   = {8},
  pages    = {1189-1202},
  volume   = {24},
  doi      = {https://doi.org/10.1016/j.mechatronics.2014.09.005},
  url      = {https://www.sciencedirect.com/science/article/pii/S0957415814001329},
}

@Article{Steurer2020,
  author   = {Mikael Steurer and Andrey Morozov and Klaus Janschek and Klaus-Peter Neitzke},
  journal  = {IFAC-PapersOnLine},
  title    = {Model-Based Dependability Assessment of Phased-Mission Unmanned Aerial Vehicles},
  year     = {2020},
  issn     = {2405-8963},
  note     = {21st IFAC World Congress},
  number   = {2},
  pages    = {8915-8922},
  volume   = {53},
  doi      = {https://doi.org/10.1016/j.ifacol.2020.12.1416},
  url      = {https://www.sciencedirect.com/science/article/pii/S2405896320318267},
}

@InProceedings{Gleirscher2023,
  author    = {Gleirscher, Mario and Haxthausen, Anne E. and Peleska, Jan},
  booktitle = {Proceedings of the 9th ACM SIGPLAN International Workshop on Formal Techniques for Safety-Critical Systems},
  title     = {{Probabilistic Risk Assessment of an Obstacle Detection System for GoA 4 Freight Trains}},
  year      = {2023},
  address   = {New York, NY, USA},
  pages     = {26–36},
  publisher = {Association for Computing Machinery},
  series    = {FTSCS 2023},
  doi       = {10.1145/3623503.3623533},
  isbn      = {9798400703980},
  location  = {Cascais, Portugal},
  numpages  = {11},

}

@InBook{Kwiatkowska2018,
  author    = {Kwiatkowska, Marta and Norman, Gethin and Parker, David},
  editor    = {Drechsler, Rolf},
  journal={Formal System Verification: State-of the-Art and Future Trends},
  pages     = {73--121},
  publisher = {Springer International Publishing},
  title     = {{Probabilistic Model Checking: Advances and Applications}},
  year      = {2018},
  address   = {Cham},
  isbn      = {978-3-319-57685-5},
  booktitle = {Formal System Verification: State-of the-Art and Future Trends},
  doi       = {10.1007/978-3-319-57685-5_3},

}

@Book{Baier2008,
  author    = {Baier, Christel and Katoen, Joost{-}Pieter},
  publisher = {The MIT Press},
  title     = {Principles of model checking},
  year      = {2008},
  address   = {Cambridge, Massachusetts},
  isbn      = {9780262267564},

  pagetotal = {1975},
  ppn_gvk   = {1657965430},
}

@Article{Kwiatkowska2012,
  author    = {Kwiatkowska, Marta and Norman, Gethin and Parker, David},
  journal   = {Formal Aspects of Computing},
  title     = {{Probabilistic verification of Herman’s self-stabilisation algorithm}},
  year      = {2012},
  pages     = {661--670},
  volume    = {24},
  publisher = {Springer},
}

@InProceedings{Kwiatkowska2001,
  author       = {Kwiatkowska, Marta and Norman, Gethin and Segala, Roberto},
  booktitle    = {{Computer Aided Verification: 13th International Conference, CAV 2001 Paris, France, July 18--22, 2001 Proceedings}},
  title        = {{Automated Verification of a Randomized Distributed Consensus Protocol Using Cadence SMV and PRISM}},
  year         = {2001},
  organization = {Springer},
  pages        = {194--206},
}

@PhdThesis{Fruth2011,
  author = {Fruth, Matthias},
  school = {University of Oxford},
  title  = {Formal methods for the analysis of wireless network protocols},
  year   = {2011},
}

@InProceedings{Dombrowski2016,
  author          = {Christian Dombrowski and Sebastian Junges and Joost-Pieter Katoen and James Gross},
  title           = {Model-Checking Assisted Protocol Design for Ultra-reliable Low-Latency Wireless Networks},
  booktitle={2016 IEEE 35th Symposium on Reliable Distributed Systems (SRDS)},
  year            = {2016},
  address         = {Budapest, Hungary},
  pages           = {307--316},
  publisher       = {IEEE},
  date            = {26-29 Sept. 2016},
  doi             = {10.1109/SRDS.2016.048},
  eventdate       = {26-29 Sept. 2016},
  eventtitleaddon = {Budapest, Hungary},
  isbn            = {978-1-5090-3514-4},
  issn            = {1060-9857},
  journal         = {2016 IEEE 35th Symposium on Reliable Distributed Systems (SRDS)},
}

@InProceedings{Mohammad2017,
  author    = {Mohammad, Nazeeruddin and Muhammad, Shahabuddin and Bashar, Abul and Khan, Majid Ali},
  booktitle = {2017 Military Communications and Information Systems Conference (MilCIS)},
  title     = {Design and modeling of energy efficient WSN architecture for tactical applications},
  year      = {2017},
  pages     = {1-6},
  doi       = {10.1109/MilCIS.2017.8190425},
}

@Article{Petridou2013,
  author  = {Petridou, Sophia and Basagiannis, Stylianos and Roumeliotis, Manos},
  journal = {IEEE Systems Journal},
  title   = {Survivability Analysis Using Probabilistic Model Checking: A Study on Wireless Sensor Networks},
  year    = {2013},
  number  = {1},
  pages   = {4-12},
  volume  = {7},
  doi     = {10.1109/JSYST.2012.2224612},
}

@Article{Basagiannis2009,
  author    = {Basagiannis, Stylianos and Katsaros, Panagiotis and Pombortsis, Andrew and Alexiou, Nikolaos},
  journal   = {Computers \& Security},
  title     = {Probabilistic model checking for the quantification of DoS security threats},
  year      = {2009},
  number    = {6},
  pages     = {450--465},
  volume    = {28},
  publisher = {Elsevier},
}

@Article{Elboukhari2010,
  author    = {Elboukhari, Mohamed and Azizi, Mostafa and Azizi, Abdelmalek},
  journal   = {International journal of Network Security \& Its Applications},
  title     = {Analysis of the Security of BB84 by Model Checking},
  year      = {2010},
  issn      = {0975-2307},
  month     = apr,
  number    = {2},
  pages     = {87--98},
  volume    = {2},
  doi       = {10.5121/ijnsa.2010.2207},
  publisher = {Academy and Industry Research Collaboration Center (AIRCC)},
}

@inproceedings{Feng2015,
  author = {Feng, Lu and Wiltsche, Clemens and Humphrey, Laura and Topcu, Ufuk},
  title = {Controller synthesis for autonomous systems interacting with human operators},
  year = {2015},
  isbn = {9781450334556},
  publisher = {Association for Computing Machinery},
  address = {New York, NY, USA},
  doi = {10.1145/2735960.2735973},
  booktitle = {Proceedings of the ACM/IEEE Sixth International Conference on Cyber-Physical Systems},
  pages = {70–79},
  numpages = {10},
  location = {Seattle, Washington},
  series = {ICCPS '15}
}

@INPROCEEDINGS{Lahijanian2010,
  author={Lahijanian, M. and Wasniewski, J. and Andersson, S. B. and Belta, C.},
  booktitle={2010 IEEE International Conference on Robotics and Automation}, 
  title={Motion planning and control from temporal logic specifications with probabilistic satisfaction guarantees}, 
  year={2010},
  volume={},
  number={},
  pages={3227-3232},
  doi={10.1109/ROBOT.2010.5509686}
}

@INPROCEEDINGS {Gleirscher2020,
  author = { Gleirscher, Mario and Calinescu, Radu },
  booktitle = { 2020 25th International Conference on Engineering of Complex Computer Systems (ICECCS) },
  title = {{ Safety Controller Synthesis for Collaborative Robots }},
  year = {2020},
  volume = {},
  ISSN = {},
  pages = {83-92},
  doi = {10.1109/ICECCS51672.2020.00017},
  url = {https://doi.ieeecomputersociety.org/10.1109/ICECCS51672.2020.00017},
  publisher = {IEEE Computer Society},
  address = {Los Alamitos, CA, USA},
  month =Oct}

@InProceedings{Kikuchi2011,
  author    = {Kikuchi, Shinji and Matsumoto, Yasuhide},
  booktitle = {2011 IEEE 4th International Conference on Cloud Computing},
  title     = {{Performance Modeling of Concurrent Live Migration Operations in Cloud Computing Systems Using PRISM Probabilistic Model Checker}},
  year      = {2011},
  pages     = {49-56},
  doi       = {10.1109/CLOUD.2011.48},
}

@Article{Baouya2019,
  author    = {Baouya, Abdelhakim and Ait Mohamed, Otmane and Bennouar, Djamal and Ouchani, Samir},
  journal   = {Computers in Industry},
  title     = {Safety analysis of train control system based on model-driven design methodology},
  year      = {2019},
  issn      = {0166-3615},
  month     = feb,
  pages     = {1--16},
  volume    = {105},
  doi       = {10.1016/j.compind.2018.10.007},
  publisher = {Elsevier BV},
}

@inproceedings{Gomes2010,
  doi       = {10.1007/978-3-642-16558-0_50},
  issn      = {1611-3349},
  title={Systematic model-based safety assessment via probabilistic model checking},
  author={Gomes, Adriano and Mota, Alexandre and Sampaio, Augusto and Ferri, Felipe and Buzzi, Julio},
  booktitle={Leveraging Applications of Formal Methods, Verification, and Validation: 4th International Symposium on Leveraging Applications, ISoLA 2010, Heraklion, Crete, Greece, October 18-21, 2010, Proceedings, Part I 4},
  pages={625--639},
  year={2010},
  organization={Springer},
}

@BOOK{Kemeny1976,
  title     = {Denumerable Markov Chains},
  author    = {Kemeny, John G and Griffeath, D S and Snell, J Laurie and Knapp, Anthony W},
  publisher = {Springer},
  series    = {Graduate texts in mathematics},
  month     =  sep,
  year      =  1976,
  address   = {New York, NY},
  language  = {en}
}

@Book{Puterman1994,
  author    = {Puterman, Martin L.},
  publisher = {John Wiley \& Sons, Inc.},
  title     = {Markov Decision Processes: Discrete Stochastic Dynamic Programming},
  year      = {1994},
  address   = {USA},
  edition   = {1st},
  isbn      = {0471619779},
}

@BOOK{Howard1971,
  title     = {Dynamic probabilistic systems: {Semi-Markov} and decision processes v. 2},
  author    = {Howard, Ronald A},
  publisher = {John Wiley \& Sons},
  series    = {Decision \& Control S.},
  month     =  aug,
  year      =  1971,
  address   = {Nashville, TN},
  language  = {en}
}

@BOOK{Anderson1991,
  isbn      = {9781461230380},
  doi       = {10.1007/978-1-4612-3038-0},
  title     = {Continuous-time Markov chains},
  author    = {Anderson, William J},
  publisher = {Springer},
  series    = {Springer series in statistics},
  edition   =  1991,
  month     =  dec,
  year      =  1991,
  address   = {New York, NY},
  language  = {en}
}

@INPROCEEDINGS{Calinescu2013,
  author={Calinescu, Radu and Johnson, Kenneth and Rafiq, Yasmin},
  booktitle={2013 28th IEEE/ACM International Conference on Automated Software Engineering (ASE)}, 
  title={Developing self-verifying service-based systems}, 
  year={2013},
  volume={},
  number={},
  pages={734-737},
  doi={10.1109/ASE.2013.6693145}
}

@InProceedings{Katoen2016,
  author    = {Katoen, Joost-Pieter},
  booktitle = {Proceedings of the 31st Annual ACM/IEEE Symposium on Logic in Computer Science},
  title     = {The Probabilistic Model Checking Landscape},
  year      = {2016},
  address   = {New York, NY, USA},
  pages     = {31–45},
  publisher = {Association for Computing Machinery},
  series    = {LICS '16},
  doi       = {10.1145/2933575.2934574},
  isbn      = {9781450343916},
  location  = {New York, NY, USA},
  numpages  = {15},
}

@Article{Norman2012,
  author    = {Norman, Gethin and Parker, David and Sproston, Jeremy},
  journal   = {Formal Methods in System Design},
  title     = {Model checking for probabilistic timed automata},
  year      = {2012},
  issn      = {1572-8102},
  month     = oct,
  number    = {2},
  pages     = {164--190},
  volume    = {43},
  doi       = {10.1007/s10703-012-0177-x},
  publisher = {Springer Science and Business Media LLC},
}

@Article{Garavel2012,
  author    = {Garavel, Hubert and Lang, Frédéric and Mateescu, Radu and Serwe, Wendelin},
  journal   = {International Journal on Software Tools for Technology Transfer},
  title     = {CADP 2011: a toolbox for the construction and analysis of distributed processes},
  year      = {2012},
  issn      = {1433-2787},
  month     = jun,
  number    = {2},
  pages     = {89--107},
  volume    = {15},
  doi       = {10.1007/s10009-012-0244-z},
  publisher = {Springer Science and Business Media LLC},
}

@InProceedings{Gnesi2000,
  author    = {Gnesi, S. and Latella, D. and Massink, M.},
  booktitle = {Proceedings. Fifth IEEE International Symposium on High Assurance Systems Engineering (HASE 2000)},
  title     = {{A stochastic extension of a behavioural subset of UML statechart diagrams}},
  year      = {2000},
  pages     = {55-64},
  doi       = {10.1109/HASE.2000.895442},
}

@PhdThesis{Argenio1999,
  author = {Pedro R. D’Argenio},
  school = {University of Twente},
  title  = {{Algebras and Automata for Timed and Stochastic Systems}},
  year   = {1999},
}

@inproceedings{gerasimou2021evolutionary,
  title={Evolutionary-guided synthesis of verified pareto-optimal MDP policies},
  author={Gerasimou, Simos and C{\'a}mara, Javier and Calinescu, Radu and Alasmari, Naif and Alhwikem, Faisal and Fang, Xinwei},
  booktitle={2021 36th IEEE/ACM International Conference on Automated Software Engineering (ASE)},
  pages={842--853},
  year={2021},
  organization={IEEE}
}

@Misc{PRISM2008,
  author = {{PRISM Team}},
  title  = {{The PRISM Language - Semantics}},
  year   = {2008},
  url    = {www.prismmodelchecker.org/doc/semantics.pdf},
}

@ARTICLE{Hahn2011,
  title     = "Probabilistic reachability for parametric {M}arkov models",
  author    = "Hahn, Ernst Moritz and Hermanns, Holger and Zhang, Lijun",
  journal   = "Int. J. Softw. Tools Technol. Transf.",
  publisher = "Springer Science and Business Media LLC",
  volume    =  13,
  number    =  1,
  pages     = "3--19",
  month     =  jan,
  year      =  2011,
  language  = "en"
}

@INPROCEEDINGS{Chen2013,
  author={Chen, Taolue and Hahn, Ernst Moritz and Han, Tingting and Kwiatkowska, Marta and Qu, Hongyang and Zhang, Lijun},
  booktitle={2013 International Symposium on Theoretical Aspects of Software Engineering}, 
  title={{Model Repair for Markov Decision Processes}}, 
  year={2013},
  volume={},
  number={},
  pages={85-92},
  doi={10.1109/TASE.2013.20}
}

@ARTICLE{Junges2024,
  title     = "{Parameter synthesis for Markov models: covering the parameter space}",
  author    = "Junges, Sebastian and {\'A}brah{\'a}m, Erika and Hensel, Christian and Jansen, Nils and Katoen, Joost-Pieter and Quatmann, Tim and Volk, Matthias",
  journal   = "Form. Methods Syst. Des.",
  publisher = "Springer Science and Business Media LLC",
  volume    =  62,
  number    = "1-3",
  pages     = "181--259",
  month     =  jun,
  year      =  2024,
  copyright = "https://creativecommons.org/licenses/by/4.0",
  language  = "en"
}

@article{Baier2020,
  title = {{Parametric Markov chains: PCTL complexity and fraction-free Gaussian elimination}},
  journal = {Information and Computation},
  volume = {272},
  pages = {104504},
  year = {2020},
  note = {GandALF 2017},
  issn = {0890-5401},
  doi = {https://doi.org/10.1016/j.ic.2019.104504},
  url = {https://www.sciencedirect.com/science/article/pii/S0890540119301208},
  author = {Christel Baier and Christian Hensel and Lisa Hutschenreiter and Sebastian Junges and Joost-Pieter Katoen and Joachim Klein},
}

@InProceedings{Eisentraut2010,
  author={Eisentraut, Christian and Hermanns, Holger and Zhang, Lijun},
  booktitle={2010 25th Annual IEEE Symposium on Logic in Computer Science}, 
  title={On Probabilistic Automata in Continuous Time}, 
  year={2010},
  volume={},
  number={},
  pages={342-351},
  doi={10.1109/LICS.2010.41}
}

@InProceedings{Christian2017,
  author="Dehnert, Christian
  and Junges, Sebastian
  and Katoen, Joost-Pieter
  and Volk, Matthias",
  editor="Majumdar, Rupak
  and Kun{\v{c}}ak, Viktor",
  title="A Storm is Coming: A Modern Probabilistic Model Checker",
  booktitle="Computer Aided Verification",
  year="2017",
  publisher="Springer International Publishing",
  address="Cham",
  pages="592--600",
  isbn="978-3-319-63390-9"
}

@inproceedings{horvath2020model,
  title={Model checking as a service: towards pragmatic hidden formal methods},
  author={Horv{\'a}th, Benedek and Graics, Bence and Hajdu, {\'A}kos and Micskei, Zolt{\'a}n and Moln{\'a}r, Vince and R{\'a}th, Istv{\'a}n and Andolfato, Luigi and Gomes, Ivan and Karban, Robert},
  booktitle={Proceedings of the 23rd ACM/IEEE International Conference on Model Driven Engineering Languages and Systems: Companion Proceedings},
  pages={1--5},
  year={2020}
}

@Book{Woodcock1996,
  title     = {{Using Z---Specification, Refinement, and Proof}},
  publisher = {Prentice-Hall},
  year      = {1996},
  author    = {J.~C.~P.~Woodcock and J.~Davies},
  address   = {Cambridge},
}

@Book{Spivey1992,
  title     = {{The Z Notation: A Reference Manual}},
  publisher = {Prentice-Hall},
  year      = {1992},
  author    = {J.~M.~Spivey},
  address   = {London},
  series    = {2nd},
}

\ifdefined \CHANGES \indexprologue{%
  This index lists for each comment the pages where the text has been modified to address the comment. Since the same page may contain multiple changes, the page number contains the index of the change in superscript to identify different changes. Finally, the page number contains a hyperlink that takes the reader to corresponding change.%
}%
\printindex[changes] \fi

\end{document}